\documentclass[11pt,a4paper]{article}
\usepackage[T1]{fontenc}
\usepackage[utf8]{inputenc}
\usepackage{lmodern}
\usepackage{amsmath,amssymb,amsthm,mathtools,bm}
\usepackage[margin=1in]{geometry}
\usepackage{graphicx,booktabs,array,alphalph}
\usepackage{microtype}
\usepackage[authoryear,round]{natbib}
\usepackage{xcolor}
\usepackage{hyperref}
\hypersetup{colorlinks=true,linkcolor=blue,citecolor=blue,urlcolor=blue,
 pdftitle={Finite-Sample Calibration and Power Benchmarks in Two-Sample Random-Integration Tests},
 pdfauthor={Pengfei Wang}}
\numberwithin{equation}{section}
\theoremstyle{plain}
\newtheorem{theorem}{Theorem}[section]
\newtheorem{proposition}[theorem]{Proposition}
\newtheorem{lemma}[theorem]{Lemma}
\newtheorem{corollary}[theorem]{Corollary}
\theoremstyle{definition}

\theoremstyle{remark}

\allowdisplaybreaks[1]
\title{Finite-Sample Calibration and Power Benchmarks in\protect\\Two-Sample Random-Integration Tests}
\author{Pengfei Wang\thanks{Email: \href{mailto:nie23.wp8738@e.ntu.edu.sg}{\texttt{nie23.wp8738@e.ntu.edu.sg}}.}\\[0.3em]
\small Mathematics and Mathematics Education\\
\small National Institute of Education, Nanyang Technological University\\
\small 1 Nanyang Walk, Singapore 637616}
\date{October 26, 2026}
\begin{document}
\maketitle
\begin{abstract}
In high-dimensional two-group studies, a supplied variable set can be
used directly or to learn a projection. We study how much attainable
power these choices retain at fixed observation and calibration budgets.
For complete unbiased random-integration statistics under Gaussian
common shape, a fixed-spectrum comparison orders direction-wise power
and gives an explicit criterion balancing an oracle advantage against
learning loss. Joint regeneration controls level for a specified learner
with unknown diagonal shape. A paired construction permits observation
reuse and finite-reference calibration of fitted scores and menus.
In the paired diagonal experiment, an invariant sparse likelihood-ratio
benchmark provides an attainable efficiency target. Finite-amplitude
bounds connect a classical bounded quadratic to this target; conditioning
on the coordinate maximum yields informative complementary bounds at
moderate and strong signals. Both comparisons include finite reference
budgets, and null-only simulations certify local comparisons with
specified fitted rules. Equal-budget experiments and a leukemia
application compare aggregation, projection and fitted quadratic scores.
For correlated paired data, classical sign randomization supplies a
separate calibration under central symmetry. The efficiency guarantees
apply to the stated experiments and signal classes; general small-budget
shape efficiency and adaptation to unknown signal allocation remain open.
\end{abstract}

\section{Introduction}
\label{sec:introduction}

The leukemia study of \citet{golub1999molecular} illustrates a common
difficulty in comparing two groups: thousands of expression measurements
are available for each subject, but the number of subjects is limited.
Its original training cohort contains 27 patients with acute lymphoblastic
leukemia and 11 with acute myeloid leukemia. The scientific question may be whether the groups have the same mean
expression profile over a specified gene set, rather than a separate
question for each gene. \citet{chen2010} studied high-dimensional two-sample
testing with applications to gene-set analysis.
We use three externally annotated Hallmark gene sets to make this
question concrete; these sets summarize biological processes
\citep{liberzon2015hallmark}. For a supplied variable set, there are
several ways to combine the measurements.
One can aggregate differences over the set, focus on its largest
coordinate difference, or estimate a direction onto which the
observations are projected. These choices retain different features
of the signal.

This application setting raises a statistical question about the use
of a limited sample. A projection can concentrate signal, but its
direction is generally unknown and must be estimated. Splitting the observations separates direction learning from the
testing observations, while leaving fewer observations for testing. Reusing observations requires
a calibration that accounts for the fit. A supplied variable set can
also be used directly without estimating an axis. At the same
observation budget, when does learning a direction retain more power
than such a direct use of the available structure?

Existing methods combine high-dimensional measurements in different
ways. The covariance inverse in the classical test of
\citet{hotelling1931} becomes unavailable when dimension exceeds
sample covariance rank. \citet{dempster1958} developed an early
high-dimensional alternative, and \citet{bai1996} studied a
squared-distance test. The U-statistic of \citet{chen2010}
accommodates unequal covariance matrices; \citet{srivastava2013unequal}
study coordinatewise standardization for that problem. For sparse
signals, \citet{cai2014} develop a maximum-type test that uses a
precision-matrix transformation. Its treatment of dependence differs
from the raw-coordinate maximum in our diagonal comparison.

Projection methods reduce the dimension before testing.
\citet{lopes2011random} study a randomly projected two-sample
statistic, while \citet{thulin2014subspaces} combines tests over random
subspaces. \citet{srivastava2016raptt} construct RAPTT from random
projections, with exact calibration under normality.
These random constructions differ from learning a signal direction.
For the one-sample mean problem, \citet{liu2024projection} estimate
an optimal projection direction, establish nonasymptotic estimation
bounds, and obtain an exact split-sample $t$-test under Gaussianity.
Their online procedure updates the direction as observations arrive
and has asymptotic calibration and power guarantees. Its exact
Gaussian split test does not require a diagonal covariance. These
results make estimation and sample allocation central to projection
testing; they also motivate asking how much attainable power remains
after the whole procedure has been fitted and calibrated.

Random integration (RI) provides a quadratic framework in which to
examine that distinction. \citet{jiang2024ri} construct a complete
unbiased two-sample RI statistic and establish its asymptotic
calibration and power properties under a general multivariate model,
allowing unequal covariance matrices. The weight is the second moment of
an integration distribution, so a baseline can be augmented along a
chosen direction. The baseline retains weight outside that direction.
The covariance subtraction in the unbiased statistic is itself random;
maximizing the statistic's expectation alone does not compare calibrated
rejection probabilities. We retain this subtraction when assessing
direction choice and fitted procedures. Our finite-sample analysis
uses a Gaussian common-shape model to make these comparisons explicit.
For general linear hypotheses, \citet{cao2024} study a nonnegative
sample-mean quadratic and give local asymptotic power for specified
weights, leaving their choice for further work. The two-sample
comparison here concerns the complete unbiased quadratic.

Power enhancement addresses a related choice between aggregation
rules. \citet{yu2023power} develop enhanced mean and covariance
components and a simultaneous test, with asymptotic size, power and
efficiency guarantees. Their mean component is relevant to our
mean-testing problem; their final joint null additionally requires
equal covariance matrices. The present efficiency question fixes the
observation and reference budgets and asks for an absolute power gap
to an attainable target. It complements asymptotic power enhancement
within a more restricted experiment.

Calibration is a separate part of the comparison.
\citet{zhang2020l2} study a chi-square approximation for an
$L^2$-norm statistic; \citet{zhang2021bf} develop a normal-reference
approach to the two-sample Behrens--Fisher problem, and
\citet{zhu2024ftype} propose an $F$-type test. Our question further
requires accounting for estimated weights and assessing actual power
at fixed observation and simulation budgets. An optimal population
axis and a valid calibration, taken separately, do not establish
that fitting the axis is worthwhile.

Our main contribution is a finite-budget comparison of fitted and
simple rules, with the comparison class stated at each step. In the
paired diagonal Gaussian experiment, an invariant sparse
likelihood-ratio (LR) test supplies an attainable target. A classical
bounded quadratic uses the supplied coordinate set directly, building
on the local-combination principle of \citet{monti1976local}; an
affine refinement controls its finite-amplitude loss. For stronger
sparse signals, conditioning LR on the largest coordinate yields
computable bounds on the power discarded by this reduction. These
results cover both ideal and finite-reference calibration.
Table~\ref{tab:u-v17-efficiency-main} gives bounds below 1.81 and 0.55
percentage points in the evaluated moderate and strong designs,
respectively. The maximum rule needs no signal-amplitude input;
the specified amplitude defines the class used to assess its efficiency.

Two further comparisons connect this target to direction learning.
First, under known common shape, fixing the whitened weight spectrum
isolates direction substitution in the complete unbiased statistic.
The resulting ordering and loss bound give an oracle-margin criterion:
a population-axis advantage exceeding the learning-loss bound is
sufficient for an advantage over a specified comparator. Second, with
unknown diagonal shape, joint regeneration gives finite-sample level
control for a specified split learner. A paired construction permits
observation reuse and joint null simulation for fitted scores and
finite menus. Its finite-reference efficiency benchmark uses the same
reference budget. Null-only simulations also provide local comparison
certificates for specified fitted procedures. These guarantees make
estimation, calibration and actual power part of one comparison; the
paired construction remains a different experiment from split RI.

The simulations examine how these comparisons change with signal
concentration, subspace coverage, estimation budget and menu adjustment.
A separate external comparison gives every method the same measured
coordinate panel and retains each competitor's stated calibration.
It separates comparisons with published procedures from matched
projection controls used to isolate a component of RI.
A mixed-signal follow-up gives block competitors the same specified
allocation information as the likelihood-ratio benchmark.
A leukemia application compares simple and fitted scores on the same
predefined gene sets, with a separate evaluation on the original test
cohort. Because the expression coordinates are correlated, it uses
paired sign randomization rather than the diagonal canonical reference.
This classical construction controls level under joint row-sign
invariance, including arbitrary correlated Gaussian paired differences.
The empirical analysis is conditional on that symmetry assumption.
Data-based simulations then measure level and power under specified
noise models, where the injected signal is known. These comparisons
do not extend the diagonal likelihood-ratio efficiency bounds to
correlated data.
General small-budget shape efficiency and adaptation to unknown
signal allocation remain unresolved.

Section~\ref{sec:setup} gives the model and direction comparison.
Section~\ref{sec:u-calibration} develops the calibration methods.
Section~\ref{sec:jasa-power} presents the paired efficiency benchmarks
and then the split-statistic baseline mechanism.
Section~\ref{sec:u-numerical-focused} reports simulations.
Section~\ref{sec:v20-application} gives the leukemia analysis and
data-based experiments; Section~\ref{sec:discussion} discusses their scope.
Proofs and additional results are in the supplementary material.

\section{Model and direction selection}
\label{sec:setup}

We examine the choice between fitting a direction and directly using
supplied structure in a Gaussian common-shape model. This model permits
the effects of direction, estimation and calibration to be distinguished.
Let the two groups be independent, with
$\bm X_{ij}\sim N_p(\bm\mu_i,\sigma_i^2\bm H)$, where
$p\ge2$, $\bm H\succ0$ and $\sigma_i^2>0$. We test
$H_0:\bm\delta=\bm0$, where $\bm\delta=\bm\mu_1-\bm\mu_2$.
A split fixed independently of the observations allocates
$N_i=m_i+n_i$ observations to training and testing, with
$m_i,n_i\ge2$. Write $\bm D_a$ for the sample-mean contrast in
block $a\in\{\mathrm{tr},\mathrm{te}\}$ and
$\bm S_{i,a}$ for its unbiased sample covariance.
For a positive-definite, training-measurable weight $\bm W$, define
\begin{align}
 Q_{\bm W}&=\bm D_{\mathrm{te}}^\top\bm W\bm D_{\mathrm{te}},
 \label{eq:setup-Q}\\
 U_{\bm W}&=Q_{\bm W}-\sum_{i=1}^2
       \frac{\operatorname{tr}(\bm W\bm S_{i,\mathrm{te}})}{n_i}.
 \label{eq:setup-U}
\end{align}
Conditional on training, $U_{\bm W}$ has expectation
$\bm\delta^\top\bm W\bm\delta$. This identifies the signal
target, but does not order rejection probabilities: the random
covariance correction and the calibrated cutoff also enter the test.
We therefore compare actual power while retaining that correction. In RI,
$\bm W=\mathbb E(\bm\omega\bm\omega^\top)$ is the second moment
of the integration distribution. To assess direction choice alone, we
hold the eigenvalues fixed after whitening. The isotropic baseline
keeps weight on every direction, while the rank-one increment
concentrates additional weight on a chosen axis. This gives the family
\begin{equation}\label{eq:setup-U-class}
 \bm W_{\bm v}=\bm H^{-1/2}
       (b\bm I_p+\eta\bm v\bm v^\top)\bm H^{-1/2},
 \quad \|\bm v\|_2=1,\quad b>0,\quad\eta\ge0.
\end{equation}

When $\bm H$ is known, set
$s^2=\sigma_1^2/n_1+\sigma_2^2/n_2$ and
$\bm\theta=\bm H^{-1/2}\bm\delta/s$. The noncentrality missed
by direction $\bm v$ is
$r_{\bm v}=\|\bm\theta\|_2^2-(\bm v^\top\bm\theta)^2$.
The signed chi-square representation in Supplementary
Section~\ref{sec:u-oracle} shows that actual power increases as
$r_{\bm v}$ decreases. At a common ideal cutoff, or under the
known-shape endpoint calibration of Supplementary
Section~\ref{sec:u-feasible}, for $\eta>0$ and $\bm\delta\ne\bm0$
the maximizing directions are
$\pm\bm\theta/\|\bm\theta\|_2$ and
\begin{equation}\label{eq:jasa-direction-bound}
 0\le\pi_* -\pi(\bm v)\le1-\exp(-r_{\bm v}/2).
\end{equation}
This compares directions at fixed $b,\eta$ within
\eqref{eq:setup-U-class}; it does not optimize all tests or spectra.
For a learned direction, averaging this bound connects missed
noncentrality to power loss. A correctly specified fixed subspace can
reduce that learning cost; exact risk and angular-minimax results are
given in Supplementary Section~\ref{sec:u-learning}.

\subsection{An explicit criterion for the cost of learning}
\label{sec:u-v12-learning-criterion}

The known-shape comparison also separates the gain available from an
axis in a supplied subspace from the cost of learning that axis.
Write $\mathcal B_{b,\eta}(x)=\min\{1-e^{-x/2},
\sqrt{2\eta x/[\pi(b+\eta)]}\}$.

\begin{proposition}[An oracle-margin criterion]
\label{prop:u-v12-learning-margin}
Assume the known-common-shape Gaussian model, a fixed rank-$q$
projector $\bm P$, and fixed $b>0$, $\eta>0$. Learn
$\widehat{\bm v}_{\bm P}$ from independent training means as in
\eqref{eq:u-subspace-learner}. Put $\bm d=\bm H^{-1/2}\bm\delta$,
$s_{\rm tr}^2=\sigma_1^2/m_1+\sigma_2^2/m_2$,
$\Lambda_\parallel=\|\bm P\bm d\|_2^2/s^2$,
$\tau=\|\bm P\bm d\|_2^2/s_{\rm tr}^2$, and
\begin{equation}\label{eq:u-v12-learning-risk}
 R_{\bm P}=\Lambda_\parallel(q-1)
       \mathbb E\frac1{q+2K},\qquad K\sim\operatorname{Poisson}(\tau/2).
\end{equation}
Let $\pi_{\bm P,*}$ use the population axis
$\bm P\bm d/\|\bm P\bm d\|_2$, with the same testing allocation,
spectrum, residual estimator and calibration as the learned test.
For the known-scale oracle cutoff, or either known-shape endpoint
calibration, including a finite independent reference bank,
\begin{equation}\label{eq:u-v12-learning-margin}
 0\le\pi_{\bm P,*}-\pi_{\rm learn}
 \le\mathcal B_{b,\eta}(R_{\bm P}).
\end{equation}
For any prespecified comparator $C$ with its own valid level-$\alpha$
calibration and the same total observations $N_i=m_i+n_i$,
\begin{equation}\label{eq:u-v12-learning-advantage}
 \pi_{\rm learn}-\pi_C
 \ge (\pi_{\bm P,*}-\pi_C)-\mathcal B_{b,\eta}(R_{\bm P}).
\end{equation}
Thus an oracle advantage exceeding the explicit learning-loss bound
is sufficient for a power advantage. If $q=1$ or
$\bm P\bm d=\bm0$, then $R_{\bm P}=0$; in the latter case any
unit axis in the subspace defines $\pi_{\bm P,*}$.
\end{proposition}

This is a population and design criterion, not an observable tuning
rule: both the signal and the oracle advantage must be specified or
bounded in a stated parameter class. The comparator may use all
observations for testing, so training is not treated as free.
The condition need not hold, especially against a strong subspace
test. It neither proves superiority to such tests nor covers
estimated shape or overlapping leave-pair-out learners.

\section{Finite-sample calibration}
\label{sec:u-calibration}
\label{sec:u-unknown-shape}

This section establishes finite-sample level control under unknown
shape. The upper reference for split RI and the exact paired reference
are distinct constructions. Efficiency comparisons are developed
separately in Section~\ref{sec:jasa-power}.
For the canonical references in the following three subsections, assume that $\bm H$ is diagonal in a specified basis.
This diagonal structure is a model assumption; coordinatewise
standardization does not remove arbitrary correlations. The basis
is fixed independently of the tested observations.
Training supplies both a positive diagonal estimate
$\widehat{\bm H}$ and a unit direction $\widehat{\bm v}$.
The testing weight and numerator are
\begin{align}
 \widehat{\bm W}_{\bm v}
 &=\widehat{\bm H}^{-1/2}(b\bm I_p+\eta\bm v\bm v^\top)
                                      \widehat{\bm H}^{-1/2},
 \label{eq:u-shape-weight}\\
 \widehat U_{\bm v}
 &=\bm D_{\mathrm{te}}^\top\widehat{\bm W}_{\bm v}\bm D_{\mathrm{te}}
 -\sum_{i=1}^2\frac{\operatorname{tr}(\widehat{\bm W}_{\bm v}
                      \bm S_{i,\mathrm{te}})}{n_i}.
 \label{eq:u-shape-statistic}
\end{align}
We form the studentizer from testing residuals:
\begin{equation}\label{eq:u-shape-scale}
 \widehat s_{\mathrm{te}}^2
 =\sum_{i=1}^2\frac{\operatorname{tr}(\widehat{\bm H}^{-1}
                                  \bm S_{i,\mathrm{te}})}{pn_i},
 \qquad \widehat T_{\bm v}=\widehat U_{\bm v}/\widehat s_{\mathrm{te}}^2.
\end{equation}
The scale estimate and numerator share testing residuals. Their
dependence, and the training randomness, enter the reference below.
\subsection{A null-invariant reference for an unknown diagonal shape}
\label{sec:u-diag-invariant}

Suppose that $\bm H$ is diagonal in a specified coordinate system.
Let $\bm P=\bm P^\top=\bm P^2$ be a fixed orthogonal projection of
rank $q\ge1$ in the corresponding whitened coordinates.
We designate the first training group for shape estimation before
observing the data. Its coordinate sample variances have independent
chi-square errors after division by their population variances.
The group's common scale cancels from the normalized direction and
the studentized testing statistic. This gives a reference with known
residual degrees of freedom. A pooled estimator would require a
different reference because its normalized errors generally depend on
the unknown group-scale ratio. The choice here specifies a tractable
learner; its efficiency is assessed separately. Estimate the shape and
learn the direction by
\begin{equation}\label{eq:u-diag-learner}
 \widehat{\bm H}
 =\operatorname{diag}(S_{1,\mathrm{tr},11},\ldots,
                       S_{1,\mathrm{tr},pp}),\qquad
 \widehat{\bm v}
 =\frac{\bm P\widehat{\bm H}^{-1/2}\bm D_{\mathrm{tr}}}
        {\|\bm P\widehat{\bm H}^{-1/2}
                         \bm D_{\mathrm{tr}}\|_2}.
\end{equation}
The diagonal estimate and direction are well defined almost surely.
The projection, $b$, $\eta$, and the split sizes are fixed in advance;
projection onto the first $q$ specified coordinates is one example.
No relation between $p$ and the training sample size is required.

The estimated shape enters both the learned direction and the testing
weight. An unconditional null reference must retain this shared
randomness, so each reference draw repeats the shape and direction
fits. Put $r=m_1-1$ and $k_i=n_i-1$. Generate mutually independent
variables
\begin{equation}\label{eq:u-diag-reference-inputs}
 R_j\sim\chi_r^2/r\quad(j=1,\ldots,p),\qquad
 \bm Z_{\mathrm{tr}},\bm Z_0\sim N_p(\bm0,\bm I_p),\qquad
 \bm G_i\in\mathbb R^{k_i\times p}\quad(i=1,2),
\end{equation}
where the entries of each $\bm G_i$ are independent standard normal
variables. Define
\begin{equation}\label{eq:u-diag-reference-matrices}
 \begin{split}
 \bm R&=\operatorname{diag}(R_1,\ldots,R_p),\qquad
 \bm v_R=\frac{\bm P\bm R^{-1/2}\bm Z_{\mathrm{tr}}}
                   {\|\bm P\bm R^{-1/2}\bm Z_{\mathrm{tr}}\|_2},\\
 \bm B_R&=\bm R^{-1/2}
       (b\bm I_p+\eta\bm v_R\bm v_R^\top)\bm R^{-1/2},
 \qquad \bm\Psi_i=\bm G_i^\top\bm G_i/k_i,\\
 Q_R&=\bm Z_0^\top\bm B_R\bm Z_0,\\
 A_{Ri}&=\operatorname{tr}(\bm B_R\bm\Psi_i),\qquad
 V_{Ri}=\operatorname{tr}(\bm R^{-1}\bm\Psi_i)/p.
 \end{split}
\end{equation}
Using the same $\bm\Psi_i$ in $A_{Ri}$ and $V_{Ri}$ retains the
dependence between covariance subtraction and scale estimation.
Under the null, the remaining variance ratio enters only as a mixing
parameter after this standardization. The statistic is a convex combination of two
endpoint ratios, with positive denominator weights. Their maximum
therefore supplies an upper reference for every group-scale ratio:
\begin{equation}\label{eq:u-diag-envelope}
 M_{\mathrm{diag}}
 =\max_{i=1,2}\frac{Q_R-A_{Ri}}{V_{Ri}}.
\end{equation}

\begin{theorem}[Finite-sample calibration for the diagonal-shape learner]
\label{thm:u-diag-invariant}
Assume the Gaussian common-shape model with positive diagonal $\bm H$,
$m_i,n_i\ge2$, and the training rule \eqref{eq:u-diag-learner}. Under
the null, the statistic in \eqref{eq:u-shape-scale} has the representation
\begin{equation}\label{eq:u-diag-null-representation}
 \widehat T_{\widehat{\bm v}}\overset{d}{=}T_\kappa
 =\frac{Q_R-\kappa A_{R1}-(1-\kappa)A_{R2}}
        {\kappa V_{R1}+(1-\kappa)V_{R2}},\qquad
 \kappa=\frac{\sigma_1^2/n_1}
                  {\sigma_1^2/n_1+\sigma_2^2/n_2}.
\end{equation}
For the common variables in
\eqref{eq:u-diag-reference-inputs}--\eqref{eq:u-diag-reference-matrices},
$T_\kappa\le M_{\mathrm{diag}}$ almost surely for every
$\kappa\in[0,1]$. The law of $M_{\mathrm{diag}}$ is continuous and
depends only on $p,m_1,n_1,n_2,\bm P,b,\eta$.

Independently of all observations, generate $L\ge1$ independent copies
$M_{\mathrm{diag}}^{(1)},\ldots,M_{\mathrm{diag}}^{(L)}$ by repeating
the entire construction above. For $0<\alpha<1$, reject when
\begin{equation}\label{eq:u-diag-pvalue}
 \widehat p_{\mathrm{diag}}
 =\frac{1+\sum_{j=1}^L\bm1\{
            M_{\mathrm{diag}}^{(j)}\ge
                         \widehat T_{\widehat{\bm v}}\}}{L+1}
 \le\alpha.
\end{equation}
Uniformly over all positive diagonal common shapes, positive group
scales, and common means,
\begin{equation}\label{eq:u-diag-level}
 \Pr_{\bm0}(\widehat p_{\mathrm{diag}}\le\alpha)
 \le\frac{\lfloor\alpha(L+1)\rfloor}{L+1}\le\alpha.
\end{equation}
The probability is over training, testing, and reference simulation.
\end{theorem}

The finite-reference step follows a Monte Carlo rank argument
\citep{dufour2006}. The resulting procedure is as follows.
\begin{enumerate}
\item Compute \eqref{eq:u-diag-learner} from the training blocks and
the complete statistic \eqref{eq:u-shape-statistic} with the scale
\eqref{eq:u-shape-scale} from the testing blocks.
\item In each of the $L$ independent reference draws, regenerate
$\bm R$, the training vector $\bm Z_{\mathrm{tr}}$, and the resulting
direction $\bm v_R$; then generate $\bm Z_0,\bm G_1,\bm G_2$ and
evaluate \eqref{eq:u-diag-envelope}.
\item Evaluate \eqref{eq:u-diag-pvalue} and reject at level $\alpha$.
\end{enumerate}
For a nontrivial finite-reference test, $\alpha(L+1)\ge1$ is needed.
The rule avoids shape-confidence endpoints and their error allocation
$\gamma$. The second training size $m_2$ affects the learned direction
under alternatives but cancels from the null reference distribution.

The level guarantee averages over training, testing and simulation.
It requires regenerating the fitted shape and direction in every
reference draw. A data-dependent choice of $\bm P$, $q$, or $\eta$
would also require a separate invariance and joint calibration
argument. Thus the theorem establishes validity for the specified
learner; it leaves direction optimality under estimated shape open.
Although joint regeneration removes the shape-confidence envelope,
the endpoint maximum in \eqref{eq:u-diag-envelope} can remain
conservative.

\begingroup
\let\section\subsection
\section{Reusing observations: calibration and interpretation}
\label{sec:u-observation-reuse}

We next consider using the same observations for learning and testing.
Suppose both groups have size $N$, and pair their observations by a
rule fixed before seeing the data. Then
$\bm Z_i=\bm X_i-\bm Y_i$ are independent
$N_p(\bm\delta,\bm C)$ vectors. For this section it suffices that
$\bm C=\bm\Sigma_1+\bm\Sigma_2$ is positive diagonal; a common
covariance shape is unnecessary. Since this construction changes the
statistic, the earlier direction ordering does not apply automatically.

First fit the shape and direction using all $N$ differences. Put
$\bm D=\operatorname{diag}(\bm S_Z)$,
$\bm y=\sqrt N\bm D^{-1/2}\overline{\bm Z}$,
$\bm R=\bm D^{-1/2}\bm S_Z\bm D^{-1/2}$ and
$\bm v=\bm P\bm y/\|\bm P\bm y\|_2$, where $\bm P$ projects onto
a specified coordinate set $J$. The resulting mean-axis score is
\begin{equation}\label{eq:reuse-mean-axis}
 T_{\rm ma}=b(\|\bm y\|_2^2-p)
       +\eta\{\|\bm P\bm y\|_2^2-\bm v^\top\bm R\bm v\}.
\end{equation}
Diagonal scaling cancels from its null law. Regenerating the complete
fit on independent standard Gaussian differences therefore gives an
exact Monte Carlo rank test (Theorem~\ref{thm:u-nosplit-level}).
Fitting the weight to the tested mean changes its target and gives the
raw score a positive null expectation. Gaussian residual directions
are independent of their lengths and sample means, yielding
\begin{equation}\label{eq:reuse-classical-reduction}
 \mathbb E(T_{\rm ma}\mid\bm y)
 =b(\|\bm y\|_2^2-p)+\eta(\|\bm P\bm y\|_2^2-1).
\end{equation}
Thus its conditional mean is a classical weighted sum of squared
coordinate $t$ statistics. The conditional variance is
$2\eta^2\{1-\sum_{j\in J}v_j^4\}/(N-1)$.
These identities motivate comparison with classical diagonal tests
\citep{srivastava2008,hu2019}; their proofs are given in
Proposition~\ref{prop:u-nosplit-reduction}. A power comparison also
requires the respective calibrated critical values. For $r=N-1>2$ and $a_r=r/(r-2)$, put
$c_0=bp(a_r-1)+\eta(qa_r-1)$. An exact centering identity is
\begin{equation}\label{eq:reuse-centered-target}
 \mathbb E\frac{T_{\rm ma}-c_0}{Na_r}
 =\bm\delta^\top\bm C^{-1/2}
     (b\bm I_p+\eta\bm P)\bm C^{-1/2}\bm\delta.
\end{equation}
Thus centering targets a fixed block quadratic. Applying the same
affine transformation to all references preserves the test. A
leave-out fit serves a different purpose: it retains the conditional
target of an independently learned rank-one weight.
Proposition~\ref{prop:u-centered-block} gives the associated variance
reduction and ancillary-randomness interpretation.
Maximizing the observed score replaces its rank-one term by
$\eta\lambda_{\max}(\bm y_J\bm y_J^\top-\bm R_{JJ})$;
the maximum must also be recomputed in every reference draw.

A second construction retains independence locally. For every pair
$i<j$, fit the diagonal shape and direction using only the other
$N-2$ differences, giving a weight $\bm W_{-ij}$. Define
\begin{equation}\label{eq:reuse-lpo-summary}
 U_{\rm LPO}=\frac{2}{N(N-1)}\sum_{i<j}
                  \bm Z_i^\top\bm W_{-ij}\bm Z_j.
\end{equation}
The two held-out observations are independent of their fitted weight.
Consequently, for $N\ge6$ and the inverse-variance learner specified
in Section~\ref{sec:u-lpo},
$\mathbb E U_{\rm LPO}=\bm\delta^\top
\mathbb E(\bm W_{-12})\bm\delta$ and the null mean is zero.
Diagonal equivariance again gives an exact finite-reference rank test
when all leave-pair-out fits are regenerated in every null draw.
The summands share training observations and are dependent; their
expectation involves the random fitted weight rather than a fixed
oracle quadratic.
At $\eta=0$ this is the leave-two-out diagonal construction of
\citet{park2013}, up to positive scaling; see also \citet{hu2024}.

Both constructions use all observations, but pairing leaves $N-1$
residual degrees of freedom and changes the covariance subtraction
relative to the original unpaired statistic. The numerical comparisons
assess their power against classical tests at the same observation
budget.

\section{A calibrated finite menu without a validation split}
\label{sec:u-menu-summary}

To accommodate signal both inside and outside the supplied subspace,
consider $K$ candidate score functions fixed before observing the data. Each score includes its
entire variance fit, direction learner and tuning constants, and is
invariant under positive diagonal scaling of the paired observations.
Generate $L$ independent canonical Gaussian data sets and evaluate
every candidate on each data set and on the observations. This gives
$M=L+1$ jointly exchangeable score rows under the null, with row $0$
denoting the observations, without reserving a validation sample.

For scores $T_{ij}$ with larger values indicating evidence, define
\begin{equation}\label{eq:u-menu-summary-ranks}
 R_{ij}=\sum_{h=0}^{L}\bm1\{T_{hj}\ge T_{ij}\},\qquad
 S_i=\min_{1\le j\le K}R_{ij},\qquad p_j=R_{0j}/M.
\end{equation}
The observed row enters the component ranks of every reference row.
The menu $p$-value is the ascending rank of $S_0$ among all $S_i$,
divided by $M$. Equal $S_i$ values are ordered using independent
uniform variables attached to every row. Under the diagonal Gaussian
null, exchangeability gives
\begin{equation}\label{eq:u-menu-summary-level}
 \Pr(p_{\rm menu}\le\alpha)=\frac{\lfloor\alpha M\rfloor}{M}.
\end{equation}
This follows the extreme-rank Monte Carlo construction
\citep{mrkvicka2017combined,myllymaki2017global}. Adaptive
high-dimensional two-sample tests are also studied by
\citet{xu2016adaptive}. Here, exact joint null regeneration of all
fitted scores supplies the required exchangeability.

Theorem~\ref{thm:u-menu} gives a weighted version and its proof.
For the equal weights used in the experiment, it implies the
pathwise inequality and power bound
\begin{equation}\label{eq:u-menu-summary-power}
 p_{\rm menu}\le\min\{1,K\min_jp_j\},\qquad
 \pi_{\rm menu}(\alpha)\ge\max_j\pi_j(\alpha/K).
\end{equation}
Thus the menu contains the Bonferroni union on the same reference
bank. The possible loss relative to the best component at nominal
level $\alpha$ is bounded by
$\max_j\Pr_{\bm\delta}(\alpha/K<p_j\le\alpha)$, which need not
be small. The bound therefore compares the menu with components at
an adjusted level. The menu is an omnibus test: its selected index
describes the observed evidence, while inference uses the adjusted
$p$-value $p_{\rm menu}$.

The candidate list may include prespecified leave-pair-out spectra
and classical diagonal scores, recomputed on the same canonical data.
Known-shape and population-direction diagnostics are infeasible and
are excluded. The split-RI endpoint envelope supplies only a
conservative marginal reference, so it is also excluded from this
exact joint calibration. A menu containing classical fallbacks is
a combined procedure; its power does not establish dominance of
its learned component. The subspace, candidate family and weights
are fixed in advance. Choosing them from the tested observations
would require calibrating that additional selection.

\endgroup
\begingroup
\let\subsection\subsubsection
\subsection{Implementation and interpretation}
\label{sec:u-practical-protocol}

The paired-menu test can be implemented as follows.
\begin{enumerate}
\item Fix the pairing, coordinate basis, subspace, candidate list and
weights before examining the tested observations. Choose a small
list using independently available structural information. Our list
includes RI scores and classical subspace, outside and all-coordinate
scores.
\item Generate complete canonical Gaussian difference datasets.
Refit every variance estimate, leave-pair-out learner and score on
every dataset, including the observed one. Use all-row component
ranks and one row-wise tie mark for the final extreme rank.
\item At the specified nominal level, report the calibrated menu
$p$-value. The smallest component $p$-value and its index may be
reported descriptively; neither identifies the power-maximizing
spectrum nor supplies an unadjusted test.
\end{enumerate}
The candidate list determines the scope and cost of adaptation.
Our experiments support retaining block and all-coordinate controls;
adding learned scores can reduce power through selection adjustment.
No single menu size or spike is preferred throughout the reported
designs. We next assess efficiency in this same paired experiment,
starting with an attainable benchmark for a prespecified axis-sparse
alternative class.

\endgroup

\subsection{A sign reference for correlated paired data}
\label{sec:v20-signflip}

The canonical reference above removes diagonal scale parameters, but
does not preserve arbitrary coordinate dependence. For the application
in Section~\ref{sec:v20-application}, we instead retain each observed
paired vector and change only its overall sign. This is a specialization
of the finite-group randomization principle of
\citet{hemerik2018}. Its purpose is to accommodate dependence in
the paired experiment; the diagonal efficiency benchmarks in Section~\ref{sec:jasa-power}
are not transferred to this new reference.

\begin{proposition}[Paired sign-reference validity]
\label{prop:v20-signflip}
Let $\bm D=(\bm D_1,\ldots,\bm D_N)^\top$ be the differences from
$N$ disjoint pairs fixed independently of the observations. Suppose
that, under the null, $g\bm D\stackrel{d}=\bm D$ for every
$g=\operatorname{diag}(\epsilon_1,\ldots,\epsilon_N)$ with
$\epsilon_i\in\{-1,1\}$. Let $T$ be a prespecified measurable score,
including all direction and variance fitting, defined on every
transformed dataset. Independently draw $g_1,\ldots,g_B$ uniformly
from the sign group and put $g_0=\bm I_N$, $M=B+1$. Then
\begin{equation}\label{eq:v20-sign-p}
 p_T=\frac{1+\sum_{b=1}^B
       \mathbf 1\{T(g_b\bm D)\ge T(\bm D)\}}{M}
\end{equation}
satisfies
\[
 \Pr\{p_T\le\alpha\mid\mathcal O(\bm D)\}
 \le \lfloor M\alpha\rfloor/M\le\alpha,
 \qquad \mathcal O(\bm D)=\{g\bm D:g\text{ is a sign matrix}\}.
\]
For a complete enumeration, take $M=2^N$ and use the
conservative upper-tail rank divided by $M$; the same level bound holds.
\end{proposition}

Independent differences symmetric about zero satisfy the assumption.
In particular, independent Gaussian groups with equal means give
$\bm D_i\sim N_p(\bm0,\bm\Sigma_1+\bm\Sigma_2)$, allowing arbitrary
correlation and unequal covariance matrices. Each transformation
requires refitting the entire score. The argument applies to paired
LPO, block, maximum and projection scores; it does not justify
swapping heteroscedastic raw observations while retaining the original
two-group split statistic.

The proof conditions on the sign orbit, on which the observation and
uniform random transforms are exchangeable. Conservative ranks then
give the result; ties may make the test conservative. Section~\ref{app:v20-application}
gives the proof and the corresponding finite-menu construction.
Equality of means alone does not imply sign invariance. Thus this
calibration supplies a clear symmetry-based interpretation for an
application, rather than a distribution-free guarantee for arbitrary
expression data.

\section{Power and efficiency}
\label{sec:jasa-power}

The power question has two parts: what a specified experiment can
attain, and how much of that power an implemented rule retains. We
begin with the paired experiment of
Section~\ref{sec:u-observation-reuse}. Its sparse invariant LR
benchmark gives an attainable target. The bounded-quadratic and
maximum comparisons then bound an absolute difference of rejection
probabilities, at fixed sample size and, where applicable, fixed
reference budget. This is a stronger numerical conclusion within the
specified experiment than a limiting statement that power tends to
one, but it uses a narrower alternative class.

These comparisons also clarify when fitting an axis is unnecessary:
a simple rule with a small benchmark gap leaves little power available
for any competing rule in the same class. The null-only certificate
can additionally compare a specified fitted procedure with bounded
aggregation. Neither calculation attributes every loss to direction
estimation alone. A fixed-allocation mixed comparison examines the
role of supplied signal information. Only in
Sections~\ref{sec:u-mixed-limit}--\ref{sec:u-finite-dimension} do we
return to split RI, to explain its retained baseline and assess the
associated approximation.

\begingroup
\let\section\subsection
\section{Restricted finite-sample efficiency benchmarks}
\label{sec:u-v12-efficiency}

A level guarantee controls false rejections but does not tell us whether
learning a direction is worthwhile. We therefore compare procedures
with an attainable power target in the same paired experiment.
We start with one active coordinate in a prespecified set: this isolates
uncertainty about its location and sign while allowing an exact
benchmark. Classical likelihood-ratio and invariant permutation
optimality principles give the comparison
\citep{neyman1933,runger1992invariant,roach2018permutation}.
Consider the paired experiment
$\bm Z_i\sim N_p(\bm\delta,\bm C)$, $i=1,\ldots,N$, where
$N\ge2$ and $\bm C$ is unknown positive diagonal. Fix a coordinate
set $J$ of size $q\ge1$ and a minimum signal strength $a_0>0$ before
examining the data, and fix $0<\alpha<1$. Write $r=N-1$ and
\begin{equation}\label{eq:u-v12-axis-class}
 \bm\theta=\sqrt N\bm C^{-1/2}\bm\delta,\qquad
 \Theta_{a_0}=\{\pm a\bm e_j:j\in J,\ a\ge a_0\}.
\end{equation}
Thus only one coordinate carries signal; its location, sign and actual
strength are unknown. Restricting tests to be unchanged by positive
coordinate rescaling removes the unknown marginal scales from the
comparison. With $S_j$ the sample standard deviation, put
$Y_j=\sqrt N\overline Z_j/S_j$. Let $f_{r,a}$ denote the noncentral
$t_r$ density with noncentrality $a$, and define
\begin{equation}\label{eq:u-v12-mixture-score}
 \ell(\bm Y)=\frac1{2q}\sum_{j\in J}
  \frac{f_{r,a_0}(Y_j)+f_{r,-a_0}(Y_j)}{f_{r,0}(Y_j)}.
\end{equation}

\begin{theorem}[An invariant maximin benchmark]
\label{thm:u-v12-ideal-maximin}
Among tests based on the paired observations that are invariant under
positive diagonal scaling and have null rejection probability at most
$\alpha$ for every positive diagonal $\bm C$, the upper-tail
level-$\alpha$ test of \eqref{eq:u-v12-mixture-score} maximizes the
minimum power over \eqref{eq:u-v12-axis-class} and all such $\bm C$.
Its minimum is attained at $a=a_0$, with equal power at all
$2q$ boundary alternatives.
\end{theorem}

For implemented Monte Carlo rules, the comparison must also account
for the reference budget. Take $M=L+1$
data-set rows: the observed paired data and $L$ independent canonical
null data sets. Put $k=\lfloor\alpha M\rfloor$ and suppose $k\ge1$.
An algorithm supplies a decision probability $\psi_i\in[0,1]$ for
each row viewed as observed. Require positive diagonal scaling
invariance within each row, row-permutation equivariance, and
\begin{equation}\label{eq:u-v12-row-budget}
 \sum_{i=0}^{L}\psi_i\le k
 \quad\text{for every row matrix, almost surely in its random marks.}
\end{equation}
Independent identically distributed row marks may break ties.
Row equivariance treats each row in the same way, and this constraint
limits the total number of hypothetical rejections as in a rank test.
It is not conditional level control given a realized reference bank.

\begin{samepage}
\begin{theorem}[A finite-reference maximin benchmark]
\label{thm:u-v12-bank-maximin}
Within the preceding algorithm class, rejecting precisely the $k$
rows with largest values of \eqref{eq:u-v12-mixture-score} maximizes
the minimum observed-row power over $\Theta_{a_0}$ and unknown
positive diagonal $\bm C$. Its null level is $k/M$, averaged over
the observations and references. The jointly ranked menus of
Section~\ref{sec:u-menu-summary} satisfy
\eqref{eq:u-v12-row-budget}; randomized final ranks give equality.
\end{theorem}
\end{samepage}

The proofs and an exact identity for the average boundary-power gap
appear in Section~\ref{app:u-v12-efficiency}. They use the maximal
invariant, a boundary mixture and a deterministic allocation argument.
These are efficiency benchmarks for a specified sparse alternative
class, not a proof that a learned RI direction or the existing menu
attains them. For $q\ge2$, even the ideal benchmark falls strictly
below the coordinate oracle (Corollary~\ref{cor:u-v12-coordinate-oracle-gap}). The finite-reference theorem imposes more than
unconditional level control. Neither theorem optimizes tests using
the original unpaired observations, unknown signal structures, or
data-selected $J$ and $a_0$.

\endgroup
\subsection{A direction-free rule near the sparse benchmark}
\label{sec:u-v13-bounded-quadratic}

The benchmark suggests a simple alternative to estimating an axis.
Its positive likelihood-ratio expansion begins with a constant and
a positive multiple of a bounded quadratic score. Since neither a
constant nor positive scaling changes the ranking, this score retains
the first data-dependent term while using the supplied coordinate set
directly. In the experiment of Section~\ref{sec:u-v12-efficiency}, define
\begin{equation}\label{eq:u-v13-bounded-score}
 B(\bm Y)=\frac1q\sum_{j\in J}
       \frac{(r+1)Y_j^2}{r+Y_j^2},\qquad r=N-1.
\end{equation}
This beta-transformed quadratic belongs to the classical family
of locally optimal combinations of independent test statistics
\citep{monti1975thesis,monti1976local}.
It uses every paired observation, fits no direction and does not
require a signal-strength specification.
Under the null, its summands divided by $r+1$ are independent
$\operatorname{Beta}(1/2,r/2)$ variables. Thus its ideal cutoff is
nuisance-free, and its finite-reference version simply ranks
\eqref{eq:u-v13-bounded-score} on the common data-set rows.

\begin{theorem}[A finite-sample bridge to a bounded quadratic rule]
\label{thm:u-v13-bounded-regret}
Fix $a_0>0$, put $s=a_0^2/2$, and let
$\Delta(s)=1-(1+s)e^{-s}$. Compare either the ideal tests at
level $\alpha$, or the finite-reference tests with the same $M,k$
as Theorem~\ref{thm:u-v12-bank-maximin}. Let $\beta_*$ and
$\beta_B$ be their respective worst-case powers on
$\Theta_{a_0}$. Then
\begin{equation}\label{eq:u-v13-bounded-regret}
 0\le\beta_*-\beta_B\le\Delta(s)\le\frac{a_0^4}{8}.
\end{equation}
The bound holds for every $N\ge2$ and $q\ge1$; when $q=1$ the
two tests coincide. \end{theorem}

An additional bound based on a one-dimensional Beta integral appears
in Section~\ref{app:u-v13-bounded-quadratic}.
The result converts the exact sparse benchmark into an implementable
rule with an explicit absolute power guarantee. Its proof uses the
positive likelihood-ratio expansion, not an estimate of the covariance
shape. For example, the first bound is $0.00719$ at $a_0=0.5$
and $0.09021$ at $a_0=1$, without a growing-sample argument.
These are bounds on the difference of worst-case powers for the
specified sparse class. Small regret can coexist with power close to
the null level, and the theorem does not cover mixtures of sparse
and diffuse signals or arbitrary correlated covariance matrices.

There is also a direct criterion for the cost of a competing fitted
score or menu. In the ideal case let
$D_\psi=\mathbb E_0[B(\bm Y)(\phi_B-\psi)]$, where
$\mathbb E_0\psi=\alpha$. In the finite-reference case let
\begin{equation}\label{eq:u-v13-null-allocation-cost}
 D_\psi=\frac1M\mathbb E_0\sum_{i=0}^{M-1}
       B(\bm Y_i)(\psi_i^B-\psi_i),
 \qquad \sum_i\psi_i=k.
\end{equation}
Here the competitor satisfies the corresponding invariant comparison
class and, in the finite-reference case, row equivariance. Both
quantities are nonnegative and depend only on the canonical null
experiment and the prespecified algorithms. If
\begin{equation}\label{eq:u-v13-weak-signal-margin}
       se^{-s}D_\psi>\Delta(s),
\end{equation}
then the bounded quadratic rule has strictly larger worst-case power
on $\Theta_{a_0}$ than the competitor. For a fixed competitor with
$D_\psi>0$, the condition holds for every sufficiently small $a_0$.
This is a class-level power comparison: a Monte Carlo estimate of
$D_\psi$ requires its own uncertainty assessment, and it is not an
observed-data tuning certificate. The score and its local optimality
principle have classical precedents. The present comparison uses an
explicit finite-amplitude likelihood-ratio remainder to connect them
to the finite reference budget and fitted RI allocations.

\begingroup
\let\subsection\subsubsection
\subsection{An affine refinement and a finite-amplitude comparison}
\label{sec:u-v14-affine-bound}

A positive affine transformation of the bounded score has the same
rejection ordering. To sharpen its comparison with LR, we subtract
the best affine approximation to each coordinate likelihood ratio
and bound the contribution of the remaining residual. The resulting bound reflects the residual degrees of freedom
and the coordinate-set size. Let $X\sim\operatorname{Beta}(1/2,r/2)$,
$V=(r+1)X$, $s=a_0^2/2$, and define
\begin{align}
 G(X)&=e^{-s}{}_1F_1\!\left((r+1)/2;1/2;sX\right),
 \qquad v_V=\frac{2r}{r+3},\nonumber\\
 c_{r,s}&=\operatorname{Cov}(G(X),V),\qquad
 \gamma_{r,s}=c_{r,s}/v_V,\qquad
 v_{r,s}^{\perp}=\operatorname{Var}(G(X))-c_{r,s}^2/v_V.
 \label{eq:u-v14-affine-moments}
\end{align}
These one-dimensional Beta integrals satisfy $\gamma_{r,s}>0$ and
$v_{r,s}^{\perp}\ge0$. They depend on the specified experiment and
signal amplitude and are evaluated independently of the tested data.

\begin{proposition}[Affine residual bound]
\label{prop:u-v14-affine-regret}
Under Theorem~\ref{thm:u-v13-bounded-regret}, put $\tau=\alpha$
for ideal calibration and $\tau=k/M$ for finite references. Then
\begin{equation}\label{eq:u-v14-affine-regret}
 0\le\beta_*-\beta_B\le
 \min\left\{\Delta(s),
 \sqrt{\frac{2\min\{\tau,1-\tau\}}q\,v_{r,s}^{\perp}}\right\}.
\end{equation}
When $q=1$, the gap is zero. The variance term is no larger than
that in \eqref{eq:u-v13-variance-regret}.
\end{proposition}

The same residual bounds the boundary-power difference between $B$
and a fixed fitted rule or menu through its null allocation cost and
disagreement probability (Corollary~\ref{cor:u-v14-affine-learning-cost}).
Estimating these null quantities requires joint Monte Carlo uncertainty
control. The argument uses linear projection and Cauchy--Schwarz;
the resulting design bound can be uninformative when the residual
variance is large.

At $a_0=1$, evaluation of \eqref{eq:u-v14-affine-regret} gives
bounds of 1.40--2.37 percentage points for $N=20,40$ and $q=5,10$,
compared with 9.02 for the remainder-mass bound. The gains remain
limited at $a_0=2$ and disappear at $a_0=4$; Section~\ref{app:u-v14-bounds-evaluation}
reports all evaluations and numerical diagnostics.

\subsection{Null-only certificates for a specified learning rule}
\label{sec:u-v14-learning-cost}

The null allocation cost can be assessed before specifying an alternative
amplitude. This gives a design calculation for a fixed learner, rather than
an estimate of power from the application observations. For the
finite-reference experiment, put $\tau=k/M$ and, on an all-null matrix,
\[
 A_\psi=\frac1M\sum_{i=0}^{M-1}B_i(\psi_i^B-\psi_i),
 \qquad
 H_\psi=\frac1M\sum_{i=0}^{M-1}|\psi_i^B-\psi_i|.
\]
Their expectations are $D_\psi$ and $d_\psi$, respectively.

\begin{proposition}[A null-simulation certificate]\label{prop:u-v14-null-certificate}
Fix $N,q,M,k$ and a finite collection of invariant, row-equivariant
competitors with $\sum_i\psi_i=k$. Then
$0\le A_\psi\le N\min\{\tau,1-\tau\}$ and
$0\le H_\psi\le2\min\{\tau,1-\tau\}$ pointwise.
Independent all-null matrices therefore give simultaneous bounded-variable
confidence limits $D_\psi\ge D_\psi^-$ and $d_\psi\le d_\psi^+$.
On this simultaneous event, for every $a_0>0$,
\begin{equation}\label{eq:u-v14-certified-margin}
 \beta_B^{\min}-\beta_\psi^{\min}
 \ge \max\left\{
 se^{-s}D_\psi^- -\Delta(s),\;
 \gamma_{r,s}D_\psi^--\sqrt{d_\psi^+v_{r,s}^{\perp}/q}\right\},
 \qquad s=a_0^2/2,
\end{equation}
where $\gamma_{r,s},v_{r,s}^{\perp}$ are the Beta moments in
\eqref{eq:u-v14-affine-moments}, and minima are over $\Theta_{a_0}$.
If $D_\psi^->0$, the first expression is positive throughout
$0<a_0<\sqrt{2s_\psi}$, where $s_\psi$ is the unique positive solution of
$(e^{s_\psi}-1-s_\psi)/s_\psi=D_\psi^-$.
\end{proposition}

The proof and implementation are in Section~\ref{app:u-v14-null-certificate}.
The same null confidence event supports the entire amplitude interval;
no application data are used to choose a method. The certificate measures
the specified procedure's rejection allocation, including its spectrum
and any menu adjustment. It does not identify a direction-estimation-only
cost or establish a universal ordering of learned and classical tests.

A null experiment with $p=30$, $M=200$ and 3000 independent matrices
per $(N,q)$ design evaluates this criterion jointly for nine competitors.
At $N=20,q=5,a_0=1$, the affine margin is $0.174$ percentage points
against the fixed LPO score with $\eta=\sqrt p$, using simultaneous
95\% null-simulation confidence limits and the checked Beta quadrature.
This is a small local advantage, not a high-power guarantee; no positive
affine certificate is obtained at $a_0=2$ or $4$ for that score in the
four designs. Tables~\ref{tab:u-v14-allocation-cost}--\ref{tab:u-v14-allocation-margins}
report the design dependence and uncertainty.

\endgroup
\subsection{A conditional power bound for the maximum statistic}
\label{sec:u-v16-conditional-power}

The maximum rule retains one coordinate summary whereas LR uses all
coordinates. To quantify the resulting loss, we condition LR on the
maximum. Its conditional mean is the likelihood ratio for this reduced
experiment; its conditional variance measures the residual LR
variation after the other coordinates are discarded. The power bound
below weighs that variation by the distance to the rejection boundary,
where residual fluctuations can change a decision. This gives a
comparison beyond the weak-signal expansion.

In the paired diagonal experiment of
Section~\ref{sec:u-v12-efficiency}, put $r=N-1$, $s=a_0^2/2$ and
\[
 X_j=\frac{Y_j^2}{r+Y_j^2},\qquad
 G_{a_0}(x)=e^{-s}{}_1F_1(N/2;1/2;sx),\qquad
 \mathsf S=\max_{j\in J}X_j.
\]
Under the null the $X_j$ are independent
$\operatorname{Beta}(1/2,r/2)$ variables, with cdf $F_r$, and the
boundary likelihood ratio is $\ell=q^{-1}\sum_{j\in J}G_{a_0}(X_j)$.
For a generic such $X$, define
\begin{align}
 h_{a_0}(x)&=\frac{G_{a_0}(x)+(q-1)
       \mathbb E_0\{G_{a_0}(X)\mid X\le x\}}q,\nonumber\\
 v_{a_0}(x)&=\frac{q-1}{q^2}
       \operatorname{Var}_0\{G_{a_0}(X)\mid X\le x\},\qquad
 \mathcal R(v,d)=\frac{\sqrt{v+d^2}-|d|}{2}.
 \label{eq:u-v16-conditional-moments}
\end{align}
Thus $h_{a_0}(\mathsf S)=\mathbb E_0(\ell\mid\mathsf S)$ and
$v_{a_0}(\mathsf S)=\operatorname{Var}_0(\ell\mid\mathsf S)$.
Since $h_{a_0}$ increases, ranking it is equivalent to ranking the
maximum-coordinate score $\max_{j\in J}|Y_j|$.

\begin{theorem}[A conditional maximum-to-benchmark comparison]
\label{thm:u-v16-conditional-power}
Fix $N\ge2$, $J$ with $q\ge1$, and $a_0>0$ in
\eqref{eq:u-v12-axis-class}. Let $\beta_*$ and $\beta_{\max}$ denote
the worst-case powers of the LR and maximum-coordinate rules on
$\Theta_{a_0}$. For ideal level-$\alpha$ calibration with
$0<\alpha<1$, set
$c_\alpha=h_{a_0}\{F_r^{-1}((1-\alpha)^{1/q})\}$. Then
\begin{equation}\label{eq:u-v16-ideal-bound}
 0\le\beta_*-\beta_{\max}
 \le\mathbb E_0\mathcal R\{v_{a_0}(\mathsf S),
                              h_{a_0}(\mathsf S)-c_\alpha\}.
\end{equation}
For the finite-reference comparison of
Theorem~\ref{thm:u-v12-bank-maximin}, let $1\le k<M$ and let $c$ be
the $k$th largest $h_{a_0}(\mathsf S_i)$ on $M$ independent null rows.
For the corresponding top-$k$ rules,
\begin{equation}\label{eq:u-v16-finite-bound}
 0\le\beta_*^{M,k}-\beta_{\max}^{M,k}
 \le\mathbb E_0\frac1M\sum_{i=0}^{M-1}
       \mathcal R\{v_{a_0}(\mathsf S_i),h_{a_0}(\mathsf S_i)-c\}.
\end{equation}
The finite-reference powers average over the observed data and the
reference simulation, at level $k/M$, in the pathwise row-budget class
of Theorem~\ref{thm:u-v12-bank-maximin}. Both gaps vanish when $q=1$.
\end{theorem}

The proof first identifies the maximum rule as the LR test in the
reduced experiment. It then evaluates a variational upper bound for
full LR power at the reduced test's threshold, and controls the
remaining difference by a conditional two-moment bound
\citep{rockafellar2002cvar,gallego1993distribution}.
For finite references the threshold is random, so the argument
conditions jointly on all row maxima. Section~\ref{app:u-v16-conditional-power} reduces the
finite-reference bound to a two-dimensional integral. Table~\ref{tab:u-v17-efficiency-main} gives the principal finite-reference
comparisons at $M=200$, $k=10$, $N=20,40$ and $q=5,10$.
The evaluated bounds remain informative both when benchmark power is
moderate and when it is high. For example, at $N=40$, $q=5$ and
$a_0=4$, the simulated LR power is $89.80\%$ and the evaluated
maximum-to-LR bound is below $0.25$ percentage points.
The power estimate and the deterministic bound have different numerical
sources; subtracting the latter from the former does not by itself
give a confidence lower bound on power.

These are efficiency bounds for the maximum rule, complementary to
the preceding bounded-quadratic bounds. The maximum test needs no
value of $a_0$; its efficiency comparison does. They justify a strong
simple comparator for a fitted procedure in the stated sparse class.
They do not choose adaptively between aggregation and maximum tests
or bound the loss of an arbitrary learned RI direction. The full
grid and numerical integration checks appear in Supplementary
Section~\ref{app:u-v16-conditional-evaluation}.

\begin{table}[b]
\centering\small
\setlength{\tabcolsep}{6pt}
\caption{Finite-reference efficiency of the maximum-coordinate rule at moderate and strong sparse signals. All designs use $M=200$, $k=10$ and $p=30$. LR power is a percentage; the simulated maximum-to-LR power gap and its conditional-moment upper bound are in percentage points.}
\label{tab:u-v17-efficiency-main}
\begin{tabular}{rrr rrr}
\toprule
$N$ & $q$ & $a_0$ & LR power & Simulated gap & Evaluated bound\\
\midrule
20 & 5 & 2 & 26.10 & 0.48 & 1.50\\
20 & 5 & 4 & 85.78 & 0.06 & 0.34\\
\addlinespace
20 & 10 & 2 & 19.93 & 0.58 & 1.80\\
20 & 10 & 4 & 78.47 & 0.11 & 0.54\\
\addlinespace
40 & 5 & 2 & 28.69 & 0.41 & 1.38\\
40 & 5 & 4 & 89.80 & 0.04 & 0.25\\
\addlinespace
40 & 10 & 2 & 22.34 & 0.49 & 1.63\\
40 & 10 & 4 & 84.33 & 0.07 & 0.39\\
\bottomrule
\end{tabular}
\par\smallskip\parbox{0.98\linewidth}{\footnotesize Power and gap estimates reuse 2000 independent augmented matrices per design from the earlier simulation. The last column evaluates Theorem~\ref{thm:u-v16-conditional-power} by deterministic quadrature and rounds the stored value upward to two decimals; it is not a Monte Carlo confidence limit or an interval-arithmetic enclosure. The full grid, ideal-calibration bounds and integration checks appear in Table~\ref{tab:u-v16-conditional-bounds} and Section~\ref{app:u-v16-conditional-evaluation}; simultaneous simulation intervals appear in Section~\ref{app:u-v13-benchmark}.}
\end{table}

\subsection{A benchmark at a specified mixed-signal allocation}
\label{sec:u-v14-mixed-orbit}
The sparse benchmark can also be extended to a fixed mixed-signal
allocation. Let $h=p-q\ge1$, fix $a_0>0,d_0>0$, and consider
$\bm\theta_J=s a\bm e_j$, $j\in J$, $s\in\{-1,1\}$,
$a\ge a_0$, with $|\theta_k|\ge d_0/\sqrt h$ for every $k\notin J$.
These are coordinatewise lower bounds, not merely a bound on the total
outside energy. With $g_a$ as in \eqref{eq:u-v12-symmetric-t-ratio}, put
\begin{equation}\label{eq:u-v14-mixed-lr}
 \ell_{a_0,d_0}(\bm Y)
 =\left\{\frac1q\sum_{j\in J}g_{a_0}(Y_j)\right\}
   \prod_{k\notin J}g_{d_0/\sqrt h}(Y_k).
\end{equation}
\begin{proposition}[A fixed-allocation mixed benchmark]
\label{prop:u-v14-mixed-maximin}
The upper-tail test of \eqref{eq:u-v14-mixed-lr} is maximin over this
class among the scale-invariant level-$\alpha$ tests of
Theorem~\ref{thm:u-v12-ideal-maximin}. Its top-$k$ finite-reference
version is maximin in the comparison class of
Theorem~\ref{thm:u-v12-bank-maximin}. Both minima occur at
$a=a_0$ and $|\theta_k|=d_0/\sqrt h$ for all $k\notin J$.
\end{proposition}
This applies the same classical likelihood-ratio argument to a mixed
boundary orbit; the allocation $(a_0,d_0)$ remains specified.
The product is evaluated by summing logarithms, so exponentially many
sign patterns need not be enumerated. Section~\ref{app:u-v14-mixed-orbit}
gives the proof and Section~\ref{app:u-v14-mixed-experiment} a matched
comparison. Setting $d_0=0$ and $g_0=1$ recovers the sparse benchmark
when the outside mean is zero. No optimization over an unknown
sparse/diffuse allocation is asserted.

A follow-up gives block competitors the same specified allocation as
LR. It uses three fixed outside-energy fractions, $N=20$, $p=30$,
$q=5$, total standardized energy $16$, and 2000 fresh independent
augmented matrices per design. At outside fraction $0.5$, the
boundary powers are $61.07\%$ for mixed LR, $54.57\%$ for the
fixed block quadratic, $55.76\%$ for the allocation-informed
quadratic, and $43.95\%$ for LPO. At fraction $0.9$, an
allocation-informed bounded score reaches $51.00\%$, compared
with $51.62\%$ for LR and $20.93\%$ for the fixed block.
Thus the earlier fixed-weight comparison partly reflected the use
of allocation information. The follow-up weights are prespecified
energy-based choices; they do not optimize every block test.
Supplementary Section~\ref{app:u-v17-informed-mixed} gives the
formulas, all designs, uncertainty intervals and separate simulation
provenance. These remain procedure-specific comparisons.

\subsection{Why retain a baseline under mixed signals}
\label{sec:u-mixed-limit}

The preceding comparisons use the paired invariant experiment. We now
return to the original split RI statistic of
Section~\ref{sec:u-unknown-shape} and ask a different question:
why can retaining a baseline improve on projection when the supplied
learning subspace omits signal? Holding the observation budget fixed
and letting dimension increase separates the diffuse baseline from
the fitted low-dimensional component. This explains a mechanism;
the finite-sample benchmark results above require no such limit.

Normalize $\bm\Sigma_1=\bm H$, $\bm\Sigma_2=\rho\bm H$ and put
\[
 s^2=n_1^{-1}+\rho n_2^{-1},\quad
 s_{\mathrm{tr}}^2=m_1^{-1}+\rho m_2^{-1},\quad
 \bm\theta=\bm H^{-1/2}\bm\delta/s,\quad
 \kappa=n_1^{-1}/s^2.
\]
The matched known-shape procedure replaces $\widehat{\bm H}$ by
$\bm H$ in both learning and testing and sets $\bm R=\bm I_p$
in the reference. It retains the same split, testing-only scale,
endpoint maximum and finite reference-bank size $L$. It still learns
its direction and therefore differs from the population-direction
oracle. Supplementary Section~\ref{sec:u-invariant-power} gives a
power-transfer condition with increasing training information and a
fixed-budget limit for $\eta=p$. In the latter limit the baseline
vanishes; increasing dimension alone does not remove shape randomness.

For the mixed-signal regime below, fix the first $q$ coordinates as
the learning subspace. With independent
$R_j\sim\chi^2_r/r$ and $\bm Z_{\mathrm{tr},q}\sim N_q(\bm0,\bm I_q)$,
write $\bm R_q=\operatorname{diag}(R_1,\ldots,R_q)$ and
\begin{equation}\label{eq:jasa-training-factors}
 \bm X=(s/s_{\mathrm{tr}})\bm\theta_q+\bm Z_{\mathrm{tr},q},\qquad
 \bm a_R=\frac{\bm R_q^{-1}\bm X}
                  {(\bm X^\top\bm R_q^{-1}\bm X)^{1/2}},\qquad
 W_R=\|\bm a_R\|_2^2.
\end{equation}
Let $W_R^0$ have the same construction with
$\bm X=\bm Z_{\mathrm{tr},q}$. These factors retain the effect of
estimated whitening in the fixed-dimensional learned component.
The scaling $\eta=p$ in Proposition~\ref{prop:u-diag-fixed-budget-limit}
lets the spike dominate the baseline. The centered baseline fluctuates
on the $\sqrt p$ scale, so choosing $\eta=c\sqrt p$ keeps both
terms at the same order. Diffuse omitted energy of order $\sqrt p$
then contributes a nonvanishing shift on that scale. The condition
$r>4$ supplies the second inverse-variance moment used by this limit;
the fourth-power condition prevents a few omitted coordinates from
dominating the diffuse contribution.

\begin{proposition}[A mixed-signal fixed-budget limit]
\label{prop:u-mixed-limit}
Use the diagonal model, shape estimator and learning rule of
Theorem~\ref{thm:u-diag-invariant}.
Fix $q,m_i,n_i,b,c,\rho,L,\alpha$, with $q\ge1$, $m_i,n_i\ge2$,
$b,c,\rho>0$, $L\ge1$, $0<\alpha<1$, and $r=m_1-1>4$.
Let $p>q$ tend to infinity, $\eta=c\sqrt p$, and
$\bm P_p=\operatorname{diag}(\bm I_q,\bm0_{p-q})$. Suppose that
$\bm\theta_p=(\bm\theta_q^\top,\bm\theta_{\perp,p}^\top)^\top$,
where the signal sequence is deterministic, $\bm\theta_q$ is fixed and
\begin{equation}\label{eq:u-mixed-signal}
 \frac{\|\bm\theta_{\perp,p}\|_2^2}{\sqrt p}\longrightarrow\tau\ge0,
 \qquad
 \frac1p\sum_{j=q+1}^p\theta_{p,j}^4\longrightarrow0.
\end{equation}
The positive diagonal shapes may vary with $p$. Use $\bm a_R,W_R$
and $W_R^0$ as in Proposition~\ref{prop:u-diag-fixed-budget-limit},
and define
\[
 \mu_r=\frac r{r-2},\qquad
 \nu_r=\frac{r^2}{(r-2)(r-4)},\qquad
 \zeta_R=\frac{\bm a_R^\top\bm\theta_q}{\sqrt{W_R}}.
\]
Let $G_0,G_1,G_2$ be independent centered Gaussian variables with
variances $2\nu_r,2\nu_r/k_1,2\nu_r/k_2$, respectively, independent
of the training quantities, where
$k_i=n_i-1$. Independently, let $Z\sim N(0,1)$ and
$E_i\sim\chi^2_{k_i}$ be mutually independent and independent of
the training quantities. With $\kappa=n_1^{-1}/s^2$,
\begin{align}
 \frac{\widehat T_{\widehat{\bm v}}}{\sqrt p}
 &\ \Rightarrow\ T_c(\tau)
 =b\tau+\frac b{\mu_r}\{G_0-\kappa G_1-(1-\kappa)G_2\}
 \notag\\[-0.2em]
 &\hspace{1.1cm}
 +\frac{cW_R}{\mu_r}
 \left\{(Z+\zeta_R)^2-\kappa\frac{E_1}{k_1}
                         -(1-\kappa)\frac{E_2}{k_2}\right\}.
 \label{eq:u-mixed-T}
\end{align}
Each independently regenerated reference satisfies
\begin{equation}\label{eq:u-mixed-M}
 \frac{M_{\mathrm{diag}}}{\sqrt p}
 \ \Rightarrow\ M_c
 =\frac1{\mu_r}\max_{i=1,2}
 \left[b(G_0-G_i)+cW_R^0
                  \left\{Z^2-\frac{E_i}{k_i}\right\}\right].
\end{equation}
The two endpoints in \eqref{eq:u-mixed-M} share $G_0,Z,W_R^0$.
The observed limit and its $L$ reference limits are independent.
If $k_\alpha=\lfloor\alpha(L+1)\rfloor\ge1$ and
$j=L-k_\alpha+1$, the actual rank-test power converges to
\begin{equation}\label{eq:u-mixed-power}
 \pi_c(\tau)=\Pr\{T_c(\tau)>M_{c,(j)}\},
\end{equation}
where $M_{c,(j)}$ is the $j$th ascending order statistic of the
$L$ reference limits. The matched known-shape limit is obtained by
setting $\mu_r=\nu_r=W_R=W_R^0=1$ and
$\bm a_R=\bm X/\|\bm X\|_2$.
\end{proposition}

The baseline retains a Gaussian component and the outside-signal
shift $b\tau$. Estimated shape inflates its variance through
$\nu_r/\mu_r^2=(r-2)/(r-4)$ and also leaves random factors in the
spike. The diffuseness condition in \eqref{eq:u-mixed-signal} is
substantive: concentrating energy of order $\sqrt p$ in one omitted
coordinate leaves its individual inverse-variance error in the limit.
This proposition does not cover that concentrated alternative.

\begin{corollary}[Power from omitted diffuse signal]
\label{cor:u-mixed-projection}
Under Proposition~\ref{prop:u-mixed-limit}, fix $\bm\theta_q$ and
the other design parameters, and assume $k_\alpha\ge1$. Then
$\pi_c(\tau)$ is continuous and strictly increasing in $\tau$ and
tends to one as $\tau\to\infty$. If $n_1=n_2=n$, the equally
informed two-sided level-$\alpha$ paired projection $t$ test has
power independent of $\tau$
and strictly below one. Consequently there is a finite $\tau$ for
which the limiting RI power is greater; along the corresponding
sequence, the same strict inequality holds for all sufficiently
large $p$.
\end{corollary}

This is an existence result for a specified mixed-signal regime.
The finite-dimensional bound in Section~\ref{sec:u-finite-dimension}
quantifies its approximation error under stronger moment conditions;
neither result asserts superiority over the no-spike or all-data tests.
The value $c$ and the subspace remain fixed before the observations
are used. In particular, the corollary does not select a spike or
resolve the fixed-budget cost of estimating the shape.

\subsection{Finite-dimensional error control}
\label{sec:u-finite-dimension}

To interpret the preceding limit at finite dimension, we must control
errors in both the observed statistic and its simulated critical value.
The argument first bounds their distribution functions, then transfers
those errors to rejection probability. We first record the required
stability property of the finite-reference test.
It compares the distributions of one observed statistic and one
reference, and does not require a simultaneous coupling of all $L$
references.

\begin{lemma}[Finite-reference power stability]
\label{lem:u-rank-cdf-stability}
Let $L\ge1$ be an integer, $0<\alpha<1$, and let $G,F$ be the continuous distribution functions of an observed
statistic and each of $L$ independent, identically distributed references,
independent of the observation. Let $G_0,F_0$ be another such pair.
For $k=\lfloor\alpha(L+1)\rfloor\ge1$ and $j=L-k+1$, define
\[
 C_{L,k}=\max_{0\le u\le1}
 L\binom{L-1}{j-1}u^{j-1}(1-u)^{L-j}.
\]
If $\pi(G,F)$ denotes the probability that the observation exceeds
the $j$th ascending reference order statistic, then
\begin{equation}\label{eq:u-rank-cdf-stability}
 |\pi(G,F)-\pi(G_0,F_0)|
 \le \min\{1,\|G-G_0\|_\infty
                  +C_{L,k}\|F-F_0\|_\infty\}.
\end{equation}
For $L>1$ the maximum occurs at $u=(j-1)/(L-1)$,
with $0^0=1$; $C_{1,1}=1$.
\end{lemma}

The approximation below strengthens $r>4$ to $r>6$ to obtain the
third inverse-variance moment needed for the quantitative Gaussian
approximation. The supplement groups its constants by the numerator,
remainder and denominator errors that they control.

\begin{theorem}[An explicit mixed-signal approximation bound]
\label{thm:u-mixed-finite}
Use the model and fixed design parameters of
Proposition~\ref{prop:u-mixed-limit}, strengthen $r>4$ to $r>6$,
and let $p>q$ be finite. Put
$\tau_p=\|\bm\theta_{\perp,p}\|^2/\sqrt p$ and
$k=\lfloor\alpha(L+1)\rfloor\ge1$.
There are fully specified nonnegative quantities $\Delta_{T,p}$ and
$\Delta_{M,p}$, defined in
Section~\ref{app:u-finite-dimension} using only the design parameters
and the signal coordinates, for which
\begin{equation}\label{eq:u-mixed-finite-power}
 |\pi_p-\pi_c(\tau_p)|
 \le \min\{1,\Delta_{T,p}+C_{L,k}\Delta_{M,p}\}.
\end{equation}
Here $\pi_p$ is the actual finite-$p$ rank-test power, with the full
unbiased statistic and the testing-only studentizer, and
$\pi_c(\tau_p)$ uses exactly $L$ reference draws from
\eqref{eq:u-mixed-M}. No limit in $L$ is taken.
For a fixed target $\tau$, the right-hand side of
\eqref{eq:u-mixed-finite-power} additionally contains
$h_Tb|\tau_p-\tau|$, where
\[
 h_T=\frac{\mu_r}
 {2b\sqrt{\pi\nu_r\{1+\kappa^2/k_1+(1-\kappa)^2/k_2\}}}.
\]
With all design parameters and $\bm\theta_q$ fixed and $\tau_p$
bounded, the untruncated bound has order
\begin{equation}\label{eq:u-mixed-finite-rate}
 O\left(p^{-1/6}
       +\left\{\frac1p\sum_{j=q+1}^p\theta_{p,j}^4\right\}^{1/3}
       +|\tau_p-\tau|\right)
\end{equation}
when compared with $\pi_c(\tau)$.
\end{theorem}

The bound separates a Gaussian approximation, the omitted baseline
and signal terms, and the random denominator. The reference calculation
retains the dependence between its two endpoints.
All its constants can be evaluated without fitting an asymptotic
error rate. Nevertheless, the universal Gaussian approximation
constants and the smoothing steps can make the bound equal to one
at practical dimensions. Thus the theorem quantifies convergence;
it does not establish an accurate approximation at a fixed small
observation budget or optimality of a chosen spike. The bound is also
an oracle comparison involving the unknown signal, rather than a
data-dependent certificate. If $k=0$, both rank procedures have zero
rejection probability and the comparison is exact.

An independent numerical study uses $3\,000$ reference banks per
law to assess the approximation at six specified designs. The
resulting upper bounds on absolute finite-dimensional versus limiting
power differences range from $1.77$ to $3.02$ percentage points,
simultaneously with $95\%$ simulation coverage (Supplementary
Section~\ref{app:u-v9-limit-certificate}). These are bounds for the
six fixed designs, not observed errors or a uniform model guarantee.

\section{Simulation studies}
\label{sec:u-numerical-focused}

The simulations examine how the choice between a fitted axis and direct
use of supplied structure depends on signal concentration and coverage.
Coordinate sets are fixed; signal coverage is varied through the
generating model, rather than by selecting variables from the tested
data. The experiments address three questions. We first measure the gap
between simple or fitted paired rules and their attainable benchmark.
We then compare split RI with projection and all-data competitors at
equal observation budgets. Finally, controlled comparisons examine
which differences arise from shape estimation, direction fitting,
weight allocation and menu adjustment. Each study states its
information and calibration conditions.

\subsection{Comparison with an attainable sparse benchmark}
\label{sec:u-v13-benchmark-experiment}

The restricted benchmark can be compared directly with the implemented
procedures. We fix $p=30$, $N\in\{20,40\}$, $q\in\{5,10\}$ and
$a_0\in\{1,2,4\}$, giving 12 prespecified settings. The alternatives
are the boundary mixture $\bm\theta=\pm a_0\bm e_j$, $j\in J$.
Every finite-reference procedure uses the same paired observations and
$M=200$ data-set rows, with $k=10$. The scores and all three candidate
menus are fixed before simulation. Coordinate-scale invariance permits
canonical Gaussian generation despite the unknown diagonal covariance.

For each of 2000 independent matrices, we place one boundary-mixture
row at a uniformly random index and generate the other rows under the
null. If $\ell_i$ is the sparse likelihood ratio and $\psi_i$ is a
rule's hypothetical row decision, the contribution
$\sum_i\ell_i\psi_i/\sum_i\ell_i$ lies in $[0,1]$ and has expectation
equal to its average boundary power. Subtracting a competitor's
contribution from the top-$k$ LR contribution gives a nonnegative,
bounded estimate of the allocation gap in
Proposition~\ref{prop:u-v12-regret-identity}; Proposition~\ref{prop:u-v13-tilted-gap}
justifies the bounded estimator and its intervals. The change of measure is
on maximal invariants; it does not assert that $\ell_i$ is a likelihood
ratio for the original unstandardized Gaussian observations. We record
matrix-level MCSEs and simultaneous bounded-variable Chernoff intervals.
The interval statements concern the specified simulation expectations.

\begin{table}[t]
\centering
\small
\setlength{\tabcolsep}{4pt}
\caption{Comparison with the attainable finite-reference sparse benchmark. The LR column gives boundary power (\%); the remaining columns give its average boundary-power advantage over each rule (percentage points). All 12 prespecified settings use $p=30$, $L=199$ and 2000 independent augmented matrices. Bounded denotes the bounded block score, quadratic denotes $\sum_{j\in J}Y_j^2$, and maximum denotes $\max_{j\in J}|Y_j|$. The LPO score has spike $\eta=\sqrt p$; combined denotes the 13-component menu.}
\label{tab:u-v13-benchmark}
\begin{tabular}{rrr r rrrrr}
\toprule
$N$ & $q$ & $a_0$ & LR power & Bounded & Quadratic & Maximum & LPO & Combined\\
\midrule
20 & 5 & 1 & 9.12 & 0.12 & 0.00 & 0.41 & 2.23 & 2.36\\
20 & 5 & 2 & 26.10 & 3.20 & 1.10 & 0.48 & 8.77 & 10.13\\
20 & 5 & 4 & 85.78 & 12.27 & 4.46 & 0.06 & 14.35 & 18.14\\
\addlinespace
20 & 10 & 1 & 7.73 & 0.10 & 0.00 & 0.47 & 1.25 & 1.57\\
20 & 10 & 2 & 19.93 & 3.58 & 1.35 & 0.58 & 5.78 & 7.64\\
20 & 10 & 4 & 78.47 & 27.18 & 9.76 & 0.11 & 17.62 & 23.82\\
\addlinespace
40 & 5 & 1 & 9.48 & 0.12 & 0.04 & 0.36 & 2.53 & 2.64\\
40 & 5 & 2 & 28.69 & 2.90 & 1.75 & 0.41 & 10.25 & 11.54\\
40 & 5 & 4 & 89.80 & 6.99 & 4.27 & 0.04 & 12.83 & 16.10\\
\addlinespace
40 & 10 & 1 & 7.99 & 0.13 & 0.04 & 0.42 & 1.50 & 1.77\\
40 & 10 & 2 & 22.34 & 3.84 & 2.36 & 0.49 & 7.50 & 9.36\\
40 & 10 & 4 & 84.33 & 17.89 & 10.16 & 0.07 & 18.22 & 23.70\\
\bottomrule
\end{tabular}
\par\smallskip\parbox{0.98\linewidth}{\footnotesize Maximum matrix-level MCSE: 0.67 percentage points for power and 0.48 for a gap. Simultaneous confidence bounds, all nine methods and the independently approximated ideal benchmark appear in Section~\ref{app:u-v13-benchmark}.}
\end{table}

Table~\ref{tab:u-v13-benchmark} shows two different regimes. At $a_0=1$, the bounded and ordinary quadratic block scores are close to the benchmark, whose absolute power is only 7.73\%--9.48\%. At $a_0=4$, the maximum-coordinate score has an estimated gap of at most 0.11 percentage points; its simultaneous Monte Carlo upper bound is at most 0.81 percentage points across these four settings. The learned LPO score loses 12.83--18.22 percentage points, and the combined menu loses 16.10--23.82 percentage points.

Theorem~\ref{thm:u-v16-conditional-power} complements these maximum-rule
comparisons with analytic bounds evaluated at the same finite reference
budget; Table~\ref{tab:u-v17-efficiency-main} separates those
calculations from the archived Monte Carlo estimates.

Thus the sparse experiment supplies an attainable target and a measurable
cost for departing from it. It does not support a preference for learned
RI directions in this class. Simple block aggregation is competitive at
weak boundary signals, whereas coordinate maxima are competitive at the
strong boundaries studied. These design comparisons do not identify an
observed-data rule for choosing between the regimes: $J$ and $a_0$ were
prespecified, and the alternative class contains only one active coordinate.
Mixed sparse and diffuse signals remain a separate comparison.

Holding $a_0$ fixed while varying $N$ holds total standardized noncentrality
fixed, so the raw effect changes as $N^{-1/2}$. These comparisons isolate
finite-residual effects at matched noncentrality; they do not describe
collecting more observations at a fixed population mean difference.

\subsection{Comparisons on a common coordinate panel}
\label{sec:u-v22-external-comparison}
The attainable benchmarks above concern a specified paired experiment.
We next ask how the fitted scores compare with published procedures when
all methods receive the same raw observations and coordinate panel.
This comparison is needed because an advantage over one projection rule
does not establish an advantage over methods that aggregate coordinates
without estimating a direction.

We generate two independent Gaussian samples of size 40 on a fixed panel
of $p=q=60$ coordinates, with diagonal covariances $H$ and $\rho H$,
$\rho\in\{1,4\}$. The diagonal entries of $H$ range from $0.5$ to $2$.
Every method receives the whole panel; neither the active coordinate nor
the signal type is supplied. In addition to the null, we consider a
single-coordinate signal and a same-sign diffuse signal with
$\Lambda_{\rm all}=40\delta^\top H^{-1}\delta/(1+\rho)\in\{16,64\}$.
Each of the ten settings has 10,000 simulated datasets. The signal settings
share noise, and the four paired scores share 199 Gaussian references
within each of 500 independent simulation banks, with 20 datasets per bank. Standard errors account
for this reuse; the largest is 0.60 percentage points.

\begin{table}[t]
\centering\small
\caption{Same-panel comparisons at a fixed total observation budget. Each group has $N=40$ observations and every method receives the same $p=q=60$ coordinates. Entries are rejection percentages at nominal level $5\%$; nonnull signal labels give $\Lambda_{\rm all}$. The last two columns retain the published asymptotic calibrations.}
\label{tab:v22-jasa-comparison}
\begin{tabular}{clrrrrrr}
\toprule
$\rho$ & Signal, $\Lambda_{\rm all}$ & Bounded & Quadratic & Maximum & LPO & Jiang RI & Yu PE \\
\midrule
1 & Null & 4.71 & 4.84 & 4.95 & 4.94 & 6.02 & 9.37 \\
1 & Sparse, 16 & 23.92 & 35.52 & 65.95 & 31.94 & 8.39 & 73.16 \\
1 & Sparse, 64 & 66.77 & 98.34 & 100.00 & 97.92 & 20.79 & 100.00 \\
1 & Diffuse, 16 & 36.39 & 35.69 & 12.33 & 31.92 & 78.81 & 41.59 \\
1 & Diffuse, 64 & 99.35 & 99.21 & 40.72 & 98.91 & 100.00 & 99.12 \\
\midrule
4 & Null & 5.11 & 5.06 & 4.86 & 4.78 & 6.35 & 9.91 \\
4 & Sparse, 16 & 24.91 & 35.71 & 66.20 & 32.05 & 8.97 & 73.55 \\
4 & Sparse, 64 & 66.94 & 98.23 & 100.00 & 97.95 & 21.02 & 100.00 \\
4 & Diffuse, 16 & 37.20 & 36.30 & 12.27 & 32.17 & 78.30 & 42.56 \\
4 & Diffuse, 64 & 99.08 & 98.92 & 39.57 & 98.64 & 100.00 & 99.05 \\
\bottomrule
\end{tabular}
\begin{minipage}{0.98\linewidth}\footnotesize
Each cell uses 10000 datasets grouped in 500 independent reference banks. The largest bank-cluster Monte Carlo standard error is 0.60 percentage points. 
The four paired scores share 199 canonical reference draws per bank; Jiang RI and Yu PE use all original observations with their own calibrations. All methods use the same supplied panel. No external-method evaluation was invalid. 
Full standard errors, paired differences, implementation checks and timings are given in Supplementary Section~\ref{app:v22-jasa-comparison}. The paired likelihood-ratio optimality statements do not extend to this comparison with unpaired procedures.
\end{minipage}
\end{table}

The four paired scores use the same fixed differences and are refitted
on every reference dataset. LPO uses $b=1$ and $\eta=\sqrt{60}$.
The RI procedure of \citet{jiang2024ri} and the mean component of
\citet{yu2023power} use the original two samples and their published
asymptotic calibrations. The former is independently implemented from
the published formulas; the latter is checked against the authors'
\texttt{PEtests} R implementation. Supplementary
Section~\ref{app:v22-jasa-comparison} gives the protocol, implementation
checks, uncertainty estimates and computation times.

Table~\ref{tab:v22-jasa-comparison} shows why a strong simple comparator
is necessary. At $\Lambda_{\rm all}=16$, the maximum rule has power
65.95--66.20\% against a single-coordinate signal, compared with
31.94--32.05\% for LPO. Against the same-sign diffuse signal, Jiang RI
has power 78.30--78.81\%, compared with 31.92--32.17\% for LPO.
At nominal level 5\%, the four paired rules have null rejection rates
between 4.71\% and 5.11\%. The corresponding rates are 6.02\% and
6.35\% for Jiang RI, and 9.37\% and 9.91\% for Yu PE. These
external-method power comparisons are therefore at the same nominal level,
not at matched attained size. No external method is recalibrated after
seeing the outcomes.
Thus the benefit of a procedure depends on the signal configuration,
and the present fitted RI score is not uniformly preferable.

These comparisons do not extend paired-experiment maximin claims to
all tests using the original two samples. Nor do they establish a
comparison with the online projection test of \citet{liu2024projection}:
a verified implementation of its complete tuning procedure is not yet
included, and a generic split projection is not used as a substitute.

\subsection{Other fixed-budget comparisons}

Every method receives $N$ raw observations per group; reference
simulation does not consume observations. Unless specified otherwise,
$\rho=4$, $b=1$, $q=5$, the training/testing split is equal,
$L=199$, and the nominal level is $0.05$. The common shape is diagonal,
with entries ranging geometrically from $1/2$ to $2$. Its values are
withheld from estimated-shape methods. We index alternatives by
$\Lambda_{\mathrm{all}}=N\bm\delta^\top\bm H^{-1}\bm\delta/(1+\rho)$.
To vary omitted diffuse signal while holding its total whitened
magnitude fixed, we set
\[
 \bm H^{-1/2}\bm\delta=
 \sqrt{\frac{\Lambda_{\mathrm{all}}(1+\rho)}{N}}
 \left\{\sqrt{1-\epsilon}\,\bm e_1+
 \sqrt{\frac{\epsilon}{p-q}}\sum_{j=q+1}^{p}\bm e_j\right\}.
\]
All splits, subspaces and spike values are prespecified. The grid uses
null and alternative draws, so power differences can be interpreted
alongside attained sizes. Supplementary material records complete
grids, seeds, bank counts and timings.

The equally informed projection comparator uses the same training
shape estimate and learned axis as RI. For balanced testing samples,
it applies a two-sided paired $t$ test to
$a_j=\widehat{\bm v}^\top\widehat{\bm H}^{-1/2}
(\bm X_{1j}-\bm X_{2j})$. Conditional on training its null law is
$t_{n-1}$; alternative rejection probabilities are integrated using
the noncentral $t$ law in the earlier retained-baseline experiments.
Those Monte Carlo errors concern training only; the expanded raw-data
comparison uses actual rejection indicators. The all-data methods
use the full raw samples. Among them, a paired Hotelling test receives
the same specified $q$-dimensional subspace; the remaining all-data
comparators do not receive this signal-subspace information. Thus
equal observation budgets do not imply identical structural information.

\subsection{Stronger external benchmarks}
\label{sec:u-v9-external-focused}

An expanded raw-data experiment compares ten procedures at
$N=40$ per group, $p\in\{30,300\}$ and $\rho\in\{1,4\}$.
RI and its no-spike version use the joint calibration; the same-axis
projection uses a two-sided Gaussian $t$ reference. All-data comparators
include CQ \citep{chen2010}, the $F$-type procedure of
\citet{zhu2024ftype}, the $U$--$t$ statistic of
\citet{li2023finite}, coordinatewise paired $t$ tests with Bonferroni
correction, and Hotelling's test in the specified $q=5$ subspace.
The cross-$U$ statistic of \citet{kim2024cross} uses a full-dimensional
training mean and is reported with both its normal reference and the
exact one-sided Gaussian $t$ correction. The subspace Hotelling test
receives the same structural information as RI and uses all observations.
The other all-data procedures do not receive that subspace.

The initial grid used the null and $\Lambda_{\mathrm{all}}=64$.
After inspecting the high-strength outcomes, a separately recorded
follow-up added $\Lambda_{\mathrm{all}}=16$, keeping methods and
tuning parameters unchanged. All 44 cells use $10\,000$ datasets,
with $200$ independent reference-bank clusters of $50$. Each RI bank
contains $L=199$ references. Full protocols, formulas, per-cell standard
errors, paired contrasts and timings are in
Section~\ref{supp:v9external} and the numerical archive.

\begin{table}[tbp]
\centering\small
\caption{Same-budget raw-data comparisons at $p=300$, $\rho=4$,
$N=40$ per group. Entries are rejection percentages at nominal
$5\%$. ``Inside'' has $\epsilon=0$; ``Conc.'' and ``Dense'' have
$\epsilon=1$, with concentrated and diffuse omitted signal.
The largest rate Monte Carlo standard error over the full grids is
$0.87$ percentage points. Exact Gaussian and approximate references
are distinguished in the text; these are nominal-level comparisons.}
\label{tab:u-v9-external}
\setlength{\tabcolsep}{3pt}
\begin{tabular}{lrrrrrrr}
\toprule
 & Null & \multicolumn{3}{c}{$\Lambda_{\rm all}=16$} & \multicolumn{3}{c}{$\Lambda_{\rm all}=64$}\\
\cmidrule(lr){3-5}\cmidrule(lr){6-8}
Method & & Inside & Conc. & Dense & Inside & Conc. & Dense\\
\midrule
RI, $\eta=\sqrt p$ & 3.99 & 49.87 & 6.29 & 6.07 & 99.10 & 13.84 & 12.47\\
RI, $\eta=0$ & 3.87 & 7.14 & 7.01 & 6.76 & 27.97 & 27.19 & 26.56\\
Projection $t$ & 4.73 & 54.92 & 5.21 & 5.21 & 99.53 & 4.73 & 4.73\\
CQ & 5.88 & 9.55 & 9.76 & 16.55 & 29.96 & 30.71 & 74.39\\
$F$-type & 5.32 & 8.72 & 8.65 & 15.44 & 28.38 & 29.12 & 72.83\\
Li $U$--$t$ & 5.84 & 9.52 & 9.73 & 16.44 & 29.63 & 30.33 & 73.88\\
Cross-$U$ normal & 6.06 & 9.14 & 9.06 & 13.25 & 22.46 & 22.62 & 52.04\\
Cross-$U$ $t$ & 4.91 & 7.43 & 7.44 & 11.00 & 19.13 & 19.25 & 47.33\\
Bonferroni $t$ & 4.89 & 48.20 & 48.63 & 7.04 & 99.97 & 99.96 & 11.85\\
Subspace $T^2$ & 4.62 & 83.15 & 5.35 & 5.35 & 100.00 & 4.62 & 4.62\\
\bottomrule
\end{tabular}

\end{table}

The new comparisons materially restrict the efficiency claim.
For the moderate inside signal in Table~\ref{tab:u-v9-external},
RI has power $49.87\%$, versus $54.92\%$ for the matched projection
and $83.15\%$ for subspace Hotelling. Its paired losses are
$5.05$ and $33.28$ percentage points, with standard errors $0.46$
and $0.57$. The latter comparator has an exact Gaussian null law
and uses the same subspace information. Thus the gain over an
uninformed quadratic does not establish competitiveness with an
all-data, structure-aware test.

For strong entirely omitted diffuse signal, RI has power $12.47\%$,
compared with $72.83\%$ for the approximate $F$-type procedure and
$47.33\%$ for the exact Gaussian cross-$U$ $t$ test. Concentrated
omission instead favors Bonferroni. At leakage $0.9$, the retained
baseline can still improve on the same-axis projection and no spike;
the complete panels retain these gains and the losses at leakage one.
Equal Mahalanobis signal strengths do not imply equal Euclidean
norms under the nonconstant shape, so the two omission geometries
are not a pure sparsity comparison for unweighted tests.

Attained sizes also differ. The normal-reference cross-$U$ statistic
has analytic null rejection probability $0.06269$ at $n=20$ and
nominal $0.05$; its $t$ correction is exact under Gaussian sampling.
CQ and Li's test are also somewhat liberal in the displayed design.
We do not interpret their extra rejections as size-controlled gains.
These findings support a restricted account of the full quadratic's
role, rather than a claim of general efficiency under a fixed budget.

The retained-baseline experiment compares three separate procedures,
$\eta\in\{0,\sqrt p,p\}$, using $400$ independent reference banks
and $100$ datasets per bank. Data and reference draws are coupled
across spike values. At $p=300$, $N=40$,
$\Lambda_{\mathrm{all}}=64$ and $\epsilon=0.9$, the powers are
$26.70\%$, $29.39\%$ and $20.72\%$, respectively, compared with
$20.63\%$ for projection. The $\sqrt p$ spike exceeds projection
by $8.76$ percentage points (paired standard error $0.33$) and
no spike by $2.69$ points (standard error $0.39$). The ordering
changes elsewhere: no spike is preferable for entirely omitted
signal at $p=30$, whereas projection is preferable for the weaker
entirely inside signal. Complete results and an infeasible
size-adjustment diagnostic appear in Supplementary
Section~\ref{app:u-v8-experiments}. These outcomes do not define a
data-selected spike rule.

Earlier calibration experiments isolate a separate gain. At
$p=30$, $N=80$ and $\Lambda_{\mathrm{all}}=16$, the joint reference
has power $57.39\%$, compared with $0.68\%$ for a shape-confidence
envelope and $59.65\%$ for a matched known-shape learner.
At larger dimensions, the shape cost persists: with $\eta=p$,
the inside-signal powers at $(p,N)=(300,40)$ and $(1000,80)$ trail
the matched known-shape learner by about $5.43$ and $2.68$ percentage
points. These comparisons establish neither oracle efficiency nor
uniform superiority. The diagonal assumption is also substantive:
correlations $0.7^{|j-k|}$ produce $8.88\%$ null rejection in an
archived nominal-$5\%$ diagnostic. These diagnostic data are simulated.

\subsection{Controlled shape, direction and spectrum comparisons}

The next experiment separates losses that the earlier split and reuse
comparisons changed simultaneously. It fixes $N=40$, $q=5$, $\rho=4$
and $p\in\{30,300\}$, with seven null or signal settings at each
dimension and $39$ prespecified procedures. The $2\times2$ comparison
varies true versus estimated paired diagonal covariance and fixed versus learned
axis while retaining every leave-pair-out product. Each version has
its own canonical null calibration using common reference data.
The true covariance and the correctly aligned fixed axis are infeasible
controls. Fixed axes are specified in population-whitened coordinates;
they are not claimed optimal conditional on estimated whitening.
Every cell uses $200$ independent banks of $L=199$ reference draws,
each shared by $50$ datasets. Bank-based uncertainty, all grid members,
raw features and binary outcomes are recorded in
Section~\ref{supp:v11ablation}.

\begin{table}[tbp]
\centering\small
\caption{Controlled ablations at $p=300$. Rejection percentages use
the same $10\,000$ raw datasets in each column. Inside 1 and Inside
$q$ have strength $16$, concentrated on the first coordinate and
diffuse within the supplied subspace, respectively. Mix .9 has
strength $64$ with $90\%$ diffuse omitted signal; Out 64 is entirely
diffuse outside. The displayed learned spikes use $\eta=\sqrt p$; no-spike rows use
$\eta=0$. Rank-only and classical subspace/all-coordinate scores
have no full baseline-plus-spike decomposition.
Known-shape and fixed-axis controls are diagnostics, not selectable
candidates. The largest bank MCSE across the full study is $1.06$
percentage points.}
\label{tab:u-v11-ablation}
\setlength{\tabcolsep}{4pt}
\begin{tabular}{lrrrrr}
\toprule
Method & Null & Inside 1 & Inside $q$ & Mix .9 & Out 64\\
\midrule
LPO, estimated shape & 4.97 & 72.97 & 74.96 & 47.34 & 15.33\\
LPO, known shape & 4.96 & 77.03 & 77.25 & 50.75 & 16.41\\
Estimated shape, fixed $e_1$ & 4.97 & 95.14 & 44.11 & 87.41 & 43.64\\
Known shape, fixed $e_1$ & 5.11 & 96.22 & 46.29 & 88.90 & 46.06\\
LPO, $q$ baseline & 5.14 & 74.88 & 76.63 & 29.54 & 5.14\\
Estimated shape, rank only & 5.06 & 73.36 & 75.11 & 27.90 & 5.06\\
LPO, full baseline, $c=0$ & 4.89 & 15.61 & 15.56 & 73.90 & 74.00\\
LPO, $q$ baseline, $c=0$ & 4.75 & 84.99 & 85.96 & 42.21 & 4.75\\
Estimated shape, LPO block & 4.82 & 83.33 & 84.24 & 60.67 & 16.16\\
Classical block $t^2$ & 4.90 & 83.86 & 84.72 & 61.41 & 16.44\\
Subspace diagonal $T^2$ & 4.87 & 85.29 & 86.11 & 42.77 & 4.87\\
All-coordinate $t^2$ & 4.90 & 15.49 & 15.54 & 74.37 & 74.11\\
\bottomrule
\end{tabular}

\end{table}

For the concentrated inside signal, replacing the fitted paired
diagonal covariance by its true value
in learned LPO raises power from $72.97\%$ to $77.03\%$; the paired
estimated-minus-known contrast is $-4.06$ points (MCSE $0.40$).
Replacing the learned axis by the correctly aligned fixed axis at
estimated shape gives $95.14\%$, a learned-minus-fixed contrast of
$-22.17$ points (MCSE $0.60$). These are controlled comparisons,
not additive components of a universal power-loss decomposition.

The scalar-spike grid provides a more direct diagnostic for tuning.
Among $c\in\{0,.25,.5,1,2,4\}$ with $\eta=c\sqrt p$ and a full
baseline, plus the rank-only limit, the largest reported inside power
is $73.59\%$. This is an infeasible, post-outcome grid maximum, not a
selected test or an upper bound over every possible spike. By contrast,
the prespecified no-spike $q$-baseline LPO score attains $84.99\%$,
and subspace diagonal $T^2$ attains $85.29\%$. Retaining the learned
rank term with a $q$ baseline gives only $74.88\%$. Thus restricting
the baseline alone does not close this particular gap while the
learned spike remains. For strength-$64$ diffuse omitted signal,
the original LPO score has $15.33\%$ power, whereas its no-spike
full-baseline version has $74.00\%$. The next experiment calibrates
a fixed family that includes these competing allocations.

\subsection{Selection without reserving validation observations}

A separately seeded experiment evaluates four menus fixed before
outcomes: seven full-baseline/rank-only LPO spectra; those seven plus
$q$-only and outside-only LPO baselines; four classical $q$, all,
outside and block sums; and the combined thirteen candidates.
The rank-only and coordinate-restricted endpoints extend the original
positive-definite weight class to positive-semidefinite weights.
They are not covered by its fixed-spectrum direction comparison.
The experiment retains all fourteen settings and uses $200$ banks,
$L=999$ and $50$ datasets per bank. All reference and observed
candidates are refitted, and all component ranks include the observed
row. Independent uniforms break outer-rank ties. Theorem~\ref{thm:u-menu}
gives exact $5\%$ level averaged over data and bank. No new observed
validation sample or pilot bank is used.
Section~\ref{sec:v11-adaptive-experiments} gives all candidates,
Bonferroni controls, selected-index diagnostics and timings.

\begin{table}[tbp]
\centering\small
\caption{Frozen-menu comparisons at $p=300$, $N=40$, $q=5$ and
$\rho=4$. Entries are rejection percentages. Inside 16 is concentrated
inside the subspace at strength $16$; Mix .9 has strength $64$ and
$90\%$ diffuse omitted signal; Outside 64 is entirely diffuse outside.
This experiment has a separate random stream and $L=999$; its entries
are not paired with Table~\ref{tab:u-v11-ablation}. All candidates
receive the same observations and structural information. The largest
bank MCSE over the complete experiment is $0.76$ percentage points.}
\label{tab:u-v11-menu-main}
\setlength{\tabcolsep}{5pt}
\begin{tabular}{lrrrr}
\toprule
Method & Null & Inside 16 & Mix .9 & Outside 64\\
\midrule
Fixed LPO, $c=1$ & 5.13 & 74.19 & 48.92 & 16.01\\
No-spike LPO, $q$ only & 5.12 & 85.31 & 42.46 & 5.12\\
Classical $q$ sum & 5.11 & 85.34 & 42.77 & 5.11\\
Classical all sum & 5.18 & 16.07 & 74.98 & 75.18\\
LPO menu, $K=7$ & 4.89 & 63.30 & 70.02 & 64.25\\
Expanded LPO, $K=9$ & 5.05 & 73.35 & 67.78 & 61.66\\
Classical menu, $K=4$ & 5.39 & 78.22 & 73.78 & 65.67\\
Combined menu, $K=13$ & 5.14 & 74.04 & 69.24 & 62.05\\
\bottomrule
\end{tabular}

\end{table}

The seven-candidate LPO menu raises wholly omitted-signal power from
$16.01\%$ to $64.25\%$, a paired gain of $48.24$ points (MCSE $0.60$).
It lowers concentrated inside-signal power from $74.19\%$ to $63.30\%$,
a difference of $-10.89$ points (MCSE $0.40$). Thus valid adaptation
reduces sensitivity to the particular spike choice but does not make
its cost negligible. Null rates of the four menus across both dimensions
range from $4.89\%$ to $5.39\%$.

Adding all thirteen candidates does not improve on the smaller
classical menu in these two alternatives. The paired combined-minus-
classical-menu differences are $-4.18$ points (MCSE $0.24$) inside
and $-3.62$ points (MCSE $0.25$) outside. Classical fallbacks therefore
provide a stronger benchmark than a poorly chosen fixed RI spike.
The reported empirical best candidate in each setting is retained only
as an infeasible diagnostic. Across all $560\,000$ menu evaluations, every corresponding
Bonferroni rejection implies a menu rejection, consistent with
the pathwise theorem; this is not dominance at the full nominal
component level. At $p=300$, the shared implementation takes about
$1.17$ seconds for a fresh $999$-dataset bank and $1.12$ milliseconds
for all candidate scores and menu ranks on one observed dataset.

\section{A leukemia application and data-based experiments}
\label{sec:v20-application}

\subsection{Scientific question and analysis design}

The expression study of \citet{golub1999molecular} compared acute
lymphoblastic leukemia (ALL) with acute myeloid leukemia (AML).
We ask whether the groups differ in their mean transformed expression
profile on each of three supplied biological sets. This is a
multivariate mean question, not a patient-classification exercise or a
claim about a causal treatment effect.
The original training cohort has 27 ALL and 11 AML observations;
the original test cohort has 20 ALL and 14 AML observations. Both
provide measurements on 7\,129 probes. We analyze the cohorts
separately, without tuning the scores after examining either result.
The historical test cohort includes both bone-marrow and peripheral-blood
specimens and differs in institutional composition. It therefore
provides a second analysis, not a controlled replication under an
identical sampling distribution.

We fixed the Hallmark G2M checkpoint, apoptosis and interferon-gamma
response sets before calculating test results. These represent cell-cycle,
cell-death and immune-response processes in the collection of
\citet{liberzon2015hallmark}. Probe mappings from the official
\texttt{hu6800.db} annotation were intersected with MSigDB 2024.1
Entrez identifiers. We retain all mapped probes, including multiple
probes for a gene, giving 156, 169 and 176 coordinates, respectively.
The target is therefore the mean vector on the measured probes,
without treating probes as independent biological replicates.
Expression intensities undergo the fixed elementwise transformation
$x\mapsto\log_{10}\{\min(16000,\max(100,x))\}$.
There is no expression-based gene screening or across-subject
standardization. Section~\ref{app:v20-application} documents the inputs and procedures.

All methods receive the same observations and the same pathway.
A frozen random seed selects 11 ALL observations for the 11 AML
observations in the training cohort, and 14 ALL observations for the
14 AML observations in the test cohort. Pairing is computational,
not biological matching; the remaining 16 and 6 ALL observations are
excluded from every compared method. Within each pathway we compare
the bounded quadratic, the sum of squared coordinate $t$ statistics,
the coordinate maximum, a paired LPO score and a split-projection
score. The LPO weight has $b=1$ and $\eta=\sqrt q$ and uses only
the $q$ supplied coordinates. All methods test the same pathway
null; no score can obtain significance from outside coordinates.

For the training cohort we enumerate all $2^{11}$ row-sign changes.
For the test cohort we use 1\,999 independent random sign changes
and the identity. Every transformed dataset repeats all variance
and direction fitting. We report conservative ranks, and apply Holm
adjustment across the three pathways separately for each method
and cohort. This does not authorize selecting the smallest
$p$-value across methods. A fixed variance floor specifies the scores
even when intensity clipping gives a zero fitted variance.

The sign-reference interpretation assumes central symmetry of the
independent paired differences under the null. Coordinate correlation
is retained, but this assumption is not verified by the expression
data or by a nonsignificant diagnostic. Accordingly, the application
is a model-based illustration of the competing scores. Its $p$-values
do not measure power or establish the diagonal efficiency bounds.

\subsection{Results in the two cohorts}

\begin{table}[tbp]
\centering\small
\caption{Leukemia pathway comparisons: Holm-adjusted sign-reference $p$-values. Adjustment is across the three pathways within each method and cohort.}\label{tab:v20-real}
\setlength{\tabcolsep}{4pt}
\begin{tabular}{llrrrrrr}
\toprule
Cohort & Set & $q$ & Bounded & Quadratic & Maximum & LPO & Projection\\
\midrule
Training & G2M & 156 & 0.0205 & 0.0156 & 0.0264 & 1.0000 & 0.0254\\
Training & Apoptosis & 169 & 0.0029 & 0.0029 & 0.0264 & 1.0000 & 0.0156\\
Training & IFN-$\gamma$ & 176 & 0.0029 & 0.0029 & 0.0264 & 0.9404 & 0.0059\\
\midrule
Test & G2M & 156 & 0.0110 & 0.0090 & 0.0560 & 0.6540 & 0.0385\\
Test & Apoptosis & 169 & 0.0030 & 0.0015 & 0.0030 & 0.6945 & 0.0210\\
Test & IFN-$\gamma$ & 176 & 0.0030 & 0.0030 & 0.0380 & 0.1665 & 0.0210\\
\bottomrule
\end{tabular}
\par\smallskip\begin{minipage}{.97\linewidth}\footnotesize
All methods use only the indicated pathway and the same 11 training-cohort or 14 test-cohort pairs. The references use all 2,048 signs or 1,999 random signs plus the identity, respectively. Projection uses a fixed training/testing split. Interpretation requires the symmetry assumption; smaller values do not establish higher power.
\end{minipage}
\end{table}

Table~\ref{tab:v20-real} reports the adjusted $p$-values. Bounded
aggregation, the quadratic sum and the split projection reject all
three pathway nulls at 5\% in both cohorts under the sign-reference
interpretation. The maximum also rejects all three in the training
cohort, but its adjusted G2M value is 0.056 in the test cohort. The
specified LPO implementation rejects none: its adjusted values range
from 0.1665 to 1. These results are retained even though they do not
favor the fitted LPO score.

The distinction is substantive for the motivating choice. Direct
use of the supplied set yields evidence of group differences without
estimating an axis, while one fitted procedure reaches different
conclusions on the same observations. Nevertheless, the split projection
also detects the differences, so this application cannot support a
general objection to learning directions. The comparison concerns
whole procedures, including variance fitting, the fixed floor,
weighting and calibration. Smaller $p$-values are not estimates of
larger power, and the historical cohort differences prevent a causal
or clinical interpretation. The complete unadjusted values, descriptive
dependence summaries and numerical-floor counts are reported in
Section~\ref{app:v20-application}.

\subsection{What can be checked with known injected signals}

To assess level and power separately, we freeze a noise model using
the within-class centered residuals of the original training cohort.
The panel contains the 156 G2M probes followed by 100 externally
annotated probes outside that set, selected by probe identifier rather
than observed expression. Thus $p=256$, $q=156$ and $N=11$ paired
differences. We use diagonal Gaussian, correlated Gaussian and
correlated $t_5$ noise. The correlated covariance retains 90\% of
the empirical paired-difference covariance and a diagonal remainder
that preserves its marginal variances, subject to the fixed variance floor;
the $t_5$ construction has the same covariance and central symmetry.
These are declared data-based generators, not the unknown true
expression distribution.

For each noise model, we consider zero signal, a single signal
coordinate inside the set, diffuse signal inside the set, and diffuse
signal outside it. Each nonzero mean obeys
$N\sum_{j=1}^p\delta_j^2/\Sigma_{jj}=16$; this is marginally
standardized energy, not Mahalanobis energy. All procedures now test
the global zero-mean null on the full panel. A pathway-only score is
valid for that null but can miss signal outside the supplied set.
The seven prespecified scores add a full-baseline LPO and a full-panel
quadratic to the five application scores.
Each of the 12 cells uses 500 independent datasets, with 199 random
sign changes plus the identity per dataset. Directions and variances
are refitted throughout. The shared datasets and signs permit paired
comparisons of rejection indicators.

\begin{table}[tbp]
\centering\small
\caption{Data-based rejection percentages at nominal 5\%, with $N=11$, $q=156$ and $p=256$. Each cell uses 500 independent datasets and 199 random signs plus the identity.}\label{tab:v20-semi}
\setlength{\tabcolsep}{3pt}
\begin{tabular}{llrrrrrrr}
\toprule
Noise & Signal & Bounded & Quad.$_q$ & Max.$_q$ & LPO$_q$ & LPO$_{\rm all}$ & Quad.$_{\rm all}$ & Proj.\\
\midrule
Diagonal Gaussian & Null & 2.6 & 3.8 & 3.8 & 4.0 & 4.0 & 4.4 & 7.0\\
 & Inside sparse & 8.4 & 15.6 & 20.8 & 12.4 & 12.6 & 12.2 & 7.0\\
 & Inside diffuse & 15.0 & 12.2 & 5.6 & 10.4 & 10.4 & 12.0 & 6.4\\
 & Outside diffuse & 2.6 & 3.8 & 3.8 & 4.0 & 5.8 & 10.4 & 7.0\\
\midrule
Correlated Gaussian & Null & 5.4 & 5.8 & 6.4 & 6.2 & 6.0 & 5.8 & 4.0\\
 & Inside sparse & 6.0 & 7.8 & 26.8 & 6.6 & 6.2 & 7.2 & 5.4\\
 & Inside diffuse & 10.4 & 10.0 & 6.8 & 8.2 & 8.0 & 7.8 & 5.6\\
 & Outside diffuse & 5.4 & 5.8 & 6.4 & 6.2 & 6.2 & 8.4 & 4.0\\
\midrule
Correlated $t_5$ & Null & 7.8 & 7.6 & 5.6 & 7.0 & 6.8 & 7.6 & 5.0\\
 & Inside sparse & 9.0 & 11.6 & 39.0 & 9.4 & 10.0 & 10.2 & 6.8\\
 & Inside diffuse & 12.4 & 13.0 & 8.6 & 10.4 & 10.4 & 11.4 & 6.6\\
 & Outside diffuse & 7.8 & 7.6 & 5.6 & 7.0 & 7.0 & 10.8 & 5.0\\
\bottomrule
\end{tabular}
\par\smallskip\begin{minipage}{.98\linewidth}\footnotesize
Maximum Monte Carlo standard error: 2.18 percentage points. Every nonzero signal has marginally standardized energy 16. All procedures test the global zero-mean null on the full panel; pathway-only scores can miss outside signal. The covariance model and its constant-probe variance floor are fixed before simulated rejection outcomes. These rates concern the declared generators, not the biological population.
\end{minipage}
\end{table}

Table~\ref{tab:v20-semi} separates the observed null rejection rates
from alternative rejection rates. The former range from 2.6\% to
7.8\%; 500 replications leave appreciable Monte Carlo uncertainty.
For example, the 7.8\% rate has standard error 1.20 percentage points.
Bonferroni--Clopper--Pearson intervals with 95\% simultaneous coverage
over the 21 null method--noise combinations all include 5\%
(Section~\ref{app:v20-application}). This is an uncertainty statement,
not an empirical proof of exact calibration. Level control follows
from sign invariance of the specified generators.

For the correlated Gaussian sparse signal, the maximum rejects in
26.8\% of datasets, compared with 6.2\% for the full-baseline LPO.
For the same sparse signal under $t_5$ noise, the corresponding rates
are 39.0\% and 10.0\%. Diffuse signals are difficult at this budget:
under correlated Gaussian noise, inside-signal rates are 10.4\% for
bounded aggregation and 8.0\% for full-baseline LPO, while the
full-panel quadratic reaches only 8.4\% for outside signal.
Pathway-only scores have exactly the same rejection indicators under
the coupled null and outside-only designs, because their input
coordinates are unchanged.

These results link the motivating data structure to a controlled
experiment. They support strong direct comparators and reveal limited
detection ability when only 11 pairs inform a large panel. They do not
establish a universal ordering, isolate an unavoidable direction-learning
cost, or solve fixed-small-sample efficiency. The full arrays and paired
Monte Carlo errors are retained, including the unfavorable outcomes.

\section{Discussion}
\label{sec:discussion}

The motivating application question is whether a supplied variable set
should be used directly or to fit a projection. The central comparison
is between the gain available from a population
axis and the power retained after learning the direction, choosing the
weight and calibrating the test. These quantities need not agree at a fixed
observation budget. The known-shape result isolates direction
substitution with spectrum and calibration held fixed. The paired
benchmarks then assess complete procedures against simple tests that
use the same supplied coordinate set.

The sparse results show that a fitted axis is not necessary to retain
most benchmark power in the studied designs. Bounded aggregation
provides a weak-signal comparison, while the maximum rule has
informative finite-sample bounds at moderate and strong signals and
does not require the signal amplitude. Some fitted RI rules and larger
menus lose appreciably more power. These findings motivate direct
structure-aware controls when evaluating a learner; they do not
establish an unavoidable cost shared by all direction-learning methods.
The separate split-statistic analysis explains why retaining a
baseline can help when diffuse signal is omitted from the learned
subspace.

Relative to the RI framework of \citet{jiang2024ri}, the
fixed-spectrum comparison gives a finite-sample power interpretation
of direction choice for the complete statistic. Relative to the
projection procedures of \citet{liu2024projection}, the benchmarks
ask how much power remains available after the entire fitted rule is
specified, and how much can be retained by direct use of the supplied
coordinate set. The complementary power-enhancement analysis of
\citet{yu2023power} expands high-power regions asymptotically.
Our absolute gap bounds instead quantify a restricted experiment at
its stated observation and reference sizes. This distinction in the
object of the guarantee is the methodological contribution; a broader
model or uniformly higher power has not been established.

The direct comparison on a common 60-coordinate panel reinforces this
distinction. The maximum rule is strong for a single-coordinate signal,
whereas the original Jiang RI procedure is strong for the same-sign
diffuse signal. The fitted LPO rule does not dominate either.
The original Jiang RI and power-enhanced mean calibrations both yield
null rejection rates above the nominal level in these designs, with a
larger discrepancy for the latter. Their reported powers therefore
compare procedures at a common nominal level, rather than at matched
attained sizes. Reporting size together with power is essential to
interpreting the external comparisons. The online projection procedure remains an
uncompleted numerical comparator.

The two simple-rule bounds are complementary. The bounded-quadratic
bound is useful near the null but can be wide for strong signals.
The conditional maximum bound remains informative in the evaluated
moderate and strong designs. This supplies quantitative comparisons
at the specified signal classes, rather than a theorem covering every
intermediate signal strength or an observed-data choice between
aggregation and maximum tests. The null-only certificates similarly
apply to specified fitted rules and design classes. They can assess a
design without searching alternative outcomes, but are not an
application-data tuning rule. The likelihood-ratio and local-optimality
principles are established; the added content is their explicit
finite-budget connection to actual power losses.

The main guarantee for unknown-shape split RI is finite-sample level
control. The separate shape-transfer and mixed-limit approximation
bounds identify error terms and provide asymptotic comparisons, but
their evaluated upper bounds can equal one at small training budgets.
They therefore give no useful uniform small-budget bound on the loss
from estimating shape. The maximum-to-LR result concerns a different
invariant experiment and does not resolve this shape-estimation
problem. The mixed-signal LR benchmark also prespecifies the allocation.
Our allocation-informed block follow-up gives competitors that same
information; its weights are explicit design choices, not solutions
of an optimization over all block weights.

The Gaussian and diagonal assumptions permit canonical regeneration of
the original null experiment. The application uses a different route:
row-sign randomization preserves correlation in paired differences,
under a central-symmetry assumption. This classical extension supplies
level control for consistently refitted scores; it does not transfer
the diagonal likelihood-ratio efficiency bounds or establish validity
under a general equal-means null.

In the leukemia cohorts, simple aggregation and the split projection
produce more consistent evidence across the three fixed pathways than
the specified LPO implementation. These are comparisons of complete
procedures, including clipping, variance regularization and calibration.
They do not identify the population power or an intrinsic cost of
estimating any direction. The original test cohort also differs in
specimen and institution mix. The data-based simulations complement
this illustration by fixing the noise law and injecting known signals,
so that attained level and power can be evaluated separately.
Their conclusions concern those generators and the fixed probe panel.

Efficient adaptation to an unknown sparse and diffuse allocation remains
unresolved. A supplied variable set gives a scientific reason to compare
aggregation with direction fitting, but does not establish Gaussianity,
diagonal covariance or central symmetry. The remaining methodological
question is to retain useful calibration and efficiency guarantees when
these structural conditions are weakened, without spending a large
additional sample on learning or validation.

\section*{Data and code availability}
The leukemia application uses public data from \citet{golub1999molecular}
and external Hallmark annotations. The reproducibility archive includes
the prepared inputs, raw-data snapshot, probe mapping, source URLs,
versions, licenses, hashes and a reconstruction script in
\texttt{data/golub}. Both raw cohorts were checked against the official
Bioconductor distribution. Ordinary analysis can run offline on the
bundled inputs. An optional download check revalidates the official
source snapshots.

The archive also contains generators, competitor implementations,
seeds, numerical arrays, environment requirements and manuscript
sources. \texttt{REPRODUCIBILITY\_MAP.md} maps tables and figures
to their inputs and scripts. The common \texttt{run\_workflow.py}
entry point separates archived-result checks, table reconstruction
and new numerical runs. The new application protocol, the documented
pre-outcome generator correction and all reported data-based cells
are retained. No pathway or pairing was selected for favorable results.

The conditional-maximum and affine bounds use deterministic quadrature,
with resolution checks distinguished from interval-arithmetic
certification. Earlier simulation archives are preserved. The current external-method comparison is a separate numerical study;
its source provenance, numerical conventions and frozen protocol are
provided with its results. Official code and independent
implementations are distinguished. The leukemia analysis and earlier
simulations are retained, rather than presented as new reruns. The complete reproducibility archive is available from the author
on request.

\clearpage
\begingroup
\small
\setlength{\bibsep}{4pt plus 1pt}
\interlinepenalty=10000
\bibliographystyle{plainnat}
\bibliography{references}
\endgroup

\clearpage
\appendix
\renewcommand{\thesection}{\AlphAlph{\value{section}}}
\newenvironment{vFifteenLowerHeadings}{%
  \let\section\subsection
  \let\subsection\subsubsection
  \let\subsubsection\paragraph
}{}

\section{Known-shape direction comparisons}
\label{app:v15-known-shape}
\subsection{Relation to projection, random integration and small-sample inference}
\label{sec:u-related}

The covariance-inverse signal direction is established in projection
testing \citep{huang2015optimal,liu2022multiple}.
\citet{liu2024projection} develop regularized direction estimation
with nonasymptotic $L_1$ and $L_2$ error bounds, an exact split-sample
$t$-test under normality, and asymptotic power guarantees. Its main
formulation is a one-sample problem, with a two-sample extension
discussed by the authors.
Regularized covariance-weighted quadratics adapt the baseline
\citep{li2020adaptable}, while anchored projections address classifier
degeneracy under the global null \citep{zhang2025adaptive}.
These methods establish the importance of learning and calibrating a
direction; our comparisons concern a complete covariance-subtracted
quadratic with a fixed whitened spectrum.

The orientation benchmark follows the stochastic-ordering framework
of \citet{mathew1997ordering}. Gaussian mean--Wishart independence
makes the subtraction law common across orientations, and independent
convolution preserves the classical order. Our self-contained coupling
records strictness and bounds the loss in actual rejection probability.
The population direction and this ordering principle are established
ingredients of the feasible analysis.

The two-sample RI construction of \citet{jiang2024ri} already uses
an unbiased statistic. For a fixed weight $\bm W$, its within-group
cross-products and between-group contrast give exactly
\[
 (\overline{\bm X}_1-\overline{\bm X}_2)^\top\bm W
 (\overline{\bm X}_1-\overline{\bm X}_2)
 -\frac{\operatorname{tr}(\bm W\bm S_1)}{n_1}
 -\frac{\operatorname{tr}(\bm W\bm S_2)}{n_2},
\]
which is \eqref{eq:setup-U}. Their nonnegative population RI contrast
must be distinguished from this sample statistic. Their Theorems~2--4
give asymptotic calibration and power under a general multivariate
model. We study finite-sample orientation comparisons for this
unbiased statistic under Gaussian common shape and a fixed whitened
spectrum, together with calibration of specified learned weights.
Equation~(2.5) of \citet{cao2024} uses a nonnegative sample-mean
quadratic; its null law differs from that of \eqref{eq:setup-U},
and their Theorem~2.6 gives local asymptotic power for a specified weight.
\citet{huang2024projection} add a projection term to a Euclidean
two-sample statistic, and \citet{wang2026bfprojection} study
Behrens--Fisher projection testing. \citet{jiang2025double} combine
RI and an extreme-type statistic by allocating significance levels.
That allocation differs from choosing the orientation of a fixed RI
weight spectrum.

Unknown covariance and small samples also have substantial precedents.
\citet{blanchard2023nonasymptotic} study nonasymptotic mean testing
and detection separation using unbiased quadratics. Its detailed
unknown-covariance implementation is a one-sample
result with sample-size and effective-dimension conditions; it does
not cover every fixed small two-sample budget.
\citet{li2023finite} allows fixed sample sizes with unknown, unequal
covariances in the two-sample problem, but obtains a $t$ limit as
$p\to\infty$ under latent-coordinate and spectral-trace assumptions.
\citet{dong2016shrinkage} address unstable diagonal variance
estimates at small sample sizes through shrinkage. Its two-sample
model has a common diagonal covariance, its shrinkage optimality
concerns estimation loss, and its test calibration is approximate.
These results distinguish nonasymptotic detection rates, fixed-sample
dimension asymptotics and variance-estimation efficiency.

Cross-statistics provide another use of independent splits.
\citet{kim2024cross} develop dimension-agnostic inference and
asymptotic minimax-rate power; \citet{shekhar2022cross} study
two-sample kernel testing. Gaussian limiting references are not the
only calibration available: Section~2.5 of \citet{kim2024cross}
also gives finite-sample-valid sign-flip and self-normalized variants
under null symmetry. The associated minimax-rate power guarantee
still uses increasing sample size. Finite-sample validity and
rate-optimal detection therefore need not imply negligible actual
power loss at a fixed small observation budget.

Our unknown-diagonal-shape constructions jointly simulate estimated
whitening, the specified learner and the testing residuals. The split
procedure uses a nuisance-free upper reference; the paired-difference
procedure gives an exact null reference for coordinatewise scale-invariant
scores. Invariance and Monte Carlo rank calibration are standard
\citep{dufour2006}. The upper-reference endpoint maximum can be
conservative, whereas an exact reference addresses level without
itself establishing high power. The matched known-shape comparator
uses the same observations and learning rule. A nontrivial uniform
power-loss guarantee for this comparison at fixed small budgets
remains unproved here. The finite-sample maximin benchmark developed
below instead specifies an axis-signal class and a class of invariant
tests; it uses the classical likelihood-ratio principle
\citep{neyman1933} to distinguish attainable power from an oracle
that knows the active coordinate. Power comparisons between Monte
Carlo tests and their ideal critical-value counterparts also have
a classical precedent in \citet{hope1968}.
The confidence-envelope extension follows the general nuisance-set
strategy of \citet{berger1994}; its fitted-family bounds and the
local general-baseline expansion have separate comparison classes
(Appendices~\ref{sec:u-shape-envelope-extension} and
\ref{sec:u-general}).

Adaptive testing by combining candidate statistics is also established.
\citet{xu2016adaptive} combines sum-of-powers tests for high-dimensional
two-sample means and studies their asymptotic properties. For a simulated
null reference, the componentwise ranks of the observed and simulated
statistic vectors can be reduced to their row minima and ranked again.
This is the extreme-rank Monte Carlo construction of
\citet{myllymaki2017global,mrkvicka2017combined}; the latter explicitly
allows a finite vector of one-sided component tests. Exchangeability
and randomized tie breaking follow the general Monte Carlo framework
of \citet{dufour2006}.
Our finite-candidate procedure applies this established construction
to jointly simulated diagonal-invariant scores. It removes the need
for an additional observational validation split within the stated
candidate family. The rank construction itself is not a new principle,
and its multiplicity-adjusted power bound does not establish closeness
to the best candidate tested at the full nominal level.

The classical multivariate mean problem and its high-dimensional
extensions have a long history. The $T^2$ statistic of
\citet{hotelling1931} supplies the benchmark when covariance estimation
is feasible. \citet{dempster1958} studied a two-sample significance test
in high dimension, and \citet{srivastava2013unequal} developed a
coordinate-scale-invariant test for unequal covariance matrices with
an asymptotic calibration. These results motivate the distinction
between covariance estimation, scale invariance and finite-sample
calibration in the present analysis.

Random projections provide another way to reduce the dimension before
testing. \citet{lopes2011random} combined random projection with
Hotelling's statistic and studied asymptotic power.
\citet{thulin2014subspaces} used random coordinate subspaces and
permutation calibration, whereas \citet{srivastava2016raptt}
combined random-projection tests under a Gaussian common-covariance
model, with its reference distribution evaluated by simulation.
Thus dimension reduction and exact Gaussian calibration are established
ideas. Our comparisons concern the complete RI quadratic, its fitted
rank-one component and its actual power under a specified calibration.
They do not replace the optimality or validity analyses of those
projection procedures.

Signal sparsity also affects the appropriate aggregation rule.
\citet{chen2019threshold} used thresholding and data transformation
for high-dimensional mean tests. In a broader testing framework,
\citet{fan2015power} combined an asymptotically pivotal statistic with
a screening component to improve power against sparse alternatives.
These approaches complement the adaptable tests discussed in the main
article. Our single-coordinate and fixed-allocation mixed classes give
restricted finite-sample comparison targets; they do not establish
adaptation over an unknown sparse/dense allocation.

Combining low-dimensional tests is itself well established.
\citet{frostig2022hotelling} combined random-subspace Hotelling tests
using Simes' procedure and established asymptotic validity. Calibration
assumptions remain essential: \citet{chung2013permutation} distinguish
finite-sample permutation exactness under equality of distributions
from asymptotic validity for a weaker parameter null. The present
Gaussian simulation reference regenerates the complete fitted rule
under a nuisance-free null law. It is not obtained by treating the
original heteroscedastic observations as exchangeable. The classical
likelihood-ratio and local-combination precedents cited in the main
article likewise remain the starting point for the stated restricted
efficiency bounds.

\subsection{What the finite-budget comparison adds}
\label{sec:u-v21-positioning}

The comparison target in this article is the actual rejection
probability of a fully specified procedure. This differs from an
estimation error, a limiting detection rate and a local efficiency
comparison. Each answers a useful question, but only the first directly
measures how much power is retained at the stated observation and
reference budgets. Table~\ref{tab:u-v21-positioning} records the relation
to the closest strands of work. The restrictions of the present
experiments are part of this comparison: stronger Gaussian and
invariance structure permits a more explicit finite-sample answer.

\begin{table}[tbp]
\centering
\caption{Established results and the finite-budget questions studied here.
The entries distinguish the objects of the guarantees; they do not
assert an ordering of the methods' power over general models.}
\label{tab:u-v21-positioning}
\small
\setlength{\tabcolsep}{4pt}
\begin{tabular}{p{0.19\linewidth}p{0.35\linewidth}p{0.39\linewidth}}
\toprule
Work & Established contribution & Present comparison \\
\midrule
\citet{jiang2024ri} & Complete unbiased two-sample RI statistic;
asymptotic calibration and power under a general multivariate model.
& Actual direction-wise power for that complete statistic under
Gaussian common shape and a fixed whitened spectrum. \\
\addlinespace
\citet{liu2024projection} & Nonasymptotic direction-estimation
bounds, an exact Gaussian split $t$-test, and online projection
with asymptotic calibration and power.
& An explicit oracle-margin condition and finite-budget power
comparisons with rules that use the same supplied coordinate set
without fitting an axis. \\
\addlinespace
\citet{yu2023power} & Power-enhanced mean and covariance
components, a simultaneous test, and asymptotic size, consistency
and efficiency results.
& Absolute power gaps to an attainable benchmark in a specified
sparse paired experiment. Unknown sparse/dense adaptation remains
a distinct problem. \\
\addlinespace
\citet{monti1976local} & Locally optimal combinations of
transformations of independent test statistics.
& Finite-amplitude bounds for a bounded quadratic, complementary
maximum-to-LR bounds, and finite-reference comparisons. \\
\bottomrule
\end{tabular}
\end{table}

Three distinctions are useful when interpreting these results.
First, the covariance correction in \eqref{eq:setup-U} already belongs
to the RI statistic of \citet{jiang2024ri}; the present direction
comparison retains it rather than introducing it. Second, exact
Gaussian projection calibration already appears in
\citet{liu2024projection}. The joint reference here is for the
specified fitted quadratic or menu and includes its shape estimates,
directions and testing residuals. Third, the attainable benchmark is
defined in a paired, diagonal, scale-invariant experiment. Its
finite-reference version further restricts row-wise rejection
allocations. The benchmark therefore does not optimize over all
tests based on the original unpaired samples.

The useful numerical content comes from connecting these targets to
simple procedures. The bounded-quadratic comparison controls a
finite-amplitude remainder, while the maximum comparison conditions
the full likelihood ratio on a reduced statistic. Its conditional
variance measures evidence discarded by that reduction, and the
distance to the rejection boundary determines whether that discarded
evidence can change the decision. These steps yield computable
absolute power bounds at the stated sample and reference sizes.
The likelihood-ratio, conditioning and Monte Carlo rank principles
themselves are classical.

The interpretation of external simulations requires an equally precise
comparison. A projection using our diagonal training estimate is a
matched structural control, not the online algorithm of
\citet{liu2024projection}. Likewise, the final simultaneous test of
\citet{yu2023power} tests equality of both means and covariance
matrices. Only its mean-test component addresses the mean null
considered here. Comparisons at a common nominal level are reported
together with attained null rejection rates, since an asymptotic
calibration and a finite-sample calibration can attain different sizes.

\begin{vFifteenLowerHeadings}
\section{A classical orientation benchmark for the unbiased statistic}
\label{sec:u-oracle}

We now retain the sample covariance terms in the unbiased two-sample
quadratic. Their distributions are part of the null calibration
and cannot be discarded. Under a common covariance shape, however,
their law is independent of orientation and independent of the mean
contrast. The comparison below therefore specializes classical
noncentral chi-square ordering \citep{mathew1997ordering} and uses
preservation of stochastic order under independent convolution.

Suppose the two testing samples are independent and
\begin{equation}\label{eq:u-model}
 \bm X_{ij}\sim N_p(\bm\mu_i,\sigma_i^2\bm H),
 \qquad j=1,\ldots,n_i,\quad i=1,2,
 \qquad n_i\ge2,
\end{equation}
where $\bm H\succ0$ and $\sigma_i^2>0$. The two scales may differ.
Let $\bm\delta=\bm\mu_1-\bm\mu_2$, and define
\[
 \bm D=\overline{\bm X}_1-\overline{\bm X}_2,
 \qquad
 \bm S_i=\frac1{n_i-1}\sum_{j=1}^{n_i}
  (\bm X_{ij}-\overline{\bm X}_i)
  (\bm X_{ij}-\overline{\bm X}_i)^\top.
\]
In this section the covariance parameters are treated as known for oracle
calibration. A procedure with unknown scales is developed separately.

\subsection{The unbiased statistic and its exact distribution}

Fix $b>0$ and $\eta>0$. For $\|\bm v\|_2=1$, put
\begin{equation}\label{eq:u-weight}
 \bm A_{\bm v}=b\bm I_p+\eta\bm v\bm v^\top,
 \qquad
 \bm W_{\bm v}=\bm H^{-1/2}\bm A_{\bm v}\bm H^{-1/2}.
\end{equation}
Equivalently, $\bm W_{\bm v}=b\bm H^{-1}+\eta\bm a\bm a^\top$
with $\bm a=\bm H^{-1/2}\bm v$ and
$\bm a^\top\bm H\bm a=1$. The statistic of interest is
\begin{align}
 U_{\bm v}
 &=\frac1{n_1(n_1-1)}\sum_{j\ne k}
        \bm X_{1j}^\top\bm W_{\bm v}\bm X_{1k}
   +\frac1{n_2(n_2-1)}\sum_{j\ne k}
        \bm X_{2j}^\top\bm W_{\bm v}\bm X_{2k}\notag\\
 &\hspace{1em}
   -\frac2{n_1n_2}\sum_{j=1}^{n_1}\sum_{k=1}^{n_2}
        \bm X_{1j}^\top\bm W_{\bm v}\bm X_{2k}\notag\\
 &=\bm D^\top\bm W_{\bm v}\bm D
     -\frac{\operatorname{tr}(\bm W_{\bm v}\bm S_1)}{n_1}
     -\frac{\operatorname{tr}(\bm W_{\bm v}\bm S_2)}{n_2}.
 \label{eq:u-statistic}
\end{align}
Thus $\mathbb E_{\bm\delta}U_{\bm v}
=\bm\delta^\top\bm W_{\bm v}\bm\delta$.
If a random integration vector has mean
$\sqrt\eta\,\bm a$ and covariance $b\bm H^{-1}$, integrating the
corresponding scalar two-sample unbiased statistic gives
\eqref{eq:u-statistic}. Both the positive-definite baseline and the
rank-one contribution are retained.

Write
\begin{equation}\label{eq:u-parameters}
 s^2=\frac{\sigma_1^2}{n_1}+\frac{\sigma_2^2}{n_2},
 \qquad
 \kappa_i=\frac{\sigma_i^2/n_i}{s^2},
 \qquad d_i=n_i-1,
 \qquad \kappa_1+\kappa_2=1,
\end{equation}
and set
\begin{equation}\label{eq:u-signal}
 \bm\theta=s^{-1}\bm H^{-1/2}\bm\delta,
 \qquad \Lambda=\|\bm\theta\|_2^2,
 \qquad r_{\bm v}=\Lambda-(\bm v^\top\bm\theta)^2.
\end{equation}
For $p=1$, chi-square variables with zero degrees of freedom and zero
noncentrality below are interpreted as identically zero.

\begin{proposition}[A signed chi-square reference]
\label{prop:u-reference}
Under \eqref{eq:u-model},
\begin{align}
 \frac{U_{\bm v}}{s^2}\ \overset d={}&
 (b+\eta)Y_1+bY_\perp
 -\sum_{i=1}^2\frac{\kappa_i}{d_i}
       \{(b+\eta)V_{i1}+bV_{i\perp}\},
 \label{eq:u-alternative-law}\\
 Y_1&\sim\chi^2_1(\Lambda-r_{\bm v}),
 &Y_\perp&\sim\chi^2_{p-1}(r_{\bm v}),\notag\\
 V_{i1}&\sim\chi^2_{d_i},
 &V_{i\perp}&\sim\chi^2_{(p-1)d_i},\notag
\end{align}
where all variables on the right are independent. Under the null,
the law is independent of $\bm v$. Denote this central signed law by
$G_0(\kappa_1;b,\eta,p,d_1,d_2)$. It is continuous, has support
$\mathbb R$, and its distribution function is strictly increasing.

The first two moments are
\begin{align}
 \mathbb E(U_{\bm v}/s^2)
 &=b\Lambda+\eta(\Lambda-r_{\bm v}),
 \label{eq:u-mean}\\
 \operatorname{Var}(U_{\bm v}/s^2)
 &=2\{(b+\eta)^2+(p-1)b^2\}
           \left(1+\sum_{i=1}^2\frac{\kappa_i^2}{d_i}\right)
   +4\{(b+\eta)^2(\Lambda-r_{\bm v})+b^2r_{\bm v}\}.
 \label{eq:u-variance}
\end{align}
\end{proposition}

The null reference includes the negative Wishart contributions through
the variables $V_{i1}$ and $V_{i\perp}$. In particular, centering the
positive mean quadratic by its population expectation would not give the
same distribution: the variance inflation in
\eqref{eq:u-variance} would be omitted. The reference also depends on
the scale ratio through $\kappa_1$, as well as on the testing block sizes.

Let $c^U_\alpha(\kappa_1)$ be the $(1-\alpha)$ quantile of $G_0$,
with the other arguments suppressed. For known covariance parameters,
the test
\begin{equation}\label{eq:u-oracle-test}
 U_{\bm v}/s^2>c^U_\alpha(\kappa_1)
\end{equation}
has size exactly $\alpha$. All directions in \eqref{eq:u-weight}
use the same null quantile. Define
\[
 \pi_U(\bm v;\bm\theta)
 =\mathbb P_{\bm\theta}
       \{U_{\bm v}/s^2>c^U_\alpha(\kappa_1)\}.
\]

\subsection{Classical power ordering and a coupling bound}

\begin{corollary}[Orientation ordering from noncentral chi-square comparison]
\label{thm:u-order}
Suppose \eqref{eq:u-model} holds and $\bm\delta\ne\bm0$.
If unit vectors $\bm v_1,\bm v_2$ satisfy
$r_{\bm v_1}\le r_{\bm v_2}$, then
\begin{equation}\label{eq:u-stochastic-order}
 \mathbb P_{\bm\theta}(U_{\bm v_1}/s^2>t)
 \ge\mathbb P_{\bm\theta}(U_{\bm v_2}/s^2>t)
 \quad\text{for every }t\in\mathbb R.
\end{equation}
If $r_{\bm v_1}<r_{\bm v_2}$, the inequality is strict for every
finite $t$. Consequently, the power of \eqref{eq:u-oracle-test} is
maximized over all unit directions precisely by
\begin{equation}\label{eq:u-optimal-direction}
 \bm v_* =\pm\frac{\bm H^{-1/2}\bm\delta}
                    {\|\bm H^{-1/2}\bm\delta\|_2},
 \qquad
 \bm a_* =\pm\frac{\bm H^{-1}\bm\delta}
              {(\bm\delta^\top\bm H^{-1}\bm\delta)^{1/2}}.
\end{equation}
The signs give the same rank-one weight. This conclusion holds for
every fixed $b>0$, $\eta>0$, $0<\alpha<1$, and all allowed
sample sizes and unequal scales.
\end{corollary}

The corollary concerns the unbiased two-sample RI statistic in
\eqref{eq:u-statistic}. The subtraction terms are included in both
the statistic and its null quantile. Their joint distribution is invariant
under rotation in the whitened coordinates and independent of the mean
contrast. They can therefore be held fixed in a coupling that reallocates
the noncentrality between the positive coefficients $b$ and $b+\eta$.
For example, if $Q_1\ge_{\mathrm{st}}Q_2$ and an independent
$R$ has the common subtraction law, then
$\Pr(Q_1-R>t)=\mathbb E\{\Pr(Q_1>t+R\mid R)\}
\ge\Pr(Q_2-R>t)$.
This explains why retaining the random subtraction changes the null
reference without requiring a new ordering principle. The proof in
Appendix~\ref{app:u-oracle-proofs} verifies strictness directly.

For $r\ge0$, define the bound
\[
 \mathcal B_{b,\eta}(r)
 =\min\left\{1-e^{-r/2},
       \sqrt{\frac{2\eta r}{\pi(b+\eta)}}\right\}.
\]

\begin{proposition}[A coupling bound for a misspecified axis]
\label{thm:u-regret}
Under the conditions of Corollary~\ref{thm:u-order}, let
$\pi_{U,*}=\pi_U(\bm v_*;\bm\theta)$. Every unit vector
$\bm v$ satisfies
\begin{equation}\label{eq:u-regret}
 0\le\pi_{U,*}-\pi_U(\bm v;\bm\theta)
 \le\mathcal B_{b,\eta}(r_{\bm v})
 \le\min\{1,r_{\bm v}/2\}.
\end{equation}
If $\phi_{\bm v}\in[0,\pi/2]$ is the angle between the
unoriented axes of $\bm v$ and $\bm\theta$, then
$r_{\bm v}=\Lambda\sin^2\phi_{\bm v}$. Equivalently, for a
nonzero direction $\bm a$ in the original coordinates,
\begin{equation}\label{eq:u-missed-signal}
 r_{\bm v}
 =\frac1{s^2}\left\{
  \bm\delta^\top\bm H^{-1}\bm\delta
  -\frac{(\bm a^\top\bm\delta)^2}{\bm a^\top\bm H\bm a}
  \right\},
 \qquad
 \bm v=\frac{\bm H^{1/2}\bm a}
                {(\bm a^\top\bm H\bm a)^{1/2}}.
\end{equation}
\end{proposition}

The first term in $\mathcal B_{b,\eta}$ is the probability of at
least one transferred Poisson increment. The second uses the
concentration function of a central $\chi^2_1$ component. It records
that, at fixed missed noncentrality, the effect of orientation vanishes
as $\eta/b\to0$. This is a quantitative refinement of the same
classical coupling, not a new general stochastic-order principle.

\begin{corollary}[An independently learned direction]
\label{cor:u-learned}
Let $\mathcal G$ be independent of both testing samples, and let
$\widehat{\bm v}$ be a $\mathcal G$-measurable unit vector.
For the oracle-calibrated test,
\begin{equation}\label{eq:u-learned-level}
 \mathbb P_{\bm0}\{U_{\widehat{\bm v}}/s^2
                  >c^U_\alpha(\kappa_1)\mid\mathcal G\}=\alpha
 \quad\text{almost surely}.
\end{equation}
At a nonzero alternative, with
$\widehat r=\Lambda-(\widehat{\bm v}^{\top}\bm\theta)^2$,
\begin{align}
 0&\le\pi_{U,*}
  -\mathbb P_{\bm\theta}\{U_{\widehat{\bm v}}/s^2
                              >c^U_\alpha(\kappa_1)\}\notag\\
 &\le\mathbb E\{\mathcal B_{b,\eta}(\widehat r)\}
 \le\mathcal B_{b,\eta}(\mathbb E\widehat r).
 \label{eq:u-learned-regret}
\end{align}
The corresponding conditional bound holds with
$\mathcal B_{b,\eta}(\widehat r)$ on the right. In a sequence of experiments,
$\widehat r=o_p(1)$ is sufficient for the unconditional power gap
to vanish.
\end{corollary}

The oracle and learned procedures here use the same testing sample sizes.
An independent training block provides one choice of $\mathcal G$;
its observations must be included in total-budget comparisons. The
corollary identifies the scale of direction estimation error needed for
power recovery. It does not imply that this scale is attainable without
restrictions on the signal or the training budget. Moreover, the exact
level statement still uses the true $s^2$ and $\kappa_1$. Substituting
estimated scales requires a separate argument.

\subsection{Fixed spectra and the scope of the result}

\begin{proposition}[Fixed-spectrum application of classical ordering]
\label{prop:u-fixed-spectrum}
Under \eqref{eq:u-model}, fix
$\tau_1\ge\cdots\ge\tau_p\ge0$ with $\tau_1>0$, and let
\[
 \bm W_{\bm O}=\bm H^{-1/2}\bm O
       \operatorname{diag}(\tau_1,\ldots,\tau_p)
       \bm O^\top\bm H^{-1/2},
 \qquad \bm O^\top\bm O=\bm I_p.
\]
Use this weight in the unbiased statistic \eqref{eq:u-statistic}.
All orientations have the same null distribution. An orientation whose
largest-eigenvalue eigenspace in whitened coordinates contains
$\bm H^{-1/2}\bm\delta$ stochastically dominates every member of
this class and therefore maximizes power at each common null quantile.
\end{proposition}

To make the connection explicit, put
$\xi_j=\{(\bm O^\top\bm\theta)_j\}^2$. The mean quadratic has
law $\sum_j\tau_j\chi_1^2(\xi_j)$, with
$\sum_j\xi_j=\Lambda$. The noncentrality vector
$(\Lambda,0,\ldots,0)$ has every leading partial sum at least as
large as that of $(\xi_1,\ldots,\xi_p)$. The classical
noncentral chi-square comparison therefore applies; the same
partial-sum criterion is used in a quadratic-test power analysis by
\citet{yajnik2020power}. The independent residual law then transfers
the ordering to $U$. A direct proof is retained in the appendix so
that this specialization does not depend on a change of notation or
noncentrality convention in another source.

The common-shape condition is what permits the entire mean--Wishart
reference distribution to be shared across directions. With arbitrary
unequal covariance shapes, whitening the covariance of the mean contrast
alone need not make the two sample covariance contributions rotationally
invariant. Likewise, a general fixed baseline $\bm B$ with a Euclidean
unit-direction constraint need not produce a fixed spectrum after
whitening. Corollary~\ref{thm:u-order} supplies the corresponding
finite-sample comparison for the specified full-rank family under
\eqref{eq:u-model}; it does not
optimize all positive-definite weights or all covariance models.

\section{Learning a direction in a specified subspace}
\label{sec:u-learning}

The power-loss bounds identify missed noncentrality as the relevant error
of a direction estimator. We now give an explicit learner and a
finite-sample risk calculation when a subspace is specified before the
data are observed. This structure can reduce the dimension of the
learning problem while retaining the full-rank RI statistic in the
testing sample.

Let $\bm P$ be a fixed orthogonal projector of rank
$q\in\{1,\ldots,p\}$ in the whitened coordinates. Its range represents
the directions in which the mean difference is expected to lie. Write
\begin{equation}\label{eq:u-learning-model}
 \bm d=\bm H^{-1/2}\bm\delta,\qquad
 \bm X_{\mathrm{tr}}=\bm H^{-1/2}\bm D_{\mathrm{tr}}
      \sim N_p(\bm d,s_{\mathrm{tr}}^2\bm I_p),\qquad
 s_{\mathrm{tr}}^2=\frac{\sigma_1^2}{m_1}
                         +\frac{\sigma_2^2}{m_2}.
\end{equation}
The testing scale remains
$s^2=\sigma_1^2/n_1+\sigma_2^2/n_2$. Define
\begin{equation}\label{eq:u-subspace-learner}
 \widehat{\bm v}_{\bm P}
 =\frac{\bm P\bm X_{\mathrm{tr}}}
         {\|\bm P\bm X_{\mathrm{tr}}\|_2}.
\end{equation}
The denominator is nonzero almost surely; a fixed unit vector in the
range of $\bm P$ may be assigned on the null event where it vanishes.
The rule needs no estimate of either scalar variance. It is measurable
with respect to the training means, as required by the calibration in
Section~\ref{sec:u-feasible}. The testing weight remains
$\bm H^{-1/2}(b\bm I_p+\eta\widehat{\bm v}_{\bm P}
\widehat{\bm v}_{\bm P}^{\top})\bm H^{-1/2}$; projecting the
learner does not discard the baseline contribution outside the subspace.

\begin{theorem}[Exact missed-noncentrality risk]
\label{thm:u-subspace-risk}
Under \eqref{eq:u-learning-model}, put
\begin{equation}\label{eq:u-subspace-signals}
 \Lambda_{\parallel}=\frac{\|\bm P\bm d\|_2^2}{s^2},\qquad
 \Lambda_{\perp}=\frac{\|(\bm I_p-\bm P)\bm d\|_2^2}{s^2},
 \qquad
 K\sim\operatorname{Poisson}
       \left(\frac{\|\bm P\bm d\|_2^2}{2s_{\mathrm{tr}}^2}\right).
\end{equation}
For
$\widehat r=\|\bm d\|_2^2/s^2
      -(\widehat{\bm v}_{\bm P}^{\top}\bm d)^2/s^2$,
\begin{equation}\label{eq:u-subspace-risk}
 \mathbb E\widehat r
 =\Lambda_{\perp}
   +\Lambda_{\parallel}(q-1)
                         \mathbb E\frac1{q+2K}.
\end{equation}
This identity also holds for $q=1$ and for $\bm P\bm d=\bm0$.
For a nonzero alternative, the power loss relative to the unrestricted
direction oracle of the same feasible procedure satisfies
\begin{equation}\label{eq:u-subspace-power}
 0\le\pi_{\mathrm U,*}-\pi_{\mathrm U}(\widehat{\bm v}_{\bm P})
 \le 1-\exp\left[-\frac12\left\{
    \Lambda_{\perp}
   +\Lambda_{\parallel}(q-1)\mathbb E\frac1{q+2K}
                  \right\}\right].
\end{equation}
Here both rejection probabilities include the independent reference
simulation if the Monte Carlo calibration is used, and the oracle uses
the same testing allocation, residual-scale estimator, and calibration
mechanism as the learned procedure.
\end{theorem}

The two terms in \eqref{eq:u-subspace-risk} have distinct meanings.
The first is the signal missed by the specified subspace. The second is
the cost of estimating an axis inside it. If $\bm d$ belongs to that
subspace, the second term depends on $q$ rather than the ambient
dimension $p$. If the subspace is misspecified, the first term records
the resulting approximation error. Choosing $\bm P=\bm I_p$ recovers
the full-dimensional mean learner. Neither the identity nor the bound
asserts that the right-hand side is observable: it is a performance
guarantee at the population signal strength.

The subspace restriction also gives a direct comparison of actual powers
when it contains the signal.

\begin{corollary}[Power improvement from a correctly specified subspace]
\label{cor:u-subspace-dominance}
Suppose $\bm d\ne\bm0$ and $\bm P\bm d=\bm d$. Let
$\widehat{\bm v}_{\bm I_p}=\bm X_{\mathrm{tr}}/
\|\bm X_{\mathrm{tr}}\|_2$ be the full-dimensional mean learner.
For either feasible calibration in Section~\ref{sec:u-feasible},
\begin{equation}\label{eq:u-subspace-dominance}
 \pi_{\mathrm U}(\widehat{\bm v}_{\bm P})
 \ge \pi_{\mathrm U}(\widehat{\bm v}_{\bm I_p}).
\end{equation}
Both procedures use the same training and testing allocations, the same
full $p$-dimensional weight spectrum, the same residual-scale estimator,
and the same reference calibration mechanism. Under the null they have
the same rejection probability.
\end{corollary}

This comparison does not follow merely by comparing the upper bounds in
\eqref{eq:u-subspace-power}. It uses a pointwise alignment coupling and
the actual-power ordering. Its assumption that the subspace contains the
signal is essential; the term $\Lambda_\perp$ records the cost when
that assumption fails.

The mean axis is also minimax for squared angular loss in the
fixed-radius Gaussian learning experiment
(Theorem~\ref{thm:u-angular-minimax},
Appendix~\ref{app:u-learning}). This is distinct from minimax testing
power; the upper bound in \eqref{eq:u-subspace-power} need not be sharp.

The subspace and its rank are fixed before the training and testing data
are examined. Selecting them from the same training contrast does not
preserve \eqref{eq:u-subspace-risk} without accounting for that selection.
Likewise, a small rank does not remove all learning cost at weak signal.
The result explains how a justified, specified direction class can reduce
that cost without estimating candidate powers on a validation sample.

\section{Proofs for the unbiased two-sample RI statistic}
\label{app:u-oracle-proofs}

\begin{proof}[Proof of Proposition~\ref{prop:u-reference}]
For a symmetric weight $\bm W$, expansion around the sample mean gives
\[
 \sum_{j\ne k}\bm X_{ij}^\top\bm W\bm X_{ik}
 =n_i(n_i-1)\overline{\bm X}_i^\top\bm W\overline{\bm X}_i
    -(n_i-1)\operatorname{tr}(\bm W\bm S_i).
\]
The cross-group sum divided by $n_1n_2$ is
$\overline{\bm X}_1^\top\bm W\overline{\bm X}_2$.
These identities prove the equality in \eqref{eq:u-statistic}.
Independence of distinct observations proves unbiasedness directly;
equivalently, subtracting the two covariance traces removes the
covariance contribution from the expectation of the mean quadratic.

Whiten the observations by $\bm H^{-1/2}$. Gaussian
mean--residual independence yields
\[
 \bm Z=s^{-1}\bm H^{-1/2}\bm D
       \sim N_p(\bm\theta,\bm I_p),
 \qquad
 \bm R_i=\frac{d_i}{\sigma_i^2}
             \bm H^{-1/2}\bm S_i\bm H^{-1/2}
       \sim\mathcal W_p(\bm I_p,d_i).
\]
The vector $\bm Z$ and the two Wishart matrices are mutually
independent. This representation remains valid when $d_i<p$ and the
Wishart matrices are singular. Indeed, each $\bm R_i$ can be
represented as
$\sum_{\ell=1}^{d_i}\bm G_{i\ell}\bm G_{i\ell}^\top$
with independent $N_p(\bm0,\bm I_p)$ vectors. It follows that
\begin{equation}\label{eq:proof-u-wishart}
 \frac{U_{\bm v}}{s^2}
 =\bm Z^\top\bm A_{\bm v}\bm Z
    -\sum_{i=1}^2\frac{\kappa_i}{d_i}
                    \operatorname{tr}(\bm A_{\bm v}\bm R_i)
 \quad\text{in distribution}.
\end{equation}

Choose an orthonormal basis with first vector $\bm v$. In that
basis, the coordinates of $\bm Z$ are independent with unit
variances. Their squared first-coordinate mean is
$\Lambda-r_{\bm v}$, and their remaining squared mean norm is
$r_{\bm v}$. The independent Wishart representations in the same
basis give
\[
 \operatorname{tr}(\bm A_{\bm v}\bm R_i)
 \overset d=(b+\eta)V_{i1}+bV_{i\perp},
 \quad V_{i1}\sim\chi^2_{d_i},\quad
 V_{i\perp}\sim\chi^2_{(p-1)d_i},
\]
with the displayed independence in the proposition. This proves
\eqref{eq:u-alternative-law}. Its null law is independent of the
chosen basis and hence of $\bm v$.

To verify the quantile assertions, write the central reference as
$Q_0-L_0$, where $Q_0$ is the positive central mean quadratic and
$L_0$ is the positive sum of residual terms. Each has an absolutely
continuous distribution on $(0,\infty)$, with strictly positive
probability in every nonempty open subinterval of $(0,\infty)$.
The two variables are independent. Their difference is absolutely
continuous and assigns strictly positive probability to every nonempty
open interval of $\mathbb R$: for an interval $(a,b)$, choose a
sufficiently large $\ell>0$ and small neighborhoods of $\ell$
and $\ell+(a+b)/2$ for $L_0$ and $Q_0$. This also proves that
the difference has full real support and a strictly increasing
distribution function. Its quantiles at levels in $(0,1)$ are therefore
finite and unique. The argument includes $p=1$; the zero-dimensional
terms are simply omitted.

Finally, the mean and variance of $\chi^2_d(\lambda)$ are
$d+\lambda$ and $2d+4\lambda$. Apply these formulas to the
independent terms in \eqref{eq:u-alternative-law}, and use
$\kappa_1+\kappa_2=1$, to obtain
\eqref{eq:u-mean}--\eqref{eq:u-variance}.
\end{proof}

For the remaining proofs we use the classical noncentral chi-square
Poisson representation to verify the required special case of the
ordering framework of \citet{mathew1997ordering}. We give the
construction explicitly, including the independent
residuals needed by the present statistic.
If $H\sim\chi^2_d$, $N\sim\operatorname{Poisson}(\lambda/2)$,
and $E_1,E_2,\ldots$ are independent $\chi^2_2$ variables,
all mutually independent, then
\begin{equation}\label{eq:u-poisson-representation}
 H+\sum_{j=1}^N E_j\ \overset d=\ \chi^2_d(\lambda).
\end{equation}
An empty sum is zero. The identity follows by multiplying the
chi-square moment-generating function by the compound Poisson
moment-generating function:
\[
 (1-2t)^{-d/2}
 \exp\left[\frac\lambda2\{(1-2t)^{-1}-1\}\right]
 =(1-2t)^{-d/2}\exp\left(\frac{\lambda t}{1-2t}\right),
 \qquad t<1/2.
\]

\begin{proof}[Proof of Corollary~\ref{thm:u-order}]
Put $w_1=b+\eta$, $w_0=b$, and write
$r_j=r_{\bm v_j}$ with $r_1\le r_2$.
Let $G_0$ have the central signed distribution in
Proposition~\ref{prop:u-reference}. Independently, take
\[
 N_A\sim\operatorname{Poisson}\{(\Lambda-r_2)/2\},
 \quad N_B\sim\operatorname{Poisson}(r_1/2),
 \quad N_H\sim\operatorname{Poisson}\{(r_2-r_1)/2\},
\]
and independent sequences $\{A_j\}$, $\{B_j\}$, $\{H_j\}$
of $\chi^2_2$ variables. Define
\begin{align*}
 \widetilde U_1
 &=G_0+w_1\left(\sum_{j=1}^{N_A}A_j+
                         \sum_{j=1}^{N_H}H_j\right)
                 +w_0\sum_{j=1}^{N_B}B_j,\\
 \widetilde U_2
 &=G_0+w_1\sum_{j=1}^{N_A}A_j
                 +w_0\left(\sum_{j=1}^{N_B}B_j+
                                   \sum_{j=1}^{N_H}H_j\right).
\end{align*}
The Poisson representation and addition of independent Poisson counts
give $\widetilde U_j\overset d=U_{\bm v_j}/s^2$. Their
central mean quadratic and both negative Wishart contributions are
shared through $G_0$. Moreover,
\[
 \widetilde U_1-\widetilde U_2
 =\eta\sum_{j=1}^{N_H}H_j\ge0
 \quad\text{almost surely}.
\]
This proves the stochastic ordering.

If $r_1<r_2$, the event $N_A=N_B=0$, $N_H=1$ has positive
probability. Conditional on this event and on $H_1=h>0$, the interval
$(t-w_1h,t-w_0h)$ is nonempty for every real $t$. The reference
$G_0$ assigns positive probability to this interval. Consequently,
$\mathbb P(\widetilde U_2\le t<\widetilde U_1)>0$ for every
finite $t$, proving strictness. Finally,
$r_{\bm v}\ge0$ with equality precisely when
$\bm v=\pm\bm\theta/\|\bm\theta\|_2$ by
Cauchy--Schwarz. The null critical value is finite, so strictness
identifies exactly these maximizing axes. Transforming by
$\bm a=\bm H^{-1/2}\bm v$ gives
\eqref{eq:u-optimal-direction}.
\end{proof}

\begin{proof}[Proof of Proposition~\ref{thm:u-regret}]
Use the preceding coupling with $r_1=0$ and
$r_2=r_{\bm v}$. The coupled variables are equal whenever
$N_H=0$, and the oracle variable is never smaller. At their common
critical value, their rejection indicators therefore differ by at most
$\bm1\{N_H>0\}$. Taking expectations gives
\[
 0\le\pi_{U,*}-\pi_U(\bm v;\bm\theta)
 \le\mathbb P(N_H>0)=1-e^{-r_{\bm v}/2}.
\]
For the second bound, put $r=r_{\bm v}$ and $w=b+\eta$.
Realize the common central signed variable as
$G_0=wX+B_0$, where $X\sim\chi^2_1$ is independent of $B_0$.
Here $B_0$ contains the perpendicular central mean quadratic and
the negative residual terms. With
$S=\sum_{j=1}^{N_H}H_j$ and with the shared noncentral increments
absorbed into $B$, the coupled variables are
\[
 \widetilde U_*=wX+B+wS,\qquad
 \widetilde U_{\bm v}=wX+B+bS.
\]
The variable $X$ is independent of $(B,S)$ and $\mathbb ES=r$.
Conditioning on $(B,S)$, the rejection indicators at any common
threshold can differ only when $X$ lies in an interval of length
$\eta S/w$. The density of $\chi^2_1$ is decreasing on its
positive support. Thus the probability of any real interval of length
$d\ge0$ is at most
\[
 F_{\chi^2_1}(d)=2\Phi(\sqrt d)-1
 \le\sqrt{2d/\pi}.
\]
Taking expectations and using Jensen's inequality gives
\[
 \pi_{U,*}-\pi_U(\bm v;\bm\theta)
 \le\sqrt{\frac{2\eta}{\pi(b+\eta)}}\,\mathbb E\sqrt S
 \le\sqrt{\frac{2\eta r}{\pi(b+\eta)}}.
\]
Combining the two bounds proves the assertion; the inequality
$1-e^{-x}\le\min\{1,x\}$ gives its final comparison.
The angle identity follows from its definition. Substituting the
normalized direction in \eqref{eq:u-missed-signal} into
$\Lambda-(\bm v^\top\bm\theta)^2$ proves that formula.
\end{proof}

\begin{proof}[Proof of Corollary~\ref{cor:u-learned}]
Conditionally on $\mathcal G$, the learned direction is fixed and
the two testing samples retain their distributions in
\eqref{eq:u-model}. The common null reference proves
\eqref{eq:u-learned-level}. Proposition~\ref{thm:u-regret} applies
conditionally and gives the bound
$\mathcal B_{b,\eta}(\widehat r)$. Taking expectations yields the
first unconditional bound. Bound the expectation of the minimum
by the expectation of each term separately, and apply Jensen's
inequality to $1-e^{-x/2}$ and $\sqrt x$, to obtain the second. If
$\widehat r=o_p(1)$, then $1-e^{-\widehat r/2}\to0$ in
probability and these variables lie in $[0,1]$. Their expectations
therefore tend to zero.
\end{proof}

\begin{proof}[Proof of Proposition~\ref{prop:u-fixed-spectrum}]
This is a direct specialization of weighted noncentral chi-square
ordering followed by independent convolution. The following coupling
also verifies the claim without invoking the general theorem.
Put $\lambda_j=\{(\bm O^\top\bm\theta)_j\}^2$.
By the same Gaussian and Wishart decomposition as before, the
standardized statistic has law
\[
 \sum_{j=1}^p\tau_j\left\{
    \chi^2_1(\lambda_j)
       -\sum_{i=1}^2\frac{\kappa_i}{d_i}\chi^2_{d_i}
                         \right\},
 \qquad \sum_{j=1}^p\lambda_j=\Lambda,
\]
where the central residual variables indexed by $(i,j)$ and all
mean variables are independent. Its central signed reference is
independent of $\bm O$. On one probability space, take this
reference $G_0$, independent
$N_j\sim\operatorname{Poisson}(\lambda_j/2)$, and independent
$E_{jk}\sim\chi^2_2$. The candidate statistic is represented by
\[
 G_0+\sum_{j=1}^p\tau_j\sum_{k=1}^{N_j}E_{jk}.
\]
Replacing every $\tau_j$ in the second term by $\tau_1$
increases this variable almost surely. The total Poisson count has
mean $\Lambda/2$, so the resulting law is exactly that of an
orientation assigning all noncentrality to a largest-eigenvalue
coordinate. Such an orientation exists by completing the standardized
mean direction to an orthonormal basis. This proves both the
stochastic-order and power assertions.
\end{proof}

\section{Proofs for structured direction learning}
\label{app:u-learning}

There is also a decision-theoretic justification for normalizing the
projected mean. The following result concerns angular loss in the
Gaussian learning experiment, not direct optimization of the final
test's power.

\begin{theorem}[Minimax angular risk of the mean axis]
\label{thm:u-angular-minimax}
Fix $q\ge1$, $\rho>0$, and $\tau>0$. In the experiment
$\bm X\sim N_q(\rho\bm u,\tau^2\bm I_q)$ with
$\bm u\in\mathbb S^{q-1}$, consider all measurable, possibly
randomized unit-vector estimators $\widetilde{\bm v}(\bm X)$ and the
loss $1-(\widetilde{\bm v}^{\top}\bm u)^2$. Then
\begin{align}
 \inf_{\widetilde{\bm v}}
 \sup_{\bm u\in\mathbb S^{q-1}}
 \mathbb E_{\bm u}\{1-(\widetilde{\bm v}^{\top}\bm u)^2\}
 &=(q-1)\mathbb E\frac1{q+2K_\rho},
 &K_\rho&\sim\operatorname{Poisson}
                          \left(\frac{\rho^2}{2\tau^2}\right).
 \label{eq:u-angular-minimax}
\end{align}
The estimator $\widehat{\bm v}=\bm X/\|\bm X\|_2$ attains this
value. The same estimator is minimax for every fixed positive
$\rho$ and $\tau$; its definition does not require their values.
\end{theorem}

The theorem follows from rotational invariance and a uniform-sphere
prior. It identifies an optimal learner for squared angular loss within
the specified model. Combining it with
Theorem~\ref{thm:u-subspace-risk} gives a power guarantee, but does not
make the learner minimax for actual rejection probability. The power
bound need not be sharp.

\begin{proof}[Proof of Theorem~\ref{thm:u-subspace-risk}]
Since $\widehat{\bm v}_{\bm P}$ lies in the range of $\bm P$,
orthogonality gives
\[
 \widehat r=\Lambda_{\perp}
 +\frac{\|\bm P\bm d\|_2^2
       -(\widehat{\bm v}_{\bm P}^{\top}\bm P\bm d)^2}{s^2}.
\]
If $\bm P\bm d=\bm0$, the second term is zero and the claimed
formula follows immediately. Suppose $\bm P\bm d\ne\bm0$ and
choose an orthonormal basis of the range of $\bm P$ with first axis
$\bm P\bm d/\|\bm P\bm d\|_2$. In these coordinates,
\[
 s_{\mathrm{tr}}^{-1}\bm P\bm X_{\mathrm{tr}}
 \overset d=
 \left(\frac{\|\bm P\bm d\|_2}{s_{\mathrm{tr}}}+G_1,
              G_2,\ldots,G_q\right)^\top,
\]
where the $G_j$ are independent standard normals. For $q\ge2$, the
squared sine of its angle with the first axis therefore has distribution
\[
 \frac{B}{A+B},\qquad
 A\sim\chi^2_1\left(\frac{\|\bm P\bm d\|_2^2}
                             {s_{\mathrm{tr}}^2}\right),\qquad
 B\sim\chi^2_{q-1},
\]
with independent $A$ and $B$. Under the Poisson representation,
conditionally on $K$ in \eqref{eq:u-subspace-signals}, the variable
$A$ is central chi-square with $1+2K$ degrees of freedom. Hence
\[
 \left.\frac{B}{A+B}\right|K
 \sim\operatorname{Beta}\left(\frac{q-1}{2},\frac{1+2K}{2}\right),
 \qquad
 \mathbb E\left[\left.\frac{B}{A+B}\right|K\right]
 =\frac{q-1}{q+2K}.
\]
Taking expectations proves \eqref{eq:u-subspace-risk}. For $q=1$,
the learner always selects the only axis in the subspace, so its
within-subspace angular error is zero and the same formula holds.

The learned-direction bound in
Corollary~\ref{thm:u-feasible-regret} gives
$0\le\pi_{\mathrm U,*}-\pi_{\mathrm U}(\widehat{\bm v}_{\bm P})
 \le\mathbb E(1-e^{-\widehat r/2})$.
Since $x\mapsto1-e^{-x/2}$ is concave on $[0,\infty)$, Jensen's
inequality and \eqref{eq:u-subspace-risk} give
\eqref{eq:u-subspace-power}. The reference simulation is independent
of the learning means, so its use does not alter this argument.
\end{proof}

\begin{proof}[Proof of Corollary~\ref{cor:u-subspace-dominance}]
Choose an orthonormal basis whose first vector is
$\bm d/\|\bm d\|_2$ and whose first $q$ vectors span the range of
$\bm P$. In the standardized training contrast, write
\[
 A=(\|\bm d\|_2/s_{\mathrm{tr}}+G_1)^2,\qquad
 V=\sum_{j=2}^{q}G_j^2,\qquad
 W=\sum_{j=q+1}^{p}G_j^2,
\]
where the $G_j$ are independent standard normals and empty sums are
zero. The squared alignments with the signal satisfy, almost surely,
\[
 \frac{(\widehat{\bm v}_{\bm P}^{\top}\bm d)^2}
            {\|\bm d\|_2^2}
 =\frac{A}{A+V}
 \ge\frac{A}{A+V+W}
 =\frac{(\widehat{\bm v}_{\bm I_p}^{\top}\bm d)^2}
            {\|\bm d\|_2^2}.
\]
Conditional on the training means, both selected directions are fixed.
Under the common-shape Gaussian model, the calibrated rejection
probability depends on the direction through this squared alignment
and is nondecreasing in it by Corollary~\ref{thm:u-feasible-regret}.
The preceding coupling therefore gives the conditional power ordering;
integration proves \eqref{eq:u-subspace-dominance}. Under the null,
the conditional reference law is independent of the selected direction,
so the null rejection probabilities are equal. For Monte Carlo
calibration these probabilities include its independent reference
simulation.
\end{proof}

\begin{proof}[Proof of Theorem~\ref{thm:u-angular-minimax}]
For $q=1$, every unit vector determines the same axis and the loss is
identically zero. Suppose $q\ge2$ and put a uniform prior on
$\bm u\in\mathbb S^{q-1}$. Conditional on a nonzero observation
$\bm X=\bm x$, the posterior density with respect to uniform surface
measure is proportional to
\[
 \exp(\kappa\bm e^\top\bm u),\qquad
 \bm e=\bm x/\|\bm x\|_2,\qquad
 \kappa=\rho\|\bm x\|_2/\tau^2.
\]
Invariance under orthogonal transformations fixing $\bm e$ implies
\begin{equation}\label{eq:u-learning-posterior-moment}
 \mathbb E(\bm u\bm u^\top\mid\bm X=\bm x)
 =a(\bm I_q-\bm e\bm e^\top)+c\bm e\bm e^\top,
 \qquad (q-1)a+c=1.
\end{equation}
To compare its eigenvalues, let $T$ denote the first coordinate of a
uniform unit vector. Symmetry gives
\[
 c=\frac{\mathbb E\{T^2\cosh(\kappa T)\}}
          {\mathbb E\{\cosh(\kappa T)\}}.
\]
Set $A=T^2$ and $g(A)=\cosh(\kappa\sqrt A)$. The function $g$ is
nondecreasing on $[0,1]$. If $A'$ is an independent copy of $A$, then
\[
 \operatorname{Cov}(A,g(A))
 =\tfrac12\mathbb E\{(A-A')[g(A)-g(A')]\}\ge0.
\]
Since $\mathbb E A=1/q$, it follows that $c\ge1/q$ and hence
$c\ge a$. The conditional expected loss of a unit direction $\bm v$
is
\[
 1-\bm v^\top
       \mathbb E(\bm u\bm u^\top\mid\bm X=\bm x)\bm v.
\]
It is minimized by $\bm v=\bm e$ in
\eqref{eq:u-learning-posterior-moment}. Randomization cannot reduce
the minimum, because conditional expected loss is linear in the
randomization distribution. The zero observation has probability zero
and can be assigned any axis. Thus $\bm X/\|\bm X\|_2$ is a Bayes
rule for the uniform prior.

Rotational invariance of the experiment and the loss makes the risk of
this rule constant over $\mathbb S^{q-1}$. The same noncentral
chi-square calculation as in the preceding proof gives that constant as
$(q-1)\mathbb E\{(q+2K_\rho)^{-1}\}$. Every estimator's maximum
risk is at least its average risk under the uniform prior, which is at
least the Bayes risk just obtained. The normalized-mean rule attains
this bound at every parameter value. It is therefore minimax, proving
\eqref{eq:u-angular-minimax}.
\end{proof}

\section{Proof of the learning-cost criterion}
\label{app:u-v12-learning-criterion}

\begin{proof}[Proof of Proposition~\ref{prop:u-v12-learning-margin}]
Fix the population signal, covariance scales and testing design, and
write $\psi(r)$ for actual rejection probability at missed
noncentrality $r\in[0,\Lambda]$. The chosen calibration is kept the
same for every direction. We first record the incremental version
of the coupling bound: for $0\le r_0\le r_1\le\Lambda$,
\begin{equation}\label{eq:u-v12-incremental-regret}
 0\le\psi(r_0)-\psi(r_1)
 \le\mathcal B_{b,\eta}(r_1-r_0).
\end{equation}

To verify it directly, put $\Delta=r_1-r_0$, $a=b+\eta$, and
take independent variables
\[
 Y\sim\chi_1^2(\Lambda-r_1),\qquad
 Z\sim\chi_{p-1}^2(r_0),\qquad
 K_\Delta\sim\operatorname{Poisson}(\Delta/2),\qquad
 V\mid K_\Delta\sim\chi_{2K_\Delta}^2.
\]
The convention for $K_\Delta=0$ is $V=0$. The two positive mean
quadratics can be coupled as
\[
 Q_{r_0}=a(Y+V)+bZ,\qquad
 Q_{r_1}=aY+b(Z+V),\qquad Q_{r_0}-Q_{r_1}=\eta V\ge0.
\]
They have the required marginal distributions, since addition of the
Poisson chi-square increment adds noncentrality $\Delta$.
Keep the entire subtraction and studentization variables common in
this coupling, independent of the positive mean variables. The
rejection event can then be written $Q_r>A+SC$, where $S>0$.
For the known-scale test $S=1$ and $C$ is its fixed cutoff. For
the endpoint procedures, $A$ and $S$ are the dependent subtraction
and residual-scale variables in \eqref{eq:u-feasible-null-law};
$C$ is either the fixed endpoint quantile or the order statistic of
one common independent reference bank. Thus this coupling also
preserves the finite-reference rejection rule. If its rank threshold
is zero, both rejection probabilities vanish and the result is
immediate.

The coupled decisions agree whenever $K_\Delta=0$, giving the bound
$1-e^{-\Delta/2}$. For the second bound, split $Y$ into an independent
central $\chi_1^2$ variable and its noncentral Poisson increments.
Condition on every variable except that central component. The
probability that $a\chi_1^2$ belongs to an interval of length $h$
is at most $\sqrt{2h/(\pi a)}$: its decreasing density makes the
interval starting at zero maximal, and
$\Pr(\chi_1^2\le h/a)\le\sqrt{2h/(\pi a)}$.
Take $h=\eta V$ and use $\mathbb E V=\Delta$ and Jensen's
inequality. This proves the second bound
$\sqrt{2\eta\Delta/\{\pi(b+\eta)\}}$ and hence
\eqref{eq:u-v12-incremental-regret}. In particular, the argument
requires no independence between the subtraction and studentizer.

The best axis constrained to the range of $\bm P$ has missed
noncentrality $r_0=\Lambda_\perp$, where
$\Lambda_\perp=\|(\bm I-\bm P)\bm d\|_2^2/s^2$.
Every learned axis in that range has $\widehat r\ge r_0$.
Independence of the training means and the Gaussian residuals permits
the preceding comparison conditional on the training means, for the
same reason as in Corollary~\ref{thm:u-feasible-regret}. The exact
risk calculation in Theorem~\ref{thm:u-subspace-risk} gives
\[
 \mathbb E(\widehat r-\Lambda_\perp)
 =\Lambda_\parallel(q-1)\mathbb E\frac1{q+2K}=R_{\bm P}.
\]
The function $\mathcal B_{b,\eta}$ is increasing and concave on
$[0,\infty)$, as the minimum of the two increasing concave functions
in its definition. Averaging \eqref{eq:u-v12-incremental-regret}
and applying Jensen's inequality proves
\eqref{eq:u-v12-learning-margin}. Subtracting $\pi_C$ proves
\eqref{eq:u-v12-learning-advantage}; no coupling with the comparator
is required. Its use of the same total budget is part of the stated
comparison, not a claim that its allocation or statistic is identical.
For $q=1$ the unoriented axis is already determined by the subspace.
If $\Lambda_\parallel=0$, every subspace axis has missed
noncentrality $\Lambda$, and the same conclusions hold with zero
learning risk.
\end{proof}

For interpreting the design condition, when $q\ge2$ and $\tau>0$,
the Poisson expectation also gives the simple bound
\begin{align*}
 \mathbb E\frac1{q+2K}
 &=\frac12\int_0^1 u^{q/2-1}e^{-\tau(1-u)/2}\,du
 \le\min\left\{\frac1q,\frac{1-e^{-\tau/2}}{\tau}\right\},\\
 R_{\bm P}
 &\le(q-1)\min\left\{
       \frac{\Lambda_\parallel}{q},\frac{s_{\rm tr}^2}{s^2}\right\}.
\end{align*}
With equal training sizes $m_1=m_2=m$ and testing sizes
$n_1=n_2=n$, the second ratio is $n/m$. Increasing the training
allocation can reduce this bound while reducing the testing
noncentrality. The oracle advantage in
\eqref{eq:u-v12-learning-advantage} must therefore be recomputed
for that allocation; it cannot be held fixed when claiming a
same-budget benefit.

\end{vFifteenLowerHeadings}

\section{Calibration and power with estimated shape}
\label{app:v15-shape}
\begin{vFifteenLowerHeadings}
\section{Feasible calibration of the unbiased RI statistic}
\label{sec:u-feasible}

The signed reference law in Section~\ref{sec:u-oracle} contains the
unknown relative variance contribution. We now remove this nuisance
parameter while retaining the covariance-subtraction terms of the
unbiased two-sample statistic. The construction uses an upper reference distribution
that can be simulated from independent chi-square variables. Simulation
replaces numerical evaluation of a reference distribution; it requires no
additional observations for estimating or comparing candidate powers.

Assume independent Gaussian observations
\[
 \bm X_{ij}\sim N_p(\bm\mu_i,\sigma_i^2\bm H),\qquad
 i=1,2,\quad j=1,\ldots,N_i,
\]
where \(p\ge2\), the positive-definite common shape \(\bm H\) is known,
and both positive scales are unknown. Split group \(i\) into training and
testing blocks of sizes \(m_i,n_i\ge2\), with \(N_i=m_i+n_i\).
The split is fixed independently of the observations. Let
\(\widehat{\bm v}\) be a unit vector measurable with respect to the two
training means, and put
\[
 \bm A(\bm v)=b\bm I_p+\eta\bm v\bm v^\top,\qquad
 \bm W(\bm v)=\bm H^{-1/2}\bm A(\bm v)\bm H^{-1/2},
 \qquad b>0,\quad\eta\ge0.
\]
The constants \(b,\eta\) are fixed in advance. Write
\(\bm D_{\mathrm{te}}\) for the testing mean difference and
\(\bm S_{i,\mathrm{te}}\) for the usual unbiased testing sample covariance.
The statistic before scale standardization is exactly
\begin{equation}
 U_{\bm v}
 =\bm D_{\mathrm{te}}^\top\bm W(\bm v)\bm D_{\mathrm{te}}
 -\sum_{i=1}^2\frac{\operatorname{tr}\{\bm W(\bm v)
                              \bm S_{i,\mathrm{te}}\}}{n_i}.
 \label{eq:u-feasible-U}
\end{equation}
The conditional expectation is
\[
 \mathbb E\{U_{\widehat{\bm v}}\mid\text{training means}\}
 =\bm\delta^\top\bm W(\widehat{\bm v})\bm\delta.
\]
Thus learning the weight does not alter the conditional unbiasedness
of the testing statistic.
The value \(\eta=0\) is included for the norm-quadratic benchmark;
in that case the direction has no effect on the statistic.

Both blocks contribute to the scalar variance estimates
\begin{equation}
 \begin{gathered}
 \widehat\sigma_i^2
 =\frac{1}{\nu_i}
 \sum_{a\in\{\mathrm{tr},\mathrm{te}\}}\ 
 \sum_{j\in\mathcal I_{i,a}}
 \left\|\bm H^{-1/2}
       (\bm X_{ij}-\overline{\bm X}_{i,a})\right\|_2^2,\\
 \nu_i=p(N_i-2).
 \end{gathered}
 \label{eq:u-feasible-scale}
\end{equation}
Define
\begin{equation}
 T_{\bm v}=\frac{U_{\bm v}}{\widehat s^2},\qquad
 \widehat s^2=\frac{\widehat\sigma_1^2}{n_1}
               +\frac{\widehat\sigma_2^2}{n_2},\qquad
 s^2=\frac{\sigma_1^2}{n_1}+\frac{\sigma_2^2}{n_2}.
 \label{eq:u-feasible-T}
\end{equation}
The testing residuals appear in both \(U_{\bm v}\) and
\(\widehat s^2\). Their dependence is retained in the following reference
construction; they are not treated as independent estimates.

\subsection{A reference distribution free of the variance ratio}

Put \(k_i=n_i-1\), \(\ell_i=m_i-1\), and
\[
 \kappa=\frac{\sigma_1^2/n_1}{s^2}\in(0,1).
\]
For each \(i=1,2\), generate independent variables
\begin{equation}
 \begin{gathered}
 E_i\sim\chi^2_{k_i},\quad F_i\sim\chi^2_{(p-1)k_i},\quad G_i\sim\chi^2_{p\ell_i},\\
 A_i=\frac{(b+\eta)E_i+bF_i}{k_i},\qquad V_i=\frac{E_i+F_i+G_i}{p(k_i+\ell_i)}.
 \end{gathered}
 \label{eq:u-feasible-reference-residuals}
\end{equation}
The triples are independent between groups. Independently generate
\begin{equation}
 Q_0=(b+\eta)Z_1^2+bZ_{p-1},
 \qquad Z_1\sim N(0,1),\qquad Z_{p-1}\sim\chi^2_{p-1},
 \label{eq:u-feasible-reference-Q}
\end{equation}
with independent terms. All quantities in
\eqref{eq:u-feasible-reference-residuals}--\eqref{eq:u-feasible-reference-Q}
have known distributions.

\begin{proposition}[A nuisance-free upper reference]
\label{prop:u-envelope}
Conditional on the training means and under \(H_0\), the statistic
\(T_{\widehat{\bm v}}\) has the distribution
\begin{equation}
 T_\kappa^0
 =\frac{Q_0-\kappa A_1-(1-\kappa)A_2}
        {\kappa V_1+(1-\kappa)V_2}.
 \label{eq:u-feasible-null-law}
\end{equation}
On the reference probability space, for every \(0\le\kappa\le1\),
\begin{equation}
 T_\kappa^0\le
 M:=\max\left\{\frac{Q_0-A_1}{V_1},
                \frac{Q_0-A_2}{V_2}\right\}
 \quad\text{almost surely}.
 \label{eq:u-feasible-envelope}
\end{equation}
The distribution of \(M\) is continuous and depends only on
\(p,m_i,n_i,b,\eta\). In particular, it depends on neither the unknown
scales nor the selected direction.
\end{proposition}

Let \(c_\alpha^M\) be the \((1-\alpha)\)-quantile of \(M\).
Proposition~\ref{prop:u-envelope} immediately gives a test of level at most
\(\alpha\) by rejecting when \(T_{\widehat{\bm v}}>c_\alpha^M\).
The endpoint maximum is a calibration device: the reported testing
statistic remains the unbiased RI statistic
\eqref{eq:u-feasible-U}. No maximum over candidate projection directions
is taken.

\subsection{Finite-reference Monte Carlo calibration}

A simulation-based rank calculation avoids treating an estimated
reference quantile as exact. Fix an integer \(L\ge1\), independently
generate \(M_1,\ldots,M_L\) from the distribution in
\eqref{eq:u-feasible-envelope}, and calculate
\begin{equation}
 \widehat p
 =\frac{1+\sum_{j=1}^L
              \bm1\{M_j\ge T_{\widehat{\bm v}}\}}{L+1}.
 \label{eq:u-feasible-MC-p}
\end{equation}
All reference draws are independent of the observations. The addition of
one is essential to the finite-reference rank argument, as in standard
Monte Carlo testing \citep{phipson2010,hemerik2018}.

\begin{theorem}[Finite-sample level with unknown unequal scales]
\label{thm:u-feasible-level}
Under the common-shape Gaussian model, the rule
\(\widehat p\le\alpha\) satisfies
\begin{equation}
 \Pr_{\bm0}\{\widehat p\le\alpha
             \mid\text{training means}\}
 \le \frac{\lfloor\alpha(L+1)\rfloor}{L+1}\le\alpha.
 \label{eq:u-feasible-MC-level}
\end{equation}
The probability includes the independent reference simulation. The bound
holds for every \(p,m_i,n_i\), every pair of positive unknown scales,
and every training-mean-measurable unit direction.
\end{theorem}

If \(h=\lfloor\alpha(L+1)\rfloor\ge1\), the rule is equivalently
\begin{equation}
 T_{\widehat{\bm v}}>M_{(L-h+1)},
 \label{eq:u-feasible-order-cutoff}
\end{equation}
where the reference sample is ordered increasingly. If \(h=0\), the rule
never rejects. For example, \(L=199\) and \(\alpha=0.05\) use the
190th order statistic of the reference sample. The reference sample size
is a computational choice, not an additional observational sample size.
The guarantee in \eqref{eq:u-feasible-MC-level} is not a conditional
guarantee for every realized reference sample. In particular, when one
reference sample is reused in a simulation study, its additional
calibration randomness must be distinguished from the conditional
Monte Carlo variation of the observed data.

If the weight is specified without direction-learning observations, the
same calibration applies with all \(N_i\) observations used for testing.
In that case set \(n_i=N_i\), \(\ell_i=0\), \(G_i=0\), and
\(\nu_i=p(N_i-1)\). There is no training block in this version.
The proof uses only the testing mean--residual decomposition and is
otherwise unchanged. In particular, this gives the all-data norm-RI
benchmark when \(\eta=0\).

\begin{proposition}[Fixed-dimension calibration limit]
\label{prop:u-feasible-calibration-limit}
Keep \(p,b,\eta,\alpha\) fixed and let
\(\min(n_1,n_2)\to\infty\). Then the ideal envelope-quantile rule has
null rejection probability tending to \(\alpha\). For fixed \(L\),
the null rejection probability of \eqref{eq:u-feasible-MC-p} tends to
\(\lfloor\alpha(L+1)\rfloor/(L+1)\).
These conclusions hold for any sequence of positive scales and
training block sizes satisfying the stated assumptions.
\end{proposition}

\subsection{Power ordering after feasible calibration}

The orientation comparison is inherited from the classical coupling.
More generally, if $Q_1\ge_{\mathrm{st}}Q_2$ are independent of a
common triple $(A,V,C)$ with $V>0$, then conditioning on that triple
orders the probabilities of $Q_j>A+VC$. Dependence among $A,V,C$
is allowed. Here $A$ and $V$ represent the jointly generated
subtraction and studentizer, while $C$ is the common ideal or
independently simulated cutoff. This is a transfer of the existing
ordering, not a separate optimal-direction principle.

For a nonzero alternative, put
\[
 \bm\theta=s^{-1}\bm H^{-1/2}\bm\delta,\qquad
 \Lambda=\|\bm\theta\|_2^2,\qquad
 \bm v_*=\bm\theta/\|\bm\theta\|_2,\qquad
 r_{\bm v}=\Lambda-(\bm v^\top\bm\theta)^2.
\]
Choose either the ideal envelope-quantile rule or the finite-reference
rule, and let \(\pi_{\mathrm U}(\bm v)\) denote its rejection
probability for a fixed direction. The latter probability includes
reference simulation when that rule is used. Define
\(\pi_{\mathrm U,*}=\pi_{\mathrm U}(\bm v_*)\).
This oracle has the same testing allocation, residual estimator,
augmentation strength, and calibration mechanism as the learned test.

\begin{corollary}[Inherited power ordering and learning loss]
\label{thm:u-feasible-regret}
For either calibration rule and every unit direction,
\begin{equation}
 0\le\pi_{\mathrm U,*}-\pi_{\mathrm U}(\bm v)
 \le\mathcal B_{b,\eta}(r_{\bm v}).
 \label{eq:u-feasible-fixed-regret}
\end{equation}
More generally, increasing \((\bm v^\top\bm\theta)^2\) cannot decrease
the calibrated power. If the direction is learned from the training
means, then
\begin{equation}
 \begin{split}
 0&\le\pi_{\mathrm U,*}-\Pr_{\bm\delta}
                     \{\text{reject using }\widehat{\bm v}\}\\
  &\le\mathbb E_{\bm\delta}
       \{\mathcal B_{b,\eta}(r_{\widehat{\bm v}})\}\\
  &\le\mathcal B_{b,\eta}
       (\mathbb E_{\bm\delta}r_{\widehat{\bm v}})
   \le\min\{1,\mathbb E_{\bm\delta}r_{\widehat{\bm v}}/2\}.
 \end{split}
 \label{eq:u-feasible-learned-regret}
\end{equation}
\end{corollary}

Thus the covariance-subtraction terms and their dependence on the
estimated denominator preserve the optimal population axis in this
model. The corollary supplies an error-to-power guarantee for any
learner based on the independent training means. It does not make the unknown direction
available, nor remove the observations allocated to learning it.
Conditions that make \(r_{\widehat{\bm v}}\) small are needed to obtain
an oracle-power approximation. The construction assumes a known common
covariance shape; replacing that shape by an arbitrary estimate is not
covered by the finite-sample calibration.

\section{Proofs for feasible calibration of the RI statistic}
\label{app:u-feasible}

\begin{proof}[Proof of Proposition~\ref{prop:u-envelope}]
Condition on the training means, so the selected unit direction is
fixed. After whitening by \(\bm H^{-1/2}\), rotate coordinates to make
this direction the first axis. Gaussian mean--residual independence
separates all four block means from the within-block residuals.
An orthogonal transformation in the testing observation index gives
\(k_i\) independent standard normal residual coordinates in the first
axis, and \((p-1)k_i\) in its orthogonal complement, after division by
\(\sigma_i\). Their squared sums are respectively \(E_i,F_i\) in
\eqref{eq:u-feasible-reference-residuals}. The independent training
residual sum is \(G_i\) with \(p\ell_i\) degrees of freedom.
Consequently,
\[
 \frac{\operatorname{tr}\{\bm W(\widehat{\bm v})
                          \bm S_{i,\mathrm{te}}\}}{\sigma_i^2}
 \ \overset d=A_i,\qquad
 \frac{\widehat\sigma_i^2}{\sigma_i^2}
 \ \overset d=V_i,
\]
jointly, with the dependence represented by their shared \(E_i,F_i\).
The group triples are mutually independent. They are also independent
of the standardized testing mean contrast, whose null law is
\(N_p(\bm0,\bm I_p)\). Its weighted squared norm therefore has law
\eqref{eq:u-feasible-reference-Q}. Dividing both numerator and
denominator of \eqref{eq:u-feasible-T} by \(s^2\) proves
\eqref{eq:u-feasible-null-law}. Rotation leaves the stated joint laws
unchanged, establishing their independence of the selected direction.

For any \(q\in\mathbb R\), \(a_1,a_2\in\mathbb R\),
\(v_1,v_2>0\), and \(0\le t\le1\), the algebraic identity
\[
 \frac{q-ta_1-(1-t)a_2}{tv_1+(1-t)v_2}
 =
 \frac{tv_1}{tv_1+(1-t)v_2}\frac{q-a_1}{v_1}
 +\frac{(1-t)v_2}{tv_1+(1-t)v_2}\frac{q-a_2}{v_2}
\]
expresses the left side as a convex combination of the two endpoints.
The variables \(V_i\) are positive almost surely, so the maximum of
these endpoints gives \eqref{eq:u-feasible-envelope}.

Conditional on \(A_1,A_2,V_1,V_2\), the envelope is a continuous,
strictly increasing function of \(Q_0\), the maximum of two affine
functions having positive slopes. Since \(Q_0\) has a continuous
distribution, this conditional distribution has no atoms. Integration
over the residual variables proves continuity of the marginal
distribution of \(M\). Its defining variables involve only the listed
known parameters.
\end{proof}

\begin{proof}[Proof of Theorem~\ref{thm:u-feasible-level}]
Continue to condition on the training means. Proposition~\ref{prop:u-envelope}
provides a coupling of the observed null statistic with a reference
variable \(M_0\) such that \(T_\kappa^0\le M_0\) almost surely.
Choose the simulated \(M_1,\ldots,M_L\) independently of this coupling.
For every fixed reference sample, the function
\[
 t\longmapsto
 \frac{1+\sum_{j=1}^L\bm1\{M_j\ge t\}}{L+1}
\]
is nonincreasing. Hence the rejection event obtained by substituting
\(T_\kappa^0\) is contained in the rejection event obtained by
substituting \(M_0\). The variables \(M_0,\ldots,M_L\) are independent
and identically distributed with a continuous law. The descending
rank of \(M_0\) among them is therefore uniform on
\(\{1,\ldots,L+1\}\). Exactly
\(\lfloor\alpha(L+1)\rfloor\) of these ranks yield a value at most
\(\alpha\), proving \eqref{eq:u-feasible-MC-level}.
This proof averages over the reference sample and does not assert
conditional validity given its realized values.

For the order-statistic representation, let
\(h=\lfloor\alpha(L+1)\rfloor\). In the absence of ties, the rank
inequality is equivalent to at most \(h-1\) reference values exceeding
the observed statistic, which is equivalent to
\(T_{\widehat{\bm v}}>M_{(L-h+1)}\) when \(h\ge1\).
When \(h=0\), no rank can satisfy the inequality.
\end{proof}

\begin{proof}[Proof of Proposition~\ref{prop:u-feasible-calibration-limit}]
Let \(a=bp+\eta\). The chi-square representation gives
\[
 \mathbb E A_i=a,\qquad
 \operatorname{Var}(A_i)
 =\frac{2\{(b+\eta)^2+(p-1)b^2\}}{k_i},
 \qquad
 \mathbb E V_i=1,\qquad
 \operatorname{Var}(V_i)=\frac{2}{p(k_i+\ell_i)}.
\]
Thus \(A_i\to a\) and \(V_i\to1\) in probability. Uniformly over
\(0\le\kappa\le1\), both reference endpoints and their convex combination
converge in distribution to \(Q_0-a\). This limit has a continuous
distribution and a positive density at every interior quantile.
The envelope quantile therefore converges to the corresponding quantile
of \(Q_0-a\), and the rejection probability of \(T_\kappa^0\) at that
threshold tends to \(\alpha\).

For fixed \(L\), the observed statistic and the \(L\) independent
reference variables converge jointly to \(L+1\) independent copies of
\(Q_0-a\). Ties have probability zero in the limit. The limiting
descending rank of the observed statistic is uniform, giving the stated
grid level. These arguments are uniform in \(\kappa\), so the unknown
scales may vary arbitrarily along the sequence.
\end{proof}

\begin{proof}[Proof of Corollary~\ref{thm:u-feasible-regret}]
Fix a direction and write \(r=r_{\bm v}\). Rotation of the standardized
testing contrast, independently of all residuals, gives the positive
quadratic law
\[
 Q_{\bm v}\ \overset d=
 (b+\eta)\chi^2_1(\Lambda-r)+b\chi^2_{p-1}(r).
\]
The residual pair \((A_i,V_i)\) has the joint distribution in
\eqref{eq:u-feasible-reference-residuals} under every alternative, and
does not depend on the direction after the same rotation. The
studentized statistic consequently has the distribution
\[
 \frac{Q_{\bm v}-\kappa A_1-(1-\kappa)A_2}
      {R},\qquad R=\kappa V_1+(1-\kappa)V_2,
\]
where the positive quadratic is independent of the residual variables.

The noncentral chi-square Poisson representation supplies a coupling
of \(Q_{\bm v}\) and the oracle positive quadratic \(Q_*\) with
\[
 Q_*-Q_{\bm v}
 =\eta\sum_{j=1}^{K_r}Y_j\ge0,\qquad
 K_r\sim\operatorname{Poisson}(r/2),\qquad
 Y_j\sim\chi^2_2.
\]
The positive quadratics are identical on \(\{K_r=0\}\).
For completeness, construct their common central part as
\((b+\eta)\chi^2_1+b\chi^2_{p-1}\), add common increments with rate
\((\Lambda-r)/2\) and coefficient \(b+\eta\), and give the increments
with rate \(r/2\) coefficient \(b\) in the candidate and \(b+\eta\)
in the oracle. All these quantities may be generated independently
of the joint residual pairs and of the reference simulation.

Use the same residual variables and the same reference cutoff for the
candidate and oracle. For the ideal quantile rule this cutoff is
constant. For the finite-reference rule it is the independent order
statistic in \eqref{eq:u-feasible-order-cutoff}; when \(h=0\), both
tests never reject and the result is immediate. Since \(R>0\),
the resulting oracle statistic is at least the candidate statistic,
and they coincide whenever \(K_r=0\). Thus the difference between
their rejection indicators is nonnegative and at most
\(\bm1\{K_r>0\}\). Taking expectations gives the zero-increment
bound \(1-e^{-r/2}\).

For the other term in \(\mathcal B_{b,\eta}\), leave the central
\((b+\eta)\chi^2_1\) component unconditioned and condition jointly
on all other mean components, both residual pairs, and the common
cutoff \(C\). Rejection is equivalent to the positive quadratic
exceeding \(\kappa A_1+(1-\kappa)A_2+RC\).
The interval argument in the proof of
Proposition~\ref{thm:u-regret} therefore applies unchanged, giving
\(\sqrt{2\eta r/\{\pi(b+\eta)\}}\).
This conditioning retains the dependence between each subtraction
term and its studentizer. It does not require level validity conditional
on the reference bank. If \(\eta=0\), the two statistics coincide
and both power gaps are zero. These observations prove
\eqref{eq:u-feasible-fixed-regret}. To compare two nonoracle directions,
transfer the difference between their noncentralities from coefficient
\(b\) to coefficient \(b+\eta\) in the same construction. This proves
the asserted ordering.

For a learned direction, condition on the training means. The testing
contrast, residual variables, and reference simulation retain the
specified joint laws. The deterministic comparison then applies with
\(r=r_{\widehat{\bm v}}\), while the conditional oracle power is
constant. Integration yields the first upper bound in
\eqref{eq:u-feasible-learned-regret}. Applying Jensen's inequality
separately to the two terms in \(\mathcal B_{b,\eta}\), as in
Corollary~\ref{cor:u-learned}, gives the next bound. The elementary
inequality \(1-e^{-x}\le\min\{1,x\}\) gives the final comparison.
\end{proof}

\end{vFifteenLowerHeadings}
\subsection{Power transfer for the invariant procedure}
\label{sec:u-invariant-power}

The null calibration in Theorem~\ref{thm:u-diag-invariant} does not by
itself quantify the power cost of estimating the shape. We now compare
it with a matched known-shape procedure, including the randomness of
the reference simulation. In this subsection, $\bm P$ projects onto a
fixed set of coordinates; in particular, $\bm P=\bm I_p$ is allowed.
This restriction permits uniform control of the fitted direction even
when the training signal is weak.

Normalize $\bm\Sigma_1=\bm H$ and $\bm\Sigma_2=\rho\bm H$ and put
\[
 s^2=n_1^{-1}+\rho n_2^{-1},\qquad
 s_{\mathrm{tr}}^2=m_1^{-1}+\rho m_2^{-1},\qquad
 \bm\theta=\bm H^{-1/2}\bm\delta/s,\qquad
 \Lambda=\|\bm\theta\|_2^2.
\]
The matched procedure replaces $\widehat{\bm H}$ by $\bm H$ in both
the training rule and the testing statistic. Thus its learned direction
is
\[
 \bm v_0=
 \frac{\bm P(\bm H^{-1/2}\bm\delta/s_{\mathrm{tr}}
                         +\bm Z_{\mathrm{tr}})}
      {\|\bm P(\bm H^{-1/2}\bm\delta/s_{\mathrm{tr}}
                         +\bm Z_{\mathrm{tr}})\|_2}.
\]
It uses the same testing-only studentizer as
\eqref{eq:u-shape-scale}, with $\widehat{\bm H}=\bm H$, and the
same rank rule with $L$ independent reference values obtained by
setting $\bm R=\bm I_p$ in
\eqref{eq:u-diag-reference-matrices}--\eqref{eq:u-diag-envelope}.
It still learns the direction and still uses the endpoint maximum.
It is therefore distinct from the direction oracle and from the
pooled-scale procedure of Section~\ref{sec:u-feasible}.
Write $\pi_{\mathrm{diag}}$ and $\pi_0$ for the actual rejection
probabilities of these two finite-reference procedures.

A finite-reference coupling bound is given in
Theorem~\ref{thm:u-invariant-power} of
Appendix~\ref{app:u-invariant-power}. It controls both the observed
statistic and all regenerated reference values, yielding the following
sufficient condition.

\begin{corollary}[A high-dimensional sufficient condition]
\label{cor:u-invariant-power}
Consider any sequence of the above models and designs with fixed
$L$ and $\alpha(L+1)\ge1$. If
\begin{equation}\label{eq:u-invariant-power-rate}
 \frac{\log(p+1)}{m_1-1}\longrightarrow0,
 \qquad \sup\Lambda<\infty,
 \qquad \sup\frac{\sqrt{pb^2+2b\eta+\eta^2}}{b+\eta}<\infty,
\end{equation}
then $|\pi_{\mathrm{diag}}-\pi_0|\to0$.
In particular, the last condition holds when $\eta=cpb$ for a fixed
$c>0$. Neither $p<N_1+N_2$ nor consistency of the learned direction
is required.
\end{corollary}

This result compares two procedures with the same sample split and
the same direction-learning task. It does not assert recovery of
oracle power under weak signals. It also does not provide an
efficiency guarantee when the training sample sizes remain fixed:
in that regime the shape error need not disappear.

The explicit bound is not a quantitative small-budget certificate.
For the reported $L=199$, $m_1\in\{20,40\}$ and $p\ge10$,
its truncated right-hand side equals one: already
$(L+1)pF_r(r/2)>1$ lower-bounds its error term for every allowed
perturbation tolerance. We therefore give a separate limit that keeps
the training and testing budgets fixed.

\begin{proposition}[A fixed-budget high-dimensional limit]
\label{prop:u-diag-fixed-budget-limit}
Fix $q,m_i,n_i,b,\rho,L$ and $\alpha$, with $q\ge1$, $m_i,n_i\ge2$,
$b,\rho>0$, $L\ge1$, $0<\alpha<1$ and $r=m_1-1>2$.
Let $p\to\infty$, set
$\bm P_p=\operatorname{diag}(\bm I_q,\bm0_{p-q})$ and $\eta=p$,
and suppose that the testing noncentrality vector is
$(\bm\theta_q^\top,\bm0^\top)^\top$ for a fixed
$\bm\theta_q\in\mathbb R^q$. The positive diagonal shapes may
otherwise vary with $p$. With the normalization above, define
\[
 \bm X=(s/s_{\mathrm{tr}})\bm\theta_q+\bm Z_{\mathrm{tr},q},
 \quad \bm a_R=
 \frac{\bm R_q^{-1}\bm X}
      {(\bm X^\top\bm R_q^{-1}\bm X)^{1/2}},
 \quad W_R=\|\bm a_R\|_2^2,
 \quad \mu_r=\frac r{r-2},
\]
where $\bm R_q=\operatorname{diag}(R_1,\ldots,R_q)$,
$R_j\sim\chi_r^2/r$, and $\bm Z_{\mathrm{tr},q}\sim
N_q(\bm0,\bm I_q)$ are independent. Let $Z\sim N(0,1)$ and
$E_i\sim\chi^2_{k_i}$ be independent of each other and of these
training variables. Then
\begin{equation}\label{eq:u-diag-fixed-budget-T}
 \frac{\widehat T_{\widehat{\bm v}}}{p}
 \ \Rightarrow\ T_\infty(\bm\theta_q)
 =\frac{W_R}{\mu_r}
 \left\{\left(Z+\frac{\bm a_R^\top\bm\theta_q}{\sqrt{W_R}}\right)^2
       -\kappa\frac{E_1}{k_1}-(1-\kappa)\frac{E_2}{k_2}\right\},
 \qquad \kappa=\frac{n_1^{-1}}{s^2}.
\end{equation}
Let $W_R^0$ have the same construction with
$\bm X=\bm Z_{\mathrm{tr},q}$. Each reference draw satisfies
\begin{equation}\label{eq:u-diag-fixed-budget-M}
 \frac{M_{\mathrm{diag}}}{p}\ \Rightarrow\ M_\infty
 =\frac{W_R^0}{\mu_r}
      \max_{i=1,2}\left\{Z^2-\frac{E_i}{k_i}\right\}.
\end{equation}
Here $W_R^0$ is independent of $Z,E_1,E_2$, and both endpoints use
the same $Z^2$. The observed statistic and its $L$ independent
references converge jointly to independent variables with the
specified limit laws. For $j=L+1-\lfloor\alpha(L+1)\rfloor\le L$, the actual
finite-reference power converges to
\begin{equation}\label{eq:u-diag-fixed-budget-power}
 \Pr\{T_\infty(\bm\theta_q)>M_{\infty,(j)}\},
\end{equation}
where $M_{\infty,(j)}$ is the $j$th ascending order statistic of
$L$ independent reference limits. Its null value is at most
$\lfloor\alpha(L+1)\rfloor/(L+1)$.
\end{proposition}

This limit reduces the power calculation to $q$ training coordinates,
one Gaussian mean coordinate and two chi-square residuals. The
baseline contribution vanishes after division by $p$, but the random
covariance subtraction remains because the testing sample sizes stay
fixed. For the matched known-shape procedure, replace $\bm R_q$ by
$\bm I_q$ and $\mu_r$ by one: then $W_R=1$ and
$\bm a_R=\bm X/\|\bm X\|_2$. The constant $\mu_r$ cancels from
the unknown-shape rank comparison, whereas the random scale $W_R$
and the distorted training direction remain. Even for $q=1$,
$W_R=R_1^{-1}$ is nondegenerate. Increasing the ambient dimension
alone therefore does not eliminate shape-learning randomness.
The result neither orders the two limiting powers nor establishes
superiority to a scalar projection test. It requires a fixed
coordinate subspace containing the signal and the specified
augmentation $\eta=p$.

\begin{vFifteenLowerHeadings}
\section{Proofs for the diagonal-shape invariant calibration}
\label{app:u-diag-invariant}

\begin{proof}[Proof of Theorem~\ref{thm:u-diag-invariant}]
We first express the entire fitted statistic through canonical mean
and residual variables, which removes the diagonal shape while
retaining the shared training randomness. We then write the remaining
scale dependence as a convex combination of two endpoint ratios.
The resulting common upper reference permits an exchangeable-rank
argument after averaging over training, testing and simulation.

Because the common shape is identifiable only up to scale, absorb
$\sigma_1^2$ into $\bm H$ and write
$\bm\Sigma_1=\bm H$ and $\bm\Sigma_2=\rho\bm H$, where
$\rho=\sigma_2^2/\sigma_1^2>0$. This reparameterization changes
neither the observations nor the fitted statistic. Put $r=m_1-1$.
The diagonal Gaussian model and the Wishart law of the first training
covariance give
\begin{equation}\label{eq:u-diag-proof-training}
 \widehat{\bm H}
 =\bm H^{1/2}\bm R\bm H^{1/2},\qquad
 \bm D_{\mathrm{tr}}
 =\sqrt{m_1^{-1}+\rho m_2^{-1}}\,
                         \bm H^{1/2}\bm Z_{\mathrm{tr}}
\end{equation}
in joint distribution under the null, with the independent variables
specified in \eqref{eq:u-diag-reference-inputs}. In particular, the
$R_j$ are mutually independent because the population coordinates are
independent. The sample means are independent of the residual
covariances in a Gaussian sample; hence $\bm Z_{\mathrm{tr}}$ is
independent of $\bm R$ even though both use the first training block.
All testing variables are independent of this pair.

Since $\bm H$ and $\bm R$ are positive diagonal matrices,
\begin{equation}\label{eq:u-diag-proof-cancellation}
 \widehat{\bm H}^{-1/2}\bm D_{\mathrm{tr}}
 =\sqrt{m_1^{-1}+\rho m_2^{-1}}\,
                       \bm R^{-1/2}\bm Z_{\mathrm{tr}}.
\end{equation}
The positive scalar cancels on normalizing after projection. Thus
$\widehat{\bm v}=\bm v_R$ on this probability space. No commutation
between $\bm P$ and either diagonal matrix is required. Conditional on
$\bm R$, the projected normal vector has a positive-definite
covariance on the rank-$q$ range of $\bm P$, so its norm is positive
almost surely. Also $R_j>0$ almost surely for every $r\ge1$.

Write $s^2=n_1^{-1}+\rho n_2^{-1}$,
$a_1=1$, $a_2=\rho$, and $k_i=n_i-1$. Gaussian
mean--residual independence represents the testing contrast and sample
covariances jointly as
\begin{equation}\label{eq:u-diag-proof-testing}
 \bm D_{\mathrm{te}}=s\bm H^{1/2}\bm Z_0,\qquad
 \bm S_{i,\mathrm{te}}
   =a_i\bm H^{1/2}\bm\Psi_i\bm H^{1/2},\qquad
 \bm\Psi_i=\bm G_i^\top\bm G_i/k_i,
\end{equation}
with $\bm Z_0,\bm G_1,\bm G_2$ mutually independent and independent
of the training pair. Cyclicity of trace and diagonal commutation now
give
\[
 \widehat U_{\widehat{\bm v}}
 =s^2Q_R-\frac{A_{R1}}{n_1}-\frac{\rho A_{R2}}{n_2},
 \qquad
 \widehat s_{\mathrm{te}}^2
 =\frac{V_{R1}}{n_1}+\frac{\rho V_{R2}}{n_2}.
\]
Dividing both quantities by $s^2$ proves
\eqref{eq:u-diag-null-representation}, with
$\kappa=n_1^{-1}/s^2$. Each $V_{Ri}$ is positive almost surely.
Set
\[
 w_\kappa
 =\frac{\kappa V_{R1}}
        {\kappa V_{R1}+(1-\kappa)V_{R2}}\in[0,1].
\]
Then the exact identity
\begin{equation}\label{eq:u-diag-proof-endpoints}
 T_\kappa
 =w_\kappa\frac{Q_R-A_{R1}}{V_{R1}}
  +(1-w_\kappa)\frac{Q_R-A_{R2}}{V_{R2}}
 \le M_{\mathrm{diag}}
\end{equation}
holds for all $\kappa\in[0,1]$, including when either endpoint
numerator is negative. This establishes a common null coupling for
every variance ratio. The reference variables contain no unknown
covariance parameters, common mean, or $m_2$.

Conditional on $\bm R,\bm Z_{\mathrm{tr}},\bm G_1,\bm G_2$,
$\bm B_R\succ0$, and $Q_R$ has a continuous distribution. The map
\[
 x\longmapsto\max_{i=1,2}(x-A_{Ri})/V_{Ri}
\]
is continuous and strictly increasing, because both slopes are
positive. The conditional distribution of $M_{\mathrm{diag}}$ is
therefore atomless, and so is its unconditional distribution. No
inverse moment of $\bm R$ is needed for this argument.

Finally, couple the observed statistic with a value
$M_{\mathrm{diag}}^{(0)}$ as in
\eqref{eq:u-diag-proof-endpoints}, independently of the $L$
reference draws. These $L+1$ reference values are independent and
identically distributed, and have no ties almost surely. Since
$\widehat T_{\widehat{\bm v}}\le M_{\mathrm{diag}}^{(0)}$ in this
coupling,
\[
 \widehat p_{\mathrm{diag}}
 \ge\frac{1+\sum_{j=1}^L\bm1\{
        M_{\mathrm{diag}}^{(j)}\ge
                         M_{\mathrm{diag}}^{(0)}\}}{L+1}.
\]
By exchangeability, the numerator on the right is uniform on
$\{1,\ldots,L+1\}$. The rejection event is therefore contained in
an event of probability
$\lfloor\alpha(L+1)\rfloor/(L+1)$. This proves
\eqref{eq:u-diag-level} uniformly in the nuisance parameters.
All training variables were integrated out in this rank argument;
it makes no conditional-on-training level assertion.
\end{proof}

\begin{proof}[Proof of Proposition~\ref{prop:u-diag-ideal-dominance}]
Use the canonical variables of
\eqref{eq:u-diag-reference-inputs}--\eqref{eq:u-diag-reference-matrices}.
The diagonal confidence construction gives
\[
 \mathcal E_R=\{\ell\bm I_p\preceq\bm R^{-1}
                                      \preceq h\bm I_p\},
 \qquad \Pr(\mathcal E_R)\ge1-\gamma.
\]
This is the event \eqref{eq:u-shape-confidence} after the normalization
$\bm H=\bm\Sigma_1$ and the representation
\eqref{eq:u-diag-proof-training}. Condition on
$\bm R,\bm Z_{\mathrm{tr}}$, and choose a measurable orthogonal
eigenbasis $\bm O$ of $\bm B_R$, with eigenvalues
$\lambda_1\ge\cdots\ge\lambda_p$. Write
$a_1=b+\eta$ and $a_2=\cdots=a_p=b$. On $\mathcal E_R$,
\begin{equation}\label{eq:u-diag-comparison-eigenvalues}
 \ell a_j\le\lambda_j\le h a_j,\qquad j=1,\ldots,p.
\end{equation}
Indeed, $\bm B_R$ has the same eigenvalues as
$(b\bm I_p+\eta\bm v_R\bm v_R^\top)^{1/2}\bm R^{-1}
 (b\bm I_p+\eta\bm v_R\bm v_R^\top)^{1/2}$, which lies
between $\ell(b\bm I_p+\eta\bm v_R\bm v_R^\top)$ and
$h(b\bm I_p+\eta\bm v_R\bm v_R^\top)$ in Loewner order.

Put $\bm\zeta_0=\bm O^\top\bm Z_0$ and
$\bm\zeta_{it}=\bm O^\top\bm g_{it}$, where
$\bm g_{it}^\top$ is row $t$ of $\bm G_i$. Their conditional law is
that of independent standard normal vectors and does not depend on
$\bm R,\bm Z_{\mathrm{tr}}$. Thus this whole rotated collection is
independent of the training variables and of $\mathcal E_R$. Define
\[
 Q_0=\sum_{j=1}^p a_j\zeta_{0j}^2,\qquad
 A_i^0=\frac1{k_i}\sum_{t=1}^{k_i}\sum_{j=1}^p
                                     a_j\zeta_{itj}^2,\qquad
 V_i^0=\frac1{pk_i}\sum_{t=1}^{k_i}\|\bm\zeta_{it}\|_2^2.
\]
These variables have exactly the joint law in
\eqref{eq:u-shape-reference}. On $\mathcal E_R$,
\eqref{eq:u-diag-comparison-eigenvalues} and the bounds on
$\bm R^{-1}$ give, simultaneously,
\[
 Q_R\le hQ_0,\qquad A_{Ri}\ge\ell A_i^0,\qquad
 \ell V_i^0\le V_{Ri}\le hV_i^0.
\]
Apply the two branches of $g_{\ell,h}$ in
\eqref{eq:u-shape-envelope} separately according to the sign of
$hQ_0-\ell A_i^0$. Since $g_{\ell,h}$ is increasing,
\[
 M_{\mathrm{diag}}
 \le\max_{i=1,2}g_{\ell,h}
         \left(\frac{hQ_0-\ell A_i^0}{V_i^0}\right)
 =g_{\ell,h}\left(\max_{i=1,2}
         \frac{hQ_0-\ell A_i^0}{V_i^0}\right)
 =M_{\ell,h}
 \quad\hbox{on }\mathcal E_R.
\]
In this coupling, $M_{\ell,h}$ is independent of $\mathcal E_R$,
because the endpoints are deterministic and its Gaussian inputs have
a constant conditional law. Writing $\gamma_R=\Pr(\mathcal E_R^c)
\le\gamma$, we obtain, for every real $t$,
\begin{equation}\label{eq:u-diag-tail-comparison}
 \begin{split}
 \Pr(M_{\mathrm{diag}}>t)
 &\le\gamma_R+(1-\gamma_R)\Pr(M_{\ell,h}>t)\\
 &\le\gamma+(1-\gamma)\Pr(M_{\ell,h}>t).
 \end{split}
\end{equation}
Continuity of the confidence-envelope reference implies
$\Pr(M_{\ell,h}>c_{\mathrm{env}})=\beta$. Hence the last bound
at $t=c_{\mathrm{env}}$ is
$\gamma+(1-\gamma)\beta=\alpha$, which proves
\eqref{eq:u-diag-ideal-dominance}. Finally, the event
$\{\widehat T_{\widehat{\bm v}}>c_{\mathrm{env}}\}$ is contained in
$\{\widehat T_{\widehat{\bm v}}>c_{\mathrm{diag}}\}$ for every
realization of the data. This deterministic inclusion proves the power
comparison under each alternative without a distributional comparison
between the two fitted statistics.
\end{proof}

\section{Proofs for invariant power transfer}
\label{app:u-invariant-power}

The comparison must control two changes caused by estimating shape:
the observed statistic changes, and so does its simulated critical
value. We use the same Gaussian observations and residuals for the
fitted and known-shape procedures. On an event where the shape error
is small and the studentizers are bounded away from zero, their
statistics are close. The same coupling applies to every reference
draw, so the order-statistic critical values are also close. Rejection
can then change only when the known-shape statistic lies near its
critical value. An anti-concentration bound controls this last event.

The constants below correspond to these three steps.
The quantities $h_\varepsilon,a_\varepsilon,f_\varepsilon$ control
the perturbation of the fitted weight; $B$ converts matrix-norm
bounds into numerator bounds; and $e(\Lambda),e(0)$ control the
observed and reference statistics, respectively. The probabilities
$\beta_R,\beta_V$ account for large shape errors and small
studentizers. This division also explains why the final bound contains
both a boundary-neighborhood term and a failure-probability term.

Here is an explicit comparison. Write $k=\min(n_1-1,n_2-1)$ and
\[
 a_0=b+\eta,\qquad
 f_0=(pb^2+2b\eta+\eta^2)^{1/2}.
\]
For $0<\varepsilon<1$, define
\[
 h_\varepsilon=(\sqrt{1+\varepsilon}+1)
       \left\{\frac{1+\varepsilon}{\sqrt{1-\varepsilon}}-1\right\},
 \quad a_\varepsilon=b\varepsilon+\eta h_\varepsilon,
 \quad f_\varepsilon=b\varepsilon\sqrt p+\eta h_\varepsilon.
\]
For $a,f\ge0$, $\lambda\ge0$ and $x>0$, let
\begin{equation}\label{eq:u-invariant-B}
 B(a,f;\lambda,x)
 =a\lambda+2f(1+k^{-1/2})\sqrt x
       +2a(1+k^{-1})x+2a\sqrt{2\lambda x}.
\end{equation}
For $0<d<1$, set
\begin{equation}\label{eq:u-invariant-e}
 e(\lambda)=
 \frac{B(a_\varepsilon,f_\varepsilon;\lambda,x)
       +\varepsilon B(a_0,f_0;\lambda,x)}{(1-\varepsilon)d}.
\end{equation}
If $F_\nu$ is the distribution function of $\chi^2_\nu$, define
\[
 \beta_R=p\left\{F_r\!\left(\frac r{1+\varepsilon}\right)
                +1-F_r\!\left(\frac r{1-\varepsilon}\right)\right\},
 \qquad
 \beta_V=\sum_{i=1}^2 F_{pk_i}(pk_i d),
 \quad r=m_1-1,\quad k_i=n_i-1.
\]

\begin{theorem}[Finite-reference power transfer]
\label{thm:u-invariant-power}
Under the assumptions of Theorem~\ref{thm:u-diag-invariant}, suppose
that $\bm P$ is a fixed coordinate projection. For every alternative,
every $L$ with $\alpha(L+1)\ge1$, and every
$\varepsilon,d\in(0,1)$ and $x>0$,
\begin{equation}\label{eq:u-invariant-power-bound}
 |\pi_{\mathrm{diag}}-\pi_0|
 \le \min\left[1,\,
       2\sqrt{\frac{e(\Lambda)+e(0)}{\pi a_0}}
       +(L+1)\{\beta_R+\beta_V+16e^{-x}\}\right].
\end{equation}
The bound holds uniformly over positive diagonal $\bm H$ and all
positive variance ratios, with the alternative entering through
$\Lambda$. The term $e(\Lambda)$ controls the fitted observed
statistic; $e(0)$ controls the simulated critical value. No ideal
quantile replaces the finite-reference rank rule in this statement.
\end{theorem}

The bound exploits the cancellation between the quadratic statistic
and its random covariance subtraction. Consequently the baseline
enters through its Frobenius norm $f_0$, rather than its trace $pb$.
The bound can be conservative at small training sizes; its role is to
give a sufficient condition for power transfer, not to certify the
numerical size of the power loss in every design.

\begin{proof}[Proof of Theorem~\ref{thm:u-invariant-power}]
\emph{Step 1: couple the statistics and bound their perturbations.}
Use the normalization and independent canonical variables in the
proof of Theorem~\ref{thm:u-diag-invariant}. Under an alternative,
replace the training vector by
$\bm Y=\bm H^{-1/2}\bm\delta/s_{\mathrm{tr}}+\bm Z_{\mathrm{tr}}$
and the testing vector by $\bm Z_0+\bm\theta$. Couple the fitted
and known-shape procedures by using these same Gaussian variables
and the same testing residuals, and by replacing $\bm R$ by
$\bm I_p$ only for the known-shape procedure. Couple each of the
$L$ reference draws in the same way, independently across draws and
independently of the observed data.

First consider the event
$\mathcal E_R=\{\|\bm R^{-1}-\bm I_p\|_{\mathrm{op}}
\le\varepsilon\}$, whose complement has probability at most
$\beta_R$ by the union bound. Put $\bm D=\bm R^{-1}$ and
$\bm v=\bm P\bm Y/\|\bm P\bm Y\|_2$.
Because $\bm P$ is a coordinate projection, it commutes with
$\bm R$, and
\[
 \bm a_R=\bm R^{-1/2}\bm v_R
       =\frac{\bm D\bm v}{\sqrt{\bm v^\top\bm D\bm v}}.
\]
On $\mathcal E_R$,
\[
 \|\bm a_R\|_2\le\sqrt{1+\varepsilon},\qquad
 \|\bm a_R-\bm v\|_2
 \le\frac{\varepsilon}{\sqrt{1-\varepsilon}}
       +\frac1{\sqrt{1-\varepsilon}}-1.
\]
The Frobenius and operator norms of
$\bm a_R\bm a_R^\top-\bm v\bm v^\top$ are therefore at most
$h_\varepsilon$. With
\[
 \bm B_0=b\bm I_p+\eta\bm v\bm v^\top,\qquad
 \bm\Delta=\bm B_R-\bm B_0,
\]
we obtain
\begin{equation}\label{eq:u-invariant-perturbation}
 \|\bm\Delta\|_{\mathrm{op}}\le a_\varepsilon,
 \qquad \|\bm\Delta\|_F\le f_\varepsilon,
 \qquad \|\bm B_0\|_{\mathrm{op}}=a_0,
 \qquad \|\bm B_0\|_F=f_0.
\end{equation}
These bounds hold for every nonzero $\bm P\bm Y$, without a lower
bound on its norm.

For a symmetric matrix $\bm A$ independent of the testing
variables, define
\[
 N_\kappa(\bm A)
 = (\bm Z_0+\bm\theta)^\top\bm A(\bm Z_0+\bm\theta)
   -\kappa\operatorname{tr}(\bm A\bm\Psi_1)
   -(1-\kappa)\operatorname{tr}(\bm A\bm\Psi_2).
\]
Conditional on $\bm A$, Gaussian quadratic-form and linear-form
tail bounds give, with failure probability at most $8e^{-x}$,
\begin{equation}\label{eq:u-invariant-centered-bound}
 |N_\kappa(\bm A)|
 \le B(a,f;\Lambda,x)
 \quad\text{simultaneously for every }\kappa\in[0,1],
\end{equation}
whenever $\|\bm A\|_{\mathrm{op}}\le a$ and
$\|\bm A\|_F\le f$. To see this, bound the centered testing
quadratic form by $2f\sqrt x+2ax$, the linear term by
$2a\sqrt{2\Lambda x}$, and the two centered residual traces by
$2f\sqrt{x/k_i}+2ax/k_i$. The quadratic and each residual bound
have two-sided failure probability at most $2e^{-x}$; the linear
bound has the same failure probability. The deterministic mean is
$\bm\theta^\top\bm A\bm\theta$, whose absolute value is at most
$a\Lambda$. The traces cancel because the two residual weights sum
to one. Taking the larger residual bound proves
\eqref{eq:u-invariant-centered-bound}, including for indefinite
$\bm A$. When $\bm A$ depends on training variables, the argument
is applied conditionally on those variables. On $\mathcal E_R$,
the same conditional bound applies with the deterministic norm
bounds in \eqref{eq:u-invariant-perturbation}.

For completeness, the quadratic-form bound used here follows
directly from the Gaussian moment-generating function. If
$\bm Z\sim N(\bm0,\bm I)$ and
$|t|<(2\|\bm A\|_{\mathrm{op}})^{-1}$, diagonalization and the
series for $-\log(1-u)$ give
\[
 \log\mathbb E\exp\{t(\bm Z^\top\bm A\bm Z-
                                  \operatorname{tr}\bm A)\}
 \le\frac{t^2\|\bm A\|_F^2}
              {1-2|t|\|\bm A\|_{\mathrm{op}}}.
\]
The exponential Markov inequality for either sign yields
$2\|\bm A\|_F\sqrt x+2\|\bm A\|_{\mathrm{op}}x$ as a
two-sided threshold with failure probability at most $2e^{-x}$.
For a residual trace, apply this calculation to the block diagonal
matrix with $k_i$ copies of $\bm A/k_i$. The zero matrix is
immediate and needs no moment-generating-function restriction.

Write
\[
 V_{0i}=\operatorname{tr}(\bm\Psi_i)/p,\qquad
 V_0=\kappa V_{01}+(1-\kappa)V_{02},\qquad
 V_R=\kappa V_{R1}+(1-\kappa)V_{R2}.
\]
Except on an event of probability at most $\beta_V$, both
$V_{0i}\ge d$. On $\mathcal E_R$ we also have
$|V_R-V_0|\le\varepsilon V_0$ and
$V_R\ge(1-\varepsilon)V_0$. Apply
\eqref{eq:u-invariant-centered-bound} to $\bm\Delta$ and
$\bm B_0$. Outside an additional event of probability at most
$16e^{-x}$,
\[
 \left|\frac{N_\kappa(\bm B_R)}{V_R}
             -\frac{N_\kappa(\bm B_0)}{V_0}\right|
 \le\frac{|N_\kappa(\bm\Delta)|
                  +\varepsilon|N_\kappa(\bm B_0)|}
                {(1-\varepsilon)V_0}
 \le e(\Lambda).
\]
For each reference draw, the same argument with $\bm\theta=0$
holds for both endpoints $\kappa=0,1$ simultaneously. Taking their
maximum gives $|M_{\mathrm{diag}}-M_0|\le e(0)$, with the same
upper bound on the failure probability.

\emph{Step 2: transfer the reference errors to the critical value.}
Let $j=L+1-\lfloor\alpha(L+1)\rfloor$, and let $C_R$ and $C_0$
be the $j$th ascending order statistics of the two coupled banks.
Order statistics are Lipschitz in the maximum norm. A union bound
over the observed data and all $L$ reference draws therefore gives
an event of probability at least
\[
 1-(L+1)\{\beta_R+\beta_V+16e^{-x}\}
\]
on which $|T_R-T_0|\le e(\Lambda)$ and
$|C_R-C_0|\le e(0)$. On this event, the rejection decisions can
differ only if
$|T_0-C_0|\le e(\Lambda)+e(0)$. Both rank rules use strict
comparison with these order statistics; ties have probability zero.

\emph{Step 3: control decisions near the rejection boundary.}
It remains to bound this last probability. Conditional on the
known-shape training direction, testing residuals and the independent
reference bank, diagonalize $\bm B_0$ and condition on all but the
Gaussian coordinate corresponding to its largest eigenvalue $a_0$.
For a standard normal $Z$, any $\mu,t\in\mathbb R$ and $w\ge0$,
\[
 \Pr\{|a_0(Z+\mu)^2-t|\le w\}
 \le 2\sqrt{\frac{w}{\pi a_0}}.
\]
Indeed, the inverse image under squaring of an interval of length
$2w/a_0$ has total length at most $2\sqrt{2w/a_0}$, and the
normal density is at most $(2\pi)^{-1/2}$. Applying this with
$w=hV_0$ and then averaging, Jensen's inequality and
$\mathbb E V_0=1$ yield
\begin{equation}\label{eq:u-invariant-margin}
 \Pr\{|T_0-C_0|\le h\}
 \le 2\sqrt{\frac h{\pi a_0}}.
\end{equation}
This bound is uniform in the signal and the training direction.
Combining the coupling and \eqref{eq:u-invariant-margin} proves
\eqref{eq:u-invariant-power-bound}.
\end{proof}

\begin{proof}[Proof of Corollary~\ref{cor:u-invariant-power}]
Put $t=\log(p+1)/r$ and take
$\varepsilon=t^{1/4}$, $x=\varepsilon^{-1/4}$ and
$d=\varepsilon^{1/4}$ for sufficiently large indices.
For $0<\varepsilon\le1/2$, the chi-square Chernoff bounds give
\[
 \beta_R\le2p\exp(-r\varepsilon^2/32)\longrightarrow0.
\]
Also, for $0<d<1$ and any $\nu\ge1$,
$F_\nu(\nu d)\le\{d\exp(1-d)\}^{\nu/2}$.
It follows that $\beta_V\to0$, uniformly over $p,k_i\ge1$.
We have $h_\varepsilon=O(\varepsilon)$,
$a_\varepsilon=O(\varepsilon a_0)$ and
$f_\varepsilon=O(\varepsilon f_0)$.
The assumptions in \eqref{eq:u-invariant-power-rate} then imply
\[
 \frac{e(\Lambda)+e(0)}{a_0}
 =O\!\left(\frac{\varepsilon(1+\sqrt x+x)}d\right)
 =o(1).
\]
Since $L$ is fixed and $e^{-x}\to0$, the theorem completes the
proof. If $\eta=cpb$, then
$f_0/a_0=\sqrt{p+2cp+c^2p^2}/(1+cp)$ is bounded.
\end{proof}

\begin{proof}[Proof of Proposition~\ref{prop:u-diag-fixed-budget-limit}]
Construct the canonical representation on an infinite sequence of
independent coordinates, separately for the observed experiment and
each of the finitely many reference experiments. The first $q$
coordinates of the fitted direction are unchanged as $p$ grows, and
all remaining direction coordinates are zero. Consequently
$\bm R_p^{-1/2}\widehat{\bm v}$ equals $\bm a_R$ on the first
$q$ coordinates and zero elsewhere. Write
$\kappa_1=\kappa$ and $\kappa_2=1-\kappa$. The standardized
numerator divided by $p$ is
\begin{align*}
 &\frac b p\sum_{j=1}^p R_j^{-1}
    \left\{(Z_{0j}+\theta_j)^2
                     -\sum_{i=1}^2\kappa_i\Psi_{i,jj}\right\}\\
 &\quad+\{\bm a_R^\top(\bm Z_{0,q}+\bm\theta_q)\}^2
            -\sum_{i=1}^2\kappa_i
                         \bm a_R^\top\bm\Psi_{i,q}\bm a_R,
\end{align*}
where $\theta_j=0$ for $j>q$ and $\bm\Psi_{i,q}$ is the
leading $q\times q$ residual block. Outside the first $q$
coordinates, the summands in the first line are independent and
identically distributed, have mean zero, and have finite absolute
first moment since $r>2$ implies
$\mathbb E R_j^{-1}=\mu_r<\infty$. The strong law sends their
average to zero almost surely. Each of the finitely many remaining
terms is finite almost surely and vanishes after division by $p$.
No second inverse moment is needed.

The standardized studentizer is
\[
 \frac1p\sum_{j=1}^p R_j^{-1}
                         \sum_{i=1}^2\kappa_i\Psi_{i,jj}.
\]
Its summands are independent, identically distributed and integrable,
with expectation $\mu_r$, so it converges almost surely to this
positive constant. Conditional on the training variables, Gaussian
rotational invariance represents the surviving rank-one numerator by
\[
 W_R\left\{\left(Z+
          \frac{\bm a_R^\top\bm\theta_q}{\sqrt{W_R}}\right)^2
                    -\sum_{i=1}^2\kappa_i E_i/k_i\right\},
\]
with $Z,E_1,E_2$ having the independent laws in the statement.
This proves \eqref{eq:u-diag-fixed-budget-T}.

For each null reference the training vector is central. Apply the
same strong-law argument to each endpoint, whose denominator
converges to $\mu_r$. Their mean quadratic terms share the same
Gaussian variable; the two residual terms are independent given
training. Taking the maximum yields
\eqref{eq:u-diag-fixed-budget-M}. The construction applies
simultaneously to the observed experiment and all $L$ independent
reference experiments, proving the asserted joint convergence.

An order statistic is a continuous function of its input vector.
Conditional on its training and residual variables,
$T_\infty(\bm\theta_q)$ is a positive multiple of a squared
nondegenerate shifted normal plus a constant, since $W_R>0$
almost surely. Its law is therefore atomless. Independence from the
reference bank implies
$\Pr\{T_\infty(\bm\theta_q)=M_{\infty,(j)}\}=0$.
Joint convergence thus yields
\eqref{eq:u-diag-fixed-budget-power}. Passing to the limit in
Theorem~\ref{thm:u-diag-invariant} gives the null bound.
If $\lfloor\alpha(L+1)\rfloor=0$, the original and limiting
rank rules never reject. Finally, the same proof with
$\bm R_p=\bm I_p$ gives the stated known-shape counterpart.
\end{proof}

\end{vFifteenLowerHeadings}
\subsection{Proofs for the mixed-signal limit}
\label{app:u-mixed-limit}

\begin{proof}[Proof of Proposition~\ref{prop:u-mixed-limit}]
Use the canonical normalization from the proof of
Theorem~\ref{thm:u-diag-invariant}. Write $w_j=R_j^{-1}$, so that
$\mathbb Ew_j=\mu_r$ and $\mathbb Ew_j^2=\nu_r<\infty$.
For all coordinates let $Z_j$ be independent standard normal
variables and set $Q_{ij}=\Psi_{i,jj}\sim\chi^2_{k_i}/k_i$.
For the coordinates outside the learning subspace, the $w_j,Z_j,Q_{1j},Q_{2j}$
are independent within a coordinate and independent across
coordinates. Their common $w_j$ is used in both residual terms.
Define
\[
 A_{0p}=\frac1{\sqrt p}\sum_{j=q+1}^p w_j(Z_j^2-1),\qquad
 A_{ip}=\frac1{\sqrt p}\sum_{j=q+1}^p w_j(Q_{ij}-1),\quad i=1,2.
\]
The summands have zero means, variances
$2\nu_r,2\nu_r/k_1,2\nu_r/k_2$ and zero cross-covariances.
The multivariate central limit theorem gives
$(A_{0p},A_{1p},A_{2p})\Rightarrow(G_0,G_1,G_2)$.
The limiting Gaussian coordinates are independent. They are also
independent of all first-$q$ training and testing variables, since
the underlying coordinate collections are independent for every $p$.

The outside-signal terms satisfy
\[
 \mathbb E\left(\frac2{\sqrt p}
           \sum_{j=q+1}^p w_j\theta_{p,j}Z_j\right)^2
 =\frac{4\nu_r}{p}\|\bm\theta_{\perp,p}\|_2^2\longrightarrow0,
\]
and
\[
 \operatorname{Var}\left(\frac1{\sqrt p}
              \sum_{j=q+1}^p w_j\theta_{p,j}^2\right)
 =\frac{\nu_r-\mu_r^2}{p}
                \sum_{j=q+1}^p\theta_{p,j}^4\longrightarrow0.
\]
The mean of the latter sum tends to $\mu_r\tau$.
The baseline numerator, divided by $\sqrt p$, therefore converges to
$b\{G_0-\kappa G_1-(1-\kappa)G_2+\mu_r\tau\}$.
The contribution of its first $q$ coordinates is $o_p(1)$ under
this scaling. The cancellation of the central quadratic and residual
means is essential: no extra fluctuation of $\sum_j w_j$ remains.

The scale denominator, after normalization by $s^2$, is
\[
 V_p=\frac1p\sum_{j=1}^p
          w_j\{\kappa Q_{1j}+(1-\kappa)Q_{2j}\}.
\]
The law of large numbers gives $V_p\to\mu_r$ in probability;
the first $q$ terms do not affect this limit. The same argument gives
$p^{-1}\sum_jw_jQ_{ij}\to\mu_r$ for each reference endpoint.
For the spike, conditional Gaussian rotational invariance in the
first $q$ coordinates yields exactly
\[
 W_R\left\{(Z+\zeta_R)^2-\kappa E_1/k_1
                                      -(1-\kappa)E_2/k_2\right\}.
\]
The standardized variables $Z,E_1,E_2$ have their stated laws
independently of the training quantities. Multiplication by
$\eta/\sqrt p=c$ and division by $V_p$ prove
\eqref{eq:u-mixed-T} by Slutsky's theorem.

Under each canonical reference the training mean is null. Apply
the same joint central limit theorem to both endpoints, retaining
their shared mean quadratic and training variables. Their Gaussian
baseline components are $b(G_0-G_i)$, and their spike components
are $cW_R^0(Z^2-E_i/k_i)$. Taking their maximum proves
\eqref{eq:u-mixed-M}. Independence of the data and the regenerated
references gives joint convergence for every fixed $L$.
Since $b>0$, the observed limit contains an independent nondegenerate
Gaussian summand. Each reference limit also contains the independent
common summand $bG_0/\mu_r$. Hence all relevant laws are continuous
and the observed statistic ties the reference order statistic with
probability zero. Convergence of the rank rejection probabilities
follows. Setting every $w_j$ to one gives the matched known-shape law.
\end{proof}

\begin{proof}[Proof of Corollary~\ref{cor:u-mixed-projection}]
In the limiting construction, changing $\tau$ changes only the
deterministic shift: $T_c(\tau)=b\tau+T_c(0)$, while the reference
law remains unchanged. The random difference
$M_{c,(j)}-T_c(0)$ has a continuous density that is strictly positive
on the real line, because $T_c(0)$ contains an independent Gaussian
component of positive variance. Thus its distribution function,
evaluated at $b\tau$, is continuous and strictly increasing and tends
to one as $\tau\to\infty$.

For balanced testing groups, conditionally on training, the paired
projection statistic has a noncentral $t_{n-1}$ distribution with
noncentrality $\zeta_R$. Its critical value is
$t_{n-1,1-\alpha/2}$. The first $q$ training coordinates, and
therefore the law of $\zeta_R$, do not depend on the outside signal.
Moreover $|\zeta_R|\le\|\bm\theta_q\|_2<\infty$, so this test's
power is strictly below one. Select a finite $\tau$ at which
$\pi_c(\tau)$ is larger. Proposition~\ref{prop:u-mixed-limit}
then transfers the strict inequality to all sufficiently large
dimensions along that sequence. No quantitative threshold dimension
is obtained from this convergence argument.
\end{proof}

\begin{vFifteenLowerHeadings}
\section{Finite-dimensional error bounds}
\label{app:u-finite-dimension}

This section gives the constants in
Theorem~\ref{thm:u-mixed-finite}. All the bounds are valid for each
$p>q$; asymptotic signal assumptions are needed only for the concluding
rate statement. The proof first separates the outside-coordinate
baseline, the retained rank-one term and a remainder. It then compares
the baseline with its Gaussian limit while keeping the rank-one
randomness and the dependence between reference endpoints. Finally,
second-moment and density bounds control the remainder and random
studentizer. Lemma~\ref{lem:u-rank-cdf-stability} converts the resulting
two distribution-function errors into a power error at the stated
finite reference budget.

The constants are grouped by their role in this argument. The
$d$ terms control studentizers, the $R$ terms control remainders,
and the $J$ terms bound the leading numerators needed when perturbing
a ratio. The $\beta$ and $\gamma$ terms control Gaussian approximation;
the $h$ terms translate small perturbations into distribution-function
errors. Thus the final $\Delta$ terms collect different sources of
error rather than introduce additional approximation assumptions.

\emph{Moments and scales.} Write
\[
 \mu=\frac r{r-2},\quad
 \nu=\frac{r^2}{(r-2)(r-4)},\quad
 \xi=\frac{r^3}{(r-2)(r-4)(r-6)},\quad
 N=p-q,\quad t_p=N/p,
\]
and set
\[
 a=\frac{\kappa^2}{k_1}+\frac{(1-\kappa)^2}{k_2},\qquad
 v_T=2\nu(1+a),\qquad K=\|\bm\theta_q\|^2,\qquad
 \tau_p=\|\bm\theta_{\perp,p}\|^2/\sqrt p.
\]
The quantities $\mu,\nu,\xi$ are the first three moments of
$w=(\chi_r^2/r)^{-1}$. The condition $r>6$ ensures that the third
moment used in the Gaussian approximation is finite. Here $N=p-q$
counts outside coordinates; it is not an observation sample size.

\emph{Studentizer and remainder errors.} Define denominator variances
\begin{equation}\label{eq:u-finite-denom}
 d_T=\frac{\nu(1+2a)-\mu^2}{p},\qquad
 d_i=\frac{\nu(1+2/k_i)-\mu^2}{p}\quad(i=1,2),
\end{equation}
remainder second moments
\begin{align}
 R_T=\frac{b^2}{p}\left[
  \mu^2K^2+2\nu q(1+a)+4\nu\|\bm\theta_p\|^2
       +(\nu-\mu^2)\sum_{j=1}^p\theta_{p,j}^4\right],
 &\label{eq:u-finite-remainder-T}\\
 R_M=\frac{2b^2\nu q}{p}
           (2+k_1^{-1}+k_2^{-1}),
 &\label{eq:u-finite-remainder-M}
\end{align}
and bounds on numerator second moments
\begin{align}
 J_T&=b^2t_pv_T+
 \left[c\sqrt{q\nu\{K^2+4K+2(1+a)\}}
                       +b\mu\tau_p\right]^2,
 \label{eq:u-finite-JT}\\
 J_i&=2\nu(1+k_i^{-1})(b^2t_p+c^2q),\quad i=1,2.
 \label{eq:u-finite-JM}
\end{align}
\emph{Boundary sensitivity.} The density bounds used below are
\begin{equation}\label{eq:u-finite-density}
 h_T=\frac{\mu}{b\sqrt{2\pi v_T}},\qquad
 h_M=\frac{\mu}{2b\sqrt{\pi\nu}}.
\end{equation}

\emph{Gaussian approximation.} The constants below use
closed-form upper bounds on the third absolute moments. Let
\[
 \bm K_M=2\begin{pmatrix}1+k_1^{-1}&1\\1&1+k_2^{-1}\end{pmatrix},
 \qquad \bm A=\bm K_M^{-1},
\]
and define
\begin{align}
 B_T&=\frac{\xi}{\nu^{3/2}}
 \left[3+\frac{12\{1+\kappa^4/k_1^3+(1-\kappa)^4/k_2^3\}}{(1+a)^2}\right]^{1/2},
 \label{eq:u-finite-BT}\\
 B_M&=\frac{\xi}{\nu^{3/2}}
 \left[2\left\{8+48\left[
 (A_{11}+2A_{12}+A_{22})^2
          +\frac{A_{11}^2}{k_1^3}+\frac{A_{22}^2}{k_2^3}
 \right]\right\}\right]^{1/2}.
 \label{eq:u-finite-BM}
\end{align}
Put
\begin{align}
 \beta_T&=\min\{1,\,0.4756 B_T/\sqrt N\},&
 \beta_M&=\min\{1,\,(42\,2^{1/4}+16)B_M/\sqrt N\},
 \label{eq:u-finite-BE}\\
 \gamma_d&=\min\{1,\,\tfrac12
             \sqrt{d(t_p-1-\log t_p)}\},\quad d=1,2.
 \label{eq:u-finite-gaussian-tv}
\end{align}
The scalar Berry--Esseen constant in
\eqref{eq:u-finite-BE} is valid by \citet{shevtsova2011};
the two-dimensional convex-set bound is Theorem~1.1 of
\citet{raic2019}.

\emph{Combining the errors.} The following explicit choices complete the
statement of Theorem~\ref{thm:u-mixed-finite}:
\begin{align}
 \Delta_{T,p}=\min\Bigg\{1,\;&\beta_T+\gamma_1
       +\frac{4d_T}{\mu^2}
       +3\left(\frac{h_T^2R_T}{\mu^2}\right)^{1/3}
       +2\left(\frac{2h_T\sqrt{J_Td_T}}{\mu^2}\right)^{1/2}
                    \Bigg\},
 \label{eq:u-finite-DeltaT}\\
 \Delta_{M,p}=\min\Bigg\{1,\;&\beta_M+\gamma_2
       +\frac{4(d_1+d_2)}{\mu^2}
       +3\left(\frac{h_M^2R_M}{\mu^2}\right)^{1/3}
       +2\left(\frac{2h_M\{\sqrt{J_1d_1}+\sqrt{J_2d_2}\}}{\mu^2}\right)^{1/2}
                    \Bigg\}.
 \label{eq:u-finite-DeltaM}
\end{align}
These expressions require only scalar arithmetic and the inverse of
a specified positive-definite $2\times2$ matrix. They contain no
unestimated distributional quantities or universal constants left
unspecified.

The size of these constants is consequential. For the mixed-signal
diagnostic with $m_i=n_i=20$, $q=5$, $\rho=4$, $b=c=1$,
$L=199$, $\alpha=0.05$, $K=3.2$ and
$\tau_p=28.8/\sqrt{300}$, the rank factor is
$C_{199,10}=26.8358$. The untruncated reference Berry--Esseen term
alone is $29.6230$ at $p=300$ and $14.7182$ at $p=1200$.
Consequently \eqref{eq:u-mixed-finite-power} is one at both dimensions;
it is still one at $p=10^6$ after the rank factor is applied.
These are deterministic evaluations of the stated bound, not measured
approximation errors. They show why the numerical agreement in the
mixed-signal diagnostic must not be described as a consequence of
this bound. The accompanying evaluator
\texttt{reproducibility/finite\_dimension\_bound.py} records all
terms in \texttt{results/v9\_finite\_bound.json} and its CSV counterpart.

\begin{proof}[Proof of Lemma~\ref{lem:u-rank-cdf-stability}]
Conditioning on the observation gives
\[
 \pi(G,F)=\int h(F(x))\,dG(x),\qquad
 h(u)=\Pr\{\operatorname{Bin}(L,u)\ge j\}.
\]
The function $h$ is nondecreasing from zero to one and
$h'(u)=L\binom{L-1}{j-1}u^{j-1}(1-u)^{L-j}$.
Integration by parts bounds the change in $G$ by
$\|G-G_0\|_\infty$, since $h\circ F$ has total variation one.
The change in $F$ is bounded by $C_{L,k}\|F-F_0\|_\infty$
using the derivative bound. Summing these two bounds proves
\eqref{eq:u-rank-cdf-stability}. Maximizing the displayed beta
density gives the stated value of its maximizer, including the
endpoint cases.
\end{proof}

\begin{proof}[Proof of Theorem~\ref{thm:u-mixed-finite}]
\emph{Step 1: decompose the statistics and identify their moment bounds.}
All statistics in this proof are divided by $\sqrt p$.
Use the variables $w_j,Z_j,Q_{ij},A_{ip}$ in the proof of
Proposition~\ref{prop:u-mixed-limit}. In particular,
$A_{0p},A_{1p},A_{2p}$ use only coordinates $j>q$ and are independent
of all first-$q$ training and testing variables. Write
\[
 S=W_R\{(Z+\zeta_R)^2-\kappa E_1/k_1-(1-\kappa)E_2/k_2\}.
\]
The observed statistic has the exact decomposition
\[
 X_p=\frac{N_T+D_T}{V_p},\qquad
 N_T=b\{A_{0p}-\kappa A_{1p}-(1-\kappa)A_{2p}\}
                       +cS+b\mu\tau_p,
\]
where
\begin{align*}
 \frac{D_T}{b}=\frac{1}{\sqrt p}
 &\sum_{j\le q}w_j\{(Z_j+\theta_{p,j})^2
                       -\kappa Q_{1j}-(1-\kappa)Q_{2j}\}\\
 +\frac{1}{\sqrt p}
 &\sum_{j>q}\{2w_j\theta_{p,j}Z_j+(w_j-\mu)\theta_{p,j}^2\}.
\end{align*}
The denominator is
$V_p=p^{-1}\sum_jw_j\{\kappa Q_{1j}+(1-\kappa)Q_{2j}\}$.
Independence across coordinates and the centered normal means give
\[
 \mathbb EV_p=\mu,\quad
 \operatorname{Var}(V_p)=d_T,\quad
 \mathbb ED_T^2=R_T.
\]
For the last equality, the first sum has mean $\mu K/\sqrt p$
and variance
$p^{-1}\{2\nu q(1+a)+4\nu K
 +(\nu-\mu^2)\sum_{j\le q}\theta_{p,j}^4\}$.
The two outside-signal summands have zero covariance and their total
variance is
$p^{-1}\{4\nu\|\bm\theta_{\perp,p}\|^2
 +(\nu-\mu^2)\sum_{j>q}\theta_{p,j}^4\}$.
This also accounts for the nonzero first-$q$ baseline mean.

From the definition of the learned direction,
\[
 W_R=\frac{\sum_{j\le q}w_j^2X_j^2}{\sum_{j\le q}w_jX_j^2}
       \le\max_{j\le q}w_j,
 \qquad \mathbb EW_R^2\le q\nu.
\]
The exceptional event with a zero denominator has probability zero.
Since $\zeta_R^2\le K$, conditional Gaussian and chi-square moments
imply
\[
 \mathbb ES^2
 =\mathbb E\left[W_R^2
       \{\zeta_R^4+4\zeta_R^2+2(1+a)\}\right]
 \le q\nu\{K^2+4K+2(1+a)\}.
\]
The centered outside baseline is independent of $S$. Applying the
$L^2$ triangle inequality to $cS+b\mu\tau_p$ proves
$\mathbb EN_T^2\le J_T$.

For a reference, the two endpoint numerators and denominators can be
written, on their shared canonical probability space, as
\[
 X_{i,p}=\frac{N_i+D_i}{V_{i,p}},\quad
 N_i=b(A_{0p}-A_{ip})+cS_i^0,\quad
 D_i=\frac{b}{\sqrt p}\sum_{j\le q}w_j(Z_j^2-Q_{ij}),
\]
where $S_i^0=W_R^0(Z^2-E_i/k_i)$ and
$V_{i,p}=p^{-1}\sum_jw_jQ_{ij}$.
Thus
\[
 \operatorname{Var}(V_{i,p})=d_i,\qquad
 \sum_{i=1}^2\mathbb ED_i^2=R_M,\qquad
 \mathbb EN_i^2\le J_i.
\]
The second-moment bound follows from
$\mathbb E(S_i^0)^2=2(1+1/k_i)\mathbb E(W_R^0)^2
\le2(1+1/k_i)q\nu$.
No independence of the two endpoints is used here or below.

\emph{Step 2: approximate the outside-coordinate baseline.}
Next compare $Y_T=N_T/\mu$ and
$Y_M=\max_i N_i/\mu$ with their Gaussian counterparts.
For one outside coordinate put
$U_0=Z^2-1$, $U_i=Q_i-1$. These are independent, centered, and
have variances $2,2/k_1,2/k_2$ and fourth cumulants
$48,48/k_1^3,48/k_2^3$, respectively. For
$Y=U_0-\kappa U_1-(1-\kappa)U_2$,
\[
 \mathbb EY^2=2(1+a),\quad
 \mathbb EY^4=3\{2(1+a)\}^2
           +48\{1+\kappa^4/k_1^3+(1-\kappa)^4/k_2^3\}.
\]
The inequality
$\mathbb E|Y|^3\le(\mathbb EY^2\mathbb EY^4)^{1/2}$,
independence of $w$ and $Y$, and $\mathbb Ew^3=\xi$ show that
the standardized third absolute moment of $wY$ is at most $B_T$.
The scalar Berry--Esseen theorem therefore gives error $\beta_T$
for the outside observed baseline.

For the reference vector use
$\bm D=(U_0-U_1,U_0-U_2)^\top$, with covariance $\bm K_M$.
Its standardized squared norm is
$H=\bm D^\top\bm A\bm D$ and satisfies
\[
 \mathbb EH=2,\qquad
 \mathbb EH^2=8+48\{(A_{11}+2A_{12}+A_{22})^2
                        +A_{11}^2/k_1^3+A_{22}^2/k_2^3\}.
\]
This identity follows by adding the fourth cumulants of the three
independent scalar components, whose coefficient vectors are
$(1,1)^\top,(-1,0)^\top,(0,-1)^\top$.
Cauchy--Schwarz gives
$\mathbb EH^{3/2}\le(\mathbb EH\mathbb EH^2)^{1/2}$.
Consequently the standardized third norm moment of $w\bm D$ is
at most $B_M$. Conditional on the first-$q$ variables, the event
$Y_M\le x$ is a translated rectangle in the two-dimensional outside
baseline. A nonsingular linear transformation preserves its convexity.
The convex-set Berry--Esseen inequality of \citet{raic2019} gives
error $\beta_M$, uniformly over these conditional shifts.
Integrating the conditional bounds is legitimate because the
outside coordinates and first-$q$ quantities are independent.

These Gaussian approximations have covariance $t_p$ times the
corresponding limiting covariance. For any positive-definite
$d$-dimensional covariance matrix, Pinsker's inequality and the
normal relative-entropy formula give total variation distance at
most
$\frac12\sqrt{d(t_p-1-\log t_p)}$
between its centered normal law and the law with covariance multiplied
by $t_p$. This proves the corrections $\gamma_1,\gamma_2$.
Convolution, independent mixing and measurable transformations cannot
increase total variation. It follows that the distribution functions
of $Y_T,Y_M$ are within $\beta_T+\gamma_1$ and
$\beta_M+\gamma_2$ of the respective limits at $\tau_p$.

The limiting observation contains an independent normal summand of
variance $b^2v_T/\mu^2$, so its density is bounded by $h_T$.
The limiting reference maximum has the representation
$bG_0/\mu+\max_i\{-bG_i+cS_i^0\}/\mu$.
The first summand is independent of the second; its variance is
$2b^2\nu/\mu^2$. This proves the density bound $h_M$ while retaining
the shared $G_0$.

\emph{Step 3: control the ratio perturbations and transfer to power.}
It remains to control the denominator without asserting its
independence of the numerator. On $|V_p-\mu|\le\mu/2$,
\[
 |X_p-Y_T|\le \frac{2|D_T|}{\mu}
                 +\frac{2|N_T|\,|V_p-\mu|}{\mu^2}.
\]
For arbitrary $u,v>0$, Chebyshev's inequality, Markov's inequality
and Cauchy--Schwarz therefore give
\begin{equation}\label{eq:u-finite-ratio-coupling}
 \Pr\{|X_p-Y_T|>2u/\mu+2v/\mu^2\}
 \le \frac{4d_T}{\mu^2}+\frac{R_T}{u^2}
                         +\frac{\sqrt{J_Td_T}}{v}.
\end{equation}
If the distribution function of $Y$ is within $\beta$ of that of
a variable with density at most $h$, then a coupling satisfying
$\Pr(|X-Y|>\epsilon)\le e$ gives a distribution-function distance
at most $\beta+h\epsilon+e$ from $X$ to that target.
Apply this observation to \eqref{eq:u-finite-ratio-coupling}, then
minimize separately over $u$ and $v$, using
\[
 \inf_{u>0}\left(\frac{2hu}{\mu}+\frac{R}{u^2}\right)
       =3(h^2R/\mu^2)^{1/3},\qquad
 \inf_{v>0}\left(\frac{2hv}{\mu^2}+\frac{A}{v}\right)
       =2(2hA/\mu^2)^{1/2}.
\]
This proves \eqref{eq:u-finite-DeltaT}.
For the reference use
$|\max_i X_{i,p}-\max_iN_i/\mu|
\le\max_i|X_{i,p}-N_i/\mu|$ and a union bound over the two
endpoints. In \eqref{eq:u-finite-ratio-coupling} replace
$d_T,R_T,\sqrt{J_Td_T}$ by
$d_1+d_2,R_M,\sqrt{J_1d_1}+\sqrt{J_2d_2}$.
The same optimization proves \eqref{eq:u-finite-DeltaM}.
Lemma~\ref{lem:u-rank-cdf-stability} now proves
\eqref{eq:u-mixed-finite-power}.

Changing $\tau_p$ to $\tau$ shifts only the limiting observation
by $b(\tau_p-\tau)$. Its distribution-function distance is at most
$h_Tb|\tau_p-\tau|$, which proves the additional error term.
For the rate, with bounded $\tau_p$ and fixed other parameters,
the numerator moment bounds are bounded,
$d_T,d_i=O(p^{-1})$, $R_M=O(p^{-1})$, and
\[
 R_T=O\left(p^{-1/2}
                 +p^{-1}\sum_{j>q}\theta_{p,j}^4\right).
\]
The Berry--Esseen terms are $O(p^{-1/2})$, the normal covariance
corrections are $O(p^{-1})$, and the denominator smoothing terms
are $O(p^{-1/4})$. Substitution yields
\eqref{eq:u-mixed-finite-rate}, since $L$ is fixed.
\end{proof}

\end{vFifteenLowerHeadings}

\section{Observation reuse and centering}
\label{app:v15-reuse}
\subsection{A no-split diagnostic: exact calibration and its limits}
\label{sec:u-nosplit}

Removing the training split can improve the information available for
direction learning, but it changes the statistic and its interpretation.
The following construction separates these issues in a diagonal model.
It is a diagnostic extension, not an extension of the unbiased
direction-optimality theorem.

Suppose that the two groups have the same size $N\ge2$, and fix a pairing
before observing the data. Write $\bm Z_i=\bm X_i-\bm Y_i$, so that
$\bm Z_i\sim N_p(\bm\delta,\bm C)$ independently, where
$\bm C=\bm\Sigma_1+\bm\Sigma_2$ is positive diagonal in the specified
coordinates. Neither a common covariance shape nor equal group scales
are needed for this construction. Let $\overline{\bm Z}$ and $\bm S_Z$
be the sample mean and unbiased sample covariance of the differences,
and put
\begin{equation}\label{eq:u-nosplit-input}
 \bm D=\operatorname{diag}(S_{Z,11},\ldots,S_{Z,pp}),\qquad
 \bm y=\sqrt N\bm D^{-1/2}\overline{\bm Z},\qquad
 \bm R=\bm D^{-1/2}\bm S_Z\bm D^{-1/2}.
\end{equation}
Fix a set $J$ of $q\ge1$ coordinates and let $\bm P$ be its coordinate
projection. For fixed $b>0$ and $\eta\ge0$, define
$\bm v=\bm P\bm y/\|\bm P\bm y\|_2$ and
$\bm W=\bm D^{-1/2}(b\bm I_p+\eta\bm v\bm v^\top)\bm D^{-1/2}$.
The zero-denominator event has probability zero. The mean-axis score is
\begin{equation}\label{eq:u-nosplit-stat}
 \begin{split}
 T_{\rm ma}
 &=N\overline{\bm Z}^{\top}\bm W\overline{\bm Z}
       -\operatorname{tr}(\bm W\bm S_Z)\\
 &=b(\|\bm y\|_2^2-p)
       +\eta(\|\bm P\bm y\|_2^2-\bm v^\top\bm R\bm v).
 \end{split}
\end{equation}
The usual average diagonal studentizer is identically one because
$p^{-1}\operatorname{tr}(\bm D^{-1}\bm S_Z)=1$.

Here is an inexpensive reference generator. Put $r=N-1$. For $j\in J$,
generate independent $\bm g_j\sim N_r(\bm0,\bm I_r)$ and set
$Q_j=\|\bm g_j\|_2^2/r$ and
$\bm u_j=\bm g_j/\|\bm g_j\|_2$. For $j\notin J$, generate independent
$Q_j\sim\chi_r^2/r$. Independently generate $\bm z\sim N_p(\bm0,\bm I_p)$.
Set $y_j^0=z_j/\sqrt{Q_j}$, form
$\bm v_J^0=\bm y_J^0/\|\bm y_J^0\|_2$, and calculate
\begin{equation}\label{eq:u-nosplit-generator}
 T_{\rm ma}^{0}
 =b\left\{\sum_{j=1}^p(y_j^0)^2-p\right\}
  +\eta\left\{\|\bm y_J^0\|_2^2
       -\left\|\sum_{j\in J}v_j^0\bm u_j\right\|_2^2\right\}.
\end{equation}
Only an $r\times q$ residual matrix is needed; a $p\times p$ sample
covariance need not be formed.

\begin{theorem}[No-split diagonal null calibration]
\label{thm:u-nosplit-level}
Under the model above and $\bm\delta=\bm0$,
$T_{\rm ma}\overset{d}{=}T_{\rm ma}^{0}$. For $L\ge1$ independent reference
draws, independent of the observations, the rule
\begin{equation}\label{eq:u-nosplit-pvalue}
 \widehat p_{\rm ma}
 =\frac{1+\sum_{\ell=1}^{L}\bm1\{
                 T_{{\rm ma},\ell}^{0}\ge T_{\rm ma}\}}{L+1}
 \le\alpha
\end{equation}
has, for $0<\alpha<1$, rejection probability at most
$\lfloor\alpha(L+1)\rfloor/(L+1)\le\alpha$, uniformly over positive
diagonal $\bm C$ and common means. No restriction $p<N$ is required.
\end{theorem}

The result follows from diagonal-scale invariance and the standard
Monte Carlo rank argument \citep{dufour2006}. Every reference draw must
recompute the mean-axis direction. The coordinate set $J$, $b$, and
$\eta$ are fixed before observing the data. A fixed noncoordinate
projection in standardized coordinates also has positive-diagonal-scale
invariance, but the economical $q$-column generator and the reduction
below use the stated coordinate-subset structure.

Exact calibration alone does not recover the conditional learned-weight
identity or establish optimality. The following identities distinguish
that target from a centered fixed-block target.

\begin{proposition}[Classical reduction and residual noise]
\label{prop:u-nosplit-reduction}
Under the model above, including alternatives, $\bm R$ is independent
of $\bm y$. Consequently,
\begin{align}
 \mathbb E(T_{\rm ma}\mid\bm y)
 &=b(\|\bm y\|_2^2-p)+\eta(\|\bm P\bm y\|_2^2-1),
 \label{eq:u-nosplit-conditional-mean}\\
 \operatorname{Var}(T_{\rm ma}\mid\bm y)
 &=\frac{2\eta^2}{r}
       \left(1-\sum_{j\in J}v_j^4\right).
 \label{eq:u-nosplit-conditional-variance}
\end{align}
If $r>2$ and $a_r=r/(r-2)$, then
\begin{equation}\label{eq:u-nosplit-null-bias}
 \mathbb E_{\bm0}T_{\rm ma}
 =bp(a_r-1)+\eta(qa_r-1)>0.
\end{equation}
For $q=1$, the conditional-mean identity holds pointwise.
\end{proposition}

Each $y_j$ is the ordinary one-sample $t$ statistic for the paired
differences. Thus \eqref{eq:u-nosplit-conditional-mean} is, apart from
a constant, a weighted sum of squared coordinate $t$ statistics, with
weight $b+\eta$ on $J$ and $b$ elsewhere. Diagonal Hotelling aggregation
and its relatives are established methods
\citep{srivastava2008,hu2019}. Equation
\eqref{eq:u-nosplit-null-bias} gives the raw score's known null bias.
Subtracting this constant preserves ranks, and a further known scaling
yields a fixed-block unbiased target in Proposition~\ref{prop:u-centered-block}.
This does not recover the original conditional learned-weight identity.
Equation \eqref{eq:u-nosplit-conditional-variance} measures the additional
sample-correlation noise. Replacing the score by its conditional mean
does not, by itself, imply a power ordering between the separately
calibrated tests.

There is also a distinction between the mean axis and the maximizer of
the observed complete score. Over unit vectors supported on $J$, the
latter gives
\begin{equation}\label{eq:u-nosplit-eigen}
 T_{\rm eig}=b(\|\bm y\|_2^2-p)
       +\eta\lambda_{\max}
            (\bm y_J\bm y_J^\top-\bm R_{JJ}).
\end{equation}
The same generator, with this maximization repeated in each reference,
gives the level guarantee in Theorem~\ref{thm:u-nosplit-level}.
Maximizing the observed score also changes its null critical value;
it is not a proof of maximal power. In general the maximizing vector
is not the mean axis when $q>1$.

Finally, pairing has an information cost. If
$\bm C_{XY}=(N-1)^{-1}\sum_i
(\bm X_i-\overline{\bm X})(\bm Y_i-\overline{\bm Y})^\top$, then
\begin{equation}\label{eq:u-nosplit-pair-covariance}
 \bm S_Z=\bm S_X+\bm S_Y-\bm C_{XY}-\bm C_{XY}^\top.
\end{equation}
The covariance subtraction in \eqref{eq:u-nosplit-stat} therefore differs
from that in the original two-sample statistic. Although all observations
contribute to the mean difference, the paired covariance has $N-1$
residual degrees of freedom; the two separate group covariances have
$2(N-1)$ in total. This construction gives an exact no-split benchmark
within a specified diagonal model. It neither transfers the earlier
unbiased direction-optimality result nor resolves efficient tuning under
general unknown covariance.

\begin{vFifteenLowerHeadings}
\section{The paired no-split diagnostic}
\label{app:u-nosplit}

\subsection{Joint representation and exact level}

Write $\bm C=\operatorname{diag}(c_1,\ldots,c_p)$ and $r=N-1$.
Independence of the groups and the fixed pairing give independent
$N_p(\bm\delta,\bm C)$ differences. The usual Gaussian mean--residual
decomposition yields
\begin{equation}\label{eq:u-nosplit-proof-gaussian}
 \sqrt N\bm C^{-1/2}\overline{\bm Z}=\bm z+\bm\theta,
 \qquad
 \bm C^{-1/2}\bm S_Z\bm C^{-1/2}=\bm G^\top\bm G/r,
 \qquad
 \bm\theta=\sqrt N\bm C^{-1/2}\bm\delta,
\end{equation}
where $\bm z\sim N_p(\bm0,\bm I_p)$, $\bm G$ has $r$ rows of
independent standard normal entries, and $\bm z$ is independent of
$\bm G$. Let $\bm g_j$ denote column $j$ of $\bm G$ and set
$Q_j=\|\bm g_j\|_2^2/r$ and
$\bm u_j=\bm g_j/\|\bm g_j\|_2$. Then
\begin{equation}\label{eq:u-nosplit-proof-representation}
 D_{jj}=c_jQ_j,\qquad
 y_j=\frac{z_j+\theta_j}{\sqrt{Q_j}},\qquad
 R_{jk}=\bm u_j^\top\bm u_k.
\end{equation}
The random directions $\bm u_j$ are independent and uniform on
the unit sphere in $\mathbb R^r$, and are jointly independent of all
the radii $Q_j$ and of $\bm z$. This remains true for $r=1$, when the
sphere consists of the two signs. It follows in particular that
$\bm R$ and $\bm y$ are independent under every $\bm\delta$.

Since $\operatorname{tr}(\bm R)=p$ and
$\bm v^\top\bm y=\|\bm P\bm y\|_2$,
substituting the definition of $\bm W$ gives
\eqref{eq:u-nosplit-stat}. Its correlation term depends only on
$\bm R_{JJ}$, and
\[
 \bm v^\top\bm R\bm v
   =\left\|\sum_{j\in J}v_j\bm u_j\right\|_2^2.
\]
Under the null, $\bm\theta=\bm0$ in
\eqref{eq:u-nosplit-proof-representation}, giving exactly the reference
in \eqref{eq:u-nosplit-generator}. The radii outside $J$ may be drawn
directly as independent $\chi_r^2/r$ variables because their directions
do not enter the score. In particular, the reference does not require
forming the full correlation matrix, even when $p>N$.

For completeness, the nuisance cancellation also follows directly from
equivariance. For a positive diagonal matrix $\bm A$, replacing
every $\bm Z_i$ by $\bm A\bm Z_i$ gives
$\bm D\mapsto\bm A\bm D\bm A$,
$\bm y\mapsto\bm y$, and $\bm R\mapsto\bm R$. The fixed coordinate
projection therefore gives the same score. The representation
\eqref{eq:u-nosplit-proof-representation} additionally proves the claimed
null distribution and independence under alternatives. A fixed
noncoordinate projection applied to $\bm y$ also preserves this positive
diagonal invariance. The coordinate restriction is used here for the
$q$-column reference generator and weighted-coordinate reduction.

Under the null, the observed score and the $L$ independently generated
reference scores are exchangeable. Add independent continuous
tie-breaking variables, and let $K$ be the rank of the observed score
in descending order with this tie breaking. Then $K$ is uniform on
$\{1,\ldots,L+1\}$. The numerator in
\eqref{eq:u-nosplit-pvalue} is at least $K$, because it counts all
reference ties as exceeding the observed score. Consequently,
\[
 \Pr_{\bm0}(\widehat p_{\rm ma}\le\alpha)
 \le \Pr\{K\le\lfloor\alpha(L+1)\rfloor\}
 =\frac{\lfloor\alpha(L+1)\rfloor}{L+1}.
\]
This proves Theorem~\ref{thm:u-nosplit-level}, including the possible
presence of ties, without any moment condition or dimension restriction.

\subsection{Conditional reduction, variance and bias}

For distinct coordinates, sphere symmetry gives
\[
 \mathbb E R_{jk}=0,\qquad
 \mathbb E R_{jk}^2=1/r,\qquad R_{jj}=1.
\]
To verify the second identity, condition on $\bm u_j$ and use
$\mathbb E(\bm u_k\bm u_k^\top)=\bm I_r/r$.
Moreover, distinct unordered off-diagonal pairs are uncorrelated.
For disjoint pairs this follows from independence. For pairs sharing
one coordinate, for example $(j,k)$ and $(j,l)$ with distinct
$j,k,l$, condition on $\bm u_j$ and use independence and zero means of
$\bm u_k$ and $\bm u_l$. Since $\bm R$ is independent of $\bm y$,
the same identities hold conditionally on $\bm y$.

The vector $\bm v$ is then fixed and has unit norm. Thus
\begin{align*}
 \mathbb E(\bm v^\top\bm R\bm v\mid\bm y)&=1,\\
 \operatorname{Var}(\bm v^\top\bm R\bm v\mid\bm y)
 &=\operatorname{Var}\left(
       2\sum_{\substack{j<k\\j,k\in J}}v_jv_kR_{jk}
          \,\middle|\,\bm y\right)\\
 &=\frac4r\sum_{\substack{j<k\\j,k\in J}}v_j^2v_k^2
   =\frac2r\left(1-\sum_{j\in J}v_j^4\right).
\end{align*}
These equations prove
\eqref{eq:u-nosplit-conditional-mean}--\eqref{eq:u-nosplit-conditional-variance}.
They require no finite unconditional second moment of the $t$ statistics:
the randomness conditional on $\bm y$ is only in the correlation term.
For $q=1$, this term is identically one, so the reduction is pointwise.

If $r>2$, then $\mathbb E Q_j^{-1}=a_r=r/(r-2)$ and
\[
 \mathbb E y_j^2=a_r(1+\theta_j^2).
\]
Combining this identity with the conditional mean gives the more general
expectation formula
\begin{equation}\label{eq:u-nosplit-proof-mean-alternative}
 \mathbb E T_{\rm ma}
 =b\{a_r(p+\|\bm\theta\|_2^2)-p\}
    +\eta\{a_r(q+\|\bm P\bm\theta\|_2^2)-1\}.
\end{equation}
Its null case is \eqref{eq:u-nosplit-null-bias}. The right-hand side is
strictly positive because $b>0$, $p\ge1$, and $a_r>1$.
For $r\le2$, these unconditional expectations are not finite; the
exchangeability argument establishing level still applies.
This completes the proof of Proposition~\ref{prop:u-nosplit-reduction}.

The conditional mean is a weighted diagonal Hotelling score because
$y_j=\sqrt N\overline Z_j/\sqrt{S_{Z,jj}}$ is the coordinate $t$
statistic. This identity is a reduction of a score, not a comparison of
rejection probabilities at recalibrated critical values. If a test's
rejection indicator were integrated over the conditional law of
$\bm R$ given $\bm y$, its resulting randomized rejection probability
would have the same expectation under every $\bm\theta$. Thresholding
the conditional mean of the score is a different operation, so no
power dominance follows merely from the variance reduction.

\subsection{Observed maximization and the pairing cost}

For an arbitrary unit vector $\bm w$ supported on $J$, the same
covariance-subtracted score is
\[
 b(\|\bm y\|_2^2-p)
   +\eta\bm w_J^\top
        (\bm y_J\bm y_J^\top-\bm R_{JJ})\bm w_J.
\]
The Rayleigh variational formula gives \eqref{eq:u-nosplit-eigen}.
For $\eta>0$, any unit top eigenvector maximizes this observed score;
when $\eta=0$, the score is independent of the direction. Regenerating
$\bm y_J^0$ and $\bm R_{JJ}^0$ and taking the largest eigenvalue in
each reference draw gives exchangeable null scores. The same rank
proof therefore establishes the stated finite-sample level guarantee.
There is no need to select an eigenvector at a repeated eigenvalue,
because the score uses only the largest eigenvalue.

The mean axis is not generally an eigenvector of the full matrix:
\[
 (\bm y_J\bm y_J^\top-\bm R_{JJ})\bm y_J
   =\|\bm y_J\|_2^2\bm y_J-\bm R_{JJ}\bm y_J.
\]
It is an eigenvector only when $\bm R_{JJ}\bm y_J$ is proportional to
$\bm y_J$; even that condition alone does not ensure that it is a top
eigenvector. For $q=1$ the two scores coincide. These pointwise facts
contain no comparison of power after the two null laws are calibrated.

Lastly, expanding the centered products of
$\bm Z_i=\bm X_i-\bm Y_i$ proves
\eqref{eq:u-nosplit-pair-covariance}. Under Gaussian sampling,
$r\bm S_Z$ has a Wishart distribution with $r=N-1$ degrees of freedom
and scale $\bm C$. The two original sample covariance matrices instead
have independent Wishart distributions with $N-1$ degrees of freedom
each and their respective scales. Pairing therefore keeps all
observations in $\overline{\bm X}-\overline{\bm Y}$ but discards a
residual component. In particular, this is not an algebraic rewrite
of the original statistic using $\bm S_X+\bm S_Y$, and adaptive
same-sample weighting invalidates its former unbiasedness argument.

\end{vFifteenLowerHeadings}
\subsection{Leave-pair-out weighting: null centering without a training split}
\label{sec:u-lpo}

The bias in \eqref{eq:u-nosplit-null-bias} can be avoided by withholding
the two observations in each cross-product from its weight learner.
This produces a different adaptive statistic. It retains all paired
observations across its terms, but its overlapping learners must be
calibrated jointly.

Use the paired Gaussian diagonal model of Section~\ref{sec:u-nosplit},
and assume $N\ge6$. For each unordered pair $i<j$, compute the sample
mean $\overline{\bm Z}_{-ij}$ and diagonal sample covariance
$\bm D_{-ij}$ from the other $m=N-2$ differences, using divisor
$m-1=N-3$ for the variances. For the same fixed coordinate projection
$\bm P$ and fixed $b>0$, $\eta\ge0$, put
\begin{equation}\label{eq:u-lpo-weight}
 \bm v_{-ij}
 =\frac{\bm P\bm D_{-ij}^{-1/2}\overline{\bm Z}_{-ij}}
        {\|\bm P\bm D_{-ij}^{-1/2}\overline{\bm Z}_{-ij}\|_2},
 \qquad
 \bm W_{-ij}
 =\bm D_{-ij}^{-1/2}
      (b\bm I_p+\eta\bm v_{-ij}\bm v_{-ij}^\top)
       \bm D_{-ij}^{-1/2}.
\end{equation}
All denominators are nonzero almost surely. Define
\begin{equation}\label{eq:u-lpo-statistic}
 U_{\rm lpo}
 =\frac{2}{N(N-1)}\sum_{i<j}\bm Z_i^\top\bm W_{-ij}\bm Z_j,
 \qquad T_{\rm lpo}=N U_{\rm lpo}.
\end{equation}

\begin{theorem}[Leave-pair-out centering and calibration]
\label{thm:u-lpo}
Let $r_-=N-3$, $a_-=r_-/(r_--2)$, and
$\Lambda=N\bm\delta^\top\bm C^{-1}\bm\delta$.
Then $U_{\rm lpo}$ is integrable and
\begin{equation}\label{eq:u-lpo-expectation}
 \mathbb E U_{\rm lpo}
 =\bm\delta^\top\mathbb E(\bm W_{-12})\bm\delta,
 \qquad
 ba_-\Lambda\le\mathbb E T_{\rm lpo}
                  \le(b+\eta)a_-\Lambda.
\end{equation}
In particular, $\mathbb E_{\bm0}U_{\rm lpo}=0$.

Independently generate $L\ge1$ data sets, each containing $N$
independent $N_p(\bm0,\bm I_p)$ observations. In each data set recompute
all the learners in \eqref{eq:u-lpo-weight} and the entire score in
\eqref{eq:u-lpo-statistic}, obtaining $T_{{\rm lpo},1}^0,\ldots,
T_{{\rm lpo},L}^0$. Under $\bm\delta=\bm0$, for every
$0<\alpha<1$,
\begin{equation}\label{eq:u-lpo-rank}
 \Pr\left\{
 \frac{1+\sum_{\ell=1}^L\bm1\{
            T_{{\rm lpo},\ell}^0\ge T_{\rm lpo}\}}{L+1}
       \le\alpha\right\}
 \le\frac{\lfloor\alpha(L+1)\rfloor}{L+1}\le\alpha.
\end{equation}
The bound holds uniformly over positive diagonal $\bm C$, with no
restriction on $p/N$.
\end{theorem}

The expectation in \eqref{eq:u-lpo-expectation} has a random-weight
target: $\mathbb E(\bm W_{-12})$ depends on the signal and on the
learner's sample size. Thus the theorem establishes null centering and
finite-sample validity, rather than unbiased estimation of a prescribed
quadratic signal. The same-sample conditional-independence argument
cannot be applied simultaneously to all terms; they share observations.
The canonical simulation above retains this dependence.

When $\eta=0$, the statistic is, up to a positive deterministic scale,
the leave-two-out diagonal Hotelling statistic of \citet{park2013};
see also the explicit representation in \citet{hu2024}. Thus neither
the leave-out device nor the baseline aggregation is new. The adaptive
rank-one term must be assessed against that baseline as well as other
diagonal and projection tests. For $\eta=0$, multiplying by
$a_-^{-1}=(N-5)/(N-3)$ removes the inverse-variance mean inflation;
this positive deterministic factor does not affect Monte Carlo ranks.
Each adaptive direction learner uses $N-2$ paired observations.
Its costs include $\binom N2$ fitted weights per data set and the paired
residual-information loss described above. Raw sums and squared sums
permit each diagonal variance and mean to be updated after deleting
two rows; no $p\times p$ matrix is needed. For a coordinate projection,
the complete score costs $O(N^2p)$ arithmetic per data set, including
each null reference data set. These facts do not imply a favorable
power--cost tradeoff at a fixed small sample budget. In particular,
neither the earlier direction ordering nor power optimality transfers
to this family of overlapping adaptive cross-products.

\begin{vFifteenLowerHeadings}
\section{Leave-pair-out centering and finite-sample calibration}
\label{app:u-lpo}

Let $\mathcal F_{-ij}$ be the sigma field generated by the observations
other than $\bm Z_i,\bm Z_j$. The weight $\bm W_{-ij}$ is measurable
with respect to $\mathcal F_{-ij}$, and the two omitted observations
are independent of this sigma field and of each other. We first check
integrability, since the weights contain inverse sample variances.

There are $m=N-2$ observations in each learner. Their marginal diagonal
variances satisfy
\[
 (D_{-ij})_{kk}\overset{d}{=}c_k\chi_{r_-}^2/r_-,
 \qquad r_-=m-1=N-3>2.
\]
Therefore
\begin{equation}\label{eq:u-lpo-proof-inverse}
 \mathbb E(\bm D_{-ij}^{-1})=a_-\bm C^{-1},
 \qquad a_-=r_-/(r_--2).
\end{equation}
The positive-semidefinite inequalities
\begin{equation}\label{eq:u-lpo-proof-weight-bound}
 b\bm D_{-ij}^{-1}\preceq\bm W_{-ij}
                 \preceq(b+\eta)\bm D_{-ij}^{-1}
\end{equation}
hold because $\bm v_{-ij}$ is a unit vector. They imply
\[
 \mathbb E\|\bm W_{-ij}\|_{\rm op}
 \le(b+\eta)\mathbb E\operatorname{tr}(\bm D_{-ij}^{-1})
 <\infty.
\]
Gaussian vectors have a finite first norm moment. Independence of the
held-out observations from the learner consequently gives
\[
 \mathbb E\big|\bm Z_i^\top\bm W_{-ij}\bm Z_j\big|
 \le \mathbb E\|\bm W_{-ij}\|_{\rm op}
       \mathbb E\|\bm Z_i\|_2\,\mathbb E\|\bm Z_j\|_2<\infty.
\]
Conditional expectations are therefore justified, and
\begin{equation}\label{eq:u-lpo-proof-condition}
 \mathbb E(\bm Z_i^\top\bm W_{-ij}\bm Z_j
                       \mid\mathcal F_{-ij})
 =\bm\delta^\top\bm W_{-ij}\bm\delta.
\end{equation}
All weights have the same marginal distribution by exchangeability of
the paired observations. Taking expectations, summing over the pairs,
and using the normalization in \eqref{eq:u-lpo-statistic} proves the
expectation identity in \eqref{eq:u-lpo-expectation}.
Taking expectations in \eqref{eq:u-lpo-proof-weight-bound} and applying
\eqref{eq:u-lpo-proof-inverse} proves the two mean bounds after
multiplication by $N$. The null expectation is zero.

To establish the null law, let $\bm A$ be a positive diagonal matrix
and transform every observation to $\bm Z_i'=\bm A\bm Z_i$.
For every pair deletion,
\[
 \overline{\bm Z}_{-ij}'=\bm A\overline{\bm Z}_{-ij},\qquad
 \bm D_{-ij}'=\bm A\bm D_{-ij}\bm A,
 \qquad \bm v_{-ij}'=\bm v_{-ij}.
\]
The last equality follows because
$(\bm A\bm D_{-ij}\bm A)^{-1/2}\bm A
=\bm D_{-ij}^{-1/2}$ for positive diagonal matrices.
Hence
\[
 \bm W_{-ij}'=\bm A^{-1}\bm W_{-ij}\bm A^{-1},\qquad
 (\bm Z_i')^\top\bm W_{-ij}'\bm Z_j'
       =\bm Z_i^\top\bm W_{-ij}\bm Z_j.
\]
The whole statistic is invariant, with the original dependence among
overlapping terms retained. Taking $\bm A=\bm C^{-1/2}$ under the
null converts the observations to independent $N_p(\bm0,\bm I_p)$
vectors. This proves that the observed null score has exactly the law
of each canonical reference score. Independence of the reference data
sets from the observations and from one another makes the $L+1$
scores exchangeable. The rank argument used in the proof of
Theorem~\ref{thm:u-nosplit-level}, with conservative handling of ties,
then proves \eqref{eq:u-lpo-rank}.

The role of $N\ge6$ is the finite inverse-variance moment needed for
\eqref{eq:u-lpo-expectation}. The rank argument itself only requires
well-defined learners and invariance, and so can hold with fewer
observations; no such smaller-sample moment assertion is made here.
Likewise, positive diagonal invariance applies to a fixed projection
in standardized coordinates more generally. The coordinate restriction
is used for the stated efficient implementation and interpretation.

To verify that the implementation need not form covariance matrices,
let $\bm s=\sum_{k=1}^N\bm Z_k$ and
$\bm t=\sum_{k=1}^N\bm Z_k\circ\bm Z_k$, where $\circ$ denotes
coordinatewise multiplication. For each pair put
\[
 \bm s_{-ij}=\bm s-\bm Z_i-\bm Z_j,\qquad
 \bm t_{-ij}=\bm t-\bm Z_i\circ\bm Z_i-\bm Z_j\circ\bm Z_j.
\]
Then
\[
 \overline{\bm Z}_{-ij}=\bm s_{-ij}/m,\qquad
 \operatorname{diag}(\bm D_{-ij})
 =\frac{\bm t_{-ij}-\bm s_{-ij}\circ\bm s_{-ij}/m}{m-1}.
\]
For $\bm a_i=\bm D_{-ij}^{-1/2}\bm Z_i$ and
$\bm a_j=\bm D_{-ij}^{-1/2}\bm Z_j$, the pair's contribution is
\[
 \bm Z_i^\top\bm W_{-ij}\bm Z_j
   =b\bm a_i^\top\bm a_j
       +\eta(\bm a_i^\top\bm v_{-ij})
                  (\bm a_j^\top\bm v_{-ij}).
\]
For a coordinate projection, all these operations are $O(p)$ per pair,
after $O(Np)$ preprocessing. The overall $O(N^2p)$ count describes
arithmetic, not statistical efficiency. An implementation based on raw
second-moment subtraction must also guard against floating-point
cancellation; this numerical issue does not alter the exact formulas.

\section{Centering, the fixed block target, and ancillary randomness}
\label{sec:u-centered-reduction}

The positive null expectation of the same-data mean-axis score does
not imply that zero null mean requires a training split or leave-pair-out
fitting. Its null mean is a known constant in the specified diagonal
Gaussian model. The following refinement separates that elementary
centering operation from the conditional unbiasedness of a weight fitted
on independent observations.

Use the notation of Section~\ref{sec:u-nosplit}, with $N>3$,
$r=N-1$, $a_r=r/(r-2)$ and a coordinate projection $\bm P$ of rank
$q$. Define
\begin{equation}\label{eq:u-centering-constant}
 c_0=bp(a_r-1)+\eta(qa_r-1),\qquad
 U_{\rm ma}^{\rm cen}=\frac{T_{\rm ma}-c_0}{Na_r}.
\end{equation}
Let $T_{\rm block}=b(\|\bm y\|_2^2-p)
+\eta(\|\bm P\bm y\|_2^2-1)$ and similarly put
$U_{\rm block}^{\rm cen}=(T_{\rm block}-c_0)/(Na_r)$.

\begin{proposition}[A fixed-target unbiased reduction]
\label{prop:u-centered-block}
For every mean difference $\bm\delta$ and positive diagonal paired
covariance $\bm C$,
\begin{equation}\label{eq:u-centered-target}
 \mathbb E U_{\rm ma}^{\rm cen}
 =\mathbb E U_{\rm block}^{\rm cen}
 =\bm\delta^\top\bm C^{-1/2}
       (b\bm I_p+\eta\bm P)\bm C^{-1/2}\bm\delta.
\end{equation}
Moreover,
$\mathbb E(U_{\rm ma}^{\rm cen}\mid\bm y)=U_{\rm block}^{\rm cen}$.
For $N>5$ both variances are finite, and
\begin{equation}\label{eq:u-centered-variance}
 \operatorname{Var}(U_{\rm ma}^{\rm cen})
 -\operatorname{Var}(U_{\rm block}^{\rm cen})
 =\frac{2\eta^2}{N^2a_r^2r}
       \mathbb E\left(1-\sum_{j\in J}v_j^4\right)\ge0.
\end{equation}
Applying the same positive affine transformation to an observed score
and every reference score preserves all Monte Carlo ranks and hence
every rejection event.
\end{proposition}

\begin{proof}
Write $\bm\theta=\sqrt N\bm C^{-1/2}\bm\delta$.
The Gaussian representation in
\eqref{eq:u-nosplit-proof-representation} gives
$\mathbb E y_j^2=a_r(1+\theta_j^2)$ and independence of
$\bm R$ from $\bm y$. Proposition~\ref{prop:u-nosplit-reduction}
therefore gives the stated conditional expectation. Subtracting
$c_0$ from that conditional mean and taking expectations yields
$a_r\{b\|\bm\theta\|_2^2+\eta\|\bm P\bm\theta\|_2^2\}$.
Division by $Na_r$ proves \eqref{eq:u-centered-target}.
When $r>4$, the coordinate noncentral $t$ variables have finite fourth
moments; the residual-correlation term is bounded conditionally on
$\bm y$. The law of total variance and
\eqref{eq:u-nosplit-conditional-variance} then prove
\eqref{eq:u-centered-variance}. A common strictly increasing affine
transformation preserves every order comparison, proving the rank claim.
\end{proof}

The target in \eqref{eq:u-centered-target} has a fixed rank-$q$
augmentation $\eta\bm P$. It is not the original conditional target
of an independently learned rank-one weight. The identity does not
transfer the earlier fixed-spectrum direction ordering to this score,
and a variance reduction does not itself order the powers of tests
with separately calibrated critical values. It does show that null
centering alone cannot explain the observed gap between the
leave-pair-out and same-data procedures.

There is a related decision-theoretic reduction. Let $K_R$ be the law
of $\bm R$ under the diagonal Gaussian model. This law is independent
of both $\bm y$ and the standardized mean under every alternative.
For any measurable test $\phi(\bm y,\bm R)\in[0,1]$, set
\begin{equation}\label{eq:u-ancillary-randomization}
 \overline\phi(\bm y)
 =\int \phi(\bm y,\bm R)\,dK_R(\bm R).
\end{equation}
Independence gives $\mathbb E_{\bm\delta}\overline\phi(\bm y)
=\mathbb E_{\bm\delta}\phi(\bm y,\bm R)$ for every
$\bm\delta$ and $\bm C$. Thus a test using these standardized
means and residual correlations can be reproduced by a randomized
test using the standardized means alone, with exactly the same power
function. Independent calibration randomness may be conditioned on
in the same argument and then integrated out. This is an ancillary
randomness reduction within the stated statistic class, not a claim
that thresholding $T_{\rm block}$ dominates every such test.

\end{vFifteenLowerHeadings}

\section{Finite menus and efficiency benchmarks}
\label{app:v15-benchmarks}
\begin{vFifteenLowerHeadings}
\section{Joint calibration of a prespecified weight menu}
\label{sec:u-rank-selection}

A fixed augmentation may favour signal inside the supplied subspace
and lose power when that subspace omits signal. We therefore consider
a finite menu of complete testing procedures. The menu is fixed before
the observations are examined; only the choice among its members is
data dependent. The construction below uses the existing observations
and one simulated reference bank. It is a joint rank-calibration
device, not a new population-direction optimality principle.
For equal weights, the component-rank, row-minimum and outer-rank
construction is the established extreme-rank Monte Carlo test
\citep{mrkvicka2017combined,myllymaki2017global}. We apply that
construction to jointly simulated diagonal-Gaussian candidate scores
and record a weighted form and elementary power comparison. Neither
the generic selection principle nor its exchangeability argument is
claimed as a new method.

Let $\mathcal Z=(\bm Z_1,\ldots,\bm Z_N)$ contain independent
$N_p(\bm\delta,\bm C)$ vectors, where $\bm C$ is positive diagonal.
For example, these may be the fixed paired differences from two
independent groups of equal size. Let $t_1,\ldots,t_K$, $K\ge1$,
be prespecified measurable scores, finite almost surely, with larger
values giving stronger evidence against $\bm\delta=\bm0$.
Assume that their entire fitting rules satisfy
\begin{equation}\label{eq:u-menu-invariance}
 t_j(\bm A\mathcal Z)=t_j(\mathcal Z),\qquad j=1,\ldots,K,
 \quad\text{for every positive diagonal }\bm A.
\end{equation}
Here $\bm A\mathcal Z$ means that each observation is multiplied by
$\bm A$. The same coordinate subspace and deterministic tuning choices
are used in the observed and reference data sets. The leave-pair-out
scores with a fixed finite augmentation grid and classical paired
diagonal scores satisfy this requirement. Every fitted direction and
variance must be recomputed within its own data set.

Independently generate $L\ge1$ canonical data sets
$\mathcal Z^{(1)},\ldots,\mathcal Z^{(L)}$, each with $N$
independent $N_p(\bm0,\bm I_p)$ rows, and put
$\mathcal Z^{(0)}=\mathcal Z$ and $M=L+1$. Compute the full
$M\times K$ matrix $T_{ij}=t_j(\mathcal Z^{(i)})$.
For each column define the conservative upper-tail ranks
\begin{equation}\label{eq:u-menu-component-rank}
 R_{ij}=\sum_{h=0}^{L}\bm1\{T_{hj}\ge T_{ij}\},
 \qquad p_j=R_{0j}/M.
\end{equation}
Thus $p_j$ is the ordinary finite-bank Monte Carlo $p$-value of
candidate $j$. Crucially, the observed row is included when ranking
\emph{every} row, including the references.

Choose fixed weights $w_j>0$ with $\sum_jw_j=1$, and set
\begin{equation}\label{eq:u-menu-minrank}
 S_i=\min_{1\le j\le K}\frac{R_{ij}}{w_j},\qquad i=0,\ldots,L.
\end{equation}
Equal weights give the same ordering as $\min_jR_{ij}$.
For a nonrandomized rule, use
\begin{equation}\label{eq:u-menu-conservative}
 p_{\rm menu}^{+}
 =\frac1M\sum_{h=0}^{L}\bm1\{S_h\le S_0\}.
\end{equation}
To avoid conservatism caused by ties of these integer-based scores,
generate independent $V_0,\ldots,V_L\sim\operatorname{Unif}(0,1)$,
independent of all data, and use the ascending lexicographic rank
\begin{equation}\label{eq:u-menu-randomized}
 p_{\rm menu}
 =\frac1M\sum_{h=0}^{L}
 \left[\bm1\{S_h<S_0\}
  +\bm1\{S_h=S_0,\ V_h\le V_0\}\right].
\end{equation}
This randomization only orders equal $S_i$ values; it does not fit
additional parameters. Reject when the chosen menu $p$-value is at
most $\alpha$, where $0<\alpha<1$.

\begin{theorem}[Extreme-rank calibration and a power comparison]
\label{thm:u-menu}
Under \eqref{eq:u-menu-invariance} and $\bm\delta=\bm0$,
\begin{equation}\label{eq:u-menu-level}
 \Pr(p_{\rm menu}\le\alpha)
 =\frac{\lfloor\alpha M\rfloor}{M}\le\alpha,
 \qquad
 \Pr(p_{\rm menu}^{+}\le\alpha)
 \le\frac{\lfloor\alpha M\rfloor}{M}.
\end{equation}
These statements hold uniformly over positive diagonal $\bm C$.
For every realized score matrix, under the null or an alternative,
\begin{equation}\label{eq:u-menu-pathwise}
 p_{\rm menu}\le p_{\rm menu}^{+}
 \le\min\left\{1,\min_j\frac{p_j}{w_j}\right\}.
\end{equation}
Write $\pi_j(u)=\Pr_{\bm\delta}(p_j\le u)$ for candidate $j$ with
this same finite bank, and $\pi_{\rm menu}(\alpha)$ for either menu
rule. Then
\begin{align}
 \pi_{\rm menu}(\alpha)&\ge\max_j\pi_j(\alpha w_j),
 \label{eq:u-menu-power}\\
 \left[\max_j\pi_j(\alpha)-\pi_{\rm menu}(\alpha)\right]_+
 &\le\max_j\Pr_{\bm\delta}\{\alpha w_j<p_j\le\alpha\}.
 \label{eq:u-menu-regret}
\end{align}
All probabilities include reference simulation, and, for the
randomized rule, the independent tie-breaking variables.
\end{theorem}

The level proof uses exchangeability of the \emph{joint} score rows
and the standard Monte Carlo rank principle \citep{dufour2006}.
The pointwise bound shows that the menu rejection event contains the
weighted Bonferroni union $\bigcup_j\{p_j\le\alpha w_j\}$ on the
same bank. It does not require independence among candidates.
For equal weights, \eqref{eq:u-menu-power} compares with the best
candidate at level $\alpha/K$, not at level $\alpha$.
Equation~\eqref{eq:u-menu-regret} identifies the possible loss as
alternative probability in a strip of component $p$-values. It is
an exact finite-sample bound, but is not automatically small.
It provides neither negligible oracle regret nor optimality among
all tests or all positive-definite weights.

Finite simulation resolution is distinct from the level guarantee.
Since each $p_j\ge1/M$, an equal-weight lower-level event can be
nonempty only if $\lfloor\alpha M/K\rfloor\ge1$.
For example, with $L=199$ and $\alpha=0.05$, that bound is vacuous
when $K>10$; validity still holds. Using $L=999$ resolves more of
the lower tail, at additional computation rather than additional
observations. A large menu need not improve power, even if its
candidates are all individually valid.

The scope of the joint reference is essential. An upper-envelope
reference for a different statistic does not supply an exchangeable
score column merely because it gives a conservative individual test.
In particular, the earlier unpaired split-RI endpoint comparator is
kept outside this menu and calibrated separately. Infeasible
known-shape or population-direction diagnostics are likewise not
selectable candidates. A menu containing classical diagonal scores
is an omnibus procedure with classical fallbacks; any resulting
power gain cannot be attributed solely to a learned RI direction.
The selected index $\arg\min_j p_j/w_j$ may be reported descriptively,
but its unadjusted $p_j$ is not the calibrated menu $p$-value.

No extra validation observations or pilot reference bank are needed.
One bank gives unconditional level control for a single test, jointly
over the observed data and reference simulation; it does not generally
give conditional level control after the bank has been fixed.
Simulation studies that reuse a bank across data sets therefore use
independent banks as the units for Monte Carlo uncertainty.
Data-dependent changes to the candidate set or weights are not covered
by the theorem unless their additional selection is separately justified.

\section{Proof of the finite-bank joint-selection theorem}
\label{app:u-rank-selection}

\subsection{Joint null law and exchangeability}

Under the null, the transformed observations
$\bm C^{-1/2}\bm Z_i$ are independent
$N_p(\bm0,\bm I_p)$ vectors. Applying
\eqref{eq:u-menu-invariance} simultaneously to all $K$ scores gives
\[
 (t_1(\mathcal Z),\ldots,t_K(\mathcal Z))
 =(t_1(\bm C^{-1/2}\mathcal Z),\ldots,
   t_K(\bm C^{-1/2}\mathcal Z)).
\]
Consequently, row zero of $\bm T$ has the same joint distribution
as each reference row. Independence of the canonical data sets and
of the observed data implies that these $M$ rows are independent
and identically distributed. Their coordinates need not be independent.

Permuting the rows of $\bm T$ permutes the corresponding component
ranks $R_{ij}$ and menu scores $S_i$. This equivariance holds because
every rank is computed using the same $M$ rows, including the observed
row. Thus $(S_0,\ldots,S_L)$ is exchangeable under the null, although
its coordinates are generally dependent. Adding the independent
identically distributed marks $V_i$ preserves exchangeability, and
the pairs $(S_i,V_i)$ are almost surely distinct in lexicographic
order. The ascending rank of row zero is therefore uniform on
$\{1,\ldots,M\}$. Dividing this rank by $M$ proves the equality
in \eqref{eq:u-menu-level}.

The conservative rank counts all rows tied with $S_0$, whereas the
randomized rank counts only the preceding marked rows among those
ties. Hence $p_{\rm menu}\le p_{\rm menu}^{+}$ pointwise.
The conservative rejection event is contained in the randomized
rejection event, proving its level inequality. No continuous
distribution assumption on the candidate scores is needed.

\subsection{A deterministic rank-count bound}

Fix any score matrix, with arbitrary ties. For a fixed column $j$ and
a real $r\ge0$, conservative upper-tail ranks satisfy
\begin{equation}\label{eq:u-menu-proof-count}
 \#\{h:R_{hj}\le r\}\le\lfloor r\rfloor.
\end{equation}
When $r\ge M$, this follows from the total number of rows. For
$0\le r<M$, arrange the column values in nonincreasing order.
All members of a tied group receive the position of that group's
last member. Only rows among the first $\lfloor r\rfloor$ positions
can therefore have rank at most $r$, proving the bound.

By definition, the event $S_h\le S_0$ means that at least one column
satisfies $R_{hj}\le w_jS_0$. Applying a union bound for finite sets
and then \eqref{eq:u-menu-proof-count} gives
\begin{align*}
 M p_{\rm menu}^{+}
 &=\#\{h:S_h\le S_0\}\\
 &\le\sum_{j=1}^K\#\{h:R_{hj}\le w_jS_0\}\\
 &\le\sum_{j=1}^K\lfloor w_jS_0\rfloor
 \le S_0\sum_{j=1}^K w_j=S_0.
\end{align*}
Since $S_0/M=\min_jp_j/w_j$ and $p_{\rm menu}^{+}\le1$,
this proves \eqref{eq:u-menu-pathwise}. It is an algebraic statement
about the realized ranks and does not use the null hypothesis.
The bound remains true for randomized outer tie breaking because
that rank is no larger than the conservative outer rank.

In particular, if candidate $j$ rejects at level $\alpha w_j$,
then the menu rejects at level $\alpha$. Therefore, for every
alternative and every $j$,
\[
 \pi_{\rm menu}(\alpha)
 \ge\Pr_{\bm\delta}(p_j\le\alpha w_j)
 =\pi_j(\alpha w_j).
\]
Taking the maximum proves \eqref{eq:u-menu-power}. To obtain
\eqref{eq:u-menu-regret}, choose an index $j_*$ maximizing
$\pi_j(\alpha)$, which exists because the menu is finite. Then
\begin{align*}
 \max_j\pi_j(\alpha)-\pi_{\rm menu}(\alpha)
 &\le\pi_{j_*}(\alpha)-\pi_{j_*}(\alpha w_{j_*})\\
 &=\Pr_{\bm\delta}\{
       \alpha w_{j_*}<p_{j_*}\le\alpha\}\\
 &\le\max_j\Pr_{\bm\delta}\{\alpha w_j<p_j\le\alpha\}.
\end{align*}
The right-hand side is nonnegative, so taking the positive part of
the left-hand side preserves the inequality.

\subsection{Implementation and interpretation}

The score matrix is computed once per observed data set and its
reference bank. Candidate scores that differ only in deterministic
weights may share fitted sufficient statistics. Sorting each of its
$K$ columns gives the conservative ranks in
$O(KM\log M)$ operations, after score construction; evaluating and
ranking $S_i$ adds $O(KM+M\log M)$. The cost of generating a
leave-pair-out score itself remains $O(N^2p)$ per data set and is
not removed by the rank aggregation.

When a reference bank is reused for several simulated observations,
the inclusion of each new observed row changes the ranks of the
reference rows as well. Freezing reference ranks computed without
that row is a different algorithm and is not justified by the
exchangeability proof above. Likewise, all independent tie marks
must accompany their rows in any permutation-equivalence check.

The result provides a calibrated omnibus rejection rule and a
comparison with the fixed candidate menu at reduced component
levels. It does not produce an independent estimate of the most
powerful candidate at the unknown alternative. The oracle in
\eqref{eq:u-menu-regret} is only the best member of this prescribed
menu at level $\alpha$, with the same finite-reference resolution.
The strip probability on the right-hand side can be large; estimating
it at a fixed simulated alternative does not make it a uniform
small-sample efficiency bound over the model.

\section{Proofs for the restricted efficiency benchmarks}
\label{app:u-v12-efficiency}

Throughout, the experiment, coordinate set and signal class are those
of Section~\ref{sec:u-v12-efficiency}. Tests may be randomized; a
measurable function taking values in $[0,1]$ denotes its conditional
rejection probability. Null and alternative power include any independent
calibration randomness. Positive diagonal scaling means multiplication
of all observations in a data set by the same positive diagonal matrix.

\subsection{The maximal invariant and its null law}

For coordinate $j$, let $\bm z_j=(Z_{1j},\ldots,Z_{Nj})^\top$,
let $\overline Z_j$ and $S_j^2$ be its sample mean and unbiased
sample variance, and set
\begin{equation}\label{eq:u-v12-maximal-invariant}
 Y_j=\frac{\sqrt N\overline Z_j}{S_j},\qquad
 \bm U_j=\frac{\bm z_j-\overline Z_j\bm1_N}{\sqrt r S_j}.
\end{equation}
The sample variances are positive almost surely for $N\ge2$.
The vector $\bm U_j$ is on the unit sphere in the $r$-dimensional
space orthogonal to $\bm1_N$. When $N=2$ this sphere consists of
two points. Equation~\eqref{eq:u-v12-maximal-invariant} is unchanged
by positive scaling of coordinate $j$, and
\begin{equation}\label{eq:u-v12-orbit-reconstruction}
 \bm z_j=S_j\left(\frac{Y_j}{\sqrt N}\bm1_N+\sqrt r\bm U_j\right).
\end{equation}
Consequently two data sets with positive sample variances have the
same $(\bm Y,\bm U_1,\ldots,\bm U_p)$ if and only if one is
obtained from the other by positive coordinate scaling. This is a
maximal invariant.

Write $C_{jj}=c_j>0$. Gaussian mean--residual independence gives
independent variables
\[
 \frac{\sqrt N\overline Z_j}{\sqrt{c_j}}
       \sim N(\theta_j,1),\qquad
 \frac{rS_j^2}{c_j}\sim\chi_r^2,
\]
and a uniform residual direction $\bm U_j$ independent of both.
Coordinates are independent because $\bm C$ is diagonal.
It follows that $Y_j\sim t_r(\theta_j)$ independently, and the
joint distribution of all residual directions is independent of
$\bm Y$, $\bm\theta$ and $\bm C$. Any scale-invariant test is a
measurable function of this maximal invariant. Integrating its residual
directions therefore gives a randomized test depending on $\bm Y$
alone, with exactly the same power at every parameter point. In
particular, all scale-invariant power functions depend on
$\bm\delta,\bm C$ only through $\bm\theta$.

This reduction also applies row by row when several independent data
sets are jointly ranked. Under the null, their maximal invariants
have a common nuisance-free distribution, denoted by $P_0$. Independent
row marks have the same distribution under every parameter point.

\subsection{A monotone form of the boundary-mixture likelihood ratio}

Define the even function
\begin{equation}\label{eq:u-v12-symmetric-t-ratio}
 g_a(t)=\frac{f_{r,a}(t)+f_{r,-a}(t)}{2f_{r,0}(t)},
 \qquad a>0.
\end{equation}
Let $V\sim\chi_{r+1}$. Direct integration gives
\begin{equation}\label{eq:u-v12-cosh-ratio}
 g_a(t)=e^{-a^2/2}\,
 \mathbb E\cosh\!\left(\frac{atV}{\sqrt{r+t^2}}\right).
\end{equation}
To verify the formula, represent a noncentral $t$ variable as
$(A+a)/(B/\sqrt r)$, where $A\sim N(0,1)$ and
$B\sim\chi_r$ are independent. Its density is the integral of
$\phi(tu/\sqrt r-a)u/\sqrt r$ against the $\chi_r$ density.
After collecting factors, the density ratio to the central $t$ is
\[
 e^{-a^2/2}
 \frac{\displaystyle\int_0^\infty u^r
    e^{-(1+t^2/r)u^2/2}e^{atu/\sqrt r}\,du}
      {\displaystyle\int_0^\infty u^r
    e^{-(1+t^2/r)u^2/2}\,du}.
\]
The substitution $v=\sqrt{1+t^2/r}\,u$ makes the normalized
integrating density that of $\chi_{r+1}$. Averaging the ratios
at $a$ and $-a$ proves \eqref{eq:u-v12-cosh-ratio}. All the
exponential moments used here are finite. Equivalently,
\begin{equation}\label{eq:u-v12-hypergeometric-ratio}
 g_a(t)=e^{-a^2/2}\,
 {}_1F_1\!\left(\frac{r+1}{2};\frac12;
                    \frac{a^2t^2}{2(r+t^2)}\right).
\end{equation}
Indeed, the nonnegative cosh series, the moments
$\mathbb E V^{2m}=2^m((r+1)/2)_m$ and
$(2m)!=4^m m!(1/2)_m$ give the defining hypergeometric series;
$(b)_m$ denotes the rising factorial. Since
$|t|/\sqrt{r+t^2}$ is strictly increasing in $|t|$,
$g_a(t)$ is strictly increasing in $|t|$ for $a>0$.

Let $Q_{a_0}$ be the uniform mixture of the distributions of a
maximal-invariant row under $\bm\theta=\pm a_0\bm e_j$,
$j\in J$. Independence of the coordinates and the ancillary residual
directions gives
\begin{equation}\label{eq:u-v12-boundary-density-ratio}
 \frac{dQ_{a_0}}{dP_0}(\bm Y,\bm U)
   =\frac1q\sum_{j\in J}g_{a_0}(Y_j)=\ell(\bm Y).
\end{equation}
In particular, $\mathbb E_0\ell(\bm Y)=1$. The score is invariant
under sign changes and permutations of the coordinates in $J$,
and is increasing in each of their absolute standardized means.
Its null distribution is continuous: condition on the other
coordinates and use the strict monotonicity of $g_{a_0}$ in
$|Y_j|$, whose distribution is continuous.

Under $\bm\theta=\pm a\bm e_j$,
\begin{equation}\label{eq:u-v12-amplitude-order}
 Y_j^2\ \overset{d}{=}
 \frac{\chi_1^2(a^2)}{\chi_r^2/r},
\end{equation}
with independent numerator and denominator. The numerator is
stochastically increasing in $a\ge0$. For example, for $x\ge0$,
$\Pr(|A+a|\le x)=\Phi(x-a)-\Phi(-x-a)$ has a nonpositive
derivative in $a\ge0$. Independent division by a positive random
variable preserves this ordering. The other coordinates retain
their independent central $t$ distributions. Any rejection probability
increasing in $|Y_j|$ and depending on the other coordinates in a
fixed way therefore has nondecreasing power as $a$ increases.

\subsection{Proof of Theorem~\ref{thm:u-v12-ideal-maximin}}

Let $h_\alpha$ be the upper-tail null cutoff of $\ell$ with
$\Pr_0\{\ell(\bm Y)>h_\alpha\}=\alpha$, and let
$\phi_*(\bm Y)=\bm1\{\ell(\bm Y)>h_\alpha\}$.
The continuity just proved makes boundary randomization unnecessary;
allowing it would give the same argument. For any competing invariant
test $\phi$ with null size at most $\alpha$, the
Neyman--Pearson lemma and \eqref{eq:u-v12-boundary-density-ratio}
give
\begin{equation}\label{eq:u-v12-ideal-np-bound}
 \frac1{2q}\sum_{j\in J}\sum_{s\in\{-1,1\}}
   \mathbb E_{sa_0\bm e_j}\phi
 \le
 \frac1{2q}\sum_{j\in J}\sum_{s\in\{-1,1\}}
   \mathbb E_{sa_0\bm e_j}\phi_*.
\end{equation}
This remains true when $\phi$ uses residual directions, because
they are ancillary and the likelihood ratio does not involve them.

The sign and coordinate symmetry of $\phi_*$ makes all the powers
on the right equal; denote their common value by $\beta_*$.
Equations~\eqref{eq:u-v12-cosh-ratio} and
\eqref{eq:u-v12-amplitude-order} show that
$\mathbb E_{sa\bm e_j}\phi_*\ge\beta_*$ for every
$a\ge a_0$. Hence its minimum over $\Theta_{a_0}$ is $\beta_*$.
For every competing test,
\[
 \inf_{\bm\theta\in\Theta_{a_0}}\mathbb E_{\bm\theta}\phi
 \le \frac1{2q}\sum_{j\in J}\sum_{s\in\{-1,1\}}
      \mathbb E_{sa_0\bm e_j}\phi
 \le\beta_*.
\]
Scale invariance makes all these statements uniform over positive
diagonal $\bm C$. This proves the theorem.

\begin{corollary}[A strictly positive cost of an unknown coordinate]
\label{cor:u-v12-coordinate-oracle-gap}
Suppose $q\ge2$. Let $c_{r,\alpha}$ be the $(1-\alpha/2)$-quantile
of the central $t_r$ distribution, and let
\[
 \beta_{\rm coord}(a_0)
   =\Pr\{|T_r(a_0)|>c_{r,\alpha}\},
 \qquad T_r(a_0)\sim t_r(a_0).
\]
This is the boundary power of the coordinate oracle that knows the
active coordinate, uses its two-sided $t$ test, and knows neither
its sign nor its variance. The ideal invariant maximin boundary
power $\beta_*$ in Theorem~\ref{thm:u-v12-ideal-maximin} satisfies
\begin{equation}\label{eq:u-v12-coordinate-oracle-gap}
 \beta_{\rm coord}(a_0)-\beta_*>0.
\end{equation}
Both powers are the respective worst-case powers on their classes
with $a\ge a_0$; the strict comparison is at the common boundary
$a_0$, not a pointwise ordering of every procedure at every strength.
\end{corollary}

\begin{proof}
For a fixed coordinate $j$, the equally weighted mixture of
$\pm a_0\bm e_j$ has likelihood ratio $g_{a_0}(Y_j)$ against
$P_0$. Its strict monotonicity in $|Y_j|$ and the continuous central
$t$ law imply that the unique most powerful level-$\alpha$
rejection function, up to $P_0$-null sets, is
$\bm1_{A_j}$, where $A_j=\{|Y_j|>c_{r,\alpha}\}$.
Uniqueness also holds among randomized invariant tests using ancillary
residual directions. The two sign-specific powers coincide with
$\beta_{\rm coord}(a_0)$, so the average power for each coordinate
pair is at most this value. If the boundary-mixture optimizer
$\phi_*$ had average power $\beta_{\rm coord}(a_0)$, equality
would have to hold for every coordinate pair. Uniqueness would then
require $\phi_*=\bm1_{A_j}$ almost surely for every $j\in J$.
For distinct $j,h\in J$, independence under $P_0$ gives
$P_0(A_j\mathbin{\triangle} A_h)=2\alpha(1-\alpha)>0$,
a contradiction. The minimum for the coordinate oracle occurs at
$a_0$ by \eqref{eq:u-v12-amplitude-order}; the same property for
$\phi_*$ was established in the theorem's proof.
\end{proof}

\subsection{Proof of Theorem~\ref{thm:u-v12-bank-maximin}}

The following allocation argument specializes established orbitwise
likelihood-ratio optimality \citep{roach2018permutation}; it does not
introduce a new general permutation-testing principle.
For clarity, let $\bm X_i=(\bm Y_i,\bm U_i)$ denote the complete
maximal invariant of data-set row $i$, where $i=0,\ldots,L$.
Every reference row has distribution $P_0$. Write $\mathbb E_0$
for expectation when all $M$ rows have distribution $P_0$,
independently, including independent identically distributed row
marks. Put $\ell_i=\ell(\bm Y_i)$, and write
$\ell_{[1]}\ge\cdots\ge\ell_{[M]}$ for their descending order.
Row-permutation equivariance means that permuting the data rows
and their marks permutes the vector of decision probabilities in
the same way. The budget \eqref{eq:u-v12-row-budget} is imposed
on this entire vector, not only its observed-row component.

Under the null, exchangeability and the row budget imply
\begin{equation}\label{eq:u-v12-bank-level}
 \mathbb E_0\psi_0
 =\frac1M\mathbb E_0\sum_{i=0}^L\psi_i\le\frac{k}{M}.
\end{equation}
When row zero has distribution $Q_{a_0}$, while the other rows
remain independent null references, its average boundary power is
\begin{align}
 \overline\beta_\psi
 &=\mathbb E_{Q_{a_0}\otimes P_0^L}\psi_0
 =\mathbb E_0(\ell_0\psi_0)
 =\frac1M\mathbb E_0\sum_{i=0}^L\ell_i\psi_i
 \label{eq:u-v12-bank-change-measure}\\
 &\le\frac1M\mathbb E_0\sum_{m=1}^k\ell_{[m]}.
 \label{eq:u-v12-bank-allocation}
\end{align}
Marks have unchanged distributions, so the change of measure
multiplies only by $\ell_0$. Exchangeability and algorithm equivariance
give the second averaging identity. The last inequality holds
pointwise: allocating at most $k$ units, with each coordinate between
zero and one, maximizes $\sum_i\ell_i\psi_i$ by allocating one unit
to each of the $k$ largest nonnegative $\ell_i$.

Let $\psi_i^*$ reject exactly these $k$ rows, with independent marks
resolving ties if required. It attains the bound in
\eqref{eq:u-v12-bank-allocation}, and its null size is exactly $k/M$.
Sign changes and permutations within $J$ do not alter any $\ell_i$,
so its boundary powers are all equal to
$\overline\beta_{\psi^*}$. Conditional on the reference rows,
observed-row rejection is increasing in $\ell_0$ and therefore in
each $|Y_{0j}|$, $j\in J$. The ordering
\eqref{eq:u-v12-amplitude-order} shows that its power does not
decrease when $a$ increases from $a_0$. For any competitor,
\[
 \inf_{\bm\theta\in\Theta_{a_0}}
    \mathbb E_{\bm\theta\otimes P_0^L}\psi_0
 \le\overline\beta_\psi
 \le\overline\beta_{\psi^*}
 =\inf_{\bm\theta\in\Theta_{a_0}}
    \mathbb E_{\bm\theta\otimes P_0^L}\psi_0^*.
\]
Here the notation $\bm\theta\otimes P_0^L$ means that only the
observed row follows the alternative. The row-by-row scale reduction
makes the comparison uniform over the observed positive diagonal
covariance, proving the finite-reference theorem.

To check inclusion of the existing menus, compute every component
rank from the same $M$ rows. Permuting the rows permutes those ranks,
their row minima and their final marked ranks. Thus the resulting
decision vector is equivariant. Unique marked final ranks run from
$1$ to $M$, so rejecting ranks at most $k$ gives
$\sum_i\psi_i=k$. Conservative final ranks select at most $k$ rows.
The specified invariant candidate fits supply the required separate
scale invariance for each row. This argument does not include a
candidate set fitted specially to the observed row in a way that
breaks row equivariance.

The pointwise row budget is a substantive restriction. An ideal
fixed-threshold likelihood-ratio test applied separately to every row
can reject more than $k$ rows in a particular matrix and need not
belong to this finite-reference class. The theorem therefore does
not assert optimality among all tests with unconditional null size
at most $\alpha$, or conditional validity after a reference bank
has been fixed.

\subsection{An exact average boundary-power gap}

The allocation argument also gives a quantitative identity. It
compares the prescribed boundary mixture, not the unknown signal
distribution in an observed application.

\begin{proposition}[A finite-reference allocation identity]
\label{prop:u-v12-regret-identity}
Under the assumptions of Theorem~\ref{thm:u-v12-bank-maximin},
suppose the competitor saturates the row budget,
$\sum_{i=0}^L\psi_i=k$ almost surely. Then
\begin{equation}\label{eq:u-v12-regret-identity}
 \overline\beta_{\psi^*}-\overline\beta_\psi
 =\frac1M\mathbb E_0\sum_{i=0}^L
  (\ell_i-\ell_{[k]})(\psi_i^*-\psi_i)\ge0.
\end{equation}
More generally, when the budget need not be saturated, the right
side must include the additional nonnegative term
$M^{-1}\mathbb E_0[\ell_{[k]}(k-\sum_i\psi_i)]$.
\end{proposition}

\begin{proof}
Subtract \eqref{eq:u-v12-bank-change-measure} for the two algorithms.
For equal budgets, $\sum_i(\psi_i^*-\psi_i)=0$ pointwise, so any
common finite centering threshold can be subtracted from the
$\ell_i$ without changing the sum. Choosing $\ell_{[k]}$ makes
each product nonnegative: on a selected row both factors are
nonnegative, and on an unselected row both are nonpositive. Ties
at the threshold contribute zero. If the competing budget is
smaller, retaining the subtracted term gives exactly the additional
term displayed in the statement. Its sign is nonnegative because
$\ell_{[k]}\ge0$ and the budget is at most $k$.
\end{proof}

The expectation in \eqref{eq:u-v12-regret-identity} involves only
canonical null rows and known $a_0,J,N$. It is therefore evaluable
by fresh null simulation, with simulation uncertainty reported if
an estimate is used. This does not make it an observed-data
certificate of the unknown alternative power. It is an exact gap
in \emph{average boundary power}; without further assumptions on
the competitor it is not a uniform regret bound over all
$a\ge a_0$. In particular, it does not prove that the existing
learned RI score or its menu attains the benchmark.

\section{Proofs for the bounded quadratic comparison}
\label{app:u-v13-bounded-quadratic}

We work throughout in the invariant paired Gaussian diagonal
experiment of Section~\ref{sec:u-v12-efficiency}. Its ancillary
residual directions may be retained in a competing procedure; the
likelihood ratio below depends only on $\bm Y$.

\subsection{A positive likelihood-ratio decomposition}

For $x=t^2/(r+t^2)$ and $s=a^2/2$, the hypergeometric
representation~\eqref{eq:u-v12-hypergeometric-ratio} gives
\begin{align}
 g_a(t)
 &=e^{-s}\{1+(r+1)sx\}+R_a(t),
 \label{eq:u-v13-positive-remainder}\\
 R_a(t)
 &=e^{-s}\sum_{m=2}^\infty
   \frac{((r+1)/2)_m}{(1/2)_m\,m!}(sx)^m\ge0.
 \label{eq:u-v13-positive-series}
\end{align}
Under the null, $X=Y^2/(r+Y^2)$ has the
$\operatorname{Beta}(1/2,r/2)$ distribution. Its moments satisfy
\[
 \mathbb E_0 X^m=\frac{(1/2)_m}{((r+1)/2)_m}.
\]
Nonnegativity permits termwise integration, yielding
\begin{equation}\label{eq:u-v13-remainder-mass}
 \mathbb E_0R_a(Y)
   =e^{-s}\sum_{m=2}^\infty\frac{s^m}{m!}
   =1-(1+s)e^{-s}=\Delta(s).
\end{equation}
In particular, $\mathbb E_0 B=1$. The boundary-mixture likelihood
ratio is therefore
\begin{equation}\label{eq:u-v13-LR-affine}
 \ell_a(\bm Y)=e^{-s}\{1+sB(\bm Y)\}
                +\overline R_a(\bm Y),
 \qquad
 \overline R_a=\frac1q\sum_{j\in J}R_a(Y_j),
\end{equation}
where $\overline R_a\ge0$ and
$\mathbb E_0\overline R_a=\Delta(s)$. Since $0\le X\le1$,
the remainder is bounded for each finite $r,a$, and all its
moments exist. Independence of the selected coordinates gives
\begin{equation}\label{eq:u-v13-remainder-variance}
 \operatorname{Var}_0(\overline R_a)=\operatorname{Var}_0(R_a(Y_1))/q.
\end{equation}
Finally, $\Delta'(s)=se^{-s}\le s$ and $\Delta(0)=0$, so
$\Delta(s)\le s^2/2=a^4/8$.

\subsection{Proof of Theorem~\ref{thm:u-v13-bounded-regret}}

The null law of $B$ is continuous because it is a sum of
independent variables with continuous distributions. Its exact
upper-tail cutoff thus gives a test $\phi_B$ of size $\alpha$.
Among all tests of size $\alpha$, thresholding $B$ maximizes
$\mathbb E_0(B\phi)$: if $c_B$ is its cutoff, then
$(B-c_B)(\phi_B-\phi)\ge0$ pointwise, and the constant term
cancels on taking expectations. The same statement holds for
randomized tests depending on ancillary residual directions.

Let $\phi_*$ be the ideal likelihood-ratio test, of the same size.
The average boundary-power gap is
\begin{align*}
 \overline\beta_*-\overline\beta_B
 &=se^{-s}\mathbb E_0[B(\phi_*-\phi_B)]
    +\mathbb E_0[\overline R_a(\phi_*-\phi_B)]\\
 &\le\mathbb E_0(\overline R_a\phi_*)
 \le\Delta(s).
\end{align*}
Its nonnegativity follows from likelihood-ratio optimality. Both
tests are sign and coordinate-permutation symmetric and are
increasing in each $|Y_j|$. The amplitude ordering in
\eqref{eq:u-v12-amplitude-order} therefore makes their worst-case
powers on $\Theta_{a_0}$ equal to their average boundary powers
at $a=a_0$. This proves the first bound for ideal calibration.

For finite-reference calibration, let $\psi^B$ allocate its $k$
rejections to the largest $B_i=B(\bm Y_i)$. Under the all-null
row experiment it satisfies $\sum_i\psi_i^B=k$ pointwise and
$\mathbb E_0\psi_0^B=k/M$. The same properties hold for the
likelihood-ratio allocation $\psi^*$. The change-of-measure and
equivariance identity~\eqref{eq:u-v12-bank-change-measure} gives
\begin{align*}
 \overline\beta_*-\overline\beta_B
 &=\frac{se^{-s}}M\mathbb E_0\sum_i B_i(\psi_i^*-\psi_i^B)
   +\frac1M\mathbb E_0\sum_i\overline R_{a,i}
                              (\psi_i^*-\psi_i^B)\\
 &\le\frac1M\mathbb E_0\sum_i\overline R_{a,i}
 =\Delta(s).
\end{align*}
The first sum is nonpositive pointwise because top-$k$ allocation
maximizes the sum of selected $B_i$. Symmetry and monotonicity in
the observed row again identify both boundary averages with the
respective worst-case powers. When $q=1$, $B$ and $\ell_a$ are
strictly increasing functions of $|Y_1|$; their ideal rejection
regions and finite-reference orderings are identical.

\subsection{A variance refinement}

Let $\tau=\alpha$ for ideal calibration and $\tau=k/M$ for
finite-reference calibration. If
$X\sim\operatorname{Beta}(1/2,r/2)$, set
\[
 \sigma_R^2=\operatorname{Var}\!\left[
 e^{-s}\left\{{}_1F_1\!\left(\frac{r+1}{2};\frac12;sX\right)
                  -1-(r+1)sX\right\}\right].
\]
This finite quantity is a one-dimensional Beta integral. The bound
\begin{equation}\label{eq:u-v13-variance-regret}
 \beta_*-\beta_B\le
 \min\left\{\Delta(s),
       \sqrt{\frac{2\min\{\tau,1-\tau\}}q\,\sigma_R^2}\right\}
\end{equation}
follows from the next argument.

For the sharper bound, write $U=\phi_*-\phi_B$ in the ideal
case and $U=\psi_0^*-\psi_0^B$ in the finite-reference case.
In either case, the preceding argument implies
\[
 0\le\overline\beta_*-\overline\beta_B
 \le\mathbb E_0[\overline R_a U]
 =\mathbb E_0[(\overline R_a-\Delta(s))U],
\]
because both procedures have null rejection probability $\tau$.
For decisions in $[0,1]$,
\[
 U^2\le |U|\le \phi_*+\phi_B,
 \qquad
 |U|\le2-\phi_*-\phi_B,
\]
with the corresponding row decisions substituted in the
finite-reference case. Hence
$\mathbb E_0 U^2\le2\min\{\tau,1-\tau\}$.
Cauchy--Schwarz and \eqref{eq:u-v13-remainder-variance} prove
\eqref{eq:u-v13-variance-regret}. The reference rows can be
dependent on the rejection decision; the argument requires no
independence between $U$ and the observed-row remainder.

\subsection{The null-allocation criterion}

In the ideal case, the threshold argument above proves
$D_\psi\ge0$ for any equal-size competitor. In the finite-reference
case, top-$k$ allocation of $B_i$ proves the same assertion
pointwise before expectation. Subtracting the two boundary powers
using \eqref{eq:u-v13-LR-affine} gives the exact identity
\begin{equation}\label{eq:u-v13-block-comparator-identity}
 \overline\beta_B-\overline\beta_\psi
   =se^{-s}D_\psi+E_\psi,
 \qquad |E_\psi|\le\Delta(s).
\end{equation}
For the ideal case, $E_\psi=\mathbb E_0[
\overline R_a(\phi_B-\psi)]$; for finite references, it is
$M^{-1}\mathbb E_0\sum_i\overline R_{a,i}
(\psi_i^B-\psi_i)$. Each of the two expectations being subtracted
lies between zero and $\Delta(s)$, proving the stated bound.

Since the bounded quadratic rule has equal boundary powers and
nondecreasing power in amplitude,
\[
 \inf_{\Theta_{a_0}}\beta_B
       -\inf_{\Theta_{a_0}}\beta_\psi
 \ge\overline\beta_B-\overline\beta_\psi
 \ge se^{-s}D_\psi-\Delta(s).
\]
This establishes \eqref{eq:u-v13-weak-signal-margin}. For a fixed
competitor with $D_\psi>0$, its right-side condition is equivalent
to $D_\psi>(e^s-1-s)/s$, whose right side tends to zero as
$s\downarrow0$. If the competitor itself changes with $a_0$,
one must instead check its amplitude-dependent $D_\psi(a_0)$;
strictness does not follow merely from pointwise positivity.

For ideal equal-size tests, continuity of $B$ also shows
$D_\psi=0$ only if $\psi=\phi_B$ almost surely under the null.
For finite references the distinct $B_i$ imply the corresponding
statement for top-$k$ allocation. Thus a fixed learned-score
procedure whose null rejection allocation differs with positive
probability pays a strict local power cost in this sparse boundary
comparison: its average boundary deficit is
$a_0^2D_\psi/2+O(a_0^4)$. This is not a statement about all signal geometries,
all unknown covariance matrices, or all data-dependent learners.

\section{Bounded estimation of average boundary-power gaps}
\label{app:u-v13-gap-estimation}

The exact allocation identity in
Proposition~\ref{prop:u-v12-regret-identity} is an expectation
under canonical null data-set rows. Direct simulation of that
identity can be unstable when a large likelihood ratio occurs
rarely under the null. The following change of sampling law gives
bounded estimates of the same finite-reference target. It changes
the numerical experiment, not the test applied to observed data.

Write $\bm X_i$ for the complete scale maximal invariant of row
$i$, including its ancillary residual directions and independent
marks when the algorithm uses them. Let $P_0$ be its canonical
null law, and let $Q_{a_0}$ be the uniform mixture of the
$2q$ boundary alternatives. Then
$dQ_{a_0}/dP_0=\ell$ from
\eqref{eq:u-v12-boundary-density-ratio}. On the $M$-row
experiment, define
\begin{equation}\label{eq:u-v13-tilted-law}
 H=\frac1M\sum_{h=0}^{M-1}
     P_0^{\otimes h}\otimes Q_{a_0}
                         \otimes P_0^{\otimes(M-h-1)}.
\end{equation}
It can be generated from raw canonical Gaussian paired observations:
choose one row uniformly, choose its active coordinate uniformly
from $J$ and its sign uniformly, give that row mean
$\pm a_0\bm e_j/\sqrt N$, and leave the other rows at zero
mean. Every coordinate variance is one. Fitting each row exactly
as prescribed induces \eqref{eq:u-v13-tilted-law} on its maximal
invariants. The density ratio used below is that of the maximal
invariants, not the raw Gaussian observations.

\begin{proposition}[Bounded power and gap estimates]
\label{prop:u-v13-tilted-gap}
Let $\bm\psi$ be a prespecified scale-invariant,
row-permutation-equivariant procedure with decisions in $[0,1]$
and row budget at most $k$. Put $\ell_i=\ell(\bm Y_i)$,
$S_\ell=\sum_{i=0}^{M-1}\ell_i$, and
\begin{equation}\label{eq:u-v13-tilted-statistics}
 W_\psi=\frac{\sum_i\ell_i\psi_i}{S_\ell},\qquad
 G_\psi=W_{\psi^*}-W_\psi,
\end{equation}
where $\bm\psi^*$ allocates exactly $k$ rejections to the
largest $\ell_i$. Then
\begin{equation}\label{eq:u-v13-tilted-unbiased}
 0\le W_\psi\le1,\quad 0\le G_\psi\le1,\qquad
 \mathbb E_H W_\psi=\overline\beta_\psi,\quad
 \mathbb E_H G_\psi=
       \overline\beta_{\psi^*}-\overline\beta_\psi.
\end{equation}
The powers are the average boundary powers in the original
experiment with one designated observed row and $M-1$ null
reference rows.

For a prespecified collection of $J_0$ such powers or gaps,
generate $R_{\rm sim}$ independent whole matrices from $H$.
Let $\widehat\mu_h$ be the corresponding sample means. For
$0<\gamma<1$, with probability at least $1-\gamma$,
simultaneously for $h=1,\ldots,J_0$,
\begin{equation}\label{eq:u-v13-hoeffding-gap}
 |\widehat\mu_h-\mu_h|\le
 \sqrt{\frac{\log(2J_0/\gamma)}{2R_{\rm sim}}}.
\end{equation}
The resulting intervals may be intersected with $[0,1]$.
Alternatively, put
\[
 \operatorname{kl}(x,u)=x\log(x/u)+(1-x)\log\{(1-x)/(1-u)\},
\]
with the usual continuous endpoint conventions. The intervals
\begin{equation}\label{eq:u-v13-kl-gap}
 I_h=\left\{u\in[0,1]:
 R_{\rm sim}\operatorname{kl}(\widehat\mu_h,u)
                     \le\log(2J_0/\gamma)\right\}
\end{equation}
have the same simultaneous coverage of at least $1-\gamma$.
These Chernoff intervals apply to the bounded, generally nonbinary
matrix contributions in \eqref{eq:u-v13-tilted-statistics}.
\end{proposition}

\begin{proof}
All $\ell_i$ are strictly positive. Under $P_0^{\otimes M}$,
the density ratio of \eqref{eq:u-v13-tilted-law} is
$S_\ell/M$. Therefore
\[
 \mathbb E_H W_\psi
   =\frac1M\mathbb E_0\sum_i\ell_i\psi_i
   =\overline\beta_\psi,
\]
where the last equality is the change-of-measure and
row-equivariance identity
\eqref{eq:u-v12-bank-change-measure}. Independent marks have
the same law under both measures. The numerator of $W_\psi$
is between zero and $S_\ell$. The deterministic top-$k$
allocation inequality gives
$\sum_i\ell_i\psi_i^*\ge\sum_i\ell_i\psi_i$,
which establishes the bounds and expectation of $G_\psi$.
For each prescribed target, its independent matrix contributions
belong to $[0,1]$. The two-sided Hoeffding inequality gives
tail probability at most $2\exp(-2R_{\rm sim}\epsilon^2)$.
A union bound over $J_0$ targets proves
\eqref{eq:u-v13-hoeffding-gap}. Dependence between targets
within a matrix does not affect the argument.

For completeness, convexity gives
$e^{\lambda w}\le1-w+we^\lambda$ for every $w\in[0,1]$
and real $\lambda$. A bounded contribution of mean $\mu$
therefore has moment-generating function at most
$1-\mu+\mu e^\lambda$. Independence across matrices and
Chernoff optimization give
\[
 \Pr_\mu(\widehat\mu\ge x)
       \le e^{-R_{\rm sim}\operatorname{kl}(x,\mu)}
       \quad(x>\mu),
\]
and the analogous lower-tail inequality for $x<\mu$.
On either side of $\mu$, $\operatorname{kl}(x,\mu)$ is
monotone as $x$ moves away from $\mu$. Inverting these
two one-sided bounds at tail probability $\gamma/(2J_0)$,
then taking a union bound, proves \eqref{eq:u-v13-kl-gap}.
No Bernoulli sampling assumption is needed.
\end{proof}

Each matrix must be generated independently, including its null
reference rows; reusing a reference bank across matrices does not
meet this independence condition. The target coordinate set,
amplitude, candidate procedures and reported comparison family
are specified before the run. The procedure estimates an exact
finite-sample boundary-mixture target with simulation uncertainty.
It is not a certificate for the unknown power of an observed data
set, does not give a pointwise gap at all $a\ge a_0$, and does
not make an empirical choice of candidates or signal class valid
without accounting for that choice. Gaussian diagonal invariance
and the row budget remain essential. The change of measure is
an importance-sampling/Rao--Blackwell calculation, not a new
general optimality or concentration principle.

\end{vFifteenLowerHeadings}

\section{Refined efficiency and allocation costs}
\label{app:v15-costs}
\begin{vFifteenLowerHeadings}
\section{Proofs for the affine residual refinement}
\label{app:u-v14-affine-bound}

Throughout this section the experiment, alternative class and
comparison classes are those of Section~\ref{sec:u-v12-efficiency}.
All expectations in the auxiliary calculations are under the null.

\subsection{Moments and positivity of the affine coefficient}

Write $N=r+1$, $X\sim\operatorname{Beta}(1/2,r/2)$ and $V=NX$.
The Beta moments give
\[
 \mathbb E V=1,\qquad
 \operatorname{Var}(V)=\frac{2r}{r+3}=v_V>0.
\]
The coordinate boundary likelihood ratio is $G(X)$ in
\eqref{eq:u-v14-affine-moments}; hence $\mathbb E G=1$.
Every power-series term of $G$ is nonnegative. Termwise integration
and $\mathbb E X^{m+1}/\mathbb E X^m=(2m+1)/(N+2m)$ yield
\begin{equation}\label{eq:u-v14-poisson-covariance}
 c_{r,s}=\mathbb E\left(\frac{2K(N-1)}{N+2K}\right),
 \qquad K\sim\operatorname{Poisson}(s).
\end{equation}
The integrand is zero at $K=0$ and strictly positive for $K>0$
because $N\ge2$. Since $s>0$, this proves
$c_{r,s}>0$ and therefore $\gamma_{r,s}>0$.

Define the centered residual
\begin{equation}\label{eq:u-v14-centered-residual}
 H(X)=G(X)-1-\gamma_{r,s}(V-1).
\end{equation}
It has mean zero and variance $v_{r,s}^{\perp}$. This variance is
nonnegative because it is the minimum of
$\operatorname{Var}(G(X)-tV)$ over real $t$, attained at the
positive coefficient $t=\gamma_{r,s}$. In particular,
\begin{equation}\label{eq:u-v14-old-variance-comparison}
 v_{r,s}^{\perp}
 \le\operatorname{Var}(G(X)-se^{-s}V)
 =\operatorname{Var}(R_{a_0}(Y)),
\end{equation}
where $R_{a_0}$ is the remainder in
\eqref{eq:u-v13-positive-remainder}.

For $q$ independent selected coordinates, let
$\overline H=q^{-1}\sum_j H(X_j)$. The full boundary likelihood
ratio satisfies the exact identity
\begin{equation}\label{eq:u-v14-affine-decomposition}
 \ell(\bm Y)=1+\gamma_{r,s}\{B(\bm Y)-1\}+\overline H,
 \quad \mathbb E_0\overline H=0,
 \quad \mathbb E_0\overline H^2=v_{r,s}^{\perp}/q.
\end{equation}
The residual is a function of the observed-row maximal invariant.
A competing decision can depend on ancillary residual directions
and reference rows; no independence of that decision and
$\overline H$ is assumed.

\subsection{Proof of Proposition~\ref{prop:u-v14-affine-regret}}

For ideal tests, put $U=\phi_*-\phi_B$. The equal-size constraint
cancels the constant term in \eqref{eq:u-v14-affine-decomposition}.
Since thresholding $B$ maximizes $\mathbb E_0(B\phi)$ among tests
of the same size and $\gamma_{r,s}>0$,
\[
 0\le\overline\beta_*-\overline\beta_B
 \le\mathbb E_0(\overline H U)
 \le\sqrt{(v_{r,s}^{\perp}/q)\mathbb E_0U^2}.
\]
Because decisions lie in $[0,1]$ and both have mean $\tau$,
$\mathbb E_0U^2\le\mathbb E_0|U|\le2\min\{\tau,1-\tau\}$.

For finite references, the boundary average equals
$M^{-1}\mathbb E_0\sum_i\ell_i(\psi_i^*-\psi_i^B)$ by
\eqref{eq:u-v12-bank-change-measure}. Top-$k$ allocation of $B_i$
makes the corresponding $B_i$ sum nonpositive pointwise.
Equivariance then gives the same upper bound
$\mathbb E_0[\overline H_0(\psi_0^*-\psi_0^B)]$.
Both row decisions have null mean $\tau=k/M$, so the preceding
Cauchy--Schwarz argument applies unchanged.

Both the LR and bounded-quadratic rules are coordinate-permutation
and sign invariant and increase in each $|Y_j|$. Their worst-case
powers over $\Theta_{a_0}$ therefore equal their average boundary
powers. Combining the resulting variance inequality with
Theorem~\ref{thm:u-v13-bounded-regret} proves
\eqref{eq:u-v14-affine-regret}. Equation~\eqref{eq:u-v14-old-variance-comparison}
proves the stated improvement. The $q=1$ equality follows because
the two scores have the same strict ordering in $|Y_1|$.

\subsection{A residual-disagreement comparison}

The same residual controls a specified fitted rule or menu. For an
equal-size ideal rule define $d_\psi=\mathbb E_0|\phi_B-\psi|$;
for an equal-budget finite-reference rule define
\begin{equation}\label{eq:u-v14-null-disagreement}
 d_\psi=\frac1M\mathbb E_0\sum_i|\psi_i^B-\psi_i|.
\end{equation}
In either case $d_\psi\le2\min\{\tau,1-\tau\}$. With $D_\psi$
as in \eqref{eq:u-v13-null-allocation-cost} and its ideal analogue,
\begin{corollary}[A residual-disagreement criterion]
\label{cor:u-v14-affine-learning-cost}
For every competitor in the corresponding comparison class, with
exact size $\alpha$ or pathwise budget $k$, its average boundary
power satisfies
\begin{equation}\label{eq:u-v14-affine-cost}
 \left|\overline\beta_B-\overline\beta_\psi
                -\gamma_{r,s}D_\psi\right|
 \le\sqrt{\frac{d_\psi v_{r,s}^{\perp}}q}.
\end{equation}
Consequently $\gamma_{r,s}D_\psi>
\sqrt{d_\psi v_{r,s}^{\perp}/q}$ is sufficient for the bounded
quadratic to have strictly greater worst-case power on
$\Theta_{a_0}$ than the competitor.
\end{corollary}

The quantities $D_\psi$ and $d_\psi$ can be evaluated under the
canonical null for a fixed algorithm, including its fitting and
reference budget. A simulation-based sufficient condition must use
a joint lower confidence bound for $D_\psi$ and an upper bound for
$d_\psi$. This supplies a design comparison, not a power estimate
from the observed data. Neither the condition nor its positivity is
automatic: a learned rule may allocate rejections almost identically
to the bounded quadratic, and a large likelihood-ratio residual can
make the bound uninformative. The refinement uses ordinary linear
projection and Cauchy--Schwarz; its role is to quantify the present
comparison, not to introduce a new optimality principle.

\subsection{Proof of Corollary~\ref{cor:u-v14-affine-learning-cost}}

In the ideal case substitute $U=\phi_B-\psi$ into
\eqref{eq:u-v14-affine-decomposition}. Equal null size gives
\[
 \overline\beta_B-\overline\beta_\psi
 =\gamma_{r,s}D_\psi+\mathbb E_0(\overline H U).
\]
Here $\mathbb E_0U^2\le\mathbb E_0|U|=d_\psi$.
Cauchy--Schwarz proves \eqref{eq:u-v14-affine-cost}.
The finite-reference proof uses the same change-of-measure identity
and equal pathwise row budgets, followed by row equivariance.
Equivalently one can apply Cauchy--Schwarz to the joint experiment
formed by the all-null matrix and a uniformly sampled row index.
Its residual second moment is $v_{r,s}^{\perp}/q$ and its expected
absolute decision difference is exactly
\eqref{eq:u-v14-null-disagreement}.

The bounded-quadratic rule has worst-case power equal to its
average boundary power, whereas any competitor's minimum power
is at most its average over the boundary alternatives. Thus
\[
 \inf_{\Theta_{a_0}}\beta_B-
 \inf_{\Theta_{a_0}}\beta_\psi
 \ge\overline\beta_B-\overline\beta_\psi
 \ge\gamma_{r,s}D_\psi-
       \sqrt{d_\psi v_{r,s}^{\perp}/q}.
\]
This implication requires no monotonicity of the competitor's
power in amplitude. It does not assert a pointwise comparison
against that competitor at every $a\ge a_0$.

\section{Null-only allocation-cost certificates}
\label{app:u-v14-null-certificate}

\subsection{Proof of Proposition~\ref{prop:u-v14-null-certificate}}

Every bounded block score satisfies $0\le B_i\le N$.
Since $\psi^B$ maximizes the sum of $k$ selected $B_i$ and the
competitor has the same pathwise budget, $A_\psi\ge0$.
For decisions in $[0,1]$ with the same sum $k$, the positive
and negative parts of $\psi^B-\psi$ have equal total mass at
most $\min\{k,M-k\}$. Thus
\[
 0\le A_\psi\le\frac{N\min\{k,M-k\}}M,
 \qquad
 0\le H_\psi\le\frac{2\min\{k,M-k\}}M.
\]
This argument permits fractional decision probabilities as well as
randomized zero--one decisions, with independent marks regarded as part
of the null matrix. No independence between decisions and scores is used.

For completeness, let $X_1,\ldots,X_R$ be independent copies of a random
variable in $[0,1]$ with expectation $\mu$. Convexity gives
$\mathbb E e^{tX}\le1-\mu+\mu e^t$. Chernoff optimization consequently
yields the usual one-sided Bernoulli-relative-entropy bounds, although
the $X_h$ need not be Bernoulli. Define
\[
 \operatorname{kl}(x,u)=x\log(x/u)+(1-x)\log\{(1-x)/(1-u)\},
\]
with the continuous endpoint conventions. For $J_0$ scalar targets,
inverting
\begin{equation}\label{eq:u-v14-null-kl}
 R\operatorname{kl}(\overline X,u)
       \le\log(2J_0/\xi)
\end{equation}
gives simultaneous two-sided intervals with probability at least
$1-\xi$, by a union bound. Apply this to $A_\psi$ and $H_\psi$
after division by their respective deterministic bounds. The normalized
interval endpoints are multiplied back by those bounds. The implementation
solves the endpoint equations on the logit scale to retain accuracy for
very small endpoints.

On the resulting event, the positive-remainder identity
\eqref{eq:u-v13-block-comparator-identity} and
Corollary~\ref{cor:u-v14-affine-learning-cost} imply
\eqref{eq:u-v14-certified-margin}: $\gamma_{r,s}>0$, and replacing $D_\psi$
by a lower bound and $d_\psi$ by an upper bound preserves the direction
of both inequalities. Both statements hold for every amplitude using
the same pair of null moments, so no union bound over amplitudes is needed.
Each inequality first bounds average boundary powers; the bounded block
rule's minimum equals its boundary average, whereas the competitor's
minimum is no larger than that average. This establishes the claimed
worst-case comparison without a monotonicity assumption on the competitor.

Let $h(s)=(e^s-1-s)/s$ for $s>0$. Its power series
$h(s)=\sum_{m=2}^{\infty}s^{m-1}/m!$ shows that $h$ is strictly increasing
from zero to infinity. Hence $h(s_\psi)=D_\psi^-$ has a unique positive
root when $D_\psi^->0$, proving the strict amplitude interval.
The first guaranteed margin is
$e^{-s}\{1+s(1+D_\psi^-)\}-1$; its maximum occurs at
$s=D_\psi^-/(1+D_\psi^-)$ and equals
\begin{equation}\label{eq:u-v14-max-local-certificate}
 (1+D_\psi^-)\exp\{-D_\psi^-/(1+D_\psi^-)\}-1.
\end{equation}
A positive value may be small: this is a sufficient local comparison,
not a guarantee of high absolute power. Numerical evaluation of the
second margin at a finite grid of amplitudes does not by itself certify
an interval between grid points.

\subsection{A prespecified null experiment}

We use $p=30$, $N\in\{20,40\}$, $q\in\{5,10\}$,
$M=200$ and $k=10$. Each of the four designs has 3000 independent
all-null matrices. Nine fixed competitors are the coordinate quadratic,
coordinate maximum, three LPO scores with $\eta=c\sqrt p$ and
$c\in\{0.5,1,2\}$, the subspace-only LPO baseline, and the classical
four-score, LPO seven-score and combined thirteen-score menus.
Every variance estimate and all leave-pair-out fits are recomputed on
every row. The block score and all competitors share the same rows and
marks. No alternative data enter this study.

The original protocol prespecified the 36 allocation costs. While the
null simulation was running, before examining its cost estimates, the
protocol was extended to report disagreement intervals from the same
archived decisions. The failure probability $0.05$ is allocated jointly
over all 72 scalar targets. The original generation source and protocol
amendment are retained; the data-generation code was unchanged.
The mathematical confidence guarantee concerns independent simulated
matrices. The numerical implementation uses ordinary floating-point
arithmetic, not interval-certified arithmetic.

\begin{table}[htbp]
\centering
\caption{Null-only allocation-cost certificates. Each cell uses $p=30$, $M=200$, $k=10$ and 3000 independent all-null matrices. The lower bounds $d$ have simultaneous 95\% Monte Carlo coverage jointly with all 36 disagreement intervals. For every $0<a_0<a_{\max}$, the bounded block rule has strictly greater worst-case power than the specified competitor in the axis-sparse class on that same coverage event.}
\label{tab:u-v14-allocation-cost}
\begin{tabular}{rrlrrr}
\toprule
$N$ & $q$ & Competitor & $\widehat D$ & $d$ & $a_{\max}$ \\
\midrule
20 & 5 & LPO, $c=0.5$ & 0.0404 & 0.0277 & 0.330 \\
20 & 5 & LPO, $c=1$ & 0.0482 & 0.0342 & 0.366 \\
20 & 5 & LPO, $c=2$ & 0.0551 & 0.0400 & 0.395 \\
20 & 5 & Combined menu & 0.0482 & 0.0342 & 0.366 \\
20 & 10 & LPO, $c=0.5$ & 0.0197 & 0.0112 & 0.211 \\
20 & 10 & LPO, $c=1$ & 0.0277 & 0.0174 & 0.262 \\
20 & 10 & LPO, $c=2$ & 0.0342 & 0.0226 & 0.298 \\
20 & 10 & Combined menu & 0.0319 & 0.0207 & 0.286 \\
40 & 5 & LPO, $c=0.5$ & 0.0449 & 0.0265 & 0.323 \\
40 & 5 & LPO, $c=1$ & 0.0526 & 0.0326 & 0.357 \\
40 & 5 & LPO, $c=2$ & 0.0590 & 0.0376 & 0.383 \\
40 & 5 & Combined menu & 0.0519 & 0.0320 & 0.354 \\
40 & 10 & LPO, $c=0.5$ & 0.0221 & 0.0101 & 0.201 \\
40 & 10 & LPO, $c=1$ & 0.0308 & 0.0162 & 0.253 \\
40 & 10 & LPO, $c=2$ & 0.0378 & 0.0212 & 0.289 \\
40 & 10 & Combined menu & 0.0345 & 0.0188 & 0.273 \\
\bottomrule
\end{tabular}
\end{table}

\begin{table}[htbp]
\centering
\caption{Numerically evaluated affine lower bounds on the bounded block rule\textquotesingle s worst-case power advantage (percentage points), using the same simultaneous null confidence event as Table~\ref{tab:u-v14-allocation-cost}. Negative entries do not certify an advantage; they do not imply a power disadvantage. The Beta integrals are evaluated with the independent precision checks in Table~\ref{tab:u-v14-bounds}.}
\label{tab:u-v14-allocation-margins}
\begin{tabular}{rrlrr}
\toprule
$N$ & $q$ & Competitor & $a_0=0.5$ & $a_0=1$ \\
\midrule
20 & 5 & LPO, $c=1$ & 0.327 & 0.174 \\
20 & 5 & LPO, $c=2$ & 0.396 & 0.410 \\
20 & 5 & Combined menu & 0.319 & 0.056 \\
20 & 10 & LPO, $c=1$ & 0.151 & -0.143 \\
20 & 10 & LPO, $c=2$ & 0.212 & 0.062 \\
20 & 10 & Combined menu & 0.183 & -0.120 \\
40 & 5 & LPO, $c=1$ & 0.293 & -0.163 \\
40 & 5 & LPO, $c=2$ & 0.352 & 0.030 \\
40 & 5 & Combined menu & 0.276 & -0.336 \\
40 & 10 & LPO, $c=1$ & 0.128 & -0.354 \\
40 & 10 & LPO, $c=2$ & 0.187 & -0.173 \\
40 & 10 & Combined menu & 0.149 & -0.406 \\
\bottomrule
\end{tabular}
\par\smallskip\parbox{0.94\linewidth}{\footnotesize No positive affine margin is obtained at $a_0=2$ or $4$ for these reported competitors. The entire 144-row grid is retained in the archive. Grid evaluation does not certify intermediate amplitudes.}
\end{table}

The archive stores the block scores and all bit-packed row decisions,
allowing reconstruction of every contribution and rejection budget.
The first matrix of each design is also evaluated by the separate NumPy
implementation of the all-pair LPO calculations. Reported cost differences
reflect complete rejection allocations; they are not decompositions into
shape fitting, direction fitting and menu penalties.

The affine margin uses the one-dimensional Beta moments evaluated in
Table~\ref{tab:u-v14-bounds}. These deterministic quantities do not
consume the null Monte Carlo failure budget. Their reported numerical
values pass independent precision and quadrature checks, but no
floating-point interval enclosure is claimed. The underlying theorem
uses their exact values. The positive-remainder interval follows from
the analytic monotonicity argument; isolated positive affine-grid values
are reported only at the corresponding amplitudes.

\section{Conditional likelihood-ratio bounds for the maximum statistic}
\label{app:u-v16-conditional-power}

We work throughout in the paired diagonal Gaussian experiment of
Section~\ref{sec:u-v12-efficiency}, with fixed $N\ge2$, $J$ and
$a_0>0$. Null expectations refer to its canonical maximal invariant.
The variable $\mathsf S$ denotes the largest transformed coordinate,
distinct from the sample standard deviations $S_j$.

\subsection{Conditional moments and the reduced experiment}

Write $G=G_{a_0}$ and let $f_r$ and $F_r$ be the density and cdf of
$\operatorname{Beta}(1/2,r/2)$. The identity
\eqref{eq:u-v12-symmetric-t-ratio} gives
$\ell=q^{-1}\sum_{j\in J}G(X_j)$. For $0<x<1$, put
\begin{equation}\label{eq:u-v16-truncated-moments}
 I_b(x)=\int_0^xG(z)^b f_r(z)\,dz,\qquad b=1,2.
\end{equation}
Then
\begin{align}
 h_{a_0}(x)&=\{G(x)+(q-1)I_1(x)/F_r(x)\}/q,\nonumber\\
 v_{a_0}(x)&=\frac{q-1}{q^2}
       \{I_2(x)/F_r(x)-I_1(x)^2/F_r(x)^2\}.
 \label{eq:u-v16-explicit-hv}
\end{align}
Conditional on $\mathsf S=x$ and the index of the maximum, the
remaining $q-1$ coordinates are independent with density
$f_r(z)1\{z<x\}/F_r(x)$. Their labels do not affect their sum.
This proves the conditional-mean and conditional-variance identities
in \eqref{eq:u-v16-conditional-moments}. When $q=1$, these identities
hold with $h_{a_0}=G$ and $v_{a_0}=0$.

All nonconstant coefficients in the power series of $G$ are positive.
Thus $G$ is strictly increasing. Its truncated mean is nondecreasing,
as also follows by differentiating:
\[
 \frac{d}{dx}\frac{I_1(x)}{F_r(x)}
 =\frac{f_r(x)}{F_r(x)}
   \{G(x)-I_1(x)/F_r(x)\}\ge0.
\]
Consequently $h_{a_0}$ is strictly increasing. The likelihood ratio
in the experiment retaining only $\mathsf S$ is
$\mathbb E_0(\ell\mid\mathsf S)=h_{a_0}(\mathsf S)$.
Its likelihood-ratio test therefore ranks the same maximum statistic.
This reduction does not assert that $\mathsf S$ is sufficient for the
full experiment. The function $G$ is bounded on $[0,1]$ for fixed
$N,a_0$, so all moments used below exist.

\subsection{Proof of the ideal bound}

For a nonnegative likelihood ratio $L$ with $\mathbb E_0L=1$, the
optimal level-$\alpha$ power has the standard positive-part
variational representation \citep{rockafellar2002cvar}
\begin{equation}\label{eq:u-v16-np-dual}
 \sup_{0\le\phi\le1,\,\mathbb E_0\phi\le\alpha}
       \mathbb E_0(L\phi)
 =\inf_{t\ge0}
       \{\alpha t+\mathbb E_0(L-t)_+\}.
\end{equation}
Indeed $L\phi\le t\phi+(L-t)_+$, and the likelihood-ratio test,
with boundary randomization when necessary, attains equality at a
level-$\alpha$ cutoff. In the present continuous experiment
randomization is unnecessary.

Let $H=h_{a_0}(\mathsf S)$ and let $c_\alpha$ be as in
Theorem~\ref{thm:u-v16-conditional-power}. The maximum rule is
$\phi_{\max}=1\{H>c_\alpha\}$, with exact size $\alpha$, and
\[
 \mathbb E_0(\ell\phi_{\max})
 =\mathbb E_0(H\phi_{\max})
 =\alpha c_\alpha+\mathbb E_0(H-c_\alpha)_+.
\]
Evaluating \eqref{eq:u-v16-np-dual} for $L=\ell$ at this same
$c_\alpha$ bounds the difference of boundary-mixture powers by
\begin{equation}\label{eq:u-v16-jensen-gap}
 \mathbb E_0\{(\ell-c_\alpha)_+-(H-c_\alpha)_+\}.
\end{equation}
Condition on $\mathsf S$. The identity $z_+=(z+|z|)/2$ cancels
the conditional means in \eqref{eq:u-v16-jensen-gap}. Moreover,
\[
 \mathbb E_0(|\ell-c_\alpha|\mid\mathsf S)
 \le\{v_{a_0}(\mathsf S)+(H-c_\alpha)^2\}^{1/2}
\]
by Cauchy--Schwarz. This applies the two-moment positive-part bound
of \citet{gallego1993distribution} conditionally. The resulting upper bound is precisely the
right-hand side of \eqref{eq:u-v16-ideal-bound}.

\subsection{Proof of the finite-reference bound}

Let $V_k(z_0,\ldots,z_{M-1})$ be the sum of the $k$ largest
arguments. For every real vector,
\begin{equation}\label{eq:u-v16-topk-dual}
 V_k(\bm z)=\inf_{t\in\mathbb R}
       \left\{kt+\sum_{i=0}^{M-1}(z_i-t)_+\right\}.
\end{equation}
Every $t$ between the $(k+1)$st and $k$th largest arguments attains
the infimum. Under the all-null law the rows are independent. Put
$H_i=h_{a_0}(\mathsf S_i)$. Conditional on the entire vector of
maxima, $\ell_i$ has mean $H_i$ and variance
$v_{a_0}(\mathsf S_i)$.

The change of measure in \eqref{eq:u-v12-bank-change-measure} gives
the boundary-average observed-row powers
\begin{align}
 \overline\beta_*^{M,k}
   &=\frac1M\mathbb E_0V_k(\ell_0,\ldots,\ell_{M-1}),\nonumber\\
 \overline\beta_{\max}^{M,k}
   &=\frac1M\mathbb E_0\sum_i\ell_i\psi_i^{\max}
    =\frac1M\mathbb E_0V_k(H_0,\ldots,H_{M-1}).
 \label{eq:u-v16-conditional-topk-power}
\end{align}
Here $\psi^{\max}$ selects the $k$ largest maxima, equivalently
the $k$ largest $H_i$. Power is the probability of rejecting the
alternative-assigned row when that row has a uniformly chosen index
and a boundary-mixture distribution. It is not the probability of
rejecting at least one of the hypothetical rows.

Choose $c$ to be the $k$th largest $H_i$. It is measurable with
respect to all maxima and attains the infimum in
\eqref{eq:u-v16-topk-dual} for $\bm H$. Evaluating that formula
for $\bm\ell$ at $c$ and conditioning on all maxima yields
\[
 \mathbb E_0\{V_k(\bm\ell)-V_k(\bm H)\mid\bm{\mathsf S}\}
 \le\sum_i\mathcal R\{v_{a_0}(\mathsf S_i),H_i-c\}.
\]
Together with \eqref{eq:u-v16-conditional-topk-power}, this proves
the claimed upper bound for boundary-average powers. The same
argument allows any maxima-measurable $c$ between the $k$th and $(k+1)$st largest
order statistics. It does not condition validity on a fixed realized
reference bank; power and level average over the observations and
the reference simulation.

\subsection{Worst-case power and a simpler residual bound}

The LR and maximum scores are sign invariant, invariant to
permutations within $J$, and increasing in every $|Y_j|$. For a
noncentral $t_r$ variable, its absolute value is stochastically
increasing with nonnegative noncentrality: condition on its positive
chi-square denominator and use the corresponding property of
$|Z+a|$. Hence both ideal powers increase with the active amplitude.
For finite references, condition first on the null reference scores;
increasing the observed score increases its top-$k$ decision, so the
same conclusion follows after averaging. Both rules consequently
attain their worst-case powers at $a=a_0$, with equal power at all
$2q$ boundary alternatives. Their boundary-average powers used
above equal the minima in the theorem. The nonnegative gaps follow
from Theorems~\ref{thm:u-v12-ideal-maximin} and
\ref{thm:u-v12-bank-maximin}. For $q=1$ the two scores have identical
strict orderings and $v_{a_0}=0$, giving equality.

A convenient, although usually looser, consequence is
\begin{equation}\label{eq:u-v16-residual-bound}
 \beta_*-\beta_{\max}
 \le\sqrt{\tau(1-\tau)\mathbb E_0v_{a_0}(\mathsf S)},
 \qquad
 \tau=\alpha\ \text{or}\ k/M.
\end{equation}
To see this, write $R=\ell-H$. In the ideal experiment, thresholding
$H$ maximizes $\mathbb E_0(H\phi)$ at size $\alpha$, and
$\mathbb E_0(R\mid\mathsf S)=0$. The power gap is therefore at
most $\mathbb E_0\{R(\phi_*-\alpha)\}$. Apply Cauchy--Schwarz
and $\operatorname{Var}_0(\phi_*)=\alpha(1-\alpha)$.
For finite references, the corresponding $H_i$ allocation term is
nonpositive pointwise and the maximum allocation is measurable
from the maxima. Row equivariance reduces the remaining expectation
to $\mathbb E_0\{R_0(\psi_0^*-k/M)\}$. Its two second moments are
$\mathbb E_0v_{a_0}(\mathsf S)$ and $(k/M)(1-k/M)$.
No independence between a decision and its likelihood-ratio residual
is used in either argument.

\subsection{Exact low-dimensional integral representations}

Set $U=F_r(\mathsf S)^q$, which is uniform under the null, and define
\[
 \mathcal H(u)=h_{a_0}\{F_r^{-1}(u^{1/q})\},\qquad
 \mathcal V(u)=v_{a_0}\{F_r^{-1}(u^{1/q})\}.
\]
The ideal right-hand side in \eqref{eq:u-v16-ideal-bound} equals
\begin{equation}\label{eq:u-v16-ideal-integral}
 \int_0^1\mathcal R\{\mathcal V(u),
                    \mathcal H(u)-\mathcal H(1-\alpha)\}\,du.
\end{equation}
For numerical integration this interval is split at $1-\alpha$,
where the absolute-value term can have a change in derivative.

For finite references let $T$ be the $k$th largest of the $M$
independent uniforms. Then $T\sim\operatorname{Beta}(M-k+1,k)$.
Given $T=t$, there is one threshold row, $M-k$ unordered uniforms
below $t$, and $k-1$ above $t$. Put
$j(u,t)=\mathcal R\{\mathcal V(u),\mathcal H(u)-\mathcal H(t)\}$.
The right-hand side of \eqref{eq:u-v16-finite-bound} is exactly
\begin{equation}\label{eq:u-v16-finite-integral}
 \frac1M\mathbb E_T\left[
   \frac{\sqrt{\mathcal V(T)}}2
   +\frac{M-k}{T}\int_0^Tj(u,T)\,du
   +\frac{k-1}{1-T}\int_T^1j(u,T)\,du\right].
\end{equation}
The first term includes the threshold row. Thus finite-reference
calibration is evaluated directly rather than replaced by its ideal
limit.

The conditional moments in \eqref{eq:u-v16-truncated-moments} can
be computed by direct Beta integration or a positive series.
With $(a)_j$ denoting the rising factorial and $I_x(a,b)$ the
regularized incomplete Beta function,
\begin{align}
 G(x)&=\sum_{j\ge0}c_jx^j,\qquad
 c_j=e^{-s}\frac{s^j}{j!}\frac{(N/2)_j}{(1/2)_j},\nonumber\\
 I_1(x)&=\sum_{j\ge0}e^{-s}\frac{s^j}{j!}
                         I_x(j+1/2,r/2),\label{eq:u-v16-moment-series}\\
 I_2(x)&=\sum_{m\ge0}
     \left(\sum_{j=0}^mc_jc_{m-j}\right)
     \frac{(1/2)_m}{(N/2)_m}I_x(m+1/2,r/2).\nonumber
\end{align}
Nonnegativity justifies termwise integration. These formulas and
\eqref{eq:u-v16-finite-integral} require only deterministic model
integration; they consume no observations for tuning. The quadrature
values, convergence diagnostics and an independent direct-moment
implementation accompany Table~\ref{tab:u-v16-conditional-bounds}.

\section{A specified mixed-signal orbit}
\label{app:u-v14-mixed-orbit}

\subsection{Proof of Proposition~\ref{prop:u-v14-mixed-maximin}}
Let $\mathcal O_{a_0,d_0}$ consist of the $q2^{h+1}$ boundary
vectors with one coordinate in $J$ equal to $\pm a_0$, the other
coordinates in $J$ zero, and every outside coordinate equal to
$\pm d_0/\sqrt h$. The coordinate location and all signs are uniform
and independent. The maximal-invariant reduction in
Section~\ref{app:u-v12-efficiency} applies unchanged. Its coordinate
$t$ statistics are independent, and the residual directions are
ancillary. Averaging the likelihood ratio separately over the inside
coordinate and the outside signs gives exactly
\eqref{eq:u-v14-mixed-lr}. In particular,
$\mathbb E_0\ell_{a_0,d_0}=1$.

The likelihood ratio is unchanged by sign changes in any coordinate
and permutations within $J$. These transformations act transitively
on $\mathcal O_{a_0,d_0}$. Consequently its ideal upper-tail test and
its top-$k$ finite-reference version have the same power at every
boundary vector. The functions $g_a(t)$ are positive and increasing
in $|t|$ for $a>0$, so the product score is increasing in every
$|Y_j|$. Under a fixed alternative, the $|Y_j|$ are independent and
stochastically increase with $|\theta_j|$ by
\eqref{eq:u-v12-amplitude-order}. Coupling the coordinates separately
shows that the benchmark power cannot decrease when the active inside
amplitude or any outside magnitude increases from its boundary value.
For the finite-reference procedure, condition first on its independent
null reference rows: the observed-row rejection event is increasing
in its score, and therefore the same argument applies. Thus each
benchmark's worst-case power is its common boundary power.

For any invariant competitor, its worst-case power is at most its
average over $\mathcal O_{a_0,d_0}$. In the ideal case,
Neyman--Pearson optimality bounds that average by the corresponding
average of the likelihood-ratio test. In the finite-reference case,
replace $\ell_i$ in \eqref{eq:u-v12-bank-change-measure} by
$\ell_{a_0,d_0}(\bm Y_i)$. The pointwise top-$k$ allocation maximizes
the resulting sum under the same row budget. Both bounds equal the
benchmark's worst-case power, proving the proposition. If $d_0=0$
and the outside mean is fixed at zero, the outside factors equal one
and the original sparse argument is recovered. This proof extends a
classical invariant-mixture construction; it supplies no new general
optimality principle.

\subsection{Evaluation and scope}
Writing $\mathcal L_j(a)=\log g_a(Y_j)$ gives the stable expression
\begin{equation}\label{eq:u-v14-log-mixed-lr}
 \log\ell_{a_0,d_0}(\bm Y)
 =\operatorname{logsumexp}_{j\in J}\{\mathcal L_j(a_0)\}
     -\log q+\sum_{k\notin J}\mathcal L_k(d_0/\sqrt h).
\end{equation}
The score costs $p$ univariate likelihood-ratio evaluations.
The prescription uses independent outside signs, with identical
boundary magnitudes. In particular, total outside energy
$\|\bm\theta_{J^c}\|^2\ge d_0^2$ alone does not define the class:
it could allow concentration into too few coordinates for the stated
coordinatewise monotonicity argument to establish the same minimum.
The values $J,a_0,d_0$ are supplied before observing the tested data.
A mixture over different allocations is a likelihood ratio against
that specified mixture, but is not automatically maximin over the
union of their parameter classes.

The boundary-power identity of Proposition~\ref{prop:u-v13-tilted-gap}
also applies. To simulate one augmented matrix, choose one of its
$M$ rows uniformly to have the uniform boundary-orbit distribution;
all other rows are null. On complete maximal invariants, its density
relative to $P_0^M$ is $M^{-1}\sum_i\ell_{a_0,d_0}(\bm Y_i)$.
Consequently the normalized weights
$w_i=\ell_{a_0,d_0}(\bm Y_i)/\sum_h\ell_{a_0,d_0}(\bm Y_h)$
give a bounded power contribution $\sum_iw_i\psi_i$, and the
benchmark-minus-competitor contribution is nonnegative and at most
one. Their expectations are average boundary powers and gaps,
respectively. These identities justify simulation confidence bounds
for a fixed allocation; they do not estimate the power of an
unspecified signal from the observed data.

\end{vFifteenLowerHeadings}

\section{Simulation designs and external comparisons}
\label{app:v15-simulations}
\subsection{Additional summaries of archived comparisons}
These tables retain the compact summaries from the earlier complete manuscript.
They concern their stated experimental designs; they are not new simulation runs.

\begin{table}[tbp]
\centering\small
\caption{Selected raw-data comparisons at $p=30$. Entries are rejection
percentages (cluster Monte Carlo standard errors, in percentage points).
Inside and outside alternatives have $\Lambda_{\mathrm{all}}=16$ and
lie on the first and last true-whitened coordinates, respectively.}
\label{tab:u-focused-benchmarks}
\setlength{\tabcolsep}{4pt}
\begin{tabular}{rlrrrr}
\toprule
$N$ & Signal & RI & No spike & CQ & F-type\\
\midrule
40 & Null & 4.40 (0.07)&3.65 (0.07)&6.49 (0.08)&5.22 (0.07)\\
40 & Inside &53.75 (0.21)&19.02 (0.17)&22.65 (0.13)&19.06 (0.13)\\
40 & Outside &4.98 (0.08)&18.88 (0.17)&81.78 (0.12)&79.22 (0.13)\\
80 & Null &4.49 (0.08)&4.07 (0.07)&6.46 (0.07)&5.16 (0.07)\\
80 & Inside &57.01 (0.22)&20.89 (0.18)&22.45 (0.14)&19.00 (0.13)\\
80 & Outside &5.22 (0.09)&20.92 (0.18)&81.82 (0.12)&79.39 (0.13)\\
\bottomrule
\end{tabular}
\end{table}

\begin{table}[tbp]
\centering\small
\caption{Full-data diagnostic at $p=300$, $N=40$ and $\rho=4$.
Entries are rejection percentages. Conc. and Dense place all signal
outside the specified subspace, on one coordinate or diffusely,
respectively. Every entry uses $10\,000$ datasets. The largest
bank-based Monte Carlo standard error in the complete experiment
is $0.84$ percentage points.}
\label{tab:u-v10-full-data}
\setlength{\tabcolsep}{3pt}
\begin{tabular}{lrrrrrrr}
\toprule
 & Null & \multicolumn{3}{c}{$\Lambda_{\rm all}=16$} & \multicolumn{3}{c}{$\Lambda_{\rm all}=64$}\\
\cmidrule(lr){3-5}\cmidrule(lr){6-8}
Method & & Inside & Conc. & Dense & Inside & Conc. & Dense\\
\midrule
Full-data mean-axis score & 4.88 & 83.40 & 6.63 & 6.77 & 100.00 & 17.47 & 16.89\\
Full-data maximum score & 4.88 & 83.41 & 6.62 & 6.76 & 100.00 & 17.48 & 16.91\\
Classical block $t^2$ & 4.89 & 83.38 & 6.62 & 6.81 & 100.00 & 17.27 & 16.89\\
Subspace diagonal $T^2$ & 4.75 & 84.62 & 4.75 & 4.75 & 100.00 & 4.75 & 4.75\\
All-coordinate $t^2$ & 5.28 & 15.61 & 15.93 & 15.50 & 70.72 & 70.84 & 73.41\\
Subspace paired DLRT & 4.75 & 83.72 & 4.75 & 4.75 & 100.00 & 4.75 & 4.75\\
Split RI, $\eta=\sqrt p$ & 4.29 & 48.59 & 5.78 & 5.68 & 99.25 & 14.11 & 13.01\\
Split RI, $\eta=0$ & 3.69 & 7.04 & 7.19 & 6.91 & 27.26 & 27.44 & 27.23\\
Split projection $t$ & 5.18 & 54.33 & 5.18 & 5.18 & 99.52 & 5.18 & 5.18\\
CQ & 5.47 & 9.31 & 9.33 & 16.35 & 29.13 & 30.69 & 73.57\\
$F$-type & 5.03 & 8.45 & 8.66 & 15.24 & 27.45 & 29.01 & 72.04\\
Li $U$--$t$ & 5.47 & 9.02 & 9.25 & 16.19 & 28.89 & 30.38 & 73.28\\
Cross-$U$ $t$ & 4.71 & 6.99 & 7.16 & 10.78 & 18.66 & 19.64 & 47.71\\
Bonferroni $t$ & 4.72 & 47.63 & 48.15 & 6.40 & 99.98 & 100.00 & 12.43\\
Subspace Hotelling & 4.96 & 82.97 & 4.96 & 4.96 & 100.00 & 4.96 & 4.96\\
\bottomrule
\end{tabular}

\end{table}

\begin{vFifteenLowerHeadings}
\section{Numerical study of the unbiased two-sample RI U statistic}
\label{sec:u-numerical}

We now evaluate the unbiased two-sample RI U statistic, retaining its
sample-covariance subtraction terms. The study examines finite-sample
calibration, the cost of learning an unrestricted direction, and the gain
from a prespecified low-dimensional direction space. It does not reserve
observations for empirical power validation.

\subsection{Design and finite Monte Carlo calibration}

The observations follow the common-shape Gaussian model, with the known
shape removed by whitening. The parameter combinations are
\[
 p\in\{20,100,300\},\qquad N_1=N_2=N\in\{40,80\},\qquad
 \sigma_1^2=1,\quad \sigma_2^2\in\{1,4\},\qquad
 f\in\{1/4,1/2\}.
\]
Each group supplies $m=fN$ training and $n=(1-f)N$ testing observations.
The alternatives have a mean difference along the first whitened coordinate
and are indexed by
\[
 \Lambda_{\mathrm{all}}
 =\frac{\bm\delta^\top\bm H^{-1}\bm\delta}
        {(\sigma_1^2+\sigma_2^2)/N}
 \in\{0,8,16\}.
\]
Thus the training and testing noncentralities are
$f\Lambda_{\mathrm{all}}$ and $(1-f)\Lambda_{\mathrm{all}}$.
Holding $\Lambda_{\mathrm{all}}$ fixed while increasing $N$ preserves
these noncentralities and primarily changes the residual-estimation
accuracy. The known-shape assumption implies identical whitened laws for
every fixed positive-definite $\bm H$; different known shapes are not
counted as additional simulation scenarios.

The three RI procedures use $b=1$ and $\eta=p$. Learned U uses the
normalized full training contrast. Subspace U projects that contrast onto
the first five whitened coordinates before normalization. This space is
fixed before sampling and contains the true signal; its use represents
additional structural information. The full $p$-dimensional baseline in
the testing statistic is retained. Oracle U uses the true direction with
the same testing sample sizes, residual scale estimator, and reference
calibration. Two further benchmarks use $\eta=0$: split norm U uses the
testing contrast, whereas all-data norm U uses all $N$ observations per
group for its contrast and residuals. The latter has no training fraction.

All procedures use the nuisance-envelope calibration of
Section~\ref{sec:u-feasible}. For each outer repetition we generate
$L=199$ independent reference-envelope values and reject when the
observed studentized statistic exceeds their 190th ascending order
statistic. This is exactly the Monte Carlo rule at level $0.05$, since
$\lfloor0.05(L+1)\rfloor/(L+1)=0.05$. The reference draws are
computational random variables and consume no additional observations.
Fresh reference banks are generated independently across all outer
repetitions. The same independent cutoff vector can be reused across
alternative configurations with identical $(p,m,n)$; it is not held fixed
across repetitions. This design averages over the calibration randomness
required by Theorem~\ref{thm:u-feasible-level}.

The data are generated through their exact Gaussian sufficient statistics.
For each candidate direction, the same testing residual sums are used in
both the negative Wishart term and the pooled residual scale, preserving
their dependence. Procedure-specific observations are generated from
independent exact marginals; no paired uncertainty comparison is made.
There are $100\,000$ repetitions per procedure in each of the $72$
configurations, using seed $20261023$. The maximum Monte Carlo standard
error is below $0.159$ percentage points. The code records rejection counts,
standard errors, Wilson intervals, reference-cutoff summaries, and exact
training-alignment expectations.

\subsection{Size and power}

Table~\ref{tab:u-size} reports null rejection percentages. The learned RI
sizes range from $4.24\%$ to $4.78\%$, and the oracle RI sizes range
from $4.30\%$ to $4.83\%$. The envelope is conservative, as expected.
The norm benchmarks attain sizes between $3.65\%$ and $4.45\%$.
Consequently, comparisons across the RI and norm spectra do not establish
efficiency dominance at equal attained size. The learned, subspace, and
oracle RI procedures have the same null distribution and calibration;
their power comparison directly concerns directional information.

\begin{table}[tbp]
 \centering
 \caption{Empirical null rejection percentages for the unbiased two-sample
 statistics. Each range covers both variance ratios and both training
 fractions at the indicated $(p,N)$. The nominal level is $5\%$.}
 \label{tab:u-size}
 \small
 \setlength{\tabcolsep}{3pt}
 \begin{tabular}{rrccccc}
\toprule
$p$ & $N$ & Learned U & Subspace U & Oracle U & Split norm U & All-data norm U\\
\midrule
20 & 40 & 4.24--4.44 & 4.37--4.40 & 4.30--4.43 & 3.65--3.92 & 3.99--4.04 \\
20 & 80 & 4.57--4.60 & 4.45--4.62 & 4.46--4.59 & 4.13--4.33 & 4.20--4.33 \\
100 & 40 & 4.45--4.52 & 4.43--4.56 & 4.46--4.66 & 3.81--4.04 & 3.92--4.14 \\
100 & 80 & 4.61--4.78 & 4.60--4.69 & 4.55--4.66 & 4.08--4.27 & 4.28--4.38 \\
300 & 40 & 4.48--4.58 & 4.40--4.59 & 4.44--4.49 & 3.73--4.10 & 4.08--4.20 \\
300 & 80 & 4.59--4.77 & 4.61--4.71 & 4.42--4.83 & 4.15--4.26 & 4.24--4.45 \\
\bottomrule
\end{tabular}

\end{table}

Table~\ref{tab:u-power} gives powers for
$\Lambda_{\mathrm{all}}=16$ and variance ratio four;
Figure~\ref{fig:u-learning} also includes
$\Lambda_{\mathrm{all}}=8$. At $N=40$, $f=1/4$, and
$\Lambda_{\mathrm{all}}=16$, the unrestricted learner has powers
$33.87\%$, $11.54\%$, and $6.86\%$ at $p=20,100,300$.
The corresponding oracle powers are $92.00\%$, $92.25\%$, and
$92.46\%$. Thus using the unbiased two-sample RI U statistic does not
eliminate the statistical cost of estimating a dense weak direction.

The prespecified five-dimensional space changes this comparison
substantially. Its powers in the same configurations are $63.74\%$,
$62.30\%$, and $62.04\%$. Training uncertainty is now governed by
the dimension of that space, while the complete testing quadratic still
contains its $p$-dimensional baseline. The remaining gap from the oracle
is also visible. These results support structured direction learning as
a concrete use of the theory; they do not establish that an unknown useful
subspace can be selected at no statistical cost.

\begin{table}[tbp]
 \centering
 \caption{Empirical powers (percent) for unbiased U procedures, with
 $\Lambda_{\mathrm{all}}=16$ and variance ratio four. Subspace U uses a
 prespecified five-dimensional space containing the signal. Oracle U uses
 the same split and calibration as the learned procedures. The two
 all-data entries at different $f$ are separate Monte Carlo estimates of
 the same population power.}
 \label{tab:u-power}
 \small
 \setlength{\tabcolsep}{3pt}
 \begin{tabular}{rrrccccc}
\toprule
$p$ & $N$ & $f$ & Learned U & Subspace U & Oracle U & Split norm U & All-data norm U\\
\midrule
20 & 40 & 0.25 & 33.87 & 63.74 & 92.00 & 43.24 & 59.55 \\
20 & 40 & 0.50 & 33.80 & 59.25 & 77.78 & 26.53 & 59.78 \\
20 & 80 & 0.25 & 34.30 & 64.14 & 92.29 & 44.99 & 61.19 \\
20 & 80 & 0.50 & 34.48 & 60.10 & 78.58 & 27.86 & 61.13 \\
100 & 40 & 0.25 & 11.54 & 62.30 & 92.25 & 17.58 & 25.45 \\
100 & 40 & 0.50 & 12.04 & 58.86 & 78.54 & 11.33 & 25.13 \\
100 & 80 & 0.25 & 11.50 & 62.61 & 92.44 & 18.51 & 26.63 \\
100 & 80 & 0.50 & 12.63 & 59.59 & 78.99 & 12.10 & 26.59 \\
300 & 40 & 0.25 & 6.86 & 62.04 & 92.46 & 10.53 & 13.84 \\
300 & 40 & 0.50 & 7.13 & 58.94 & 78.40 & 7.43 & 13.72 \\
300 & 80 & 0.25 & 7.06 & 62.42 & 92.69 & 10.96 & 14.22 \\
300 & 80 & 0.50 & 7.24 & 59.38 & 79.41 & 7.97 & 14.57 \\
\bottomrule
\end{tabular}

\end{table}

\begin{figure}[tbp]
 \centering
 \includegraphics[width=0.97\textwidth]{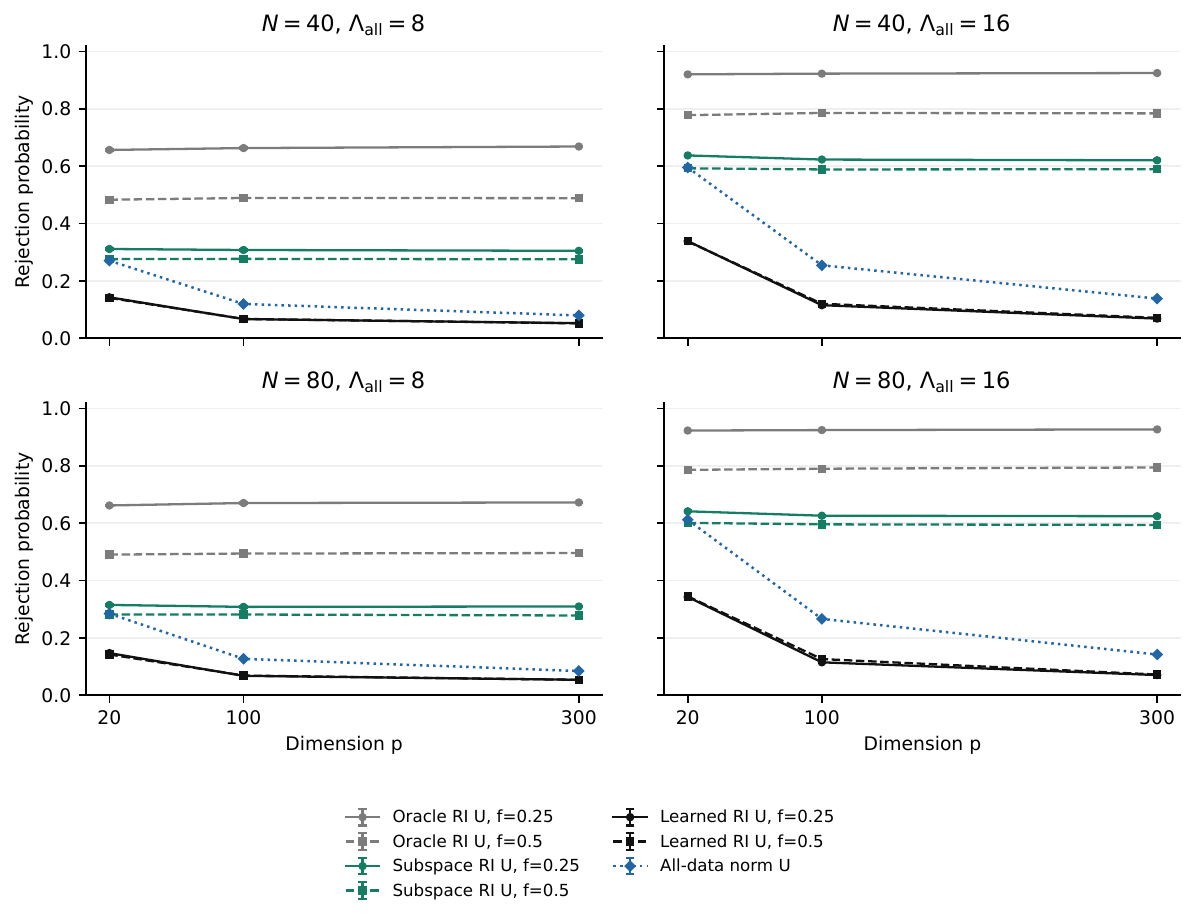}
 \caption{Power of the unbiased two-sample procedures at fixed total sample
 budgets, with variance ratio four. Error bars are $1.96$ Monte Carlo
 standard errors and are often smaller than the plotting symbols.
 The all-data norm procedure has no training fraction; its $f=1/4$
 simulation entry is displayed. All reference banks are refreshed across
 outer repetitions, so calibration randomness is included.}
 \label{fig:u-learning}
\end{figure}

\subsection{Implementation checks}

The empirical squared alignments of both learned directions agree with
their exact Poisson-mixture expectations within $2.50$ Monte Carlo standard
errors across all configurations. A separate experiment generated
$20\,000$ raw Gaussian datasets at $p=20$, $N=40$, $f=1/4$, variance
ratio four, and $\Lambda_{\mathrm{all}}=16$. Its five rejection rates
agree with the sufficient-statistic experiment within $2.20$ combined
Monte Carlo standard errors. For the first $250$ raw datasets and the full-dimensional learner,
the direct within-pair and cross-pair expression for U agrees with the mean-quadratic
minus sample-covariance expression to maximum absolute error
$2.49\times10^{-14}$. These checks explicitly cover the residual terms
that distinguish the present experiment from a mean-quadratic experiment.

\end{vFifteenLowerHeadings}
\subsection{Estimating a diagonal covariance shape}
\label{sec:u-shape-numerical}

We examine whether the invariant reference in
Theorem~\ref{thm:u-diag-invariant} avoids the power cost of the
shape-confidence envelope at a fixed sample budget. We take
\[
 \begin{gathered}
 H_{jj}=\exp\!\left\{-\log2+
       \frac{2(j-1)\log2}{p-1}\right\},\\
 \bm\Sigma_1=\bm H,\quad \bm\Sigma_2=\rho\bm H,
 \quad p\in\{10,30\},\quad N\in\{40,80\}.
 \end{gathered}
\]
The diagonal structure is specified, but its entries are unknown to the
estimated-shape procedures. Each group supplies $m=n=N/2$ training and
testing observations. The first group's training coordinate variances
form $\widehat{\bm H}$; both training means learn the direction. The
direction is the normalized projection of
$\widehat{\bm H}^{-1/2}\bm D_{\mathrm{tr}}$ onto the first $q=5$
coordinates. This subspace is fixed and contains the signal, which lies
along the first true-whitened coordinate. We set $b=1$, $\eta=p$, and
\[
 \Lambda_{\mathrm{all}}
 =\frac{N\bm\delta^\top\bm H^{-1}\bm\delta}{1+\rho}
 \in\{0,16,64\}.
\]
No additional observations are used to estimate the shape or validate
power. All four methods retain the full unbiased numerator and the
testing-residual scale in \eqref{eq:u-shape-scale}.

The invariant procedure regenerates the training variances, the null
training contrast, and the testing variables in each reference draw.
It therefore accounts for the joint randomness of the estimated shape
and learned direction without a shape-confidence event. Its level
guarantee is unconditional over training. Known shape uses the true
$\bm H$ in the direction, weight, and scale, while still learning its
direction from the training means. Naive plug-in uses the estimated
shape with the known-shape reference $M_{1,1}$. The confidence-envelope
procedure uses \eqref{eq:u-shape-diagonal-confidence} with
$\gamma=0.01$ and $M_{\ell,h}$ at
$\beta=(0.05-0.01)/0.99$. With $L=199$ draws, the invariant and
known-shape cutoffs are the 190th ascending order statistics; the
confidence cutoff is the 192nd. Its finite-reference level bound is
$0.01+0.99(8/200)=0.0496$.

For each $(p,N)$ we generate $1\,000$ independent reference banks.
Each bank calibrates $100$ independently generated datasets in each
configuration, giving $100\,000$ datasets per configuration. The
seed is $20261025$. Shared banks induce dependence within each group
of $100$ rejection indicators. We accordingly estimate the Monte Carlo
standard error by the sample standard deviation of the $1\,000$
bank-level rejection proportions divided by $\sqrt{1\,000}$. The
largest such standard error among the reported configurations is
$0.360$ percentage points. The methods share observations; cutoff
vectors are reused across alternatives and variance ratios for the
same $(p,N)$. We report marginal Monte Carlo uncertainty without
treating comparisons across methods or configurations as independent.
At an all-zero or all-one empirical rate, a normal-approximation
interval and its estimated standard error are not reported.

\begin{table}[tbp]
 \centering
 \caption{Rejection percentages (cluster Monte Carlo standard errors
 in percentage points) at $\rho=4$. Each group has $N$ observations,
 including shape and direction training. Each entry uses $1\,000$
 independent reference banks with $100$ datasets per bank and $L=199$
 draws per reference. Confidence denotes the shape-confidence envelope;
 invariant denotes Theorem~\ref{thm:u-diag-invariant}. The four confidence
 null entries are zero observed rejections, with uncertainty marked
 by a dash rather than a degenerate confidence interval.}
 \label{tab:u-invariant-power}
 \small
 \setlength{\tabcolsep}{3pt}
 \begin{tabular}{rrrcccc}
\toprule
$p$ & $N$ & $\Lambda_{\rm all}$ & Known shape & Naive plug-in & Confidence & Invariant\\
\midrule
10 & 40 & 0 & 4.18 (0.08) & 4.74 (0.08) & 0.00 (--) & 4.21 (0.08) \\
10 & 40 & 16 & 58.75 (0.23) & 55.63 (0.22) & 0.046 (0.007) & 53.73 (0.23) \\
10 & 40 & 64 & 99.92 (0.009) & 99.62 (0.02) & 10.42 (0.17) & 99.54 (0.02) \\
10 & 80 & 0 & 4.36 (0.08) & 4.62 (0.08) & 0.00 (--) & 4.44 (0.08) \\
10 & 80 & 16 & 59.99 (0.23) & 58.37 (0.22) & 1.46 (0.05) & 57.52 (0.21) \\
10 & 80 & 64 & 99.94 (0.008) & 99.87 (0.01) & 61.40 (0.35) & 99.86 (0.01) \\
30 & 40 & 0 & 4.26 (0.08) & 4.80 (0.08) & 0.00 (--) & 4.32 (0.08) \\
30 & 40 & 16 & 58.64 (0.23) & 55.49 (0.22) & 0.015 (0.004) & 53.70 (0.22) \\
30 & 40 & 64 & 99.91 (0.010) & 99.53 (0.02) & 4.83 (0.10) & 99.46 (0.02) \\
30 & 80 & 0 & 4.64 (0.08) & 4.90 (0.09) & 0.00 (--) & 4.58 (0.08) \\
30 & 80 & 16 & 59.65 (0.22) & 58.13 (0.22) & 0.68 (0.03) & 57.39 (0.22) \\
30 & 80 & 64 & 99.92 (0.009) & 99.80 (0.01) & 49.23 (0.36) & 99.79 (0.01) \\
\bottomrule
\end{tabular}

\end{table}

\begin{figure}[tbp]
 \centering
 \includegraphics[width=0.97\textwidth]{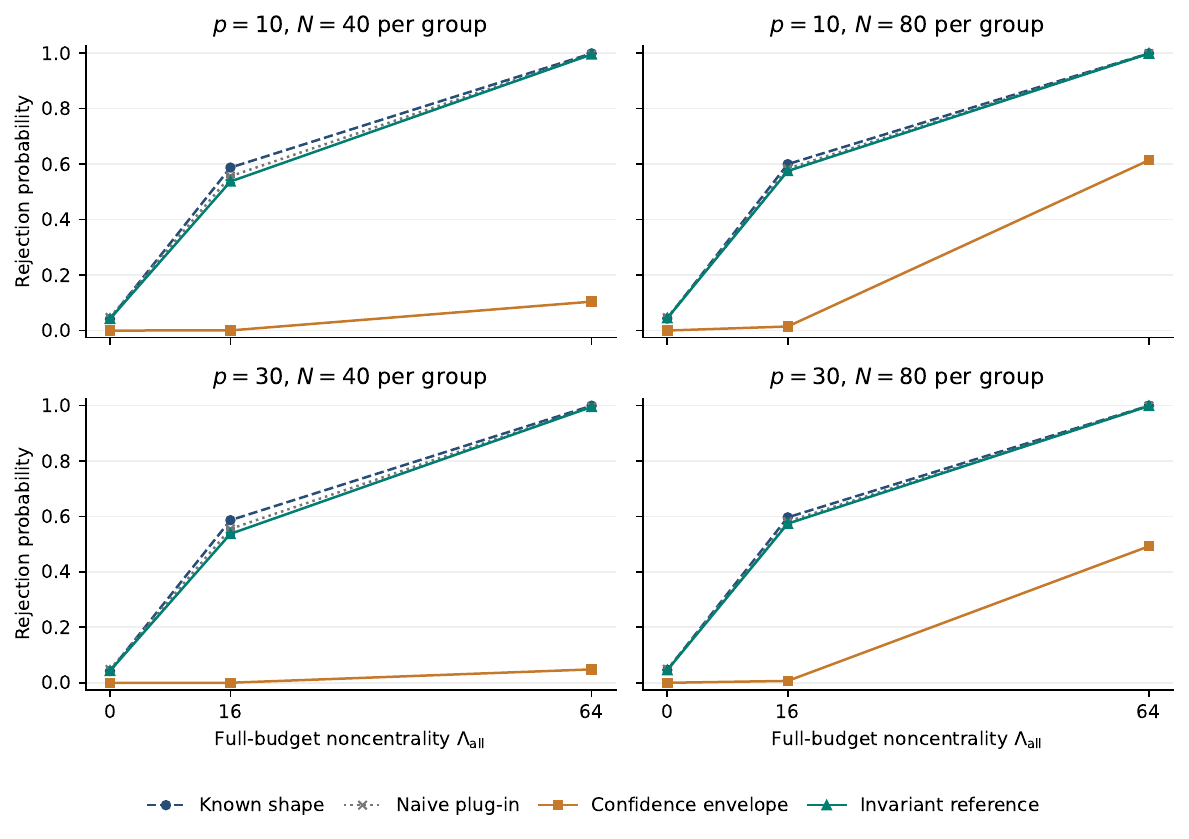}
 \caption{Rejection probabilities at $\rho=4$ under estimated diagonal
 shapes and fixed total sample budgets. The first five coordinates form
 the prespecified direction space. Bars show $1.96$ cluster Monte Carlo
 standard errors, accounting for shared finite reference banks, and are
 generally smaller than the plotting symbols. No normal-approximation
 bar is drawn for an all-zero rejection count.}
 \label{fig:u-invariant-learning}
\end{figure}

Table~\ref{tab:u-invariant-power} and
Figure~\ref{fig:u-invariant-learning} show that integrating the training
randomness into the reference removes most of the confidence envelope's
power cost in these settings. At $\Lambda_{\mathrm{all}}=16$, invariant
power ranges from $53.70\%$ to $57.52\%$, compared with $58.64\%$ to
$59.99\%$ for known shape and $0.015\%$ to $1.46\%$ for the confidence
envelope. At $\Lambda_{\mathrm{all}}=64$, invariant power ranges from
$99.46\%$ to $99.86\%$. The median invariant cutoffs lie between
$36.24$ and $105.32$, close to the corresponding known-shape cutoffs
of $35.45$ to $98.72$; confidence-envelope medians range from
$253.49$ to $1901.36$. These differences concern calibration as well
as shape estimation: the three estimated-shape procedures use exactly
the same observed statistic in every replication.

We additionally examine the null at $\rho\in\{0.1,1,10\}$, using
the same reference banks and new observation streams. Together with
the four $\rho=4$ null configurations, the sixteen invariant rejection
percentages range from $4.09\%$ to $4.58\%$
(Table~\ref{tab:u-invariant-size}). The remaining conservativeness is
consistent with using the maximum of the two variance-ratio endpoints.
These experiments do not replace the uniform level guarantee, which
follows from Theorem~\ref{thm:u-diag-invariant}. Nor does reasonable
size in these particular configurations establish validity of naive
plug-in calibration.

\begin{table}[tbp]
 \centering
 \caption{Null rejection percentages (cluster Monte Carlo standard
 errors in percentage points) for the invariant reference. Each entry
 uses $100\,000$ datasets, in $1\,000$ independent banks of $100$.
 The $\rho=4$ column repeats the corresponding entries of
 Table~\ref{tab:u-invariant-power}; the other columns use independent
 observation streams with the same bank cutoffs.}
 \label{tab:u-invariant-size}
 \small
 \begin{tabular}{rrcccc}
\toprule
$p$ & $N$ & $\rho=0.1$ & $\rho=1$ & $\rho=4$ & $\rho=10$\\
\midrule
10 & 40 & 4.20 (0.08) & 4.09 (0.07) & 4.21 (0.08) & 4.15 (0.08) \\
10 & 80 & 4.51 (0.08) & 4.37 (0.08) & 4.44 (0.08) & 4.46 (0.08) \\
30 & 40 & 4.27 (0.08) & 4.45 (0.08) & 4.32 (0.08) & 4.43 (0.08) \\
30 & 80 & 4.47 (0.08) & 4.50 (0.08) & 4.58 (0.08) & 4.55 (0.08) \\
\bottomrule
\end{tabular}

\end{table}

A separate implementation check uses $10\,000$ raw Gaussian datasets
at $p=10$, $N=40$, $\rho=4$, and $\Lambda_{\mathrm{all}}=16$, with
$1\,000$ independently generated reference banks and $10$ datasets
per bank. Its invariant rejection rate is $53.79\%$, within $0.103$
combined cluster Monte Carlo standard errors of the sufficient-statistic
result. On the first $250$ datasets, the direct pairwise U expression
agrees with the mean-quadratic minus sample-covariance expression to
maximum absolute error $5.69\times10^{-14}$. Positive coordinate
rescaling and a common translation leave the studentized statistic
unchanged to numerical precision. A further $250$ raw datasets with
unequal training and testing sizes verify these algebraic identities;
the power experiment itself uses balanced splits throughout.

Thus the invariant reference supplies a practical finite-sample
calibration in this specified diagonal Gaussian model without an
additional validation sample. The remaining power gap from known shape
includes the cost of learning the direction with estimated coordinate
scales. These experiments neither optimize the fixed subspace or spike
size nor establish such performance for arbitrary unknown covariance
shapes. The earlier confidence-envelope simulation is retained among
the reproducibility files as a separate historical experiment; all
comparisons in this subsection use the common design described above.

\subsection{Projection ablation and all-data benchmarks}
\label{sec:u-practical}

Improvement over a conservative shape envelope does not isolate the
benefit of projection. We therefore compare the invariant procedure
with a no-spike version that uses the same estimated shape, equal
training/testing split, baseline, and testing-residual scale, but sets
$\eta=0$. Its reference is regenerated at $\eta=0$; using the spiked
critical value for this ablation would not be a matched comparison.
Two further benchmarks use all $N$ observations per group on their
original coordinate scale: the test of \citet{chen2010}, denoted CQ,
with its original leave-out variance estimator, and the normal-reference
F-type test of \citet{zhu2024ftype}, with bias-corrected trace estimates
in its fitted degrees of freedom. Neither receives the true shape or
the prespecified subspace. Thus the external comparison includes both
the cost of splitting and the benefit of structural information.

We retain the diagonal shapes, $p$, $N$, and $\rho=4$ from
Section~\ref{sec:u-shape-numerical}. The full-budget noncentrality is
again $\Lambda_{\mathrm{all}}$. Inside signals lie along the first
true-whitened coordinate; outside signals lie along the last, with
$\Lambda_{\mathrm{all}}=16$. The RI learner always uses the first
five coordinates. All four methods receive the same raw observations,
with group means $\bm\delta/2$ and $-\bm\delta/2$.
Seed $20261026$ supplies new data and reference streams. Each
configuration again uses $1\,000$ independent reference banks and
$100$ datasets per bank. Uncertainty is computed from bank proportions;
paired differences and their cluster standard errors are archived.
The CQ and F-type calibrations are approximate, whereas the two RI
rules have the stated finite-reference bounds under this model.

\begin{table}[tbp]
 \centering
 \caption{Rejection percentages (cluster Monte Carlo standard errors in
 percentage points) for the projection ablation and all-data benchmarks.
 Each group supplies $N$ observations. RI uses $q=5$ and $\eta=p$;
 no spike uses $\eta=0$ with its own reference. CQ and F-type use all
 observations without splitting or true-shape information. Here
 $\Lambda=\Lambda_{\mathrm{all}}$. Each entry uses $100\,000$ datasets.}
 \label{tab:u-v5-benchmarks}
 \small
 \setlength{\tabcolsep}{3pt}
 \begin{tabular}{rrrlrrrr}
\toprule
$p$ & $N$ & $\Lambda$ & Signal & RI & No spike & CQ & F-type\\
\midrule
10 & 40 & 0 & -- & 4.14 (0.08) & 3.58 (0.07) & 6.87 (0.08) & 5.07 (0.07)\\
10 & 40 & 16 & Inside & 53.96 (0.23) & 31.96 (0.22) & 40.42 (0.15) & 33.35 (0.15)\\
10 & 40 & 64 & Inside & 99.53 (0.02) & 94.53 (0.09) & 99.78 (0.01) & 99.55 (0.02)\\
10 & 40 & 16 & Outside & 6.64 (0.11) & 31.86 (0.21) & 93.81 (0.08) & 92.16 (0.09)\\
10 & 80 & 0 & -- & 4.28 (0.08) & 3.96 (0.07) & 6.85 (0.08) & 5.06 (0.07)\\
10 & 80 & 16 & Inside & 57.71 (0.22) & 36.13 (0.23) & 40.69 (0.16) & 33.87 (0.16)\\
10 & 80 & 64 & Inside & 99.83 (0.01) & 97.09 (0.06) & 99.78 (0.01) & 99.60 (0.02)\\
10 & 80 & 16 & Outside & 7.15 (0.11) & 36.21 (0.22) & 93.95 (0.08) & 92.40 (0.08)\\
30 & 40 & 0 & -- & 4.40 (0.07) & 3.65 (0.07) & 6.49 (0.08) & 5.22 (0.07)\\
30 & 40 & 16 & Inside & 53.75 (0.21) & 19.02 (0.17) & 22.65 (0.13) & 19.06 (0.13)\\
30 & 40 & 64 & Inside & 99.45 (0.02) & 79.57 (0.17) & 92.73 (0.08) & 90.63 (0.09)\\
30 & 40 & 16 & Outside & 4.98 (0.08) & 18.88 (0.17) & 81.78 (0.12) & 79.22 (0.13)\\
30 & 80 & 0 & -- & 4.49 (0.08) & 4.07 (0.07) & 6.46 (0.07) & 5.16 (0.07)\\
30 & 80 & 16 & Inside & 57.01 (0.22) & 20.89 (0.18) & 22.45 (0.14) & 19.00 (0.13)\\
30 & 80 & 64 & Inside & 99.78 (0.01) & 85.43 (0.15) & 92.94 (0.08) & 90.85 (0.09)\\
30 & 80 & 16 & Outside & 5.22 (0.09) & 20.92 (0.18) & 81.82 (0.12) & 79.39 (0.13)\\
\bottomrule
\end{tabular}

\end{table}

Table~\ref{tab:u-v5-benchmarks} separates two effects. For inside
signals at $\Lambda_{\mathrm{all}}=16$, projection improves power
relative to the matched no-spike rule. For outside signals, that
improvement disappears and the all-data benchmarks can be substantially
more powerful. This is not a contradiction of the known-shape subspace
result, which requires the signal to belong to the specified subspace.
Moreover, moving from the first to the last coordinate changes the
signal's original-scale variance and Euclidean magnitude. It is not
a pure subspace rotation comparison for CQ or F-type. The controlled
leakage experiment below examines that distinction within the
coordinate-scale-invariant procedure.

For example, at $p=30$, $N=40$, and an inside signal with
$\Lambda_{\mathrm{all}}=16$, rejection percentages are $53.75\%$
for RI and $19.02\%$ for no spike. Their paired difference is
reported with its Monte Carlo uncertainty in the archived results.
Across the four designs, inside-signal RI power ranges from
$53.75\%$ to $57.71\%$; for the outside signal it ranges from
$4.98\%$ to $7.15\%$. No benchmark replication produced a
nonpositive variance or degrees-of-freedom estimate.

These are comparisons at a common nominal level, not at a common
attained size. In particular, CQ is liberal in the displayed low- and
moderate-dimensional configurations: its null rejection percentages
range from $6.46\%$ to $6.87\%$, compared with $5.06\%$ to $5.22\%$
for F-type and $4.14\%$ to $4.49\%$ for RI. A larger rejection rate cannot
by itself establish better size-adjusted efficiency. The comparison
does not assert uniform superiority of the structured RI rule.

\subsection{Prespecified tuning and subspace misspecification}
\label{sec:u-sensitivity}

Fix $p=30$, $\rho=4$, and $N\in\{40,80\}$. Starting from
$(q,\eta,m)=(5,p,N/2)$, vary one factor at a time: $q=p$,
$\eta\in\{p/2,2p\}$, or
$m\in\{\operatorname{round}(N/3),\operatorname{round}(2N/3)\}$.
Each group then has $n=N-m$ testing observations. We consider the
null and alternatives with
\[
 \bm H^{-1/2}\bm\delta
 =\sqrt{\frac{16(1+\rho)}{N}}
   \{\sqrt{1-\epsilon}\,\bm e_1+
      \sqrt{\epsilon}\,\bm e_6\},
 \qquad \epsilon\in\{0,1/2,1\}.
\]
Thus $\epsilon$ is the signal fraction outside the first five
coordinates; both coordinates belong to the full-space learner.
All twelve designs were fixed before simulation. Each has its own
reference banks and four signal settings, giving $48$ configurations
with $100\,000$ datasets each. Exact Gaussian sufficient statistics
are used, and the signal settings within a design share shape errors,
mean noises, residuals, and cutoffs. No outcome-based choice of a
design or seed is part of the test.

\begin{table}[tbp]
 \centering
 \caption{Prespecified one-factor sensitivity at $p=30$, $\rho=4$.
 Entries are rejection percentages (cluster Monte Carlo standard errors
 in percentage points). Non-null columns have
 $\Lambda_{\mathrm{all}}=16$ and the indicated fraction outside the
 first five coordinates. Each entry uses $1\,000$ banks of $100$
 datasets. These are separate fixed procedures, not a data-selected test.}
 \label{tab:u-v5-sensitivity}
 \small
 \setlength{\tabcolsep}{3pt}
 \begin{tabular}{rrrrrrrr}
\toprule
$N$ & $m$ & $q$ & $\eta/p$ & Null & $\epsilon=0$ & $\epsilon=1/2$ & $\epsilon=1$\\
\midrule
40 & 20 & 5 & 1 & 4.35 (0.08) & 53.71 (0.22) & 26.08 (0.19) & 5.02 (0.09)\\
40 & 20 & 30 & 1 & 4.29 (0.08) & 23.88 (0.17) & 23.85 (0.18) & 23.91 (0.18)\\
40 & 20 & 5 & 0.5 & 4.30 (0.08) & 53.81 (0.23) & 27.14 (0.20) & 5.82 (0.10)\\
40 & 20 & 5 & 2 & 4.39 (0.08) & 53.59 (0.23) & 25.66 (0.19) & 4.69 (0.08)\\
40 & 13 & 5 & 1 & 4.29 (0.08) & 55.61 (0.21) & 28.32 (0.20) & 5.22 (0.09)\\
40 & 27 & 5 & 1 & 4.09 (0.07) & 41.94 (0.23) & 20.14 (0.17) & 4.52 (0.08)\\
80 & 40 & 5 & 1 & 4.55 (0.08) & 57.14 (0.22) & 27.99 (0.20) & 5.34 (0.09)\\
80 & 40 & 30 & 1 & 4.60 (0.08) & 25.65 (0.19) & 25.74 (0.20) & 25.65 (0.19)\\
80 & 40 & 5 & 0.5 & 4.59 (0.08) & 57.10 (0.22) & 28.95 (0.20) & 6.33 (0.10)\\
80 & 40 & 5 & 2 & 4.57 (0.08) & 57.08 (0.20) & 27.50 (0.18) & 4.92 (0.08)\\
80 & 27 & 5 & 1 & 4.61 (0.08) & 60.63 (0.21) & 30.60 (0.20) & 5.63 (0.09)\\
80 & 53 & 5 & 1 & 4.24 (0.08) & 45.52 (0.23) & 21.80 (0.18) & 4.72 (0.08)\\
\bottomrule
\end{tabular}

\end{table}

Table~\ref{tab:u-v5-sensitivity} reports attained size alongside power.
At $N=40$, the reference design has powers $53.71\%$, $26.08\%$,
and $5.02\%$ as $\epsilon$ increases from zero to one; full-space
learning stays near $23.9\%$. Restricting the learner therefore helps
only when the structural information is sufficiently accurate.
For an inside signal, varying $\eta$ between $p/2$ and $2p$ changes
power little in these designs. Allocation has a larger effect:
at $N=80$, the inside-signal powers for $m=27,40,53$ are
$60.63\%$, $57.14\%$, and $45.52\%$, respectively.
The relevant tradeoff is between directional estimation, testing
allocation, and signal omitted from the learning space. A full-rank
baseline retains some response to omitted signal, but it does not
prevent the added projection component from lowering power relative
to a differently calibrated baseline. Nor does the largest entry in
this finite grid establish an optimal choice of $q$, $\eta$, or $m$.
The known-shape subspace dominance theorem does not automatically extend
to these fitted-shape learners under different calibrations.

\subsection{Executable procedure and model scope}

The reproducibility package supplies \texttt{ri\_two\_sample} for
two raw observation matrices. It returns the full studentized
statistic, numerator, scale estimate, reference exceedance count,
Monte Carlo $p$-value, and the design used. It accepts unequal block
sizes and a fixed orthonormal subspace basis. Unit checks compare the
mean/covariance form with the pairwise U expression, verify positive
coordinate-scale and common-translation invariance, and reproduce the
reference-rank decision. A separate scalar implementation checks both
external benchmark formulas. The example fixes the design and random
seeds before generating data; the defaults are not presented as
optimal tuning recommendations.
Three further $10\,000$-dataset raw checks cover a changed split,
full-space learning, and a changed spike. They agree with the respective
sufficient-statistic calculations within $1.64$ combined Monte Carlo
standard errors. All $124$ archived cluster summaries are independently
recomputed from the stored rejection counts.

The diagonal Gaussian assumption is substantive. Additional correlated
Gaussian and independent heavy-tailed null diagnostics appear in
Appendix~\ref{sec:u-misspecification}. They delimit use of the procedure
rather than extend Theorem~\ref{thm:u-diag-invariant}. In particular,
an application cannot replace a justified covariance model by an
unverified diagonal plug-in and retain the stated exact guarantee.

\begin{vFifteenLowerHeadings}
\section{External comparisons on an equal observation budget}\label{supp:v9external}
\subsection{Recorded designs and information available to each method}
Each group contributes $N=40$ independent Gaussian observations, with
$X_i\sim N_p(\delta,H)$ and $Y_i\sim N_p(0,\rho H)$.
The design grid is $p\in\{30,300\}$ and $\rho\in\{1,4\}$.
The fixed diagonal entries of $H$ are equally spaced on the logarithmic
scale from $0.5$ to $2$. The first $m=20$ observations per group train
the RI shape estimate and direction; the remaining $n=20$ observations
per group test. The supplied subspace consists of the first $q=5$
coordinate axes. RI uses $b=1$ and the prespecified choices
$\eta=\sqrt p$ and $\eta=0$ separately.

The initial protocol specified the null and $\Lambda_{\rm all}=64$,
where $\Lambda_{\rm all}=N\delta^TH^{-1}\delta/(1+\rho)$.
After inspecting these high-strength outcomes, a separate follow-up
protocol added $\Lambda_{\rm all}=16$ to examine nonsaturated power.
The follow-up retained all designs, methods, splits, directions and
calibration rules. Its protocol was frozen before its outcomes were
generated; it was not part of the original registration.
Both protocols and their separate results are retained. These are
internally frozen design records, not public-registry registrations.
The main table displays both strengths and the leakage endpoints;
the complete grid, including leakage $0.9$, is reported below.

For leakage $\epsilon\in\{0,0.9,1\}$ the whitened signal has norm
$\sqrt{\Lambda_{\rm all}(1+\rho)/N}$ and direction
\[
 \sqrt{1-\epsilon}\,e_1+\sqrt{\epsilon}\,e_{q+1}
 \quad\hbox{or}\quad
 \sqrt{1-\epsilon}\,e_1+
 \sqrt{\frac{\epsilon}{p-q}}\sum_{j=q+1}^p e_j.
\]
The first is concentrated leakage and the second is dense leakage.
At $\epsilon=0$ they coincide and are simulated only once.
The two leakage patterns have equal Mahalanobis strength, but their
Euclidean signal norms differ under this nonconstant $H$.
Consequently their difference is not a controlled sparsity comparison
for the unweighted Euclidean-norm procedures.

All procedures receive the same raw observations.
RI, the matched projection test and subspace Hotelling test receive
the specified $q$-dimensional subspace. The remaining procedures use
all coordinates without this information.
Where pairing is needed, $Z_i=X_i-Y_i$ uses the fixed original row
indices. These are independent differences from independent groups;
no matched-subject model, data-dependent matching or pairing search
is imposed. All-data procedures use all $N$ such differences.

\subsection{Benchmark statistics and calibration}
CQ uses the original leave-out variance estimator of
\citet{chen2010}; the $F$-type procedure uses the bias-corrected trace
estimators and reference degrees of freedom of \citet{zhu2024ftype}.
Both are approximate reference calibrations. The implementation
computes the original traces through observation Gram matrices and
does not change their estimands. A nonpositive variance or
degrees-of-freedom estimate is assigned $p=1$ and counted, rather than
repaired or discarded. None occurred in either experiment.

For the balanced-sample statistic of \citet{li2023finite}, write
$K=N(N-1)/2$, $a_{ij}=Z_i^TZ_j$, and let $\bar a$ and $s_a$ be
the mean and sample standard deviation of the $K$ entries with $i<j$.
The statistic and upper-tail reference are
\[
 T_{\rm Li}=\sqrt K\,\bar a/s_a,\qquad t_{K-1}.
\]
This is the paper's paired-difference two-sample construction.
The $t$ reference is justified as dimension diverges with fixed
sample size under its trace conditions; it is not an exact
finite-dimensional null law.

For the cross-$U$ construction of \citet{kim2024cross}, the training
direction is $d=m^{-1}\sum_{i=1}^m Z_i$ and the test scores are
$f_i=d^TZ_{m+i}$, $i=1,\ldots,n$. We report both
\[
 T_Z=\frac{\sqrt n\,\bar f}
 {\{n^{-1}\sum_i(f_i-\bar f)^2\}^{1/2}},\quad 1-\Phi(T_Z),
 \qquad
 T_t=\sqrt{\frac{n-1}{n}}\,T_Z,\quad 1-F_{t_{n-1}}(T_t).
\]
The normal-reference statistic follows the published convention.
The second is the exact Gaussian scalar-$t$ calibration of the same
learned direction. Conditional on training under the Gaussian null,
$T_t\sim t_{n-1}$, for arbitrary positive-definite covariance of $Z_i$.
Thus the two calibrations have different finite-sample guarantees.
At $n=20$ the exact null rejection probability of the normal version
at nominal $0.05$ is
$1-F_{t_{19}}(\sqrt{19/20}\,z_{0.95})=0.06269$.
Its extra rejections must not be credited as size-controlled power.

The coordinate Bonferroni benchmark uses
$t_j=\sqrt N\,\bar Z_j/s_j$ and
\[
 p_{\rm Bonf}=\min\{1,\;p\min_{j\le p}2[1-F_{t_{N-1}}(|t_j|)]\}.
\]
It has Gaussian finite-sample level at most $\alpha$ regardless of
the dependence among coordinates or the equality of group covariances.
For subspace Hotelling, let $\bar Z_Q,S_Q$ denote the mean and unbiased
covariance of the first $q$ coordinates of all $N$ differences. Then
\[
 F_Q=\frac{N-q}{q(N-1)}N\bar Z_Q^TS_Q^{-1}\bar Z_Q
       \sim F_{q,N-q}\quad\hbox{under the Gaussian null}.
\]
It uses the same specified subspace as RI and all observations;
$q<N$ makes its covariance invertible almost surely. It tests only
the subspace component and therefore has null-level rejection under
a purely outside-subspace signal.
The matched projection benchmark instead uses the RI training
direction and a two-sided scalar $t_{n-1}$ test on the testing pairs.

\subsection{Monte Carlo uncertainty, paired contrasts and independent audit}
Every cell contains 10,000 simulated datasets, grouped into 200
independent banks of 50. Each bank has an independent reference stream
and an independent raw-data stream. The two RI procedures share the
Gaussian endpoint-reference construction, each with $L=199$ draws.
All methods share the raw data, and signal cells share their underlying
noise to improve comparison precision. A bank's reference is reused
for its 50 datasets. Monte Carlo standard errors are therefore the
standard deviation of the 200 bank means divided by $\sqrt{200}$.
Paired contrasts use bank means of the paired rejection differences;
they do not sum marginal variances as though the methods were independent.
The maximum rate MCSE across both full grids is 0.863 percentage points.
Complete per-cell MCSEs, paired contrasts, invalid counts and bank counts
are included in the separate initial and extension CSV/NPZ outputs.

An independent script checks the optimized CQ/$F$ calculations against
covariance-matrix and explicit leave-out implementations, and all five
new benchmarks against separate scalar or matrix formulas. The largest
algebraic discrepancy is below $3\times10^{-15}$.
A further 16,000 Gaussian null datasets with unequal, correlated,
non-common-shape covariances gave rejection rates $5.156\%$ for cross-$U$
with exact $t$, $5.081\%$ for Bonferroni and $4.931\%$ for subspace
Hotelling. The corresponding normal-reference cross-$U$ rate was
$6.444\%$, consistent with its analytic finite-sample inflation.
These checks verify the implemented Gaussian references; they do not
establish uniform superiority of any procedure.

\subsection{Computational cost}
With one BLAS thread, the initial experiment took
85.2 seconds for all four designs and all six initial signal cells;
the recorded moderate-signal extension took 71.1 seconds.
The component timings below are for $p=300,\rho=4$ and exclude
raw-data generation. A paired
row reports the joint cost when methods share calculations; it is
not an isolated per-method timing. Reference simulation takes
6.11 milliseconds for a fresh pair of $199$-draw banks, or
0.122 milliseconds per dataset when amortized over the 50 datasets
used here. A single new dataset without a prespecified reusable bank
must pay the fresh-reference cost. Timing is descriptive of this
implementation and hardware, not an algorithmic complexity result.
\begin{center}
\begin{tabular}{lr}
\toprule
Component & Milliseconds per dataset\\
\midrule
Two RI statistics & 0.102\\
Matched projection $t$ & 0.008\\
CQ and $F$ jointly & 0.239\\
Li $U$--$t$ & 0.038\\
Two cross-$U$ calibrations jointly & 0.008\\
Coordinate Bonferroni $t$ & 0.090\\
Subspace Hotelling & 0.008\\
Fresh pair of RI reference banks & 6.110\\
\bottomrule
\end{tabular}
\end{center}

\subsection{Complete rejection-rate panels}
All entries below are percentages. ``Inside'' means $\epsilon=0$;
the other headings give the leakage fraction and geometry. The null
column is from the initial experiment and is shown only once per design.
The $\Lambda_{\rm all}=16$ panel is the recorded follow-up, not an
independent replication of a previously selected favorable result.

\paragraph{$p=30$, $\rho=1$.}

\noindent $\Lambda_{\rm all}=16$: \par
\begin{center}
\begin{tabular}{lrrrrr}
\toprule
Method & Inside & Conc. 0.9 & Conc. 1 & Dense 0.9 & Dense 1\\
\midrule
RI, $\eta=\sqrt p$ & 50.64 & 12.18 & 9.95 & 11.95 & 9.93\\
RI, $\eta=0$ & 19.17 & 18.78 & 18.87 & 18.74 & 18.40\\
Projection $t$ & 54.83 & 7.67 & 5.53 & 7.67 & 5.53\\
CQ & 23.14 & 29.41 & 30.09 & 55.77 & 58.96\\
$F$-type & 19.40 & 25.48 & 26.15 & 51.68 & 54.63\\
Li $U$--$t$ & 22.68 & 29.24 & 29.72 & 55.09 & 58.34\\
Cross-$U$ normal & 18.13 & 21.55 & 22.27 & 38.17 & 40.41\\
Cross-$U$ $t$ & 15.32 & 18.65 & 19.03 & 34.07 & 36.16\\
Bonferroni $t$ & 74.40 & 68.12 & 73.74 & 20.38 & 20.10\\
Subspace $T^2$ & 83.29 & 11.84 & 5.05 & 11.84 & 5.05\\
\bottomrule
\end{tabular}
\end{center}

\noindent $\Lambda_{\rm all}=64$: \par
\begin{center}
\begin{tabular}{lrrrrrr}
\toprule
Method & Null & Inside & Conc. 0.9 & Conc. 1 & Dense 0.9 & Dense 1\\
\midrule
RI, $\eta=\sqrt p$ & 3.84 & 99.36 & 63.81 & 52.93 & 65.99 & 56.36\\
RI, $\eta=0$ & 3.25 & 78.65 & 79.81 & 78.93 & 84.05 & 84.10\\
Projection $t$ & 5.12 & 99.46 & 20.67 & 5.12 & 20.67 & 5.12\\
CQ & 6.80 & 92.76 & 97.54 & 97.89 & 99.92 & 99.95\\
$F$-type & 5.22 & 90.94 & 96.52 & 96.88 & 99.88 & 99.92\\
Li $U$--$t$ & 6.55 & 92.51 & 97.30 & 97.72 & 99.88 & 99.93\\
Cross-$U$ normal & 6.61 & 75.28 & 85.08 & 85.92 & 97.75 & 98.30\\
Cross-$U$ $t$ & 5.31 & 71.57 & 82.59 & 83.35 & 97.19 & 97.75\\
Bonferroni $t$ & 4.77 & 100.00 & 100.00 & 100.00 & 74.12 & 72.72\\
Subspace $T^2$ & 4.87 & 100.00 & 40.41 & 4.87 & 40.41 & 4.87\\
\bottomrule
\end{tabular}
\end{center}

\paragraph{$p=30$, $\rho=4$.}

\noindent $\Lambda_{\rm all}=16$: \par
\begin{center}
\begin{tabular}{lrrrrr}
\toprule
Method & Inside & Conc. 0.9 & Conc. 1 & Dense 0.9 & Dense 1\\
\midrule
RI, $\eta=\sqrt p$ & 50.06 & 12.58 & 10.56 & 12.44 & 10.13\\
RI, $\eta=0$ & 18.78 & 19.47 & 19.74 & 19.26 & 19.33\\
Projection $t$ & 52.93 & 7.68 & 4.71 & 7.68 & 4.71\\
CQ & 22.68 & 28.91 & 29.85 & 54.87 & 57.96\\
$F$-type & 19.22 & 24.97 & 25.63 & 50.38 & 53.83\\
Li $U$--$t$ & 22.22 & 28.55 & 29.39 & 54.06 & 57.29\\
Cross-$U$ normal & 18.43 & 22.20 & 23.03 & 38.46 & 40.84\\
Cross-$U$ $t$ & 15.47 & 19.12 & 19.56 & 34.22 & 36.64\\
Bonferroni $t$ & 73.82 & 68.14 & 73.82 & 20.53 & 20.06\\
Subspace $T^2$ & 82.60 & 11.24 & 4.80 & 11.24 & 4.80\\
\bottomrule
\end{tabular}
\end{center}

\noindent $\Lambda_{\rm all}=64$: \par
\begin{center}
\begin{tabular}{lrrrrrr}
\toprule
Method & Null & Inside & Conc. 0.9 & Conc. 1 & Dense 0.9 & Dense 1\\
\midrule
RI, $\eta=\sqrt p$ & 3.64 & 99.40 & 63.77 & 52.97 & 65.92 & 55.43\\
RI, $\eta=0$ & 3.71 & 79.33 & 80.47 & 79.71 & 84.84 & 84.60\\
Projection $t$ & 4.79 & 99.51 & 20.33 & 4.79 & 20.33 & 4.79\\
CQ & 6.55 & 92.52 & 97.73 & 98.05 & 99.93 & 99.93\\
$F$-type & 5.12 & 90.34 & 96.87 & 97.33 & 99.86 & 99.90\\
Li $U$--$t$ & 6.35 & 92.33 & 97.55 & 97.95 & 99.90 & 99.92\\
Cross-$U$ normal & 6.75 & 74.96 & 85.12 & 85.81 & 97.74 & 98.50\\
Cross-$U$ $t$ & 5.28 & 71.08 & 82.36 & 83.16 & 97.06 & 97.84\\
Bonferroni $t$ & 4.84 & 100.00 & 100.00 & 100.00 & 74.30 & 72.92\\
Subspace $T^2$ & 4.97 & 100.00 & 39.07 & 4.97 & 39.07 & 4.97\\
\bottomrule
\end{tabular}
\end{center}

\paragraph{$p=300$, $\rho=1$.}

\noindent $\Lambda_{\rm all}=16$: \par
\begin{center}
\begin{tabular}{lrrrrr}
\toprule
Method & Inside & Conc. 0.9 & Conc. 1 & Dense 0.9 & Dense 1\\
\midrule
RI, $\eta=\sqrt p$ & 47.98 & 7.32 & 5.64 & 7.53 & 5.57\\
RI, $\eta=0$ & 6.98 & 7.08 & 7.12 & 6.58 & 6.67\\
Projection $t$ & 53.75 & 7.59 & 5.08 & 7.59 & 5.08\\
CQ & 9.45 & 9.66 & 9.71 & 15.98 & 16.88\\
$F$-type & 8.75 & 8.95 & 9.01 & 14.93 & 15.75\\
Li $U$--$t$ & 9.34 & 9.52 & 9.66 & 15.63 & 16.51\\
Cross-$U$ normal & 9.28 & 9.40 & 9.48 & 13.22 & 13.78\\
Cross-$U$ $t$ & 7.50 & 7.47 & 7.55 & 10.86 & 11.28\\
Bonferroni $t$ & 47.32 & 41.25 & 47.67 & 7.05 & 6.87\\
Subspace $T^2$ & 82.45 & 11.71 & 5.19 & 11.71 & 5.19\\
\bottomrule
\end{tabular}
\end{center}

\noindent $\Lambda_{\rm all}=64$: \par
\begin{center}
\begin{tabular}{lrrrrrr}
\toprule
Method & Null & Inside & Conc. 0.9 & Conc. 1 & Dense 0.9 & Dense 1\\
\midrule
RI, $\eta=\sqrt p$ & 3.97 & 99.10 & 29.85 & 12.67 & 29.37 & 11.91\\
RI, $\eta=0$ & 3.58 & 27.31 & 26.98 & 26.75 & 26.52 & 26.56\\
Projection $t$ & 4.72 & 99.49 & 21.36 & 4.72 & 21.36 & 4.72\\
CQ & 5.61 & 30.69 & 31.41 & 31.42 & 70.70 & 74.33\\
$F$-type & 5.17 & 28.96 & 29.93 & 29.82 & 69.12 & 72.98\\
Li $U$--$t$ & 5.52 & 30.09 & 30.93 & 30.75 & 69.95 & 73.70\\
Cross-$U$ normal & 6.32 & 22.68 & 22.82 & 22.92 & 50.08 & 53.19\\
Cross-$U$ $t$ & 5.02 & 19.55 & 19.42 & 19.56 & 45.50 & 48.70\\
Bonferroni $t$ & 4.92 & 100.00 & 99.86 & 99.96 & 17.77 & 12.01\\
Subspace $T^2$ & 5.08 & 100.00 & 40.76 & 5.08 & 40.76 & 5.08\\
\bottomrule
\end{tabular}
\end{center}

\paragraph{$p=300$, $\rho=4$.}

\noindent $\Lambda_{\rm all}=16$: \par
\begin{center}
\begin{tabular}{lrrrrr}
\toprule
Method & Inside & Conc. 0.9 & Conc. 1 & Dense 0.9 & Dense 1\\
\midrule
RI, $\eta=\sqrt p$ & 49.87 & 8.35 & 6.29 & 8.43 & 6.07\\
RI, $\eta=0$ & 7.14 & 6.87 & 7.01 & 6.81 & 6.76\\
Projection $t$ & 54.92 & 7.73 & 5.21 & 7.73 & 5.21\\
CQ & 9.55 & 9.82 & 9.76 & 15.84 & 16.55\\
$F$-type & 8.72 & 8.84 & 8.65 & 14.76 & 15.44\\
Li $U$--$t$ & 9.52 & 9.77 & 9.73 & 15.64 & 16.44\\
Cross-$U$ normal & 9.14 & 9.08 & 9.06 & 12.77 & 13.25\\
Cross-$U$ $t$ & 7.43 & 7.51 & 7.44 & 10.68 & 11.00\\
Bonferroni $t$ & 48.20 & 41.75 & 48.63 & 7.25 & 7.04\\
Subspace $T^2$ & 83.15 & 12.12 & 5.35 & 12.12 & 5.35\\
\bottomrule
\end{tabular}
\end{center}

\noindent $\Lambda_{\rm all}=64$: \par
\begin{center}
\begin{tabular}{lrrrrrr}
\toprule
Method & Null & Inside & Conc. 0.9 & Conc. 1 & Dense 0.9 & Dense 1\\
\midrule
RI, $\eta=\sqrt p$ & 3.99 & 99.10 & 29.96 & 13.84 & 29.96 & 12.47\\
RI, $\eta=0$ & 3.87 & 27.97 & 27.14 & 27.19 & 26.90 & 26.56\\
Projection $t$ & 4.73 & 99.53 & 20.46 & 4.73 & 20.46 & 4.73\\
CQ & 5.88 & 29.96 & 30.77 & 30.71 & 70.83 & 74.39\\
$F$-type & 5.32 & 28.38 & 29.06 & 29.12 & 69.28 & 72.83\\
Li $U$--$t$ & 5.84 & 29.63 & 30.36 & 30.33 & 70.04 & 73.88\\
Cross-$U$ normal & 6.06 & 22.46 & 22.73 & 22.62 & 48.88 & 52.04\\
Cross-$U$ $t$ & 4.91 & 19.13 & 19.03 & 19.25 & 44.75 & 47.33\\
Bonferroni $t$ & 4.89 & 99.97 & 99.94 & 99.96 & 17.68 & 11.85\\
Subspace $T^2$ & 4.62 & 100.00 & 40.32 & 4.62 & 40.32 & 4.62\\
\bottomrule
\end{tabular}
\end{center}

\section{Full-data diagnostic and classical diagonal benchmarks}
\label{supp:v10nosplit}
\subsection{Prospective design and fixed information}
A new experiment compares full-data scores with the split procedures and
classical tests on exactly the same raw observations. Both strengths
$\Lambda_{\rm all}\in\{16,64\}$, both dimensions $p\in\{30,300\}$,
and all leakage cases were frozen before generating this experiment's
outcomes. Each independent group has $N=40$ observations and
$\rho=4$; $q=5$, $b=1$ and $\eta=\sqrt p$ are fixed.
The diagonal shape and signal normalization are the same as in
Section~\ref{supp:v9external}. For each dimension there is one null cell
and five cells at each strength: inside the subspace, concentrated
leakage at $\epsilon=0.9,1$, and diffuse leakage at $\epsilon=0.9,1$.
All $22$ cells and all $15$ procedures are retained.

The pairing $Z_i=X_i-Y_i$ is fixed by row index. It uses all $2N$
observations but only $N$ independent differences, with $N-1$ residual
degrees of freedom. It assumes no matched-subject structure and performs
no matching search. Paired covariance $S_Z$ contains random cross terms;
it is not the original unpaired estimator $S_X+S_Y$.
The full-data scores therefore do not implement the original complete
unbiased statistic by merely omitting its split. They are different,
data-adaptive statistics whose entire fitted null law is regenerated.
The full-data mean-axis and maximum scores, classical block score,
subspace diagonal $T^2$, subspace paired DLRT, split RI and projection,
and subspace Hotelling all receive the same fixed first $q$ coordinates.
The remaining full-vector methods do not receive this subspace information.

\subsection{New scores and exact diagonal-null simulation}
Set $D=\operatorname{diag}(S_Z)$,
$y=\sqrt N D^{-1/2}\bar Z$,
$R=D^{-1/2}S_ZD^{-1/2}$, and
$v=y_Q/\|y_Q\|$. The six newly calibrated scores are
\begin{align*}
 T_{\rm mean}&=b(\|y\|^2-p)+\eta(\|y_Q\|^2-v^TR_Qv),\\
 T_{\rm max}&=b(\|y\|^2-p)+\eta\lambda_{\max}(y_Qy_Q^T-R_Q),\\
 T_{\rm block}&=b(\|y\|^2-p)+\eta(\|y_Q\|^2-1),\\
 T_Q&=\sum_{j\le q}y_j^2,\qquad T_{\rm all}=\sum_{j\le p}y_j^2,\\
 T_{\rm DLRT,Q}&=N\sum_{j\le q}\log\{1+y_j^2/(N-1)\}.
\end{align*}
The maximum score optimizes its realized Rayleigh quotient; this does
not imply power optimality after its own calibration. The mean-axis
score is not unbiased after estimating the shape and choosing its
direction from the same data. The block score is an affine transform
of a classical weighted sum of coordinate $t$ squares, and $T_Q$ is the
classical diagonal Hotelling statistic. The last score is the one-sample
diagonal likelihood-ratio construction of \citet{hu2019}, applied to
our fixed paired differences and restricted to the supplied subspace.
We do not apply that paper's equal-covariance two-sample result to
unequal original group scales.

For each null draw, generate independent Gaussian residual columns in
$\mathbb R^{N-1}$ for the $q$ coordinates and independent
$\chi^2_{N-1}$ residual lengths for the remaining coordinates.
Normalize the first columns to obtain $R_Q$ and divide independent
Gaussian mean coordinates by the corresponding residual standard
deviations to obtain $y$. Each draw regenerates both fitted means and
residual geometry. Diagonal variances and the common mean cancel under
the Gaussian null, so this is an exact canonical null construction.
Each score uses its own upper-tail Monte Carlo rank with $L=199$;
the six procedures share the underlying canonical draws.
They have unconditional level $0.05$ in the specified continuous
Gaussian model. This is not conditional validity given the fitted
shape or direction, and it does not extend to unknown correlation.

The split RI scores use their original joint-reference envelope;
projection, cross-$U$ $t$, Bonferroni and subspace Hotelling keep their
Gaussian calibrations. CQ, the $F$-type procedure and Li's $U$--$t$
retain their approximate references. Formulas for these controls are
in Section~\ref{supp:v9external}. No rejection-rate comparison to an
approximate reference is interpreted as a size-adjusted power ordering.

\subsection{Simulation uncertainty and diagnostic findings}
Each cell uses $10\,000$ datasets in $200$ independent banks of $50$.
Reference and raw-data streams are independent. All methods share the
same raw observations and signal cells share their underlying Gaussian
noise. A bank's references are reused for its $50$ datasets. Rates and
paired differences therefore use bank-cluster Monte Carlo standard
errors, not standard errors that incorrectly treat the shared-reference
decisions as independent. The maximum rate MCSE is 0.83 percentage
points. All binary outcomes, bank counts, critical values, complete
paired contrasts and timing records are archived. The design is an
internally frozen record, not a public preregistration.

At $p=300$, $\Lambda_{\rm all}=16$ and no leakage, the mean-axis score
has rejection rate $83.40\%$, versus $48.59\%$ for split RI,
$82.97\%$ for subspace Hotelling and $84.62\%$ for subspace
diagonal $T^2$. The paired gain over split RI is $34.81$
percentage points (MCSE $0.62$); its paired difference from
subspace diagonal $T^2$ is $-1.22$ points (MCSE $0.25$).
Thus the full-data route largely closes the observed power deficit in
this specified diagonal model. The statistic and calibration also change,
so this contrast does not isolate a causal effect of splitting alone.
The classical controls remain necessary for interpreting the gain.

For this same cell, the maximum-minus-mean difference is $0.01$
points (MCSE $0.02$), and the mean-minus-block difference is
$0.02$ points (MCSE $0.09$). Across all cells the
absolute mean-versus-block rejection-rate gap never exceeds
$0.28$ points, and the mean-versus-maximum gap never exceeds
$0.21$ points. This numerical proximity is not a proof of power
equivalence or dominance. Conditional on $y$, replacing $v^TR_Qv$ by
its mean $1$ removes residual directional noise, but such conditional
averaging does not generally order calibrated rejection probabilities.
Accordingly the experiment does not establish a new optimal-direction
method beyond the classical diagonal scores.

Subspace-only tests return to null-level rejection when all signal is
outside the supplied subspace, while baseline-bearing scores retain
some sensitivity. The complete panels below also show substantial
losses against outside-signal controls. Fixed $b,\eta,q$, Gaussian
coordinates and diagonal shape restrict the finding: neither unknown
correlation nor efficient data-driven tuning at a fixed small sample
budget is resolved by this experiment. Approximate-reference size
inflation is visible and is not credited as additional valid power.

\subsection{Independent implementation checks and cost}
An independent audit reconstructs the six scores from raw covariance
matrices and couples the canonical generator exactly to synthetic raw
Gaussian observations. The largest raw/canonical discrepancy through
$p=300$ is below $2\times10^{-12}$. It also checks diagonal rescaling
invariance, the $q=1$ collapse, the realized maximum-score inequality,
positive null bias of the fitted score and the residual-correlation
conditional moments. These are algebraic and calibration checks,
not evidence of optimality. No invalid CQ variance or $F$ degrees of
freedom occurred in this experiment.

The full $22$-cell run took 78.3 seconds using one BLAS thread.
The following $p=300$ timings separate fresh reference-bank generation
from data-score computation. A fresh reference bank is needed unless
a bank independent of the tested data has already been generated.
The experiment's sharing across datasets is a simulation device, not
a promise that a new dataset costs only an amortized reference fraction.
\begin{center}
\begin{tabular}{lr}
\toprule
Component & Milliseconds\\
\midrule
Six full-data scores, per dataset & 0.037\\
Two split RI scores, per dataset & 0.099\\
CQ and $F$ jointly, per dataset & 0.230\\
Subspace Hotelling, per dataset & 0.007\\
Fresh joint reference for six scores ($199$ draws) & 3.806\\
Fresh joint split RI reference ($199$ draws) & 5.782\\
\bottomrule
\end{tabular}
\end{center}

\subsection{All prespecified rejection-rate panels}
All entries are percentages at nominal $5\%$. ``Conc.'' denotes
concentrated leakage; ``Dense'' denotes diffuse leakage. The null
is repeated across the two strength panels only for readability and
comes from the same null cell. Cells at leakage zero coincide across
the two signal geometries and were simulated once.

\noindent\begin{minipage}{\textwidth}
\paragraph{$p=30$, $\rho=4$, $\Lambda_{\rm all}=16$.}
\begin{center}\small
\begin{tabular}{lrrrrrr}
\toprule
Method & Null & Inside & Conc. 0.9 & Conc. 1 & Dense 0.9 & Dense 1\\
\midrule
Full-data mean-axis score & 5.28 & 84.55 & 23.63 & 13.75 & 23.31 & 13.54\\
Full-data maximum score & 5.28 & 84.57 & 23.64 & 13.75 & 23.33 & 13.54\\
Classical block $t^2$ & 5.27 & 84.65 & 23.74 & 13.66 & 23.43 & 13.46\\
Subspace diagonal $T^2$ & 5.35 & 85.61 & 13.05 & 5.35 & 13.05 & 5.35\\
All-coordinate $t^2$ & 5.09 & 50.32 & 50.31 & 49.81 & 51.07 & 50.93\\
Subspace paired DLRT & 5.42 & 84.38 & 12.87 & 5.42 & 12.87 & 5.42\\
Split RI, $\eta=\sqrt p$ & 4.22 & 51.68 & 12.45 & 10.37 & 12.23 & 10.08\\
Split RI, $\eta=0$ & 3.65 & 18.65 & 18.72 & 18.44 & 18.99 & 19.04\\
Split projection $t$ & 5.19 & 55.56 & 7.56 & 5.19 & 7.56 & 5.19\\
CQ & 6.52 & 23.18 & 28.98 & 29.88 & 55.15 & 57.93\\
$F$-type & 5.23 & 19.32 & 24.82 & 25.64 & 50.58 & 53.98\\
Li $U$--$t$ & 6.46 & 22.96 & 28.76 & 29.31 & 54.61 & 57.50\\
Cross-$U$ $t$ & 5.04 & 15.87 & 18.99 & 19.31 & 33.93 & 36.04\\
Bonferroni $t$ & 5.27 & 73.95 & 68.18 & 73.68 & 19.60 & 19.13\\
Subspace Hotelling & 5.47 & 83.52 & 12.34 & 5.47 & 12.34 & 5.47\\
\bottomrule
\end{tabular}
\end{center}\end{minipage}\par\medskip

\noindent\begin{minipage}{\textwidth}
\paragraph{$p=30$, $\rho=4$, $\Lambda_{\rm all}=64$.}
\begin{center}\small
\begin{tabular}{lrrrrrr}
\toprule
Method & Null & Inside & Conc. 0.9 & Conc. 1 & Dense 0.9 & Dense 1\\
\midrule
Full-data mean-axis score & 5.28 & 100.00 & 90.78 & 71.86 & 92.13 & 75.51\\
Full-data maximum score & 5.28 & 100.00 & 90.82 & 72.02 & 92.20 & 75.72\\
Classical block $t^2$ & 5.27 & 100.00 & 90.86 & 71.89 & 92.11 & 75.79\\
Subspace diagonal $T^2$ & 5.35 & 100.00 & 42.54 & 5.35 & 42.54 & 5.35\\
All-coordinate $t^2$ & 5.09 & 99.72 & 99.80 & 99.79 & 99.95 & 99.96\\
Subspace paired DLRT & 5.42 & 100.00 & 41.70 & 5.42 & 41.70 & 5.42\\
Split RI, $\eta=\sqrt p$ & 4.22 & 99.50 & 64.95 & 52.92 & 67.46 & 56.94\\
Split RI, $\eta=0$ & 3.65 & 80.62 & 79.98 & 79.14 & 84.78 & 85.00\\
Split projection $t$ & 5.19 & 99.54 & 21.16 & 5.19 & 21.16 & 5.19\\
CQ & 6.52 & 92.80 & 97.33 & 97.75 & 99.91 & 99.98\\
$F$-type & 5.23 & 90.81 & 96.49 & 96.95 & 99.91 & 99.96\\
Li $U$--$t$ & 6.46 & 92.57 & 97.17 & 97.58 & 99.92 & 99.97\\
Cross-$U$ $t$ & 5.04 & 70.54 & 82.81 & 83.62 & 97.19 & 97.95\\
Bonferroni $t$ & 5.27 & 100.00 & 100.00 & 100.00 & 73.85 & 72.62\\
Subspace Hotelling & 5.47 & 100.00 & 39.73 & 5.47 & 39.73 & 5.47\\
\bottomrule
\end{tabular}
\end{center}\end{minipage}\par\medskip

\noindent\begin{minipage}{\textwidth}
\paragraph{$p=300$, $\rho=4$, $\Lambda_{\rm all}=16$.}
\begin{center}\small
\begin{tabular}{lrrrrrr}
\toprule
Method & Null & Inside & Conc. 0.9 & Conc. 1 & Dense 0.9 & Dense 1\\
\midrule
Full-data mean-axis score & 4.88 & 83.40 & 14.59 & 6.63 & 14.50 & 6.77\\
Full-data maximum score & 4.88 & 83.41 & 14.57 & 6.62 & 14.49 & 6.76\\
Classical block $t^2$ & 4.89 & 83.38 & 14.58 & 6.62 & 14.62 & 6.81\\
Subspace diagonal $T^2$ & 4.75 & 84.62 & 11.75 & 4.75 & 11.75 & 4.75\\
All-coordinate $t^2$ & 5.28 & 15.61 & 15.79 & 15.93 & 15.47 & 15.50\\
Subspace paired DLRT & 4.75 & 83.72 & 11.50 & 4.75 & 11.50 & 4.75\\
Split RI, $\eta=\sqrt p$ & 4.29 & 48.59 & 7.71 & 5.78 & 7.54 & 5.68\\
Split RI, $\eta=0$ & 3.69 & 7.04 & 7.07 & 7.19 & 6.96 & 6.91\\
Split projection $t$ & 5.18 & 54.33 & 7.43 & 5.18 & 7.43 & 5.18\\
CQ & 5.47 & 9.31 & 9.27 & 9.33 & 15.55 & 16.35\\
$F$-type & 5.03 & 8.45 & 8.58 & 8.66 & 14.37 & 15.24\\
Li $U$--$t$ & 5.47 & 9.02 & 9.18 & 9.25 & 15.19 & 16.19\\
Cross-$U$ $t$ & 4.71 & 6.99 & 7.14 & 7.16 & 10.27 & 10.78\\
Bonferroni $t$ & 4.72 & 47.63 & 41.11 & 48.15 & 6.72 & 6.40\\
Subspace Hotelling & 4.96 & 82.97 & 12.21 & 4.96 & 12.21 & 4.96\\
\bottomrule
\end{tabular}
\end{center}\end{minipage}\par\medskip

\noindent\begin{minipage}{\textwidth}
\paragraph{$p=300$, $\rho=4$, $\Lambda_{\rm all}=64$.}
\begin{center}\small
\begin{tabular}{lrrrrrr}
\toprule
Method & Null & Inside & Conc. 0.9 & Conc. 1 & Dense 0.9 & Dense 1\\
\midrule
Full-data mean-axis score & 4.88 & 100.00 & 61.04 & 17.47 & 61.02 & 16.89\\
Full-data maximum score & 4.88 & 100.00 & 61.04 & 17.48 & 60.99 & 16.91\\
Classical block $t^2$ & 4.89 & 100.00 & 61.18 & 17.27 & 60.81 & 16.89\\
Subspace diagonal $T^2$ & 4.75 & 100.00 & 41.84 & 4.75 & 41.84 & 4.75\\
All-coordinate $t^2$ & 5.28 & 70.72 & 71.54 & 70.84 & 73.46 & 73.41\\
Subspace paired DLRT & 4.75 & 100.00 & 40.72 & 4.75 & 40.72 & 4.75\\
Split RI, $\eta=\sqrt p$ & 4.29 & 99.25 & 29.56 & 14.11 & 29.08 & 13.01\\
Split RI, $\eta=0$ & 3.69 & 27.26 & 27.58 & 27.44 & 26.80 & 27.23\\
Split projection $t$ & 5.18 & 99.52 & 20.07 & 5.18 & 20.07 & 5.18\\
CQ & 5.47 & 29.13 & 30.61 & 30.69 & 70.15 & 73.57\\
$F$-type & 5.03 & 27.45 & 28.98 & 29.01 & 68.54 & 72.04\\
Li $U$--$t$ & 5.47 & 28.89 & 30.30 & 30.38 & 69.80 & 73.28\\
Cross-$U$ $t$ & 4.71 & 18.66 & 19.48 & 19.64 & 44.67 & 47.71\\
Bonferroni $t$ & 4.72 & 99.98 & 99.93 & 100.00 & 18.15 & 12.43\\
Subspace Hotelling & 4.96 & 100.00 & 39.57 & 4.96 & 39.57 & 4.96\\
\bottomrule
\end{tabular}
\end{center}\end{minipage}\par\medskip

\section{All-pair cross-fitting: a focused follow-up}
\label{sec:lpo-experiments}
The leave-pair-out construction was investigated after the external comparison
identified a substantial deficit of the split procedure.  Its purpose is to recover information
while keeping each cross-product independent of its fitted weight.  It is a
separate statistic, and the finite-sample directional ordering for the original
fixed-training statistic does not transfer to it.

\subsection{Protocol and established diagonal baseline}
We fixed $N=40$ observations per group, $q=5$, $\rho=4$, $b=1$,
$\eta=\sqrt p$, and $p\in\{30,300\}$.  The diagonal entries of $H$ are
geometrically spaced from $0.5$ to $2$.  In each dimension, the three specified
settings are the null, $\Lambda_{\rm all}=16$ with signal in the first
coordinate, and $\Lambda_{\rm all}=64$ with signal equally distributed over
coordinates $j>q$ in standardized units.  Thus the six cells are a focused
diagnostic, rather than a broad robustness study.  Pairing follows observation
index and is fixed before the data are generated; it does not assume matched
subjects.  All methods receive the same $2N$ raw observations.

For each observed or reference dataset, every one of the $\binom{40}{2}=780$
held-out pairs has its own fitted diagonal covariance and direction, trained on
the remaining $38$ paired observations.  The covariance divisor is $37$.
No pair is subsampled.  The reference regenerates the complete collection of
dependent fitted weights from an independent canonical Gaussian dataset.
Each test uses $L=199$ reference statistics and the finite-rank rule at
$\alpha=0.05$.  The original split RI comparison uses its own jointly simulated
nuisance-envelope reference.  Fixed choices of subspace and strength are not
selected from the reported outcomes.

The initial protocol was written before generating rejection outcomes.  An
independent literature audit, conducted while that experiment was running,
identified the need to include the diagonal leave-pair-out statistic of
\citet{park2013}; see also the diagonal reduction in \citet{hu2024}.
This comparison was recorded as a separate auxiliary protocol before its
outcomes were generated, and the original protocol and all outcomes were
retained.  These are internally recorded designs, not external preregistrations.
At $\eta=0$, the statistic is the Park--Ayyala diagonal statistic up to a
positive deterministic factor.  Their inverse-variance correction
$(N-5)/(N-3)$ does not affect ranks.  We calibrate this baseline with the same
finite Gaussian reference mechanism; this is not a claim that the original
asymptotic Park--Ayyala calibration is exact.

Every cell contains $10\,000$ datasets, arranged as $200$ independent reference
banks with $50$ datasets per bank.  Gaussian noise is shared across signal
settings, and the same observed data and canonical reference matrices are used
for the augmented and diagonal leave-pair-out statistics.  References are
independent of observed data.  Rates and paired differences use the standard
deviation of bank-level means divided by $\sqrt{200}$, accounting for reference
reuse.  All binary outcomes, reference thresholds, protocols, and computation
times are retained.  The subspace DLRT uses the one-sample statistic of
\citet{hu2019} on the fixed paired differences, with finite Gaussian simulation
calibration.  No equal-covariance unpaired DLRT result is invoked.

\subsection{All rejection rates and interpretation}
Tables~\ref{tab:lpo30} and~\ref{tab:lpo300} give all twelve comparisons,
including the auxiliary diagonal baseline.  Parentheses contain Monte Carlo
standard errors in percentage points; the rejection rates are percentages.
The F-type calibration is approximate.  The fitted same-data scores are not
unbiased and remain separately named.

\begin{table}[htbp]
\centering\small
\caption{Focused leave-pair-out comparison at $p=30$, $N=40$ per group, $q=5$, and $\rho=4$.}
\label{tab:lpo30}
\begin{tabular}{lrrr}
\toprule
Method & Null & \shortstack{Inside\\$\Lambda=16$} & \shortstack{Dense outside\\$\Lambda=64$} \\
\midrule
Leave-pair-out, $\eta=\sqrt p$ & 5.27 (0.24) & 76.21 (0.59) & 84.13 (0.84) \\
Park--Ayyala baseline, $\eta=0$ & 4.91 (0.22) & 50.37 (0.61) & 99.93 (0.03) \\
Same-data mean axis & 4.97 (0.22) & 83.72 (0.46) & 74.71 (0.76) \\
Same-data observed maximum & 4.95 (0.22) & 83.72 (0.46) & 74.88 (0.76) \\
Classical block $t^2$ & 4.92 (0.22) & 83.62 (0.46) & 74.81 (0.77) \\
Subspace $t^2$ & 4.83 (0.21) & 84.50 (0.44) & 4.83 (0.21) \\
All-coordinate $t^2$ & 5.11 (0.24) & 50.69 (0.61) & 99.96 (0.02) \\
Subspace DLRT & 4.90 (0.22) & 83.26 (0.46) & 4.90 (0.22) \\
Split RI, $\eta=\sqrt p$ & 4.31 (0.22) & 50.88 (0.55) & 56.98 (0.79) \\
Split projection $t$ & 4.97 (0.21) & 54.55 (0.48) & 4.97 (0.21) \\
F-type & 5.13 (0.22) & 19.46 (0.38) & 99.98 (0.01) \\
Subspace Hotelling & 4.88 (0.20) & 83.01 (0.40) & 4.88 (0.20) \\
\bottomrule
\end{tabular}
\end{table}

\begin{table}[htbp]
\centering\small
\caption{Focused leave-pair-out comparison at $p=300$, $N=40$ per group, $q=5$, and $\rho=4$.}
\label{tab:lpo300}
\begin{tabular}{lrrr}
\toprule
Method & Null & \shortstack{Inside\\$\Lambda=16$} & \shortstack{Dense outside\\$\Lambda=64$} \\
\midrule
Leave-pair-out, $\eta=\sqrt p$ & 5.32 (0.24) & 75.00 (0.52) & 16.86 (0.55) \\
Park--Ayyala baseline, $\eta=0$ & 4.92 (0.23) & 15.22 (0.43) & 73.06 (0.55) \\
Same-data mean axis & 5.07 (0.21) & 84.41 (0.42) & 17.24 (0.47) \\
Same-data observed maximum & 5.07 (0.21) & 84.39 (0.42) & 17.25 (0.47) \\
Classical block $t^2$ & 5.10 (0.22) & 84.43 (0.41) & 17.34 (0.47) \\
Subspace $t^2$ & 5.13 (0.23) & 85.88 (0.40) & 5.13 (0.23) \\
All-coordinate $t^2$ & 4.99 (0.24) & 15.55 (0.45) & 73.80 (0.57) \\
Subspace DLRT & 5.19 (0.23) & 84.88 (0.41) & 5.19 (0.23) \\
Split RI, $\eta=\sqrt p$ & 4.20 (0.22) & 50.26 (0.60) & 12.95 (0.45) \\
Split projection $t$ & 4.78 (0.20) & 56.04 (0.50) & 4.78 (0.20) \\
F-type & 4.88 (0.21) & 8.16 (0.26) & 72.25 (0.47) \\
Subspace Hotelling & 5.05 (0.21) & 83.37 (0.36) & 5.05 (0.21) \\
\bottomrule
\end{tabular}
\end{table}

For the subspace-contained signal at $p=300$, all-pair learning improves on
split RI by $+24.74$ percentage points (paired Monte Carlo standard error
$0.69$).  Nevertheless, its differences from subspace $t^2$ and subspace
Hotelling are $-10.88$ ($0.47$) and $-8.37$ ($0.47$) percentage points,
respectively.  The all-pair procedure recovers a substantial part
of the observed deficit, but does not make this fitted direction more efficient
than the established structure-informed procedures in this setting.

For fully diffuse signal outside the specified subspace, the augmented statistic
at $p=300$ differs from its Park--Ayyala diagonal baseline by $-56.20$ percentage
points (paired Monte Carlo standard error $0.74$).  This comparison directly
separates the learned rank-one addition from the existing diagonal
cross-product procedure.  The subspace-dependent addition can consume power
when the structural information is wrong.  These results do not establish
fixed-small-sample efficiency or a global advantage of the selected direction.
They instead identify the remaining methodological task: controlling the
learning and weighting cost while retaining power outside the supplied
subspace.

\subsection{Implementation and verification}
The implementation evaluates all pairs using leave-two-out sufficient
statistics.  A bundled C++ kernel accelerates the exact arithmetic; a NumPy
implementation is also provided.  An independent audit explicitly deletes
each held-out pair, refits the covariance and direction from the remaining
rows, forms the full weight matrix, and evaluates the cross-product.  It checks
agreement of both implementations, signed diagonal rescaling, row permutations,
and the $q=1$ reduction.  Maximum raw-formula errors in the stored audit are
below $1.6\times10^{-13}$.  A separate $4\,000$-test check with independent
$99$-draw reference banks gives a null rejection rate of $4.85\%$
(Monte Carlo standard error $0.345$ percentage points); a null-moment check
reports its sampling uncertainty rather than treating empirical centering as
proof.  All F-type variance estimates in this focused study were valid.

The computation is $O(N^2p)$ per observed or reference dataset.  The timing
records include the entire $199$-draw calibration bank, not only one statistic
evaluation.  The tests use no additional observed validation sample, but finite
calibration and repeated fits still incur computational cost.  This distinction
does not remove the power deficits shown above.

\end{vFifteenLowerHeadings}

\subsection{Direct same-panel comparisons with published mean tests}
\label{app:v22-jasa-comparison}
This experiment compares whole procedures at the same observation budget.
Every method receives a fixed panel of 60 coordinates; no method receives the
active coordinate or the signal type. The design avoids giving a small signal
subspace only to a learned RI method. We use 60 rather than five coordinates
because the official \texttt{PEtests} implementation warns that its
high-dimensional approximation may be inaccurate when the dimension is at
most 30. Passing that threshold does not itself establish adequate calibration.

Independently generate $X_i\sim N_{60}(\delta,H)$ and
$Y_i\sim N_{60}(0,\rho H)$, $i=1,\ldots,40$, where
$H=\operatorname{diag}(h_j)$ and the $h_j$ are equally spaced on the logarithmic
scale from $0.5$ to $2$. We fix $\rho\in\{1,4\}$.
The null has $\delta=0$. For a sparse signal set
$\delta_j=\sqrt{h_j(1+\rho)\Lambda/40}\,1\{j=1\}$;
for a same-sign diffuse signal set
$\delta_j=\sqrt{h_j(1+\rho)\Lambda/(40\cdot60)}$.
Both use $\Lambda\in\{16,64\}$, so
$\Lambda_{\rm all}=40\delta^\top H^{-1}\delta/(1+\rho)$.
This gives ten prespecified cells, all reported in
Table~\ref{tab:v22-jasa-comparison}.

The bounded, quadratic, maximum and LPO procedures use the fixed paired
differences $Z_i=X_i-Y_i$. LPO uses $b=1$, $\eta=\sqrt{60}$ and $q=p=60$;
all leave-pair-out fits are recomputed. Since these scores are invariant under
positive diagonal rescaling, their null references consist of independent
standard Gaussian matrices of size $40\times60$. Each reference dataset is
fitted anew. The observed score is ranked against 199 references using
$(1+\sum_{\ell=1}^{199}1\{T_\ell\ge T_0\})/200$.
The continuous model requires no variance floor. In contrast, the two
published competitors use all original observations directly. Thus the
procedures use the same raw observations, but the paired procedures do not
retain every feature of the original two-group experiment. The maximin claims
in the main text apply only to their specified paired experiments,
not to this cross-procedure comparison.

Jiang RI independently implements the fixed weights and leave-out trace
variance formulas of the published method, retaining its upper-tail normal
calibration. It is a formula-based implementation checked against a separate
explicit-matrix calculation; author software was not obtained. Yu PE translates
the authors' \texttt{PEtests} 0.1.0 R implementation of the mean component,
with the default threshold $2\log\log(80)\log(60)$ and the original normal
calibration. It was checked numerically against the unaltered R source.
The simultaneous mean-and-covariance procedure is not used, because it tests
a different null. Neither competitor is recalibrated to match the attained
size in these simulations. A nonfinite value or nonpositive variance estimate
is recorded as invalid and counted as a non-rejection, never deleted.

The comparison does not include a verified implementation of the online
projection test of Liu et al. A generic split projection is not substituted
for it. The accompanying source audit records the available supplementary
specification and the unresolved tuning issue. Consequently, this experiment
supports no empirical superiority claim over that online test.

Each cell contains 10000 datasets in 500 independent banks of 20 datasets.
Reference matrices are shared within a bank and across the four paired scores,
and independently generated between banks and between values of $\rho$.
Within a bank, the same raw noise is shifted across the five signal settings.
For every method, the standard error is the sample standard deviation of the
500 bank rejection proportions divided by $\sqrt{500}$. Paired contrasts use
the corresponding bank differences. This accounts for reference reuse;
treating all rejection indicators as independent would not generally do so.
The initial 1000 datasets per cell were extended to 10000 after reviewing
that preliminary run. The total of 500 banks per scale ratio was fixed
before generating the additional outcomes. The original 50 banks are
retained as the prefix, and the signal configurations, method parameters
and calibrations are unchanged. No stopping rule or tuning choice depends
on the additional results. The extension protocol and production-source
hashes are retained separately from the original protocol.
Reference scores, observed scores, p-values and decisions are retained in
the two \texttt{v22\_compare\_rho*} archives. Raw Gaussian inputs can be
regenerated exactly from the recorded seed streams; per-bank input hashes
verify regeneration without storing duplicate raw matrices. The CSV files
report every rejection rate, LPO-minus-comparator contrast, Monte Carlo error,
invalid count and timing. Production uses
\texttt{v22\_compare.py}; \texttt{v22\_render\_compare.py} reconstructs the tables.

\begin{table}[t]
\centering\scriptsize
\caption{Monte Carlo standard errors (percentage points) for every entry of Table~\ref{tab:v22-jasa-comparison}. These use independent bank means and include reference-bank reuse.}
\label{tab:v22-jasa-mcse}
\begin{tabular}{clrrrrrr}
\toprule
$\rho$ & Signal, $\Lambda_{\rm all}$ & Bounded & Quadratic & Maximum & LPO & Jiang RI & Yu PE \\
\midrule
1 & Null & 0.23 & 0.23 & 0.22 & 0.23 & 0.24 & 0.30 \\
1 & Sparse, 16 & 0.48 & 0.53 & 0.51 & 0.51 & 0.28 & 0.42 \\
1 & Sparse, 64 & 0.56 & 0.13 & 0.00 & 0.14 & 0.42 & 0.00 \\
1 & Diffuse, 16 & 0.55 & 0.55 & 0.35 & 0.54 & 0.42 & 0.51 \\
1 & Diffuse, 64 & 0.08 & 0.09 & 0.60 & 0.11 & 0.00 & 0.09 \\
\midrule
4 & Null & 0.23 & 0.23 & 0.22 & 0.22 & 0.25 & 0.29 \\
4 & Sparse, 16 & 0.46 & 0.50 & 0.51 & 0.50 & 0.29 & 0.44 \\
4 & Sparse, 64 & 0.54 & 0.14 & 0.00 & 0.15 & 0.40 & 0.00 \\
4 & Diffuse, 16 & 0.52 & 0.52 & 0.36 & 0.49 & 0.42 & 0.50 \\
4 & Diffuse, 64 & 0.09 & 0.11 & 0.60 & 0.12 & 0.00 & 0.10 \\
\bottomrule
\end{tabular}
\end{table}

\begin{table}[t]
\centering\scriptsize
\caption{Paired LPO-minus-comparator rejection-rate differences, in percentage points, with bank-cluster Monte Carlo standard errors in parentheses. Null rows compare attained sizes; nonnull rows compare nominal-level power and must be read together with those sizes.}
\label{tab:v22-jasa-paired}
\begin{tabular}{clrrrrr}
\toprule
$\rho$ & Signal, $\Lambda_{\rm all}$ & Bounded & Quadratic & Maximum & Jiang RI & Yu PE \\
\midrule
1 & Null & 0.23 (0.16) & 0.10 (0.16) & -0.01 (0.28) & -1.08 (0.27) & -4.43 (0.29) \\
1 & Sparse, 16 & 8.02 (0.32) & -3.58 (0.24) & -34.01 (0.61) & 23.55 (0.49) & -41.22 (0.56) \\
1 & Sparse, 64 & 31.15 (0.55) & -0.42 (0.07) & -2.08 (0.14) & 77.13 (0.44) & -2.08 (0.14) \\
1 & Diffuse, 16 & -4.47 (0.29) & -3.77 (0.27) & 19.59 (0.57) & -46.89 (0.54) & -9.67 (0.46) \\
1 & Diffuse, 64 & -0.44 (0.07) & -0.30 (0.06) & 58.19 (0.60) & -1.09 (0.11) & -0.21 (0.10) \\
\midrule
4 & Null & -0.33 (0.18) & -0.28 (0.17) & -0.08 (0.28) & -1.57 (0.29) & -5.13 (0.30) \\
4 & Sparse, 16 & 7.14 (0.34) & -3.66 (0.26) & -34.15 (0.61) & 23.08 (0.51) & -41.50 (0.55) \\
4 & Sparse, 64 & 31.01 (0.52) & -0.28 (0.07) & -2.05 (0.15) & 76.93 (0.42) & -2.05 (0.15) \\
4 & Diffuse, 16 & -5.03 (0.30) & -4.13 (0.28) & 19.90 (0.53) & -46.13 (0.53) & -10.39 (0.46) \\
4 & Diffuse, 64 & -0.44 (0.06) & -0.28 (0.05) & 59.07 (0.60) & -1.36 (0.12) & -0.41 (0.10) \\
\bottomrule
\end{tabular}
\end{table}

\begin{table}[t]
\centering\small
\caption{Recorded computation time in milliseconds per evaluation. The paired-score columns evaluate four scores together. Reference costs are reported separately; references are reused across 20 datasets per signal cell and the five coupled signal settings within each bank. Timings are descriptive for the recorded machine and exclude compilation and archive compression.}
\label{tab:v22-jasa-timing}
\begin{tabular}{crrrr}
\toprule
$\rho$ & Paired observation & Paired reference & Jiang RI & Yu PE \\
\midrule
1 & 0.232 & 0.246 & 0.061 & 0.100 \\
4 & 0.232 & 0.247 & 0.062 & 0.102 \\
\bottomrule
\end{tabular}
\end{table}

\section{Power benchmarks and numerical efficiency}
\label{app:v15-numerical-efficiency}
\subsection{Unknown shape when dimension exceeds sample size}
\label{sec:u-highdim}

We add $(p,N)=(300,40)$ and $(1000,80)$, retaining $q=5$ and an
equal split. For each design there are $500$ independent reference
banks, each used for $100$ datasets, giving $50\,000$ datasets per
configuration. Seed $20261027$ is fixed before simulation. The null
and three alternatives share Gaussian sufficient statistics and
reference banks within each design. For non-null cases,
\[
 \bm H^{-1/2}\bm\delta
 =\sqrt{\frac{16(1+\rho)}{N}}
   \{\sqrt{1-\epsilon}\,\bm e_1+
           \sqrt\epsilon\,\bm e_6\},
 \qquad\epsilon\in\{0,1/2,1\}.
\]
The known-shape comparator learns its direction from the same training
means, but replaces estimated whitening by true whitening in its
learner, statistic, testing-only scale, and endpoint reference. It is
the matched comparison of Section~\ref{sec:u-unknown-shape}, not the
pooled-scale procedure of Section~\ref{sec:u-feasible}.

A simpler comparator uses exactly the same estimated shape and learned
axis as RI. With a fixed pairing of the balanced testing samples, set
$\bm Z_j=\bm X_{1j}-\bm X_{2j}$ and
$a_j=\widehat{\bm v}^{\top}\widehat{\bm H}^{-1/2}\bm Z_j$.
Conditional on training, the statistic $\sqrt n\,\bar a/s_a$ has the
central $t_{n-1}$ law under the null, where $s_a^2$ is the sample
variance of the paired scalar differences. We use a two-sided $5\%$
test. Its conditional alternative law is noncentral $t$, with
noncentrality
\[
 \frac{\sqrt n\,\widehat{\bm v}^{\top}
             \widehat{\bm H}^{-1/2}\bm\delta}
 {\{\widehat{\bm v}^{\top}\widehat{\bm H}^{-1/2}
       (\bm\Sigma_1+\bm\Sigma_2)
          \widehat{\bm H}^{-1/2}\widehat{\bm v}\}^{1/2}}.
\]
We average this exact conditional rejection probability over the same
training draws. Thus projection entries integrate out testing
randomness and their Monte Carlo standard errors concern training
only; their null $5\%$ is exact, not an empirical zero-error estimate.
The comparator uses the same total observation budget and structural
information, and requires no reference simulation. It is a specified
paired projection test, not an implementation of the regularized
procedure of \citet{liu2024projection}.

\begin{table}[tbp]
\centering\small
\caption{High-dimensional diagonal-shape comparisons. Entries are
rejection percentages (Monte Carlo standard errors in percentage
points). RI, no spike and known shape use independent-bank cluster
standard errors; projection $t$ uses training-only errors after exact
integration over testing. Each configuration has $50\,000$ training
draws; $\epsilon$ is the fraction of signal outside the fixed subspace.}
\label{tab:u-v6-highdim}
\setlength{\tabcolsep}{3pt}
\begin{tabular}{rrrlrrrr}
\toprule
$p$ & $N$ & $\Lambda$ & $\epsilon$ & RI & No spike & Known shape & Projection $t$\\
\midrule
300 & 40 & 0 & -- & 4.32 (0.10) & 3.85 (0.10) & 4.38 (0.10) & 5.00 (exact)\\
300 & 40 & 16 & 0 & 53.26 (0.31) & 7.18 (0.15) & 58.69 (0.31) & 54.61 (0.08)\\
300 & 40 & 16 & 0.5 & 25.62 (0.27) & 7.26 (0.15) & 27.88 (0.28) & 26.22 (0.05)\\
300 & 40 & 16 & 1 & 4.38 (0.10) & 7.33 (0.14) & 4.47 (0.10) & 5.00 (exact)\\
1000 & 80 & 0 & -- & 4.78 (0.11) & 4.08 (0.10) & 4.70 (0.11) & 5.00 (exact)\\
1000 & 80 & 16 & 0 & 57.20 (0.31) & 5.88 (0.13) & 59.88 (0.30) & 57.98 (0.07)\\
1000 & 80 & 16 & 0.5 & 27.28 (0.27) & 5.85 (0.13) & 28.14 (0.27) & 27.85 (0.06)\\
1000 & 80 & 16 & 1 & 4.81 (0.11) & 5.92 (0.13) & 4.73 (0.11) & 5.00 (exact)\\
\bottomrule
\end{tabular}

\end{table}

Table~\ref{tab:u-v6-highdim} supports in-model high-dimensional
calibration but also shows a non-negligible shape cost at a fixed
training budget. For inside signals, RI trails the matched known-shape
learner by about $5.43$ and $2.68$ percentage points. Complete signal
omission removes the projection gain. The equally informed projection
$t$ comparator has slightly greater power in these settings. Hence
these experiments do not establish that retaining the full quadratic
improves power over a simple learned projection. The procedures also
have different attained sizes; their nominal-level comparison is not
a size-adjusted efficiency ordering. The paper's direction-optimality
theorem concerns a fixed RI spectrum, not competition among these
different tests.

The low-dimensional limit in
Proposition~\ref{prop:u-diag-fixed-budget-limit} gives inside-signal
invariant powers of $54.08\%$ and $57.37\%$ for the corresponding
fixed budgets (Monte Carlo standard errors $0.22$ percentage points).
Its matched known-shape powers are $59.14\%$ and $59.55\%$.
These separate simulations use $1\,000$ banks of $100$ datasets.
They illustrate the persistent fixed-budget shape loss; the limit
does not supply an error bound for its finite-dimensional approximation.

A separate timing experiment uses $20$ sequential calls after two
warmups on fixed datasets, with single-thread OpenBLAS. Median times
for the two designs are $10.19$ and $31.05$ milliseconds for RI with
a fresh bank, versus $0.23$ and $0.45$ for paired projection,
$3.34$ and $29.19$ for CQ, and $1.92$ and $15.19$ for F-type.
Training and calibration are included; data generation is excluded.
These are implementation timings, not complexity lower bounds:
in particular, the CQ and F-type code forms dense covariance matrices.
Timing spreads, environment and branch-equivalence checks are archived.

\subsection{When the retained baseline contributes}
\label{sec:u-baseline-value}

The next experiment fixes three distinct procedures with
$\eta\in\{0,\sqrt p,p\}$ before evaluating their powers. We use
$(p,N)=(30,40),(300,40)$, $q=5$, $\rho=4$, $b=1$, equal splits,
and $L=199$. For diffuse omitted signal, set
\[
 \bm H^{-1/2}\bm\delta
 =\sqrt{\frac{\Lambda_{\mathrm{all}}(1+\rho)}{N}}
 \left\{\sqrt{1-\epsilon}\,\bm e_1+
 \sqrt{\frac{\epsilon}{p-q}}\sum_{j=q+1}^p\bm e_j\right\}.
\]
The grid contains the null and
$\Lambda_{\mathrm{all}}\in\{16,64\}$,
$\epsilon\in\{0,1/2,0.9,1\}$. Each configuration uses
$400$ independent reference banks and $100$ datasets per bank.
The data and reference draws are coupled across the three spike
values; the projection comparator uses the same learned axis.
The main-table subset was fixed before examining rejection rates.
The complete grid and a concentrated-omission counterpart are retained
in Appendix~\ref{app:u-v8-experiments} and the numerical archive.

\begin{table}[tbp]
\centering\small
\caption{Diffuse omitted signal with $N=40$ per group. Entries are
rejection percentages (Monte Carlo standard errors in percentage
points); $\Lambda=\Lambda_{\mathrm{all}}$. RI errors account for
reference-bank sharing; projection errors concern training only.}
\label{tab:u-v8-baseline}
\setlength{\tabcolsep}{3pt}
\begin{tabular}{rrlrrrr}
\toprule
$p$ & $\Lambda$ & $\epsilon$ & $\eta=0$ & $\eta=\sqrt p$ & $\eta=p$ & Projection $t$\\
\midrule
30 & 0 & -- & 3.55 (0.11) & 3.94 (0.12) & 4.36 (0.12) & 5.00 (exact)\\
30 & 64 & 0 & 79.00 (0.29) & 99.38 (0.04) & 99.48 (0.04) & 99.50 (0.01)\\
30 & 64 & 0.5 & 83.13 (0.26) & 94.36 (0.14) & 89.87 (0.19) & 88.05 (0.06)\\
30 & 64 & 0.9 & 84.30 (0.27) & 66.76 (0.45) & 26.65 (0.33) & 20.62 (0.05)\\
30 & 64 & 1 & 84.24 (0.27) & 56.65 (0.52) & 7.93 (0.19) & 5.00 (exact)\\
300 & 0 & -- & 3.62 (0.11) & 4.12 (0.11) & 4.46 (0.11) & 5.00 (exact)\\
300 & 64 & 0 & 27.21 (0.33) & 99.09 (0.05) & 99.45 (0.04) & 99.49 (0.01)\\
300 & 64 & 0.5 & 26.84 (0.33) & 88.18 (0.19) & 88.23 (0.20) & 88.09 (0.06)\\
300 & 64 & 0.9 & 26.70 (0.33) & 29.39 (0.33) & 20.72 (0.26) & 20.63 (0.05)\\
300 & 64 & 1 & 26.75 (0.33) & 12.73 (0.26) & 4.72 (0.12) & 5.00 (exact)\\
\bottomrule
\end{tabular}

\end{table}

At $p=300$, $\Lambda_{\mathrm{all}}=64$ and $\epsilon=0.9$,
the $\sqrt p$ spike has power $29.39\%$, compared with $20.63\%$
for projection and $26.70\%$ for no spike. The paired differences
are $8.76$ and $2.69$ percentage points, with standard errors $0.33$
and $0.39$. This is a finite-dimensional example in which the retained
baseline contributes beyond the learned axis. It is consistent with
the mechanism in Proposition~\ref{prop:u-mixed-limit}, without being
a finite-dimensional consequence of its limiting formula.
The ordering changes across the grid: at $p=30$ and entirely outside
signal, no spike is preferable; for the weaker entirely inside signal,
projection is preferable to the $\sqrt p$ spike. No spike is selected
from these outcomes.

An independent sample of $100\,000$ null datasets per design estimates
each raw RI statistic's $95$th percentile at the true $\rho=4$.
This infeasible size-adjustment diagnostic is evaluated on held-out
null and alternative draws. In the $p=300$, $\epsilon=0.9$,
$\Lambda_{\mathrm{all}}=64$ example, it gives $32.81\%$, $21.87\%$
and $31.64\%$ for the $\sqrt p$, $p$ and zero spikes, respectively;
projection remains at $20.63\%$. The $\sqrt p$ versus no-spike
paired difference is $1.17$ percentage points (conditional standard
error $0.26$). These diagnostic errors exclude uncertainty in the
estimated null quantiles. The adjustment removes both endpoint
conservatism and finite-bank cutoff randomness; it is not a proposed
replacement for the nuisance-free calibration. Full held-out size
results and both signal strengths are reported in the appendix.

The experiment establishes a restricted empirical advantage over an
equally informed projection, while the limiting corollary identifies
an existence regime. The expanded raw-data comparison also evaluates
mixed signals against external methods. These experiments do not
determine an optimal spike or exhaust suitable shrinkage competitors.

\begin{vFifteenLowerHeadings}
\section{Mixed-signal experiments and limit diagnostics}
\label{app:u-v8-experiments}

The v8 protocols were fixed before inspecting their power outcomes.
The concentrated-omission protocol places the outside component on
coordinate $q+1$; a separately fixed diffuse protocol spreads that
component equally across coordinates $q+1,\ldots,p$. The latter was
added after the first run began, before inspecting its rejection
rates. Both complete grids are retained. The protocols share random
streams, so their estimates are correlated. The numerical tables
compare separate prespecified rules and do not select a rule from
their estimated powers.

\begin{figure}[tbp]
\centering
\includegraphics[width=0.98\linewidth]{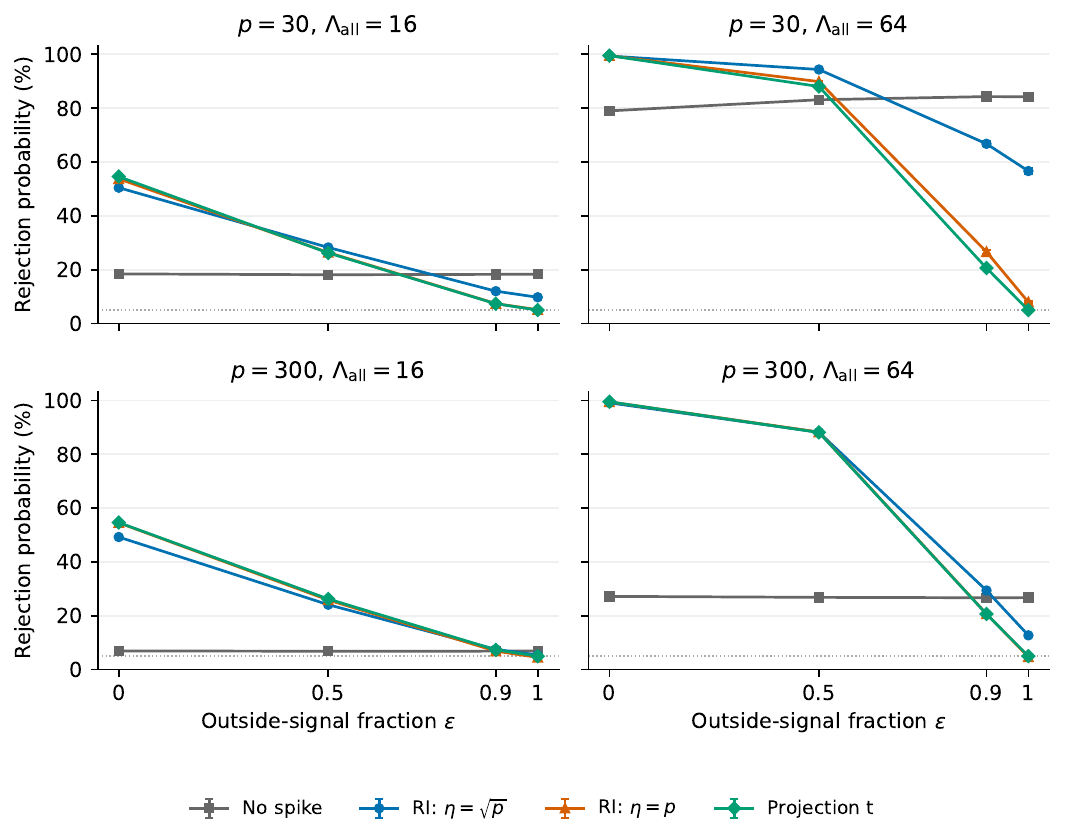}
\caption{Complete non-null grid for diffuse omitted signal at
$N=40$ per group. Error bars are plus or minus two Monte Carlo
standard errors. Each panel retains every prespecified spike and
outside-signal fraction. The dotted line is the nominal $5\%$ level.}
\label{fig:u-v8-grid}
\end{figure}

\begin{table}[tbp]
\centering\small
\caption{Concentrated omitted signal: complete nominal-level grid.
Entries are percentages (independent-bank Monte Carlo standard
errors in percentage points for RI; training-only errors for projection).
The mixed-signal limit does not apply to concentrated omitted energy
of order $\sqrt p$.}
\label{tab:u-v8-concentrated}
\setlength{\tabcolsep}{3pt}
\begin{tabular}{rrlrrrr}
\toprule
$p$ & $\Lambda_{\mathrm{all}}$ & $\epsilon$ & $\eta=0$ & $\eta=\sqrt{p}$ & $\eta=p$ & Projection $t$\\
\midrule
30 & 0 & -- & 3.55 (0.11) & 3.94 (0.12) & 4.36 (0.12) & 5.00 (exact)\\
30 & 16 & 0 & 18.45 (0.27) & 50.47 (0.35) & 53.76 (0.36) & 54.58 (0.09)\\
30 & 16 & 0.5 & 18.36 (0.27) & 28.25 (0.32) & 26.42 (0.31) & 26.23 (0.06)\\
30 & 16 & 0.9 & 18.38 (0.28) & 12.20 (0.23) & 7.49 (0.17) & 7.36 (0.01)\\
30 & 16 & 1 & 18.53 (0.27) & 10.07 (0.22) & 5.03 (0.13) & 5.00 (exact)\\
30 & 64 & 0 & 79.00 (0.29) & 99.38 (0.04) & 99.48 (0.04) & 99.50 (0.01)\\
30 & 64 & 0.5 & 81.75 (0.27) & 94.12 (0.14) & 89.89 (0.19) & 88.05 (0.06)\\
30 & 64 & 0.9 & 80.22 (0.29) & 64.47 (0.44) & 26.88 (0.34) & 20.62 (0.05)\\
30 & 64 & 1 & 79.28 (0.28) & 53.52 (0.48) & 8.24 (0.20) & 5.00 (exact)\\
300 & 0 & -- & 3.62 (0.11) & 4.12 (0.11) & 4.46 (0.11) & 5.00 (exact)\\
300 & 16 & 0 & 6.94 (0.17) & 49.26 (0.31) & 54.55 (0.32) & 54.63 (0.09)\\
300 & 16 & 0.5 & 6.95 (0.17) & 24.07 (0.27) & 25.89 (0.27) & 26.25 (0.06)\\
300 & 16 & 0.9 & 7.01 (0.17) & 7.44 (0.18) & 6.85 (0.15) & 7.36 (0.01)\\
300 & 16 & 1 & 7.02 (0.17) & 5.49 (0.14) & 4.53 (0.11) & 5.00 (exact)\\
300 & 64 & 0 & 27.21 (0.33) & 99.09 (0.05) & 99.45 (0.04) & 99.49 (0.01)\\
300 & 64 & 0.5 & 26.90 (0.33) & 88.11 (0.19) & 88.20 (0.20) & 88.09 (0.06)\\
300 & 64 & 0.9 & 27.07 (0.34) & 30.04 (0.33) & 20.74 (0.25) & 20.63 (0.05)\\
300 & 64 & 1 & 27.03 (0.33) & 13.48 (0.27) & 4.72 (0.12) & 5.00 (exact)\\
\bottomrule
\end{tabular}

\end{table}

\begin{table}[tbp]
\centering\small
\caption{Diffuse omitted signal: infeasible empirical size adjustment
at the true $\rho=4$. Independent $100\,000$-draw null samples
estimate the raw-statistic cutoffs. All displayed rates use held-out
data, and their Monte Carlo errors are conditional on the estimated
cutoffs. Cutoff-estimation uncertainty is not included. These are
diagnostics, not the proposed finite-reference tests.}
\label{tab:u-v8-adjusted}
\setlength{\tabcolsep}{3pt}
\begin{tabular}{rrlrrrr}
\toprule
$p$ & $\Lambda_{\mathrm{all}}$ & $\epsilon$ & $\eta=0$ & $\eta=\sqrt{p}$ & $\eta=p$ & Projection $t$\\
\midrule
30 & 0 & -- & 4.86 (0.11) & 4.96 (0.11) & 4.99 (0.10) & 5.00 (exact)\\
30 & 16 & 0 & 22.45 (0.20) & 54.43 (0.25) & 56.69 (0.24) & 54.58 (0.09)\\
30 & 16 & 0.5 & 22.43 (0.20) & 31.80 (0.23) & 28.68 (0.23) & 26.23 (0.06)\\
30 & 16 & 0.9 & 22.59 (0.20) & 14.62 (0.16) & 8.36 (0.13) & 7.36 (0.01)\\
30 & 16 & 1 & 22.69 (0.20) & 12.14 (0.15) & 5.82 (0.11) & 5.00 (exact)\\
30 & 64 & 0 & 83.26 (0.19) & 99.56 (0.03) & 99.59 (0.03) & 99.50 (0.01)\\
30 & 64 & 0.5 & 86.94 (0.17) & 95.73 (0.10) & 91.34 (0.14) & 88.05 (0.06)\\
30 & 64 & 0.9 & 88.17 (0.16) & 72.93 (0.22) & 29.15 (0.23) & 20.62 (0.05)\\
30 & 64 & 1 & 88.13 (0.16) & 64.19 (0.24) & 9.16 (0.14) & 5.00 (exact)\\
300 & 0 & -- & 4.81 (0.11) & 4.90 (0.10) & 4.81 (0.10) & 5.00 (exact)\\
300 & 16 & 0 & 8.98 (0.14) & 52.27 (0.25) & 56.40 (0.25) & 54.63 (0.09)\\
300 & 16 & 0.5 & 8.96 (0.13) & 26.54 (0.22) & 27.29 (0.22) & 26.25 (0.06)\\
300 & 16 & 0.9 & 8.82 (0.13) & 8.87 (0.15) & 7.36 (0.13) & 7.36 (0.01)\\
300 & 16 & 1 & 8.87 (0.13) & 6.45 (0.12) & 4.88 (0.10) & 5.00 (exact)\\
300 & 64 & 0 & 31.76 (0.24) & 99.28 (0.04) & 99.53 (0.03) & 99.49 (0.01)\\
300 & 64 & 0.5 & 31.54 (0.23) & 89.68 (0.15) & 89.05 (0.15) & 88.09 (0.06)\\
300 & 64 & 0.9 & 31.64 (0.23) & 32.81 (0.24) & 21.87 (0.20) & 20.63 (0.05)\\
300 & 64 & 1 & 31.67 (0.23) & 15.20 (0.17) & 5.06 (0.10) & 5.00 (exact)\\
\bottomrule
\end{tabular}

\end{table}

An independent diagnostic evaluates Proposition~\ref{prop:u-mixed-limit}
with $m_i=n_i=20$, $q=5$, $\rho=4$, $b=c=1$,
$\bm\theta_q=(\sqrt{3.2},0,\ldots,0)^\top$ and
$\tau=28.8/\sqrt{300}$. At each finite dimension, the outside
coordinates have squared magnitude $\tau\sqrt p/(p-q)$, while
the inside component is held fixed. Thus
$\Lambda_{\mathrm{all}}=2(3.2+\tau\sqrt p)$ varies along this
sequence; this diagnostic does not hold the total signal fixed.
It also includes null and inside-only checks.

The mixed-signal limiting power is $29.45\%$ (Monte Carlo standard
error $0.21$ percentage points), using $1\,000$ independent banks
of $100$ draws. Independent finite-dimensional evaluations give
$30.17\%$ ($0.42$) at $p=300$ and $30.25\%$ ($0.44$) at
$p=1200$, each using $250$ banks of $100$ draws. The corresponding
null rates are $4.29\%$ ($0.08$), $4.38\%$ ($0.15$) and
$4.30\%$ ($0.15$). Distribution-function checks and the full
inside-only results are archived. These comparisons do not prove
convergence or provide a bound on its finite-dimensional error;
the proof is in Appendix~\ref{app:u-mixed-limit}.

\section{Sparse-benchmark implementation and numerical verification}
\label{app:u-v13-benchmark}

The complete design is recorded in \texttt{v13\_benchmark\_design.json}.
All 12 cells and nine methods were specified before their outcomes were
generated. For each cell the observations are $N$ paired differences;
the reference budget is $L=199$, so $M=200$ and $k=10$ at
$\alpha=0.05$. A matrix has one uniformly selected row from the
uniform signed-coordinate boundary mixture and 199 independent null rows.
Its 200 fitted scores and all candidate ranks are computed together.
Independent row marks break final menu ties. Each algorithm allocates at
most ten rejections, and the implemented continuous-score and marked-menu
rules allocate exactly ten in every generated matrix. Thus their exact
unconditional null level is 0.05 under the specified Gaussian diagonal model.

The compared methods are the sparse LR benchmark; the bounded block
score $B=(N/q)\sum_{j\in J}Y_j^2/(N-1+Y_j^2)$; the quadratic score
$Q=\sum_{j\in J}Y_j^2$; the maximum $A=\max_{j\in J}|Y_j|$; the
LPO score with spike $\sqrt p$ (RI); the subspace-only LPO baseline (P);
the original classical four-component menu (C); the original seven-component
LPO menu (L); and the combined thirteen-component menu (T). No test of
an individual selected component is substituted for the calibrated menu.

For an augmented matrix let $W_\psi=\sum_i\ell_i\psi_i/\sum_i\ell_i$.
The maximal-invariant mixture density is $M^{-1}\sum_i\ell_i$ relative
to the all-null matrix law. Consequently $\mathbb E W_\psi$ is
the average boundary power, and
$W_*-W_\psi\in[0,1]$ has expectation equal to the LR power advantage.
All 2000 matrix contributions per cell are independent. The standard error
is their sample standard deviation divided by $\sqrt{2000}$; methods in
the same matrix remain paired. We also save rejection of the originally
marked alternative row. Its expectation agrees with that of $W_\psi$;
their finite-run empirical means need not agree exactly.

For $K=12(9+8)=204$ prespecified nontrivial power and gap targets, the
reported finite-reference confidence intervals invert
\[
 \operatorname{kl}(\widehat\mu\,\Vert\,\mu)
 \le \frac{\log(2K/0.025)}{2000}.
\]
Here the Bernoulli relative entropy bound applies to independent variables
in $[0,1]$, including the non-Bernoulli weighted contributions. Convexity
gives $\mathbb E e^{tW}\le 1-\mu+\mu e^t$, and the usual Chernoff
optimization and a union bound give simultaneous failure probability at
most 0.025. The zero LR-versus-itself gap is known identically and is
excluded from $K$. The corresponding Hoeffding intervals are retained in
the numerical archive as a conservative check. Neither interval is a
confidence statement for the unknown power in an observed application.

The ideal LR cutoff is approximated separately. For each cell, 400000
independent central-$t$ vectors give a binomial order-statistic bracket
for its population 0.95 quantile. Another independent sample of 200000
boundary-alternative vectors evaluates both bracket endpoints, using
one-sided Clopper--Pearson bounds. The sum of these error probabilities
over all 12 cells is at most 0.025. Together with the finite-reference
intervals, the simultaneous Monte Carlo coverage is at least 0.95 under
the exact mathematical simulation law. Evaluation uses ordinary
floating-point densities and quantiles, without interval-arithmetic
certification. The ideal powers in Table~\ref{tab:u-v13-ideal} are numerical
estimates with such brackets, not analytically evaluated powers.

\begin{table}[htbp]
\centering\small\setlength{\tabcolsep}{3pt}
\caption{Average boundary powers (percent) for all nine finite-reference rules. Abbreviations are defined in the text; all cells have $p=30$.}\label{tab:u-v13-allpowers}
\begin{tabular}{rrr rrrrrrrrr}\toprule
$N$&$q$&$a_0$&LR&B&Q&A&RI&P&C&L&T\\\midrule
20 & 5 & 1 & 9.12 & 9.01 & 9.12 & 8.72 & 6.89 & 9.04 & 7.57 & 6.35 & 6.77\\
20 & 5 & 2 & 26.10 & 22.90 & 25.00 & 25.63 & 17.33 & 24.54 & 18.74 & 14.18 & 15.97\\
20 & 5 & 4 & 85.78 & 73.51 & 81.32 & 85.72 & 71.43 & 80.29 & 71.96 & 63.41 & 67.64\\
\addlinespace
20 & 10 & 1 & 7.73 & 7.62 & 7.73 & 7.26 & 6.48 & 7.67 & 6.68 & 6.11 & 6.16\\
20 & 10 & 2 & 19.93 & 16.36 & 18.58 & 19.35 & 14.15 & 18.27 & 14.09 & 12.19 & 12.29\\
20 & 10 & 4 & 78.47 & 51.28 & 68.70 & 78.36 & 60.84 & 67.52 & 58.38 & 53.93 & 54.64\\
\addlinespace
40 & 5 & 1 & 9.48 & 9.36 & 9.44 & 9.12 & 6.96 & 9.42 & 7.82 & 6.31 & 6.85\\
40 & 5 & 2 & 28.69 & 25.80 & 26.95 & 28.29 & 18.44 & 26.82 & 20.42 & 14.70 & 17.15\\
40 & 5 & 4 & 89.80 & 82.81 & 85.54 & 89.77 & 76.98 & 85.40 & 77.84 & 68.52 & 73.71\\
\addlinespace
40 & 10 & 1 & 7.99 & 7.86 & 7.94 & 7.57 & 6.49 & 7.93 & 6.85 & 6.13 & 6.21\\
40 & 10 & 2 & 22.34 & 18.50 & 19.98 & 21.85 & 14.84 & 19.88 & 15.21 & 12.63 & 12.98\\
40 & 10 & 4 & 84.33 & 66.43 & 74.16 & 84.26 & 66.11 & 73.85 & 64.83 & 59.49 & 60.62\\
\bottomrule\end{tabular}\end{table}

\begin{table}[htbp]
\centering\small\setlength{\tabcolsep}{3pt}
\caption{Average boundary-power gaps (percentage points) from the finite-reference LR benchmark. Abbreviations are defined in the text; all cells have $p=30$.}\label{tab:u-v13-allgaps}
\begin{tabular}{rrr rrrrrrrrr}\toprule
$N$&$q$&$a_0$&LR&B&Q&A&RI&P&C&L&T\\\midrule
20 & 5 & 1 & 0.00 & 0.12 & 0.00 & 0.41 & 2.23 & 0.09 & 1.55 & 2.77 & 2.36\\
20 & 5 & 2 & 0.00 & 3.20 & 1.10 & 0.48 & 8.77 & 1.57 & 7.36 & 11.92 & 10.13\\
20 & 5 & 4 & 0.00 & 12.27 & 4.46 & 0.06 & 14.35 & 5.49 & 13.81 & 22.37 & 18.14\\
\addlinespace
20 & 10 & 1 & 0.00 & 0.10 & 0.00 & 0.47 & 1.25 & 0.06 & 1.05 & 1.62 & 1.57\\
20 & 10 & 2 & 0.00 & 3.58 & 1.35 & 0.58 & 5.78 & 1.67 & 5.84 & 7.74 & 7.64\\
20 & 10 & 4 & 0.00 & 27.18 & 9.76 & 0.11 & 17.62 & 10.94 & 20.09 & 24.53 & 23.82\\
\addlinespace
40 & 5 & 1 & 0.00 & 0.12 & 0.04 & 0.36 & 2.53 & 0.06 & 1.66 & 3.18 & 2.64\\
40 & 5 & 2 & 0.00 & 2.90 & 1.75 & 0.41 & 10.25 & 1.87 & 8.28 & 14.00 & 11.54\\
40 & 5 & 4 & 0.00 & 6.99 & 4.27 & 0.04 & 12.83 & 4.41 & 11.96 & 21.28 & 16.10\\
\addlinespace
40 & 10 & 1 & 0.00 & 0.13 & 0.04 & 0.42 & 1.50 & 0.06 & 1.14 & 1.86 & 1.77\\
40 & 10 & 2 & 0.00 & 3.84 & 2.36 & 0.49 & 7.50 & 2.46 & 7.13 & 9.71 & 9.36\\
40 & 10 & 4 & 0.00 & 17.89 & 10.16 & 0.07 & 18.22 & 10.48 & 19.49 & 24.84 & 23.70\\
\bottomrule\end{tabular}\end{table}

\begin{table}[htbp]
\centering\small\setlength{\tabcolsep}{3pt}
\caption{Simultaneous Chernoff upper bounds (percentage points) for the average boundary-power gaps. Abbreviations are defined in the text; all cells have $p=30$.}\label{tab:u-v13-gapbounds}
\begin{tabular}{rrr rrrrrrrrr}\toprule
$N$&$q$&$a_0$&LR&B&Q&A&RI&P&C&L&T\\\midrule
20 & 5 & 1 & 0.00 & 0.83 & 0.50 & 1.38 & 4.01 & 0.76 & 3.10 & 4.70 & 4.17\\
20 & 5 & 2 & 0.00 & 5.25 & 2.47 & 1.50 & 11.82 & 3.12 & 10.21 & 15.36 & 13.36\\
20 & 5 & 4 & 0.00 & 15.74 & 6.79 & 0.68 & 18.02 & 8.02 & 17.44 & 26.64 & 22.13\\
\addlinespace
20 & 10 & 1 & 0.00 & 0.80 & 0.50 & 1.49 & 2.68 & 0.68 & 2.39 & 3.18 & 3.13\\
20 & 10 & 2 & 0.00 & 5.71 & 2.82 & 1.68 & 8.36 & 3.25 & 8.44 & 10.64 & 10.53\\
20 & 10 & 4 & 0.00 & 31.70 & 12.94 & 0.81 & 21.58 & 14.27 & 24.22 & 28.93 & 28.18\\
\addlinespace
40 & 5 & 1 & 0.00 & 0.84 & 0.64 & 1.31 & 4.39 & 0.70 & 3.25 & 5.21 & 4.53\\
40 & 5 & 2 & 0.00 & 4.86 & 3.36 & 1.39 & 13.50 & 3.53 & 11.26 & 17.64 & 14.93\\
40 & 5 & 4 & 0.00 & 9.78 & 6.56 & 0.62 & 16.36 & 6.73 & 15.40 & 25.49 & 19.93\\
\addlinespace
40 & 10 & 1 & 0.00 & 0.86 & 0.65 & 1.40 & 3.03 & 0.68 & 2.51 & 3.51 & 3.40\\
40 & 10 & 2 & 0.00 & 6.04 & 4.18 & 1.52 & 10.36 & 4.31 & 9.95 & 12.89 & 12.50\\
40 & 10 & 4 & 0.00 & 21.87 & 13.40 & 0.71 & 22.22 & 13.75 & 23.58 & 29.24 & 28.05\\
\bottomrule\end{tabular}\end{table}

\begin{table}[htbp]\centering\small\setlength{\tabcolsep}{5pt}
\caption{Ideal and finite-reference sparse benchmarks. Ideal entries are independently approximated powers (percent) and simultaneous Monte Carlo brackets. The coordinate oracle knows the active coordinate but not its sign or variance, and is outside the unknown-coordinate class.}\label{tab:u-v13-ideal}
\begin{tabular}{rrr r r r r}\toprule
$N$&$q$&$a_0$&Finite LR&Ideal LR&Ideal bracket&Coordinate oracle\\\midrule
20 & 5 & 1 & 9.12 & 9.23 & [8.85, 9.65] & 15.81\\
20 & 5 & 2 & 26.10 & 26.68 & [26.01, 27.32] & 47.56\\
20 & 5 & 4 & 85.78 & 86.47 & [86.00, 86.92] & 96.64\\
20 & 10 & 1 & 7.73 & 7.76 & [7.40, 8.10] & 15.81\\
20 & 10 & 2 & 19.93 & 20.34 & [19.75, 20.92] & 47.56\\
20 & 10 & 4 & 78.47 & 79.34 & [78.74, 79.95] & 96.64\\
40 & 5 & 1 & 9.48 & 9.58 & [9.20, 9.98] & 16.41\\
40 & 5 & 2 & 28.69 & 29.04 & [28.36, 29.71] & 49.63\\
40 & 5 & 4 & 89.80 & 90.12 & [89.74, 90.52] & 97.38\\
40 & 10 & 1 & 7.99 & 8.05 & [7.72, 8.41] & 16.41\\
40 & 10 & 2 & 22.34 & 22.44 & [21.82, 23.07] & 49.63\\
40 & 10 & 4 & 84.33 & 84.86 & [84.37, 85.36] & 97.38\\
\bottomrule\end{tabular}\end{table}

The archive stores every likelihood-ratio vector, bit-packed row decision,
matrix power/gap contribution and direct marked-row rejection. The audit
reconstructs all 24000 allocations, all 108 summary rows and all interval
endpoints. It independently regenerates the first matrix in every setting
using the NumPy all-pairs LPO implementation and checks it against the
production C++ implementation. These are internal reproducibility checks;
they are distinct from a full independent rerun or external review.

The comparison estimates average power under the prescribed boundary mixture.
It does not establish that a competing procedure has its worst-case power
there, nor that the displayed choice between block and maximum scores can
be made after viewing the data without an additional calibration argument.
The fixed sparse class supplies a finite-budget efficiency benchmark;
general unknown-shape small-budget efficiency remains unresolved.

\section{Numerical evaluation of the scalar efficiency bounds}
\label{app:u-v14-bounds-evaluation}

Table~\ref{tab:u-v14-bounds} evaluates the analytic bounds for
$N\in\{20,40\}$, $q\in\{5,10\}$, $a_0\in\{0.5,1,2,4\}$ and
$\tau=0.05$. These calculations use one-dimensional Beta integrals;
they generate no new observations or reference banks. The original
remainder-mass bound is compared with its earlier variance refinement
and the optimized affine-residual bound. The latter two bounds include
their dependence on $N$ and $q$. Each reported value is the minimum
with the preceding bounds, so the sequence of reported bounds cannot
increase.

At $a_0=1$, the affine bounds range from 1.40 to 2.37 percentage
points, compared with 9.02 percentage points for the original
remainder-mass bound. At $a_0=2$, the corresponding bounds range
from 20.22 to 38.87 percentage points, compared with 59.40 percentage
points. The improvement is therefore quantitative, but it remains
far from the archived Monte Carlo gaps in the latter regime.
At $a_0=4$, the optimized residual variance is too large to improve
the remainder-mass bound, which remains 99.70 percentage points.
This bounded-quadratic calculation does not supply a useful strong-signal
efficiency guarantee. Increasing $q$ lowers the variance bound through averaging;
this does not imply that either test's absolute power increases.
The signal amplitude is standardized by $\sqrt N$, so comparing
different $N$ at fixed $a_0$ also does not hold the unstandardized
mean difference fixed.

For reproducibility, the integrals are computed after the substitution
$X=t^2$, with density
$2(1-t^2)^{r/2-1}/\mathrm B(1/2,r/2)$ on $(0,1)$.
This removes the Beta density's square-root singularity at zero.
The covariance used in the optimal slope is independently evaluated
from the positive Poisson-mixture series. All eight scalar
$(N,a_0)$ settings are evaluated with 40 and 70 decimal digits;
the largest absolute difference, divided by the larger of one and
the absolute reference value, is below $1.9\times10^{-37}$.
A separate 256-node Gauss--Jacobi calculation agrees within
$2.2\times10^{-11}$ on the same scaled criterion.
Normalization, the first Beta moment, the Poisson covariance
identity and the variance projection identity are checked separately.
These are numerical convergence diagnostics, not interval-arithmetic
certificates for the printed decimal bounds.

The final column retains the earlier bounded-rule boundary-power
gap estimates from Section~\ref{app:u-v13-benchmark}; it is included
to show the remaining looseness of the analytic bounds.
It neither changes those simulation results nor replaces their
simultaneous Monte Carlo confidence bounds. No corresponding
simulation was conducted at $a_0=0.5$.
The script \texttt{evaluate\_v14\_bounds.py} writes the complete
numerical values, high-precision diagnostics and table.

\begin{table}[t]
\centering\small
\setlength{\tabcolsep}{4pt}
\caption{Numerically evaluated analytic upper bounds for the bounded quadratic rule at $\tau=0.05$. All bound entries and archived average power gaps are in percentage points. The original bound is $\Delta(a_0^2/2)$; the variance column includes the earlier Beta-integral refinement; the affine column takes the minimum with the optimized affine-residual bound. The final column is an archived Monte Carlo estimate, not an analytic bound.}
\label{tab:u-v14-bounds}
\begin{tabular}{rrr rrrr}
\toprule
$N$ & $q$ & $a_0$ & Original & Variance & Affine & MC gap\\
\midrule
20 & 5 & 0.5 & 0.72 & 0.28 & 0.13 & ---\\
20 & 5 & 1 & 9.02 & 3.83 & 1.98 & 0.12\\
20 & 5 & 2 & 59.40 & 39.18 & 28.59 & 3.20\\
20 & 5 & 4 & 99.70 & 99.70 & 99.70 & 12.27\\
20 & 10 & 0.5 & 0.72 & 0.20 & 0.09 & ---\\
20 & 10 & 1 & 9.02 & 2.71 & 1.40 & 0.10\\
20 & 10 & 2 & 59.40 & 27.70 & 20.22 & 3.58\\
20 & 10 & 4 & 99.70 & 99.70 & 99.70 & 27.18\\
40 & 5 & 0.5 & 0.72 & 0.31 & 0.15 & ---\\
40 & 5 & 1 & 9.02 & 4.29 & 2.37 & 0.12\\
40 & 5 & 2 & 59.40 & 49.16 & 38.87 & 2.90\\
40 & 5 & 4 & 99.70 & 99.70 & 99.70 & 6.99\\
40 & 10 & 0.5 & 0.72 & 0.22 & 0.11 & ---\\
40 & 10 & 1 & 9.02 & 3.04 & 1.68 & 0.13\\
40 & 10 & 2 & 59.40 & 34.76 & 27.48 & 3.84\\
40 & 10 & 4 & 99.70 & 99.70 & 99.70 & 17.89\\
\bottomrule
\end{tabular}
\par\smallskip\parbox{0.96\linewidth}{\footnotesize The 40- and 70-decimal quadrature evaluations agree to at least 30 scaled decimal places and are independently checked by Gauss--Jacobi quadrature. These are numerical convergence checks, not certified interval enclosures. No new Monte Carlo experiment is used in this table; $a_0=0.5$ has no archived gap estimate.}
\end{table}

\section{Numerical evaluation of the conditional maximum bound}
\label{app:u-v16-conditional-evaluation}

Table~\ref{tab:u-v16-conditional-bounds} evaluates the bounds in
Theorem~\ref{thm:u-v16-conditional-power} for
$N\in\{20,40\}$, $q\in\{5,10\}$ and $a_0\in\{1,2,4\}$.
The ideal test uses $\alpha=0.05$, and the finite-reference test uses
$M=200$ rows and $k=10$ rejections. The two bounds address different
calibration experiments and are evaluated separately. Both compare
the maximum-coordinate rule with the corresponding sparse LR
benchmark. They do not replace the bounds for the bounded quadratic
rule in Table~\ref{tab:u-v14-bounds}.

The finite-reference bounds range from 1.37 to 1.80 percentage points
at $a_0=2$ and from 0.24 to 0.54 points at $a_0=4$. The earlier LR
power estimates in these designs range from 19.93\% to 28.69\%
and from 78.47\% to 89.80\%, respectively. Thus the conditional
comparison remains informative beyond the weak-signal region in
which the bounded-quadratic expansion was most accurate.
The maximum test uses the specified coordinate set $J$ but does not
require $a_0$; its efficiency comparison does require the specified
alternative class. These evaluations establish neither an adaptive
choice between maximum and aggregation tests nor a direction-learning
lower bound.

The main implementation computes the first truncated moment from
the positive Poisson--Beta series in
\eqref{eq:u-v16-moment-series}. For the second moment, it squares
the positive power series for the coordinate likelihood ratio and
integrates the resulting monomials against the Beta density.
Gauss--Legendre quadrature evaluates the ideal integral
\eqref{eq:u-v16-ideal-integral} separately on either side of
$1-\alpha$. This split accounts for the absolute-value kink at
the rejection boundary. For \eqref{eq:u-v16-finite-integral},
Gauss--Jacobi quadrature integrates the
$\operatorname{Beta}(M-k+1,k)$ cutoff, and Gauss--Legendre
quadrature integrates the rows below and above it. The threshold
row's contribution is included explicitly.

Each design is evaluated with 80 and 160 quadrature nodes, truncating
the coordinate series after indices 100 and 130, respectively.
The script records the changes in both bounds and in the full
second moment, the remaining first-moment Poisson mass, and
roundoff in the conditional variance. The exact equality case
$q=1$ is checked separately.
An independent implementation checks all four $a_0=2$ designs by
directly integrating the coordinate likelihood ratio and its square
after the substitution $x=z^2$, which removes the Beta density's
singularity at zero. It uses adaptive outer integration, with the
same boundary split for the ideal bound. This route does not use the
positive-series moment calculation.

The largest finite-bound change between the two quadrature resolutions
is below $1.1\times10^{-8}$ in probability units. At $a_0=2$,
the independent ideal and finite-reference calculations agree
with the main implementation within $10^{-8}$ and $10^{-12}$,
respectively. These are floating-point convergence and
implementation checks, not interval-arithmetic enclosures.
The independent direct-moment check targets $a_0=2$; the
$a_0=1,4$ evaluations have the stated resolution and truncation
checks.

The script \texttt{evaluate\_v16\_conditional\_bounds.py} generates
the evaluations, convergence records and table.
\texttt{audit\_v16\_conditional\_bounds.py} performs the independent
direct-moment check. Neither script generates random observations
or Monte Carlo reference banks. The last two table columns retain
the earlier simulated LR powers and maximum-to-LR gaps from
Section~\ref{app:u-v13-benchmark}, with their original uncertainty
analysis; they are not newly simulated values.

\begin{table}[t]
\centering\small
\setlength{\tabcolsep}{5pt}
\caption{Conditional-moment upper bounds for the maximum-coordinate test relative to the sparse LR benchmark. Ideal calibration uses $\alpha=0.05$; finite calibration uses $M=200$ and $k=10$. Bound evaluations and archived maximum-to-LR gaps are in percentage points; archived LR power is in percent.}
\label{tab:u-v16-conditional-bounds}
\begin{tabular}{rrr rrrr}
\toprule
$N$ & $q$ & $a_0$ & Ideal bound & Finite bound & LR power & MC gap\\
\midrule
20 & 5 & 1 & 1.11 & 1.12 & 9.12 & 0.41 \\
20 & 5 & 2 & 1.43 & 1.50 & 26.10 & 0.48 \\
20 & 5 & 4 & 0.28 & 0.34 & 85.78 & 0.06 \\
\addlinespace
20 & 10 & 1 & 1.30 & 1.30 & 7.73 & 0.47 \\
20 & 10 & 2 & 1.75 & 1.80 & 19.93 & 0.58 \\
20 & 10 & 4 & 0.45 & 0.54 & 78.47 & 0.11 \\
\addlinespace
40 & 5 & 1 & 1.04 & 1.05 & 9.48 & 0.36 \\
40 & 5 & 2 & 1.30 & 1.37 & 28.69 & 0.41 \\
40 & 5 & 4 & 0.20 & 0.24 & 89.80 & 0.04 \\
\addlinespace
40 & 10 & 1 & 1.21 & 1.21 & 7.99 & 0.42 \\
40 & 10 & 2 & 1.57 & 1.63 & 22.34 & 0.49 \\
40 & 10 & 4 & 0.32 & 0.39 & 84.33 & 0.07 \\
\bottomrule
\end{tabular}
\par\smallskip\parbox{0.98\linewidth}{\footnotesize Bounds are floating-point evaluations of analytic inequalities, with quadrature and independent-method checks; they are not interval-arithmetic enclosures. The final two columns reproduce the earlier simulated values in Table~\ref{tab:u-v13-benchmark}. No new random observations or reference banks are generated.}
\end{table}

\section{Matched comparisons at fixed mixed-signal allocations}
\label{app:u-v14-mixed-experiment}

The preceding proposition permits a finite-dimensional comparison with
both concentrated and diffuse signal. Before generating outcomes, we
fixed $N=20$, $p=30$, $q=5$, total standardized noncentrality $16$,
and outside fractions $\epsilon\in\{0,0.5,0.9\}$. Thus
$a_0=4\sqrt{1-\epsilon}$ and $d_0=4\sqrt\epsilon$.
Inside the first $q$ coordinates, one uniform coordinate has mean
noncentrality $\pm a_0$; each of the other $p-q$ coordinates has
independently signed noncentrality $d_0/\sqrt{p-q}$. These comparisons
hold the total noncentrality fixed and vary its allocation. The
outside lower bound, signs, pairing, coordinate set and allocation
belong to the stated simulation experiment; they are not estimated
from the tested observations.

For each allocation we generated 2000 independent augmented matrices,
each with one uniformly located boundary-orbit row and 199 independent
null rows. Every row contains $N$ paired Gaussian differences and is
fitted separately. The eleven procedures use the same rows and marks,
with exactly ten hypothetical rejections in every matrix. In addition
to the nine procedures in Section~\ref{sec:u-v13-benchmark-experiment},
we include the all-coordinate sum of squared $t$ statistics and the
prespecified block score
$\sum_{j=1}^pY_j^2+\sqrt p\sum_{j\in J}Y_j^2$.
The sparse LR is replaced by \eqref{eq:u-v14-mixed-lr}. None of the
RI, block or menu tuning constants is optimized after the outcomes.
The LR has the allocation information required by its parameter class;
it is an attainable comparison target when that information is supplied,
not an adaptive procedure for an unknown allocation.

Normalized likelihood-ratio weights give the bounded power and gap
contributions justified in Section~\ref{app:u-v14-mixed-orbit}.
Two-sided bounded-variable KL--Chernoff intervals allocate failure
probability $0.05$ over all 33 power and 30 nonzero-gap targets.
This simultaneous statement applies to this study; it is separate
from confidence statements for other experiments. Floating-point
likelihood-ratio evaluation is not interval-certified arithmetic.
The maximum power and gap Monte Carlo standard errors are $0.72$
and $0.60$ percentage points, respectively.

\begin{table}[htbp]
\centering\small
\caption{Fixed-allocation mixed-orbit comparisons at $N=20$, $p=30$, $q=5$ and total standardized signal energy $16$. Entries are boundary power percentages, with simultaneous upper bounds on the LR power gap in percentage points in parentheses. Each allocation uses 2000 independent matrices, $M=200$ and $k=10$. All eleven procedures receive the same data; the LR alone uses the specified allocation. The confidence allocation covers all 63 power and nonzero-gap targets in this study with probability at least $0.95$ under the simulation law.}
\label{tab:u-v14-mixed}
\begin{tabular}{lrrr}\toprule
Method & $\epsilon=0$ & $\epsilon=0.5$ & $\epsilon=0.9$ \\ \midrule
Mixed LR & 85.68 (0.00) & 61.22 (0.00) & 51.97 (0.00) \\
Bounded block & 72.75 (16.08) & 42.62 (22.20) & 11.60 (44.75) \\
Quadratic, $J$ & 81.07 (6.70) & 48.61 (15.74) & 12.25 (44.10) \\
Maximum, $J$ & 85.62 (0.58) & 51.84 (12.17) & 11.57 (44.78) \\
LPO, $c=1$ & 71.38 (17.57) & 44.09 (20.63) & 17.21 (39.05) \\
LPO, $J$ only & 80.06 (7.89) & 47.65 (16.78) & 11.88 (44.47) \\
Classical menu, $K=4$ & 71.82 (17.09) & 49.84 (14.39) & 40.72 (14.25) \\
LPO menu, $K=7$ & 62.93 (26.59) & 45.01 (19.64) & 35.72 (19.69) \\
Combined menu, $K=13$ & 67.30 (21.96) & 45.07 (19.58) & 36.59 (18.75) \\
Quadratic, all coordinates & 45.17 (44.89) & 47.53 (16.92) & 47.90 (6.06) \\
Fixed block quadratic & 79.89 (8.09) & 54.71 (8.92) & 21.46 (34.68) \\
\bottomrule\end{tabular}
\end{table}

At the sparse endpoint the maximum-coordinate rule is close to the
LR target. With half the energy outside, the LR has power $61.22\%$,
compared with $54.71\%$ for the fixed block quadratic and $44.09\%$
for the learned LPO score. Their estimated LR gaps are $6.51$ and
$17.13$ percentage points, with simultaneous intervals $[4.55,8.92]$
and $[13.97,20.63]$. With $90\%$ of energy outside, the all-coordinate
quadratic has power $47.90\%$ against $51.97\%$ for LR, whereas
LPO has power $17.21\%$; the respective gap intervals are
$[2.55,6.06]$ and $[30.63,39.05]$ percentage points.
These examples extend the efficiency comparison beyond a single
active coordinate. They continue to favor direct aggregation over
the particular learned score in the specified designs, and do not
establish a general ordering over mixed signals or unknown allocations.

The generator \texttt{simulate\_v14\_mixed.py} saves every log-likelihood
ratio, packed rejection allocation, bounded contribution and alternative
row/sign label, with a frozen design and seed. The renderer
\texttt{render\_v14\_mixed.py} rebuilds Table~\ref{tab:u-v14-mixed}
from the archived summary. All 6000 matrices passed the pathwise
budget and centered-allocation identity checks; the first matrix of
each design also passed row-permutation and positive-coordinate-scale
checks. Density-ratio checks compare the hypergeometric expression
with independently evaluated noncentral-$t$ densities.
The separate audit \texttt{audit\_v14\_mixed.py} reconstructs all
stored contributions and summary entries and regenerates the first matrix
of each design, checking its LPO features against independent NumPy all-pair
sums.

\subsection{Allocation-informed block comparisons}
\label{app:u-v17-informed-mixed}

The mixed likelihood ratio in Proposition~\ref{prop:u-v14-mixed-maximin}
uses the specified allocation $(a_0,d_0)$, whereas the block and LPO spectra
in Table~\ref{tab:u-v14-mixed} are held fixed across allocations.
We therefore conduct a follow-up comparison in which two additional
nonlearned block scores receive the same allocation information as the
likelihood ratio. Write $h=p-q$ and define
\begin{align}
 T_{\mathrm{IQ}}(Y)
 &=\frac{a_0^2}{q}\sum_{j\in J}Y_j^2
   +\frac{d_0^2}{h}\sum_{j\notin J}Y_j^2,
   \label{eq:u-v17-informed-quadratic}\\
 T_{\mathrm{IB}}(Y)
 &=\frac{a_0^2}{q}\sum_{j\in J}\frac{NY_j^2}{N-1+Y_j^2}
   +\frac{d_0^2}{h}\sum_{j\notin J}\frac{NY_j^2}{N-1+Y_j^2}.
   \label{eq:u-v17-informed-bounded}
\end{align}
These weights distribute each block's specified energy equally over its
coordinates. The bounded version is motivated by the quadratic term in
the symmetric likelihood-ratio expansion: replacing $(a_0,d_0)$ by
$(t a_0,t d_0)$ gives
$\ell_{t a_0,t d_0}(Y)
=1+\tfrac12t^2\{T_{\mathrm{IB}}(Y)-a_0^2-d_0^2\}+O(t^4)$
for fixed $Y$. We use this expansion only to specify the score; its
weights are not asserted to maximize finite-amplitude power.
The ordinary quadratic uses the same energy weights as a simpler
unbounded comparison.

The five procedures in the follow-up are the mixed LR, the original
fixed block quadratic, the original LPO score with $c=1$, and the two
scores above. All receive the same paired observations, $J$, and reference
budget. The LR and the two new scores also receive the same
$(a_0,d_0)$; the fixed block and LPO spectra remain as controls for the
earlier comparison. Each rule rejects the largest $k=10$ among $M=200$
rows, using an independent row-attached uniform to break ties. The null
level is therefore $k/M$ by exchangeability. No observations are used
to select a weight or a score from this collection.

The follow-up was motivated by the earlier results. Before generating
new outcomes, we fixed the five scores, $N=20$, $p=30$, $q=5$, total
standardized energy $16$, and outside fractions
$\epsilon\in\{0,0.5,0.9\}$. We then generated 2000 new independent
augmented matrices for each allocation, with new seeds. The boundary
mixture, likelihood-ratio weighting and bounded power estimator are the
same as in Section~\ref{app:u-v14-mixed-experiment}. The new estimates do not pool
historical contributions. A separate replay of the first 100 archived
matrices in each allocation reproduces their likelihood ratios and
alternative rows, coordinates and signs exactly; this is an
implementation check, not additional simulation evidence.

\begin{table}[htbp]
\centering\small
\caption{Fresh allocation-informed comparisons at $N=20$, $p=30$, $q=5$ and total standardized signal energy $16$. Each entry is boundary power in percent, followed by the estimated LR gap in percentage points. The LR and the two allocation-informed scores receive the same specified $(a_0,d_0,J)$. The fixed block and LPO scores retain their earlier spectra as controls. Each allocation uses 2000 new independent matrices, $M=200$ and $k=10$.}
\label{tab:u-v17-informed-mixed}
\begin{tabular}{lrrr}
\toprule Method & $\epsilon=0$ & $\epsilon=0.5$ & $\epsilon=0.9$ \\
\midrule
Mixed LR & 85.99 (0.00) & 61.07 (0.00) & 51.62 (0.00) \\
Fixed block quadratic & 80.27 (5.73) & 54.57 (6.50) & 20.93 (30.68) \\
LPO, $c=1$ & 71.79 (14.20) & 43.95 (17.13) & 16.69 (34.93) \\
Allocation-informed quadratic & 81.57 (4.43) & 55.76 (5.31) & 48.06 (3.56) \\
Allocation-informed bounded score & 73.81 (12.18) & 52.16 (8.91) & 51.00 (0.61) \\
\bottomrule\end{tabular}
\par\smallskip\parbox{0.98\linewidth}{\footnotesize These are fresh Monte Carlo estimates, not the archived values in Table~\ref{tab:u-v14-mixed}. Simultaneous intervals for the 33 power, gap and signed-contrast targets are provided in the numerical archive. Their joint coverage is at least $0.95$ under the stated fixed-design simulation law. The largest matrix-level MCSE is 0.46 percentage points for power and 0.28 for a gap.}
\end{table}

At $\epsilon=0.5$, the mixed LR has estimated power $61.07\%$,
compared with $55.76\%$ for the informed quadratic and $52.16\%$
for the informed bounded score. The fixed block and LPO controls have
powers $54.57\%$ and $43.95\%$, respectively. The informed quadratic's
estimated improvement over the fixed block is $1.19$ percentage points;
its simultaneous signed-difference interval contains zero. This design
therefore leaves a nonnegligible gap between the LR and both supplied
energy-weighted scores.

At $\epsilon=0.9$, allocating the block weights using the supplied
information changes the comparison substantially. The LR, informed
quadratic and informed bounded score have powers $51.62\%$,
$48.06\%$ and $51.00\%$, respectively, whereas the fixed block
has power $20.93\%$. The informed bounded score's estimated LR gap
is $0.61$ percentage points, with a simultaneous upper confidence
bound of $1.53$ points. Its improvement over the fixed block is
$30.07$ points, with simultaneous interval $[21.86,38.00]$ points.
Thus the earlier fixed-weight comparison does not measure an
unavoidable cost of learning a direction. The scores' use of the
specified allocation can explain a substantial part of that comparison.
The improvement is not uniform: at $\epsilon=0$, the informed bounded
score has lower estimated power than the fixed block.

The numerical archive reports simultaneous bounded-variable
KL--Chernoff intervals for all 33 targets: five powers, four nonzero
LR gaps and two informed-minus-fixed power differences in each
allocation. Signed differences lie in $[-1,1]$ and are mapped to
$[0,1]$ before applying the same inequality. The joint coverage is at
least $0.95$ under the fixed-design simulation law. Matrix-level
MCSEs are also reported. These probability statements are subject
to the documented floating-point implementation; they are not
interval-arithmetic certifications.

The script \texttt{simulate\_v17\_informed\_mixed.py} freezes the
protocol, generates the fresh matrices and reconstructs the tables.
Together with \texttt{audit\_v17\_informed\_mixed.py}, it checks
row budgets, bounded power contributions, nonnegative LR gaps and
allocation identities. The audit independently reconstructs the
informed scores and replays ten fresh matrices per allocation;
the simulation script checks positive coordinate-scale invariance
and row-permutation equivariance. Historical random archives
remain unchanged.

\end{vFifteenLowerHeadings}
\subsection{Finite-simulation certificates for the mixed approximation}
\label{app:u-v9-limit-certificate}
A separate, prespecified simulation evaluates the three signal configurations of
the mixed-limit diagnostic at $p=300,1200$ and in the limiting law.
Each of the nine power estimates uses $3\,000$ independent reference banks,
$L=199$ references per bank, and $20$ independent datasets per bank.
The seeds and outcomes are independent of the earlier diagnostic.
The observation for uncertainty quantification is the bank rejection fraction
$X_b\in[0,1]$, so that its expectation is the actual finite-reference power.
For its sample variance $S^2$ and $B=3\,000$, define
\[
 r_B=\sqrt{\frac{2S^2\log(4J/\delta)}{B}}
       +\frac{7\log(4J/\delta)}{3(B-1)},
 \qquad J=9,\quad\delta=0.05.
\]
The empirical Bernstein inequality of \citet[Theorem~4]{maurer2009},
applied to both tails and followed by a union bound, covers all nine
means simultaneously with probability at least $0.95$. No independence
between signal configurations is needed. Consequently
\[
 |\pi_p-\pi_\infty|\le
 \min\{1,|\widehat\pi_p-\widehat\pi_\infty|+r_{B,p}+r_{B,\infty}\}
\]
holds simultaneously for the six comparisons with the same coverage.
\begin{table}[tbp]
\centering
\caption{Finite-simulation certificates. Powers are percentages and error bounds are in
percentage points; the last column is a simultaneous $95\%$ upper bound
on the absolute difference of the true finite-dimensional and limiting
rank-test powers at these specified configurations.}
\label{tab:u-v9-limit-certificate}
\begin{tabular}{rlrrr}
\toprule
$p$ & Signal & Finite power & Limit power & Error bound\\
\midrule
300 & Null & 4.26 & 4.45 & 1.84\\
300 & Inside & 17.45 & 17.58 & 2.36\\
300 & Mixed & 29.36 & 29.59 & 2.76\\
1200 & Null & 4.34 & 4.45 & 1.77\\
1200 & Inside & 18.18 & 17.58 & 2.83\\
1200 & Mixed & 30.08 & 29.59 & 3.02\\
\bottomrule
\end{tabular}
\end{table}
These are probability-over-simulation certificates at six fixed designs.
They supplement the analytic bound but do not establish a uniform
finite-dimensional approximation, an optimal spike, or negligible
unknown-shape cost. The full bank counts, protocol, timings and interval
reconstruction are included in the numerical archive.

\section{Structural information and model sensitivity}
\label{sec:u-numerical-supplement}
\begin{vFifteenLowerHeadings}
\section{Controlled shape, direction and spectrum ablations}
\label{supp:v11ablation}
\subsection{Frozen design and exact feature decomposition}
The controlled experiment uses $N=40$ observations per group,
$q=5$, $\rho=4$, and $p\in\{30,300\}$. Write
$H=\operatorname{diag}(h_j)$, where
$h_j=2^{2(j-1)/(p-1)-1}$,
and $C=5H$ for the covariance of the fixed differences $Z_i=X_i-Y_i$.
Independent observations satisfy $X_i\sim N_p(\delta,H)$ and
$Y_i\sim N_p(0,4H)$. Their signal strength is
$\Lambda_{\rm all}=N\delta^TC^{-1}\delta$.
The seven prespecified settings at each dimension are the null;
$\Lambda_{\rm all}=16$ entirely inside $Q$ along either $e_1$ or the
uniform unit vector on $Q$; and diffuse outside components with
$(\Lambda_{\rm all},\epsilon)=(64,.5),(64,.9),(64,1),(16,1)$.
For the mixed cases the inside component follows $e_1$.
All $14$ settings and all $39$ procedures are retained.

For each held-out pair, the estimated-shape version uses the diagonal
sample covariance of the other $N-2$ differences, with divisor $N-3$.
The known-shape version substitutes the true $C$ and is infeasible;
here this diagnostic includes knowledge of the paired covariance scale.
Each version returns separate full-baseline, $Q$-baseline, learned-rank,
and two fixed-axis rank contributions. The learned direction is the
standardized retained mean restricted to $Q$. Every contribution is
multiplied by $2/(N-1)$ after summing all $\binom{40}{2}=780$ pairs,
so that each feature has the scale $NU$.
For either baseline, the prespecified spectrum grid is
$b=1$, $\eta=c\sqrt p$, $c\in\{0,.25,.5,1,2,4\}$;
a rank-only version uses $b=0$. No pair is subsampled.
At $c=0$, the estimated-shape score is the diagonal leave-two-out
statistic of \citet{park2013}, applied to all coordinates or restricted
to $Q$, with our finite Gaussian null calibration.
The feature decomposition is an algebraic decomposition of the score,
not an additive decomposition of power losses.

The two fixed-axis controls use $e_1$ and the uniform unit vector on $Q$
in population-whitened coordinates. Those same coordinate vectors are
held fixed when inserted into estimated whitening. They are not claimed
to optimize power conditional on a fitted shape. Under complete omission
of the signal, neither is a population signal direction in $Q$.
The additional LPO block control replaces the learned rank contribution
by the entire $Q$ baseline, giving full baseline plus $\sqrt p$ times
$Q$ baseline. It belongs to a different weight class.

\subsection{Calibration, uncertainty and interpretation}
All methods receive exactly the same raw $X,Y$ observations and fixed
pairing. Every simulated-null procedure has its own upper-tail rank
threshold from the same $L=199$ independent canonical Gaussian reference
datasets per bank, regenerating all retained-sample fits. The subspace
Hotelling control uses its exact paired Gaussian $F$ reference.
There are $200$ independent banks, each with $50$ observed datasets,
for $10\,000$ datasets per cell. Signal settings share Gaussian noise;
reference and observation streams are independent. The supplied
subspace is identical for all subspace procedures. Known-shape controls
receive additional covariance information and are explicitly diagnostic.
Every rate and paired contrast uses the standard deviation of its
$200$ bank means divided by $\sqrt{200}$. The largest rate MCSE is
1.06 percentage points. All binary decisions, bank counts,
reference and observed features, critical values, paired contrasts and
timings are archived. The design was frozen before rejection outcomes;
this was an internal record, not external preregistration.

At $p=300$, strength $16$ and signal along $e_1$, the original
estimated-shape, full-baseline LPO score with $c=1$ has power
$72.97\%$. Its known-shape counterpart has power $77.03\%$, a paired
estimated-minus-known difference of $-4.06$ percentage points
(MCSE $0.40$). The estimated-shape fixed-$e_1$ control has power
$95.14\%$, versus $96.22\%$ for its known-shape counterpart.
The learned-minus-fixed difference with estimated shape is $-22.17$ points
(MCSE $0.60$). These separately recalibrated contrasts expose costs of
the fitted shape and fitted axis within these comparisons; they do not
provide an additive or model-free attribution of the total gap.

Changing only the baseline from all coordinates to $Q$, while
retaining $c=1$ and relearning the axis for every held-out pair, gives
$74.88\%$ power; rank-only LPO gives $73.36\%$.
Replacing axis learning by the LPO block contribution gives $83.33\%$,
and the classical full-data block and subspace diagonal $T^2$ give
$83.86\%$ and $85.29\%$. The original LPO minus LPO-block difference
is $-10.36$ points (MCSE $0.51$); its difference from subspace diagonal
$T^2$ is $-12.32$ points (MCSE $0.51$).
Thus changing the spectrum or avoiding learned rank augmentation can
be more consequential than shape estimation alone in this setting.
The fixed-known-shape direction theorem does not order these different
spectra or data-adaptive statistics.

The dense inside-subspace signal gives original LPO power
$74.96\%$, known-shape learned LPO $77.25\%$, estimated-shape fixed dense
axis $95.52\%$, and LPO block $84.24\%$.
For the entirely outside diffuse signal of strength $64$, the original
LPO score gives $15.33\%$, its zero-spike full-baseline version gives
$74.00\%$, and the classical all-coordinate $t^2$ gives $74.11\%$.
The finite family therefore contains markedly different compromises
between inside and outside signals. An efficient rule that adapts to
these compromises without an additional observation split requires
its own calibration and power analysis. These diagnostics alone do not
prove uniformly efficient fixed-sample tuning.

For the concentrated inside signal, the descriptive maximum over
the seven full-baseline/rank-only candidates is only $73.59\%$, at $c=2$.
Allowing the $Q$-only baseline changes the comparison: the fixed,
zero-spike $Q$-baseline candidate has power $84.99\%$, close to the
subspace diagonal $T^2$ control in this experiment. Thus the observed
deficit persists across the prespecified scalar-spike grid, while a
different baseline with no learned rank term largely removes it.
This is evidence about a finite grid and these signal settings, not
an impossibility result for all learned-direction procedures.

For each setting, the archive also reports the largest observed rate
over the seven full-baseline/rank-only candidates and over the thirteen
full-baseline/$Q$-baseline/rank-only candidates. These are clairvoyant,
descriptive maxima after observing the outcomes. They are not feasible
selection procedures, simultaneous confidence bounds, or upper bounds
over all weights. A reported MCSE for the selected candidate does not
adjust for selection. Their purpose is to distinguish a poor fixed
spectrum from failure of every member of this specified finite family.

\subsection{Checks and cost}
An independent implementation audit reconstructs leave-pair-out scores
by literal row deletion and full matrix multiplication, including
nonconstant known covariance, fixed and learned axes, both baselines,
and all spectra. It also checks row-permutation and positive diagonal
rescaling invariance and the $q=1$ collapse. Reconstructing every LPO
decision from archived features and critical values gives exact agreement;
the archived counts, rates, bank MCSEs and paired contrasts agree to
floating-point precision. These checks establish implementation
consistency, not general power optimality.

The full experiment took 150.8 seconds in this environment.
At $p=300$, the single-pass feature computation for one raw dataset
took 1.064 milliseconds; a fresh bank of $199$ canonical datasets
took 235.509 milliseconds. These costs jointly produce all ten
features used by the $32$ LPO candidates. Applying a fresh calibration
requires the reference-bank cost in addition to the data-score cost;
reference reuse here is a simulation device.

\subsection{Complete prespecified panels}
All entries below are rejection percentages at nominal $5\%$.
``Inside 1'' and ``Inside $q$'' have strength $16$ and concentrated and
diffuse within-subspace signals, respectively. ``Mix .5'' and ``Mix .9''
have strength $64$ with the indicated outside-signal proportion.
``Out 64'' and ``Out 16'' are diffuse signals entirely outside the
subspace with the indicated strengths. All learned-rank grid panels
use $\eta=c\sqrt p$; a $Q$ baseline has zero weights outside $Q$.
The null column is repeated across panels for readability, not resimulated.

\noindent\begin{minipage}{\textwidth}
\paragraph{$p=30$: Estimated shape, learned direction.}
\begin{center}\small
\begin{tabular}{lrrrrrrr}
\toprule
Method & Null & Inside 1 & Inside $q$ & Mix .5 & Mix .9 & Out 64 & Out 16\\
\midrule
Full baseline, $c=0$ & 5.43 & 49.38 & 50.74 & 99.93 & 99.94 & 99.91 & 51.68\\
Full baseline, $c=0.25$ & 5.62 & 72.02 & 72.79 & 99.95 & 99.80 & 99.77 & 40.89\\
Full baseline, $c=0.5$ & 5.49 & 76.33 & 76.87 & 99.90 & 98.94 & 98.41 & 26.77\\
Full baseline, $c=1$ & 5.35 & 76.79 & 77.86 & 99.76 & 91.42 & 85.28 & 12.70\\
Full baseline, $c=2$ & 5.36 & 76.14 & 77.15 & 99.41 & 65.14 & 34.93 & 7.94\\
Full baseline, $c=4$ & 5.23 & 75.23 & 76.09 & 98.92 & 44.57 & 11.70 & 6.36\\
$q$ baseline, $c=0$ & 5.26 & 85.13 & 86.36 & 99.33 & 42.06 & 5.26 & 5.26\\
$q$ baseline, $c=0.25$ & 5.32 & 82.56 & 83.64 & 99.06 & 37.94 & 5.32 & 5.32\\
$q$ baseline, $c=0.5$ & 5.15 & 80.61 & 81.50 & 98.88 & 34.76 & 5.15 & 5.15\\
$q$ baseline, $c=1$ & 5.23 & 78.30 & 79.31 & 98.57 & 32.12 & 5.23 & 5.23\\
$q$ baseline, $c=2$ & 5.26 & 76.47 & 77.49 & 98.33 & 30.33 & 5.26 & 5.26\\
$q$ baseline, $c=4$ & 5.19 & 75.24 & 76.40 & 98.24 & 29.19 & 5.19 & 5.19\\
Estimated shape, rank only & 5.02 & 73.76 & 74.95 & 98.08 & 27.56 & 5.02 & 5.02\\
\bottomrule
\end{tabular}
\end{center}\end{minipage}\par\medskip

\noindent\begin{minipage}{\textwidth}
\paragraph{$p=30$: Known shape, learned direction.}
\begin{center}\small
\begin{tabular}{lrrrrrrr}
\toprule
Method & Null & Inside 1 & Inside $q$ & Mix .5 & Mix .9 & Out 64 & Out 16\\
\midrule
Full baseline, $c=0$ & 5.48 & 53.67 & 53.33 & 99.97 & 99.96 & 99.96 & 54.43\\
Full baseline, $c=0.25$ & 5.44 & 75.89 & 75.48 & 99.99 & 99.88 & 99.86 & 42.78\\
Full baseline, $c=0.5$ & 5.49 & 79.92 & 79.67 & 99.99 & 99.19 & 99.09 & 28.47\\
Full baseline, $c=1$ & 5.37 & 80.53 & 80.28 & 99.91 & 93.03 & 87.27 & 13.70\\
Full baseline, $c=2$ & 5.56 & 80.02 & 79.62 & 99.63 & 68.53 & 38.21 & 8.48\\
Full baseline, $c=4$ & 5.62 & 79.45 & 78.96 & 99.36 & 47.56 & 12.77 & 6.65\\
$q$ baseline, $c=0$ & 5.42 & 88.08 & 88.18 & 99.61 & 45.14 & 5.42 & 5.42\\
$q$ baseline, $c=0.25$ & 5.35 & 85.71 & 85.47 & 99.52 & 40.26 & 5.35 & 5.35\\
$q$ baseline, $c=0.5$ & 5.22 & 83.55 & 83.39 & 99.32 & 36.68 & 5.22 & 5.22\\
$q$ baseline, $c=1$ & 5.45 & 81.66 & 81.36 & 99.19 & 33.62 & 5.45 & 5.45\\
$q$ baseline, $c=2$ & 5.58 & 80.30 & 79.97 & 99.04 & 31.66 & 5.58 & 5.58\\
$q$ baseline, $c=4$ & 5.53 & 79.24 & 79.03 & 99.02 & 30.68 & 5.53 & 5.53\\
Known shape, rank only & 5.54 & 78.20 & 78.05 & 98.93 & 29.48 & 5.54 & 5.54\\
\bottomrule
\end{tabular}
\end{center}\end{minipage}\par\medskip

\noindent\begin{minipage}{\textwidth}
\paragraph{$p=30$: Fixed axes, block augmentation and classical controls.}
\begin{center}\small
\begin{tabular}{lrrrrrrr}
\toprule
Method & Null & Inside 1 & Inside $q$ & Mix .5 & Mix .9 & Out 64 & Out 16\\
\midrule
Estimated shape, fixed $e_1$ & 5.41 & 95.42 & 57.36 & 99.99 & 99.73 & 98.52 & 27.88\\
Estimated shape, fixed dense axis & 5.17 & 57.24 & 96.35 & 99.59 & 98.93 & 98.67 & 28.39\\
Known shape, fixed $e_1$ & 5.46 & 96.61 & 59.84 & 99.99 & 99.79 & 98.95 & 29.35\\
Known shape, fixed dense axis & 5.16 & 59.69 & 96.77 & 99.80 & 99.29 & 99.06 & 29.51\\
Estimated shape, LPO block & 5.33 & 84.12 & 85.36 & 99.84 & 92.10 & 75.64 & 12.98\\
Known shape, LPO block & 5.38 & 87.09 & 87.37 & 99.95 & 93.66 & 78.65 & 13.74\\
Full-data mean-axis score & 5.38 & 84.24 & 85.70 & 99.87 & 92.35 & 76.25 & 13.27\\
Full-data maximum score & 5.38 & 84.24 & 85.70 & 99.87 & 92.38 & 76.39 & 13.27\\
Classical block $t^2$ & 5.36 & 84.22 & 85.78 & 99.87 & 92.37 & 76.09 & 13.14\\
Subspace diagonal $T^2$ & 5.25 & 85.44 & 86.56 & 99.37 & 42.47 & 5.25 & 5.25\\
All-coordinate $t^2$ & 5.39 & 49.72 & 50.87 & 99.95 & 99.94 & 99.93 & 51.81\\
Subspace paired DLRT & 5.26 & 84.02 & 86.95 & 99.26 & 41.37 & 5.26 & 5.26\\
Subspace Hotelling & 5.36 & 83.06 & 82.55 & 98.99 & 39.88 & 5.36 & 5.36\\
\bottomrule
\end{tabular}
\end{center}\end{minipage}\par\medskip

\noindent\begin{minipage}{\textwidth}
\paragraph{$p=300$: Estimated shape, learned direction.}
\begin{center}\small
\begin{tabular}{lrrrrrrr}
\toprule
Method & Null & Inside 1 & Inside $q$ & Mix .5 & Mix .9 & Out 64 & Out 16\\
\midrule
Full baseline, $c=0$ & 4.89 & 15.61 & 15.56 & 72.99 & 73.90 & 74.00 & 15.19\\
Full baseline, $c=0.25$ & 4.84 & 58.61 & 60.52 & 97.89 & 73.18 & 61.86 & 11.78\\
Full baseline, $c=0.5$ & 5.22 & 69.91 & 71.28 & 98.88 & 64.12 & 40.88 & 8.64\\
Full baseline, $c=1$ & 4.97 & 72.97 & 74.96 & 98.90 & 47.34 & 15.33 & 6.44\\
Full baseline, $c=2$ & 5.03 & 73.59 & 75.44 & 98.68 & 36.86 & 8.40 & 5.66\\
Full baseline, $c=4$ & 5.09 & 73.53 & 75.52 & 98.47 & 32.50 & 6.42 & 5.36\\
$q$ baseline, $c=0$ & 4.75 & 84.99 & 85.96 & 99.39 & 42.21 & 4.75 & 4.75\\
$q$ baseline, $c=0.25$ & 4.96 & 78.20 & 79.63 & 98.84 & 32.82 & 4.96 & 4.96\\
$q$ baseline, $c=0.5$ & 5.06 & 76.11 & 77.59 & 98.53 & 30.80 & 5.06 & 5.06\\
$q$ baseline, $c=1$ & 5.14 & 74.88 & 76.63 & 98.36 & 29.54 & 5.14 & 5.14\\
$q$ baseline, $c=2$ & 5.05 & 74.13 & 76.04 & 98.22 & 28.72 & 5.05 & 5.05\\
$q$ baseline, $c=4$ & 5.03 & 73.77 & 75.63 & 98.12 & 28.30 & 5.03 & 5.03\\
Estimated shape, rank only & 5.06 & 73.36 & 75.11 & 98.08 & 27.90 & 5.06 & 5.06\\
\bottomrule
\end{tabular}
\end{center}\end{minipage}\par\medskip

\noindent\begin{minipage}{\textwidth}
\paragraph{$p=300$: Known shape, learned direction.}
\begin{center}\small
\begin{tabular}{lrrrrrrr}
\toprule
Method & Null & Inside 1 & Inside $q$ & Mix .5 & Mix .9 & Out 64 & Out 16\\
\midrule
Full baseline, $c=0$ & 4.86 & 15.89 & 15.62 & 75.93 & 75.88 & 76.10 & 15.85\\
Full baseline, $c=0.25$ & 4.95 & 62.73 & 62.88 & 98.70 & 75.83 & 64.88 & 12.47\\
Full baseline, $c=0.5$ & 5.02 & 73.46 & 73.55 & 99.36 & 66.68 & 42.86 & 8.58\\
Full baseline, $c=1$ & 4.96 & 77.03 & 77.25 & 99.44 & 50.75 & 16.41 & 6.42\\
Full baseline, $c=2$ & 4.99 & 77.34 & 77.81 & 99.36 & 40.07 & 8.19 & 5.66\\
Full baseline, $c=4$ & 5.04 & 77.54 & 78.04 & 99.21 & 34.90 & 6.45 & 5.40\\
$q$ baseline, $c=0$ & 4.94 & 88.08 & 88.03 & 99.74 & 45.73 & 4.94 & 4.94\\
$q$ baseline, $c=0.25$ & 4.99 & 81.37 & 81.95 & 99.36 & 35.31 & 4.99 & 4.99\\
$q$ baseline, $c=0.5$ & 5.02 & 79.71 & 80.27 & 99.26 & 32.74 & 5.02 & 5.02\\
$q$ baseline, $c=1$ & 5.04 & 78.57 & 79.18 & 99.19 & 31.09 & 5.04 & 5.04\\
$q$ baseline, $c=2$ & 5.09 & 77.96 & 78.52 & 99.08 & 30.28 & 5.09 & 5.09\\
$q$ baseline, $c=4$ & 5.10 & 77.62 & 78.17 & 99.03 & 29.84 & 5.10 & 5.10\\
Known shape, rank only & 5.05 & 77.25 & 77.82 & 99.00 & 29.46 & 5.05 & 5.05\\
\bottomrule
\end{tabular}
\end{center}\end{minipage}\par\medskip

\noindent\begin{minipage}{\textwidth}
\paragraph{$p=300$: Fixed axes, block augmentation and classical controls.}
\begin{center}\small
\begin{tabular}{lrrrrrrr}
\toprule
Method & Null & Inside 1 & Inside $q$ & Mix .5 & Mix .9 & Out 64 & Out 16\\
\midrule
Estimated shape, fixed $e_1$ & 4.97 & 95.14 & 44.11 & 99.99 & 87.41 & 43.64 & 9.18\\
Estimated shape, fixed dense axis & 4.76 & 43.78 & 95.52 & 86.88 & 57.93 & 43.72 & 8.97\\
Known shape, fixed $e_1$ & 5.11 & 96.22 & 46.29 & 99.99 & 88.90 & 46.06 & 9.29\\
Known shape, fixed dense axis & 4.91 & 45.60 & 96.20 & 88.97 & 60.16 & 45.79 & 8.95\\
Estimated shape, LPO block & 4.82 & 83.33 & 84.24 & 99.61 & 60.67 & 16.16 & 6.71\\
Known shape, LPO block & 4.97 & 86.48 & 86.69 & 99.82 & 64.81 & 17.81 & 6.93\\
Full-data mean-axis score & 4.95 & 83.82 & 84.74 & 99.64 & 61.43 & 16.56 & 6.63\\
Full-data maximum score & 4.95 & 83.80 & 84.74 & 99.64 & 61.46 & 16.57 & 6.64\\
Classical block $t^2$ & 4.90 & 83.86 & 84.72 & 99.63 & 61.41 & 16.44 & 6.70\\
Subspace diagonal $T^2$ & 4.87 & 85.29 & 86.11 & 99.40 & 42.77 & 4.87 & 4.87\\
All-coordinate $t^2$ & 4.90 & 15.49 & 15.54 & 73.16 & 74.37 & 74.11 & 15.28\\
Subspace paired DLRT & 4.94 & 83.96 & 86.67 & 99.28 & 41.66 & 4.94 & 4.94\\
Subspace Hotelling & 5.17 & 83.26 & 83.23 & 99.17 & 40.11 & 5.17 & 5.17\\
\bottomrule
\end{tabular}
\end{center}\end{minipage}\par\medskip

\end{vFifteenLowerHeadings}
\begin{figure}[tbp]
\centering
\includegraphics[width=\textwidth]{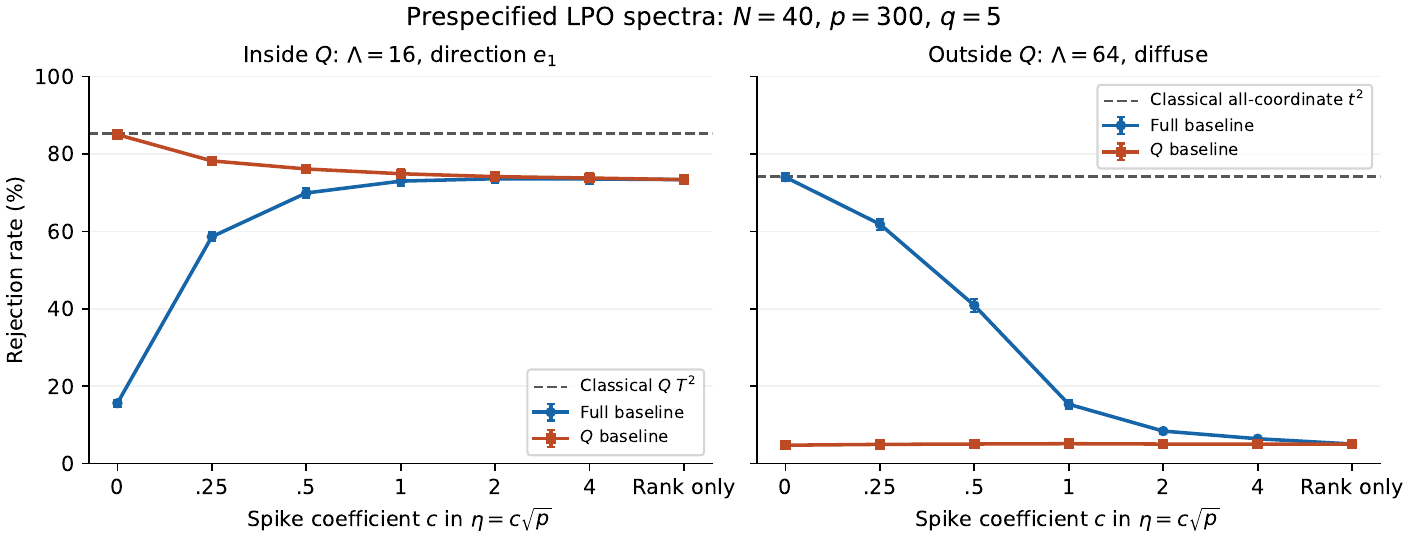}
\caption{Prespecified LPO spectra at $p=300$, $N=40$ and $q=5$.
Each tick is a separate calibrated procedure; the final tick is the
rank-only score, with no baseline. Bars are $1.96$ bank MCSEs for
individual rates, not simultaneous or selection-adjusted intervals.
The left reference line is the classical subspace diagonal $T^2$;
the right reference line is the all-coordinate $t^2$. Curves describe
fixed procedures and do not define an implementable selection rule.}
\label{fig:u-v11-spectrum}
\end{figure}

\begin{vFifteenLowerHeadings}
\section{Prespecified menu selection without a validation split}
\label{sec:v11-adaptive-experiments}
The preceding ablations motivate a different question: whether a fixed menu
can protect against a misspecified subspace or a poorly chosen spike strength
without reserving observations for validation.  We use the established
joint extreme-rank Monte Carlo construction
\citep{mrkvicka2017combined,myllymaki2017global}; its role here is to calibrate
the complete collection of fitted two-sample scores.  It does not estimate a
population-optimal strength, and we do not claim a new combination principle.

\subsection{Fixed menus and simulation protocol}
Let $E_{\rm all}$, $E_q$, and $E_v$ denote the estimated-shape all-coordinate,
subspace-coordinate, and fitted rank-one leave-pair-out features, each scaled
as $N U$.  The first menu contains
$E_{\rm all}+c\sqrt p E_v$ for
$c\in\{0,1/4,1/2,1,2,4\}$ and $E_v$ alone ($K=7$).
The expanded menu adds $E_q$ and $E_{\rm all}-E_q$ ($K=9$).
The rank-only and coordinate-restricted endpoints are positive semidefinite
extensions of the fixed positive-definite weight class, rather than an
optimization result within that original class.  The separate classical menu
uses sums of coordinate $t_j^2$ over the first $q$ coordinates, all coordinates,
and the complement, together with the all-coordinate sum plus $\sqrt p$ times
the first-$q$ sum ($K=4$).  The combined menu contains all thirteen scores.
The latter two are classical omnibus controls, not newly identified RI tests.

All four menus, all signals, and their order were frozen before rejection
outcomes were generated.  We retain both $p=30$ and $p=300$, $N=40$ per
group, $q=5$, $\rho=4$, and the same nonconstant diagonal shape as in the
ablation.  The seven cells in each dimension are the null, concentrated and
diffuse inside signal at $\Lambda_{\rm all}=16$, diffuse leakage fractions
$1/2$, $0.9$, and $1$ at $\Lambda_{\rm all}=64$, and wholly outside diffuse
signal at $\Lambda_{\rm all}=16$.  Mixed signals have standardized direction
$\sqrt{1-\epsilon}\,e_1+
\sqrt{\epsilon/(p-q)}\sum_{j>q}e_j$.

Each of $200$ independent banks contains $L=999$ canonical reference
datasets and is reused for $50$ observed datasets per cell, yielding
$10\,000$ datasets per cell.  All $780$ leave-pair-out learners are rebuilt
for every observed and reference dataset.  All methods receive identical raw
observations and the same fixed coordinate subspace.  Reference scores and
observed scores are ranked jointly, including the observed row.  Independent
continuous uniforms break final extreme-rank ties; the same uniforms are
shared across menus for paired comparisons.  At $M=1000$, the exact null
level is $0.05$, and the component levels $\alpha/K$ have nonzero finite-rank
resolution even for $K=13$.  This exact level averages over the random
reference bank.  It is not conditional on a fixed realized bank.
The matched Bonferroni union for each menu and the separately calibrated
paired subspace Hotelling test are retained as controls.

Monte Carlo standard errors use independent bank means; paired differences
use the same bank-level calculation.  We archive every binary outcome,
component rank, final rank, selected-index diagnostic, and random tie mark.
The index attaining the smallest component rank is reported descriptively
with a fixed index rule for component ties; this label does not determine the
final rejection rule.  These internally recorded designs are not external
preregistrations.

\subsection{Power, adaptation cost, and computational cost}
Table~\ref{tab:v11-adaptive} gives a focused comparison at $p=300$; the complete tables below retain all candidates and all four Bonferroni controls. Across both dimensions, menu null rejection rates range from $4.89\%$ to $5.39\%$; the largest Monte Carlo standard error over all reported rates is $0.76$ percentage points.

For concentrated inside signal at $p=300$, the seven-score LPO menu has power $63.30\%$, compared with $74.19\%$ for the fixed $c=1$ score: a paired difference of $-10.89$ percentage points (Monte Carlo standard error $0.40$). For wholly outside signal at $\Lambda_{\rm all}=64$, the corresponding powers are $64.25\%$ and $16.01\%$, a difference of $+48.24$ ($0.60$). Thus the menu reduces the sensitivity to leakage while paying for adaptation on a favorable inside alternative.

The combined thirteen-score menu differs from the four-score classical menu by $-4.18$ ($0.24$) percentage points for concentrated inside signal and $-3.62$ ($0.25$) for the wholly outside $\Lambda_{\rm all}=64$ signal. More candidates therefore do not automatically produce a more powerful test. In particular, these results do not establish uniform superiority over the structure-informed classical scores or solve fixed-small-sample oracle efficiency.

The samplewise inequality $p_{\rm menu}\leq K\min_j p_j$ was checked
on every generated dataset: there were no violations of inclusion of the
matched Bonferroni rejection event.  The archived envelope table reports the
largest nominal-level individual power in each menu, the menu's selection
gap relative to it, and the largest empirical probability of the strip
$\alpha/K<p_j\leq\alpha$.  These quantities obey the corresponding empirical
inequality on the coupled sample.  The best nominal-level individual is
selected separately for each simulated alternative after outcomes are
available; it is an infeasible diagnostic, not a test or a population oracle
estimate with a confidence guarantee.  The theoretical power comparison is
to component tests at $\alpha/K$, not to the best component at $\alpha$.

On the recorded execution at $p=300$, generating and fitting one $999$-dataset reference bank required $1.17$ seconds on average; computing all candidate scores and all four menu ranks for one observed dataset required $1.12$ milliseconds. These timings describe a shared implementation and bank construction, not the sum of thirteen separately fitted procedures. The larger simulation bank reduces rank discretization but uses no additional observed subjects.

\begin{table}[htbp]
\centering\small
\setlength{\tabcolsep}{3pt}
\caption{Prespecified joint-rank menus at $p=300$, $N=40$ per group, $q=5$, and $\rho=4$. Rejection rates are percentages. $I^{\rm c}$ and $I^{\rm d}$ denote concentrated and diffuse signal inside the first $q$ coordinates; $M^\epsilon$ denotes diffuse leakage fraction $\epsilon$; $O$ denotes wholly outside signal. Subscripts give $\Lambda_{\rm all}$. All menus were fixed before generating these outcomes.}
\label{tab:v11-adaptive}
\begin{tabular}{lrrrrrrr}
\toprule
Method & Null & $I_{16}^{\rm c}$ & $I_{16}^{\rm d}$ & $M_{64}^{.5}$ & $M_{64}^{.9}$ & $O_{64}$ & $O_{16}$ \\
\midrule
LPO $c=1$ & 5.13 & 74.19 & 76.20 & 99.01 & 48.92 & 16.01 & 6.70 \\
LPO $q$ only & 5.12 & 85.31 & 87.05 & 99.40 & 42.46 & 5.12 & 5.12 \\
Classical $q$ sum & 5.11 & 85.34 & 87.12 & 99.47 & 42.77 & 5.11 & 5.11 \\
Classical all sum & 5.18 & 16.07 & 16.00 & 74.11 & 74.98 & 75.18 & 16.30 \\
Classical outside sum & 5.35 & 5.35 & 5.35 & 34.48 & 68.24 & 75.42 & 16.31 \\
LPO menu ($K=7$) & 4.89 & 63.30 & 65.47 & 97.96 & 70.02 & 64.25 & 12.17 \\
Expanded LPO ($K=9$) & 5.05 & 73.35 & 75.57 & 98.72 & 67.78 & 61.66 & 11.21 \\
Classical menu ($K=4$) & 5.39 & 78.22 & 80.31 & 99.28 & 73.78 & 65.67 & 12.65 \\
Combined menu ($K=13$) & 5.14 & 74.04 & 76.15 & 98.87 & 69.24 & 62.05 & 11.27 \\
\bottomrule
\end{tabular}
\end{table}

\subsection{All prespecified candidate and menu rates}
Rejection rates are percentages. The compact full tables omit individual Monte Carlo standard errors to remain readable; every standard error and all paired contrasts are supplied in the accompanying CSV files. The symbol $c$ means $\eta/\sqrt p$.

\begin{table}[htbp]
\centering\small
\setlength{\tabcolsep}{3pt}
\caption{Complete candidate and menu rates at $p=30$, with the signal labels defined in Table~\ref{tab:v11-adaptive}.}
\label{tab:v11-adaptive-all-30}
\begin{tabular}{lrrrrrrr}
\toprule
Method & Null & $I_{16}^{\rm c}$ & $I_{16}^{\rm d}$ & $M_{64}^{.5}$ & $M_{64}^{.9}$ & $O_{64}$ & $O_{16}$ \\
\midrule
LPO $c=0$ & 5.16 & 51.32 & 51.07 & 99.89 & 99.93 & 99.91 & 51.96 \\
LPO $c=1/4$ & 5.13 & 72.68 & 73.79 & 99.98 & 99.76 & 99.73 & 40.84 \\
LPO $c=1/2$ & 5.15 & 76.85 & 77.85 & 99.94 & 98.92 & 98.74 & 27.11 \\
LPO $c=1$ & 4.95 & 77.57 & 78.46 & 99.86 & 92.47 & 87.02 & 12.60 \\
LPO $c=2$ & 4.93 & 76.46 & 77.55 & 99.57 & 66.57 & 35.90 & 7.48 \\
LPO $c=4$ & 4.90 & 75.72 & 76.93 & 99.11 & 45.20 & 11.63 & 5.96 \\
LPO rank only & 4.91 & 74.98 & 76.01 & 98.35 & 29.13 & 4.91 & 4.91 \\
LPO $q$ only & 5.03 & 85.74 & 86.23 & 99.46 & 42.63 & 5.03 & 5.03 \\
LPO outside only & 5.01 & 5.01 & 5.01 & 92.16 & 99.86 & 99.96 & 56.30 \\
Classical $q$ sum & 4.99 & 85.81 & 86.44 & 99.44 & 42.51 & 4.99 & 4.99 \\
Classical all sum & 5.23 & 51.44 & 51.31 & 99.89 & 99.96 & 99.93 & 52.41 \\
Classical outside sum & 5.03 & 5.03 & 5.03 & 92.12 & 99.87 & 99.97 & 56.47 \\
Classical block sum & 4.84 & 85.06 & 85.57 & 99.95 & 92.91 & 77.81 & 13.03 \\
LPO menu ($K=7$) & 5.27 & 69.17 & 69.93 & 99.88 & 99.74 & 99.78 & 41.19 \\
Expanded LPO ($K=9$) & 5.06 & 74.10 & 75.38 & 99.84 & 99.76 & 99.82 & 41.94 \\
Classical menu ($K=4$) & 5.20 & 78.76 & 79.96 & 99.93 & 99.85 & 99.87 & 46.18 \\
Combined menu ($K=13$) & 5.01 & 74.58 & 76.23 & 99.86 & 99.76 & 99.84 & 42.26 \\
Bonferroni LPO $7$ & 1.62 & 47.14 & 47.65 & 99.41 & 99.14 & 99.31 & 24.01 \\
Bonferroni LPO $9$ & 1.60 & 58.17 & 59.59 & 99.29 & 99.19 & 99.49 & 25.98 \\
Bonferroni classical $4$ & 2.84 & 71.04 & 72.23 & 99.75 & 99.69 & 99.78 & 36.60 \\
Bonferroni combined $13$ & 1.02 & 52.80 & 54.02 & 99.04 & 98.80 & 99.10 & 21.50 \\
Subspace Hotelling & 5.01 & 82.60 & 82.77 & 99.16 & 39.41 & 5.01 & 5.01 \\
\bottomrule
\end{tabular}
\end{table}

\begin{table}[htbp]
\centering\small
\setlength{\tabcolsep}{3pt}
\caption{Complete candidate and menu rates at $p=300$, with the signal labels defined in Table~\ref{tab:v11-adaptive}.}
\label{tab:v11-adaptive-all-300}
\begin{tabular}{lrrrrrrr}
\toprule
Method & Null & $I_{16}^{\rm c}$ & $I_{16}^{\rm d}$ & $M_{64}^{.5}$ & $M_{64}^{.9}$ & $O_{64}$ & $O_{16}$ \\
\midrule
LPO $c=0$ & 5.12 & 15.95 & 16.10 & 74.24 & 74.80 & 74.88 & 16.28 \\
LPO $c=1/4$ & 5.05 & 59.79 & 61.23 & 98.10 & 74.70 & 63.35 & 12.16 \\
LPO $c=1/2$ & 5.28 & 70.90 & 72.41 & 99.01 & 65.94 & 43.21 & 8.86 \\
LPO $c=1$ & 5.13 & 74.19 & 76.20 & 99.01 & 48.92 & 16.01 & 6.70 \\
LPO $c=2$ & 5.16 & 74.69 & 76.55 & 98.88 & 38.15 & 8.51 & 5.90 \\
LPO $c=4$ & 5.11 & 74.52 & 76.32 & 98.68 & 33.15 & 6.58 & 5.52 \\
LPO rank only & 5.02 & 74.17 & 76.06 & 98.33 & 28.02 & 5.02 & 5.02 \\
LPO $q$ only & 5.12 & 85.31 & 87.05 & 99.40 & 42.46 & 5.12 & 5.12 \\
LPO outside only & 5.19 & 5.19 & 5.19 & 34.03 & 67.89 & 75.33 & 16.44 \\
Classical $q$ sum & 5.11 & 85.34 & 87.12 & 99.47 & 42.77 & 5.11 & 5.11 \\
Classical all sum & 5.18 & 16.07 & 16.00 & 74.11 & 74.98 & 75.18 & 16.30 \\
Classical outside sum & 5.35 & 5.35 & 5.35 & 34.48 & 68.24 & 75.42 & 16.31 \\
Classical block sum & 5.08 & 83.66 & 85.67 & 99.70 & 61.93 & 17.01 & 6.94 \\
LPO menu ($K=7$) & 4.89 & 63.30 & 65.47 & 97.96 & 70.02 & 64.25 & 12.17 \\
Expanded LPO ($K=9$) & 5.05 & 73.35 & 75.57 & 98.72 & 67.78 & 61.66 & 11.21 \\
Classical menu ($K=4$) & 5.39 & 78.22 & 80.31 & 99.28 & 73.78 & 65.67 & 12.65 \\
Combined menu ($K=13$) & 5.14 & 74.04 & 76.15 & 98.87 & 69.24 & 62.05 & 11.27 \\
Bonferroni LPO $7$ & 1.63 & 36.28 & 37.85 & 91.23 & 51.04 & 46.31 & 4.46 \\
Bonferroni LPO $9$ & 1.64 & 56.01 & 58.17 & 95.50 & 48.24 & 43.54 & 4.22 \\
Bonferroni classical $4$ & 2.81 & 70.32 & 72.45 & 98.36 & 64.13 & 56.19 & 7.59 \\
Bonferroni combined $13$ & 1.06 & 51.37 & 53.37 & 94.33 & 43.70 & 38.46 & 3.09 \\
Subspace Hotelling & 4.99 & 82.62 & 82.78 & 99.09 & 39.31 & 4.99 & 4.99 \\
\bottomrule
\end{tabular}
\end{table}

\end{vFifteenLowerHeadings}
\subsection{Null diagnostics outside the diagonal Gaussian model}
\label{sec:u-misspecification}

We deliberately keep the diagonal-Gaussian calibration unchanged while
altering the observation model. Set $p=30$, $N=40$, $\rho=4$, and
$\bm\mu_1=\bm\mu_2=\bm0$. In the correlated Gaussian cases the common
shape is $\bm H^{1/2}\bm C\bm H^{1/2}$, with the diagonal
$\bm H$ from Section~\ref{sec:u-shape-numerical} and
$C_{jk}=a^{|j-k|}$, $a\in\{0.3,0.7\}$. In the heavy-tailed case,
each coordinate noise is an independent $\sqrt{3/5}\,t_5$ variable,
so the group covariances remain $\bm H$ and $4\bm H$ but Gaussian
mean--residual independence no longer applies. Each setting uses
$100\,000$ raw datasets with the same bank-cluster design as
Section~\ref{sec:u-practical}. Reference banks are reused from its
matching diagonal Gaussian design and are independent of these data.

\begin{table}[tbp]
 \centering
 \caption{Out-of-model null rejection percentages (cluster Monte Carlo
 standard errors in percentage points). Calibration is deliberately
 left unchanged. These diagnostics are not finite-sample guarantees.}
 \label{tab:u-v5-misspecification}
 \small
 \begin{tabular}{lrrrr}
\toprule
Noise & RI & No spike & CQ & F-type\\
\midrule
Gaussian, $a=0.3$ & 5.09 (0.08) & 4.95 (0.08) & 6.64 (0.08) & 5.23 (0.07)\\
Gaussian, $a=0.7$ & 8.88 (0.10) & 11.02 (0.11) & 7.07 (0.08) & 5.45 (0.07)\\
Independent $t_5$ & 4.52 (0.08) & 4.21 (0.07) & 6.42 (0.08) & 4.64 (0.07)\\
\bottomrule
\end{tabular}

\end{table}

\enlargethispage{\baselineskip}
The correlated null at $a=0.7$ rejects at $8.88\%$, demonstrating
failure of the diagonal calibration outside its model. The isolated
$t_5$ result supplies no guarantee for a general heavy-tailed class.

\section{General covariance and further extensions}
\label{app:v15-extensions}

For these extensions, let $\bm D=\overline{\bm X}_1-\overline{\bm X}_2$
and $\bm C=\bm\Sigma_1/n_1+\bm\Sigma_2/n_2$, so that
$\bm D\sim N_p(\bm\delta,\bm C)$. Here $\bm S_i$ is the unbiased
covariance matrix of the testing observations in group $i$, and
$\mathcal T$ is the sigma-field generated by the independent training
observations. The baseline $\bm B$ is positive definite and fixed.
This use of $\bm C$ concerns the unpaired mean contrast; the paired-data
experiment in the main article uses the covariance of one paired difference.
\subsection{The nonnegative mean quadratic}
\label{sec:u-Q-bridge}

The distinction between $Q_{\bm W}$ and $U_{\bm W}$ is relevant
to comparing RI formulations. Removing the independent covariance
subtraction terms recovers a direct instance of the classical
noncentral chi-square comparison \citep{mathew1997ordering}.
Its null quantile must be changed accordingly.

\begin{proposition}[Mean-quadratic counterpart]
\label{prop:u-Q-counterpart}
Let $\bm D\sim N_p(\bm\delta,\bm C)$ with known
$\bm C\succ0$. For fixed $b,\eta>0$ and $\|\bm v\|_2=1$,
use
\[
 \widetilde{\bm W}_{\bm v}
 =\bm C^{-1/2}(b\bm I_p+\eta\bm v\bm v^\top)\bm C^{-1/2},
 \qquad \widetilde Q_{\bm v}
 =\bm D^\top\widetilde{\bm W}_{\bm v}\bm D.
\]
Put $\bm\vartheta=\bm C^{-1/2}\bm\delta$ and
$r=\|\bm\vartheta\|_2^2-(\bm v^\top\bm\vartheta)^2$.
Then
\[
 \widetilde Q_{\bm v}\overset d=
 (b+\eta)\chi_1^2(\|\bm\vartheta\|_2^2-r)
       +b\chi_{p-1}^2(r),
\]
with independent terms. Its null law is independent of the direction.
At its own null quantile, power is maximized precisely by
$\bm v=\pm\bm\vartheta/\|\bm\vartheta\|_2$ when
$\bm\delta\ne0$, and the power loss at another direction is at
most $\mathcal B_{b,\eta}(r)$.
\end{proposition}

\begin{proof}
Orthogonally separating the coordinate along $\bm v$ gives the
displayed distribution. The Poisson coupling in the proof of
Corollary~\ref{thm:u-order} and Proposition~\ref{thm:u-regret}, with its negative
Wishart term removed, proves the ordering and power-loss bound.
Strictness holds at every positive threshold. The null quantile is
positive, and equality in the directional Cauchy--Schwarz inequality
characterizes the two optimizing signs.
\end{proof}

This proposition permits arbitrary known $\bm\Sigma_1,\bm\Sigma_2$
through $\bm C$. The analogous conclusion for $U$ requires control
of the two residual distributions as well. In particular,
Proposition~\ref{prop:u-general-noncommon-null} below demonstrates that
matching $\bm C$ alone is insufficient. The mean-quadratic and
unbiased references are not interchangeable.

\begin{vFifteenLowerHeadings}
\section{Extensions based on shape confidence bounds}
\label{sec:u-shape-envelope-extension}

This appendix retains the statistic and testing-only scale of
Section~\ref{sec:u-unknown-shape}, but permits a general positive-definite
common shape and a training-measurable direction. Its power comparisons
concern a fitted weight family and a different calibration from the
invariant procedure.

\subsection{A confidence-envelope reference}

This second route permits a general positive-definite common shape and
any training-measurable direction. Suppose finite, positive,
$\mathcal T$-measurable numbers $\ell\le h$ satisfy
\begin{equation}\label{eq:u-shape-confidence}
 \Pr(\mathcal E)\ge1-\gamma,\qquad
 \mathcal E=
 \{\ell\widehat{\bm H}\preceq\bm H
                     \preceq h\widehat{\bm H}\},
 \qquad 0\le\gamma<\alpha.
\end{equation}
The probability bound must hold under the model under consideration;
examples below make it uniform in the means and covariance parameters.
Assumption~\eqref{eq:u-shape-confidence} is used only for this
confidence-envelope route, not for Theorem~\ref{thm:u-diag-invariant}.

Set $k_i=n_i-1$. Independently generate
\begin{equation}\label{eq:u-shape-reference}
 \begin{split}
 E_i&\sim\chi^2_{k_i},\qquad
 F_i\sim\chi^2_{(p-1)k_i},\qquad
 A_i=\frac{(b+\eta)E_i+bF_i}{k_i},\qquad
 V_i=\frac{E_i+F_i}{pk_i},\quad i=1,2,\\
 Q_0&\sim(b+\eta)\chi_1^2+b\chi_{p-1}^2,
 \end{split}
\end{equation}
where the two terms in $Q_0$ and all the displayed chi-square variables
are independent. Define
\begin{equation}\label{eq:u-shape-envelope}
 g_{\ell,h}(x)=
 \begin{cases}x/\ell,&x\ge0,\\x/h,&x<0,\end{cases}
 \qquad
 M_{\ell,h}
 =g_{\ell,h}\!\left(
       \max_{i=1,2}\frac{hQ_0-\ell A_i}{V_i}\right).
\end{equation}
The two branches are needed because the unbiased numerator can be
negative. Conditional on $\mathcal T$, this reference has a known
continuous distribution and uses the observed confidence endpoints.
It is invariant under a common rescaling of the endpoints:
$M_{a\ell,ah}=M_{\ell,h}$ for every $a>0$. Only their ratio
affects the reference.

\begin{theorem}[Finite-sample calibration with shape uncertainty]
\label{thm:u-shape-level}
Under \eqref{eq:u-shape-confidence}, on the event $\mathcal E$ and
conditional on $\mathcal T$, the null distribution of
$\widehat T_{\widehat{\bm v}}$ is stochastically dominated by
$M_{\ell,h}$. Let
\[
 \beta=\frac{\alpha-\gamma}{1-\gamma},\qquad L\ge1,
\]
and, conditionally independently of the testing observations, generate
$L$ independent reference values $M_{\ell,h}^{(1)},\ldots,
M_{\ell,h}^{(L)}$. Reject when
\begin{equation}\label{eq:u-shape-pvalue}
 \widehat p_H
 =\frac{1+\sum_{j=1}^L
       \bm1\{M_{\ell,h}^{(j)}\ge
                         \widehat T_{\widehat{\bm v}}\}}{L+1}
 \le\beta.
\end{equation}
Writing $\beta_L=\lfloor\beta(L+1)\rfloor/(L+1)$, the
unconditional null rejection probability satisfies
\begin{equation}\label{eq:u-shape-level}
 \Pr_{\bm0}(\text{reject})
 \le\gamma+(1-\gamma)\beta_L\le\alpha.
\end{equation}
The probability includes training, testing, and reference simulation.
\end{theorem}

The rule is implementable once a valid shape confidence bound is
available. The reference depends on $p,n_i,b,\eta,\ell,h$ but not on
$\sigma_1^2,\sigma_2^2$ or the selected direction. The values
$b,\eta$ and the testing allocation are held fixed here. For a nontrivial
finite-reference test, $L$ must satisfy $\beta(L+1)\ge1$.
The theorem permits any training-measurable direction; it does not
require independence between the fitted shape and that direction.

\subsection{Two explicit shape confidence bounds}

The scale/shape parameterization is not identifiable without a
normalization. For the following constructions set
$\bm H=\bm\Sigma_1$ and $\sigma_1^2=1$, so the second scale is
the relative variance. This is a choice of parameterization, not an
assumption that the first covariance is known. Write $r=m_1-1$ and
let $\bm S_{1,\mathrm{tr}}$ be the first training sample covariance.

\begin{proposition}[Gaussian shape confidence constructions]
\label{prop:u-shape-confidence}
Fix $0<\gamma<1$.
\begin{enumerate}
\item For an unrestricted positive-definite $\bm H$, set
$\widehat{\bm H}=\bm S_{1,\mathrm{tr}}$ and
\begin{equation}\label{eq:u-shape-full-confidence}
 a_r=\frac{\sqrt p+\sqrt{2\log(2/\gamma)}}{\sqrt r},\qquad
 \ell=(1+a_r)^{-2},\qquad h=(1-a_r)^{-2}.
\end{equation}
If $a_r<1$, the estimate is positive definite almost surely and
\eqref{eq:u-shape-confidence} holds.
\item If $\bm H$ is diagonal in a specified coordinate system, set
$\widehat{\bm H}=\operatorname{diag}
 (S_{1,\mathrm{tr},11},\ldots,S_{1,\mathrm{tr},pp})$.
For $q_r(t)$ the $t$-quantile of $\chi_r^2$, use
\begin{equation}\label{eq:u-shape-diagonal-confidence}
 \ell=\frac{r}{q_r(1-\gamma/(2p))},\qquad
 h=\frac{r}{q_r(\gamma/(2p))}.
\end{equation}
Then \eqref{eq:u-shape-confidence} holds for every $r\ge1$,
including $p>m_1$.
\end{enumerate}
Both bounds are uniform in the group means and all admissible covariance
parameters.
\end{proposition}

The unrestricted construction requires
$r>(\sqrt p+\sqrt{2\log(2/\gamma)})^2$. It therefore does not
provide unrestricted covariance estimation when the training sample is
smaller than the dimension. The diagonal construction avoids that
requirement by imposing a known covariance structure. Its endpoints can
still be wide, particularly at small $r$. The theorem establishes valid
calibration in these cases, not efficiency at every sample budget.

\subsection{Comparison of ideal reference cutoffs}

The two diagonal-shape calibrations use the same observed statistic
when both employ the learner \eqref{eq:u-diag-learner}. Their ideal
critical values can be ordered without an asymptotic approximation.
For a real random variable $X$, write
$F_X^{-1}(u)=\inf\{t:\Pr(X\le t)\ge u\}$.

\begin{proposition}[Ideal-cutoff comparison in the diagonal model]
\label{prop:u-diag-ideal-dominance}
Under the assumptions of Theorem~\ref{thm:u-diag-invariant}, fix
$0<\gamma<\alpha$ and use the deterministic endpoints in
\eqref{eq:u-shape-diagonal-confidence}. Set
\begin{equation}\label{eq:u-diag-ideal-cutoffs}
 \beta=\frac{\alpha-\gamma}{1-\gamma},\qquad
 c_{\mathrm{diag}}=F_{M_{\mathrm{diag}}}^{-1}(1-\alpha),\qquad
 c_{\mathrm{env}}=F_{M_{\ell,h}}^{-1}(1-\beta).
\end{equation}
Then
\begin{equation}\label{eq:u-diag-ideal-dominance}
 c_{\mathrm{diag}}\le c_{\mathrm{env}}.
\end{equation}
Consequently, for the same learner, weight, split, and observed
statistic, the ideal rule
$\widehat T_{\widehat{\bm v}}>c_{\mathrm{diag}}$ has no smaller power
than the ideal rule
$\widehat T_{\widehat{\bm v}}>c_{\mathrm{env}}$ under every alternative
in the diagonal common-shape model.
\end{proposition}

This comparison concerns deterministic population quantiles of the
reference laws. It does not order the random cutoffs from finite Monte
Carlo samples, establish strict power improvement, or assert optimality
of the learned direction. The finite-reference procedures retain the
separate guarantees in Theorems~\ref{thm:u-diag-invariant}
and~\ref{thm:u-shape-level}.

\subsection{Power transfer after estimated whitening}

The results in this subsection concern the confidence-envelope rule
\eqref{eq:u-shape-pvalue}. For a realized training sample, let
$\pi_H(\bm v\mid\mathcal T)$ denote the rejection probability of
\eqref{eq:u-shape-pvalue} under the true covariance matrices and fixed
mean difference $\bm\delta$. Every direction in this comparison uses
the same realized $\widehat{\bm H},\ell,h$, sample allocation, and
reference mechanism. The comparison below allows an arbitrary scale
$c>0$, since a common covariance shape is identifiable only up to
scale. This is an auxiliary comparison scale, not a parameter required
by the test. Define
\begin{equation}\label{eq:u-shape-transfer-quantities}
 \begin{split}
 \bm K&=\widehat{\bm H}^{-1/2}\bm H
                         \widehat{\bm H}^{-1/2},\qquad
 d_0=n_1+n_2-1,\qquad
 s^2=\frac{\sigma_1^2}{n_1}+\frac{\sigma_2^2}{n_2},\\
 \bm\theta_c
   &=(s\sqrt c)^{-1}\widehat{\bm H}^{-1/2}\bm\delta,\qquad
 r_c(\bm v)
   =\|\bm\theta_c\|_2^2
                    -(\bm v^\top\bm\theta_c)^2,\\
 \Delta_{H,c}&=\min\left\{1,\frac12
       \sqrt{d_0\{\operatorname{tr}(\bm K)/c-p
                               -\log\det(\bm K/c)\}}\right\}.
 \end{split}
\end{equation}

\begin{theorem}[Conditional power loss in the fitted weight family]
\label{thm:u-shape-power}
For $\bm\delta\ne\bm0$, every unit direction and every $c>0$
satisfy, almost surely,
\begin{equation}\label{eq:u-shape-power}
 \begin{split}
 0&\le\sup_{\|\bm u\|_2=1}\pi_H(\bm u\mid\mathcal T)
                         -\pi_H(\bm v\mid\mathcal T)\\
  &\le\min\{1,\,2\Delta_{H,c}+1-\exp[-r_c(\bm v)/2]\}.
 \end{split}
\end{equation}
In particular, the fitted population axis
$\bm v_H=\bm\theta_c/\|\bm\theta_c\|_2$ does not depend on
$c$ and has conditional regret at most $\min(1,2\Delta_{H,c})$.
On $\mathcal E$, with $f(x)=x-1-\log x$, a scalar bound is
\begin{equation}\label{eq:u-shape-TV-envelope}
 \Delta_{H,c}\le
 \min\left\{1,\frac12\sqrt{d_0p
       \max\{f(\ell/c),f(h/c)\}}\right\}.
\end{equation}
\end{theorem}

The choice
\[
 c_{\ell,h}=
 \begin{cases}
 (h-\ell)/\log(h/\ell),&\ell<h,\\
 \ell,&\ell=h
 \end{cases}
\]
balances the two endpoint terms in
\eqref{eq:u-shape-TV-envelope} and minimizes their maximum.
If $\bm H=c\widehat{\bm H}$ for some $c>0$, then
$\Delta_{H,c}=0$: a scalar discrepancy introduces no loss of
direction optimality. Thus the comparison respects the scale ambiguity
of the common-shape model.

The theorem compares directions within the \emph{fitted} family
\eqref{eq:u-shape-weight}. Its oracle is not the oracle in the
true-$\bm H$ family of Section~\ref{sec:u-oracle}. The fitted
population axis remains unknown because it depends on
$\bm\delta$. Substituting $\widehat{\bm v}$ in
\eqref{eq:u-shape-power} and averaging over training gives the
corresponding unconditional learning bound.

The total-variation bound compares the full contrast--residual
experiment. Its error can therefore accumulate with the number of
residual vectors. A direct comparison of the rejection scores gives a
second bound without this factor.

\begin{proposition}[A direct statistic-level power bound]
\label{prop:u-shape-statistic-transfer}
Condition on $\mathcal T$. Suppose $p\ge2$, $\nu_i=pk_i>4$ for
$i=1,2$, and $j_L=\lfloor\beta(L+1)\rfloor\ge1$. Let $C$ be
the $(L+1-j_L)$th ascending order statistic of the $L$ reference
values in \eqref{eq:u-shape-pvalue}, and put $q_2=\mathbb E(C^2\mid
\mathcal T)$. Set
\begin{equation}\label{eq:u-shape-score-quantities}
 \begin{split}
 c&=\operatorname{tr}(\bm K)/p,\qquad
 \bm d=s^{-1}\widehat{\bm H}^{-1/2}\bm\delta,\qquad
 \kappa_i=\frac{\sigma_i^2/n_i}{s^2},\qquad
 a_k=\sum_{i=1}^2\frac{\kappa_i^2}{k_i},\\
 e&=\|\bm K^{1/2}-\sqrt c\,\bm I_p\|_{\mathrm{op}},\qquad
 a_2=(b+\eta)^2+(p-1)b^2,\\
 \overline V_U
   &=4e^2(b+\eta)^2\|\bm d\|_2^2
     +2e^2\{\sqrt{\lambda_{\max}(\bm K)}+\sqrt c\}^2
             a_2(1+a_k),\\
 V_R&=\frac{2}{p^2}
        \operatorname{tr}\{(\bm K-c\bm I_p)^2\}\,a_k,\qquad
 \rho_H=\min\left\{1,\,
       3\left(\frac{\overline V_U+q_2V_R}
                          {2c^2b(b+\eta)}\right)^{1/3}\right\}.
 \end{split}
\end{equation}
Then \eqref{eq:u-shape-power} holds at this $c$ with
$\Delta_{H,c}$ replaced by
$\tau_H=\min\{\Delta_{H,c},\rho_H\}$. The reference moment
is finite and satisfies
\begin{equation}\label{eq:u-shape-cutoff-moment}
 q_2\le L\left[
  \left(\frac h\ell\right)^2
     \{(pb+\eta)^2+2a_2\}
       \sum_{i=1}^2\frac{\nu_i^2}{(\nu_i-2)(\nu_i-4)}
       +p^2(b+\eta)^2\right].
\end{equation}
\end{proposition}

For example, keep $p,b,\eta,L$ fixed, assume
$\|\bm\theta_c\|_2^2$ and $h/\ell$ are bounded, and retain
$pk_i>4$. If
$\epsilon=\|\bm K/c-\bm I_p\|_{\mathrm{op}}\to0$, then
$\rho_H=O(\epsilon^{2/3})\to0$. This conclusion does not
require $(n_1+n_2)\epsilon^2\to0$. It concerns power transfer
for the fitted family; controlling the missed noncentrality
$r_c(\widehat{\bm v})$ remains a direction-learning problem.

Both transfer bounds can be conservative, especially with wide shape
confidence endpoints or at a fixed small training budget. The
confidence-envelope level theorem does not require either bound to be
small. These results do not
preserve the exact learning risk formula in
Section~\ref{sec:u-learning} after estimated whitening, or supply
automatic selection of a covariance structure or subspace.

\section{Proofs for estimated-shape calibration and power}
\label{app:u-unknown-shape}

\begin{proof}[Proof of Theorem~\ref{thm:u-shape-level}]
Condition on $\mathcal T$. The fitted shape, selected direction, and
confidence endpoints are then fixed. Write
$\bm A=b\bm I_p+\eta\bm v\bm v^\top$ and
$\bm K=\widehat{\bm H}^{-1/2}\bm H
\widehat{\bm H}^{-1/2}$. On $\mathcal E$,
$\ell\bm I_p\preceq\bm K\preceq h\bm I_p$.
Let $a_1=b+\eta$ and $a_2=\cdots=a_p=b$.
The eigenvalues $\lambda_1\ge\cdots\ge\lambda_p$ of
$\bm K^{1/2}\bm A\bm K^{1/2}$ satisfy
\begin{equation}\label{eq:u-shape-spectral-bracket}
 \ell a_j\le\lambda_j\le h a_j,\qquad j=1,\ldots,p.
\end{equation}
Indeed, this matrix has the same eigenvalues as
$\bm A^{1/2}\bm K\bm A^{1/2}$, which lies between
$\ell\bm A$ and $h\bm A$ in Loewner order. The eigenvalue
inequalities follow from the variational characterization of ordered
eigenvalues.

Choose an orthogonal $\bm O$ with
$\bm O^\top\bm K^{1/2}\bm A\bm K^{1/2}\bm O
 =\operatorname{diag}(\lambda_1,\ldots,\lambda_p)$.
Under the null, Gaussian mean--residual independence represents the
standardized contrast and residual covariance matrices using independent
vectors $\bm z_0,\bm z_{ij}\sim N_p(\bm0,\bm I_p)$,
$j=1,\ldots,k_i$, as
\[
 s^{-1}\widehat{\bm H}^{-1/2}\bm D_{\mathrm{te}}
   =\bm K^{1/2}\bm O\bm z_0,
 \qquad
 \sigma_i^{-2}\widehat{\bm H}^{-1/2}
       \bm S_{i,\mathrm{te}}\widehat{\bm H}^{-1/2}
   =\frac1{k_i}\sum_{j=1}^{k_i}
       \bm K^{1/2}\bm O\bm z_{ij}
                  \bm z_{ij}^{\top}\bm O^\top\bm K^{1/2}
\]
in joint distribution. Equality here specifies a common probability
space. Put
\[
 Q=\sum_{a=1}^p\lambda_a z_{0a}^2,\qquad
 A_i^K=\frac1{k_i}\sum_{j=1}^{k_i}\sum_{a=1}^p
                               \lambda_a z_{ija}^2,
 \qquad
 V_i^K=\frac1{pk_i}\sum_{j=1}^{k_i}
                    \bm z_{ij}^{\top}\bm O^\top\bm K\bm O\bm z_{ij}.
\]
Replacing the coefficients $\lambda_a$ by $a_a$ gives $Q_0,A_i$,
and putting
$V_i=(pk_i)^{-1}\sum_j\|\bm z_{ij}\|_2^2$ gives exactly the
joint reference law in \eqref{eq:u-shape-reference}. Consequently,
\begin{equation}\label{eq:u-shape-joint-bracket}
 Q\le hQ_0,\qquad A_i^K\ge\ell A_i,
 \qquad \ell V_i\le V_i^K\le hV_i.
\end{equation}
In particular, the variables bounding numerator and denominator are
constructed from the same residual vectors; they are not replaced by
independent copies.

Set $\kappa=(\sigma_1^2/n_1)/s^2$,
$A_\kappa=\kappa A_1+(1-\kappa)A_2$, and
$V_\kappa=\kappa V_1+(1-\kappa)V_2$.
The observed statistic has the joint representation
\[
 \widehat T_{\bm v}
 =\frac{Q-\kappa A_1^K-(1-\kappa)A_2^K}
        {\kappa V_1^K+(1-\kappa)V_2^K}.
\]
Let $N^+=hQ_0-\ell A_\kappa$. By
\eqref{eq:u-shape-joint-bracket}, the numerator is at most $N^+$
and the denominator lies in $[\ell V_\kappa,hV_\kappa]$.
If $N^+\ge0$, the ratio is at most
$N^+/(\ell V_\kappa)$; if $N^+<0$, it is at most
$N^+/(hV_\kappa)$. Thus
\[
 \widehat T_{\bm v}
 \le g_{\ell,h}\!\left(\frac{hQ_0-\ell A_\kappa}{V_\kappa}\right).
\]
The argument of $g_{\ell,h}$ is a convex combination of
$(hQ_0-\ell A_i)/V_i$, with weights
$\kappa V_1/V_\kappa$ and $(1-\kappa)V_2/V_\kappa$.
Since $g_{\ell,h}$ is increasing, the last display is at most
$M_{\ell,h}$. This proves the required stochastic domination.
Conditional on the residual reference variables, $M_{\ell,h}$ is a
continuous strictly increasing function of $Q_0$; its distribution is
therefore continuous.

On the event $\mathcal E$, compare the observed statistic with a
coupled dominating reference $M_{\ell,h}^{(0)}$. The finite-reference
$p$-value is nonincreasing in the observed statistic, so rejection is
contained in the event obtained by replacing that statistic by
$M_{\ell,h}^{(0)}$. Conditional on $\mathcal T$, this value and the
$L$ simulated references are independent and identically distributed,
with no ties. Their rank argument gives conditional rejection
probability at most $\beta_L$. On $\mathcal E^c$, bound that
probability by one. Therefore
\[
 \Pr_{\bm0}(\text{reject})
 \le \beta_L\Pr(\mathcal E)+\Pr(\mathcal E^c)
 \le \beta_L+(1-\beta_L)\gamma
 \le \beta+(1-\beta)\gamma=\alpha.
\]
The conditioning does not impose a fixed realized Monte Carlo bank:
reference randomness is included in each probability.
\end{proof}

\begin{proof}[Proof of Proposition~\ref{prop:u-shape-confidence}]
For the unrestricted construction,
\[
 \bm H^{-1/2}\bm S_{1,\mathrm{tr}}\bm H^{-1/2}
 \overset d=\frac{\bm Z^\top\bm Z}{r},
\]
where $\bm Z$ has $r$ rows, $p$ columns, and independent standard
normal entries. The Gaussian singular-value bounds of
\citet{davidson2001}, in the form stated by
\citet[Corollary~5.35]{vershynin2012}, give, for $t>0$ and $r\ge p$,
\[
 \Pr\{s_{\max}(\bm Z)>\sqrt r+\sqrt p+t\}\le e^{-t^2/2},
 \qquad
 \Pr\{s_{\min}(\bm Z)<\sqrt r-\sqrt p-t\}\le e^{-t^2/2}.
\]
Take $t=\sqrt{2\log(2/\gamma)}$. Since $a_r<1$, both
singular-value bounds imply, with probability at least $1-\gamma$,
\[
 (1-a_r)^2\bm H\preceq\bm S_{1,\mathrm{tr}}
                          \preceq(1+a_r)^2\bm H.
\]
Rearranging gives \eqref{eq:u-shape-confidence} with
\eqref{eq:u-shape-full-confidence}.

For the diagonal construction, each diagonal entry satisfies
$rS_{1,\mathrm{tr},jj}/H_{jj}\sim\chi_r^2$. Both chi-square
quantiles in \eqref{eq:u-shape-diagonal-confidence} are finite and
strictly positive. The union bound over the two tails and $p$
coordinates gives probability at least $1-\gamma$ that
\[
 q_r(\gamma/(2p))
 \le\frac{rS_{1,\mathrm{tr},jj}}{H_{jj}}
 \le q_r(1-\gamma/(2p)),\qquad j=1,\ldots,p.
\]
Rearranging coordinatewise gives the asserted Loewner inequalities,
because both matrices are diagonal in the specified coordinate system.
The residual distributions used in both constructions do not depend
on the group means, which proves the uniformity assertion.
\end{proof}

\begin{proof}[Proof of Theorem~\ref{thm:u-shape-power}]
Condition on $\mathcal T$ throughout and fix $c>0$. Consider a proxy testing
experiment with the same group means and covariance matrices
$c\sigma_i^2\widehat{\bm H}$, and write its conditional rejection
probability as $\pi_c(\bm v\mid\mathcal T)$. The
realized fitted weights, confidence endpoints, reference distribution,
and rejection rule are unchanged in this comparison. In particular, the
proxy experiment does not rerun training or require its confidence
endpoints to be valid for a different parameter.

The testing contrast and sample covariance matrices can be generated
from one standardized contrast vector and $k_1+k_2$ independent
standardized residual vectors. Whiten by $\widehat{\bm H}^{-1/2}$,
divide the contrast by $s\sqrt c$, and divide the group-$i$
residual vectors by $\sigma_i\sqrt c$. Their joint law under
the true model is then the product of
\[
 N_p(\bm\theta_c,\bm K/c)
 \quad\text{and}\quad
 d_0-1\text{ copies of }N_p(\bm0,\bm K/c).
\]
Under the proxy experiment, replace $\bm K/c$ by $\bm I_p$ in
every factor, leaving the mean vector unchanged. Gaussian
Kullback--Leibler divergence of the first product from the second is
\[
 \frac{d_0}{2}
       \{\operatorname{tr}(\bm K)/c-p-\log\det(\bm K/c)\}.
\]
The total-variation bound by the square root of half the divergence
therefore gives a distance of at most $\Delta_{H,c}$. Taking measurable
functions to form the sufficient statistics cannot increase this
distance. Adjoining independent reference variables with the same law
in both experiments also does not increase it. Consequently,
\begin{equation}\label{eq:u-shape-power-TV}
 \sup_{\|\bm v\|_2=1}
 |\pi_H(\bm v\mid\mathcal T)
        -\pi_c(\bm v\mid\mathcal T)|\le\Delta_{H,c}.
\end{equation}

In the proxy experiment, the covariance shape agrees with the fitted
weight metric. Its testing-residual-only scale is the
$\ell_i=0$ (and hence $G_i=0$) specialization of the joint
representation used in the proof of
Corollary~\ref{thm:u-feasible-regret}. That proof uses the same central
residual variables and the same independent reference cutoff when it
compares directions. It is valid for any such common cutoff, including
the order statistic of $M_{\ell,h}$ used here. Thus
$\bm v_H=\bm\theta_c/\|\bm\theta_c\|_2$
maximizes the proxy power and
\[
 0\le
 \pi_c(\bm v_H\mid\mathcal T)
      -\pi_c(\bm v\mid\mathcal T)
 \le1-e^{-r_c(\bm v)/2}.
\]
When the finite-reference rule never rejects, this assertion holds
trivially. Combining it with \eqref{eq:u-shape-power-TV} on the two
sides of the comparison yields
\eqref{eq:u-shape-power}; a difference between two probabilities is
also at most one.

Finally, on $\mathcal E$, every eigenvalue of $\bm K/c$ lies in
$[\ell/c,h/c]$. The convex function $f(x)=x-1-\log x$
has maximum over that interval at an endpoint. Summing its values over
the $p$ eigenvalues proves \eqref{eq:u-shape-TV-envelope}.
For $\ell<h$, the equality $f(\ell/c)=f(h/c)$ gives
$c=(h-\ell)/\log(h/\ell)\in[\ell,h]$.
On this interval $f(h/c)$ decreases with $c$ and
$f(\ell/c)$ increases, while choosing $c$ outside the interval
cannot improve the endpoint maximum. This proves the stated
minimizing property. The case $\ell=h$ is immediate.
\end{proof}

\begin{proof}[Proof of Proposition~\ref{prop:u-shape-statistic-transfer}]
All expectations below are conditional on $\mathcal T$.
Use the same independent standard Gaussian vectors
$\bm z_0,\bm z_{ij}$ to generate the true and proxy testing
experiments, where the proxy covariance matrices are
$c\sigma_i^2\widehat{\bm H}$ and $c=\operatorname{tr}(\bm K)/p$.
For a fixed unit $\bm v$, write
\[
 \bm A=b\bm I_p+\eta\bm v\bm v^\top,\qquad
 \bm E=\bm K^{1/2}-\sqrt c\,\bm I_p,\qquad
 \bm B=\bm K^{1/2}\bm A\bm K^{1/2}-c\bm A.
\]
Let $U_K,U_c$ be the corresponding complete unbiased numerators,
and let $R_K,R_c$ be their estimated squared scales divided by
$s^2$. The coupled numerator difference is
\[
 \frac{U_K-U_c}{s^2}
 =2\bm d^\top\bm A\bm E\bm z_0
  +\bm z_0^\top\bm B\bm z_0-\operatorname{tr}(\bm B)
  -\sum_{i=1}^2\frac{\kappa_i}{k_i}
      \sum_{j=1}^{k_i}
       \{\bm z_{ij}^\top\bm B\bm z_{ij}
                           -\operatorname{tr}(\bm B)\}.
\]
It has mean zero. Gaussian quadratic-form moments and independence give
\begin{equation}\label{eq:u-shape-score-variance}
 \begin{split}
 V_U(\bm v)
  &:=\mathbb E\left\{\left(\frac{U_K-U_c}{s^2}\right)^2\right\}
    =4\bm d^\top\bm A\bm E^2\bm A\bm d
                +2\operatorname{tr}(\bm B^2)(1+a_k),\\
 \mathbb E(R_K-R_c)&=0,\qquad
 \mathbb E\{(R_K-R_c)^2\}=V_R.
 \end{split}
\end{equation}
The second line follows from
\[
 R_K-R_c
  =\sum_{i=1}^2\frac{\kappa_i}{pk_i}
       \sum_{j=1}^{k_i}
         \bm z_{ij}^\top(\bm K-c\bm I_p)\bm z_{ij},
 \qquad \operatorname{tr}(\bm K-c\bm I_p)=0.
\]
Moreover,
$\bm B=\bm E\bm A\bm K^{1/2}+\sqrt c\,\bm A\bm E$,
$\|\bm A\|_{\mathrm{op}}=b+\eta$, and
$\|\bm A\|_F^2=a_2$. The Frobenius norm inequality therefore
gives $V_U(\bm v)\le\overline V_U$ uniformly in $\bm v$.

Use the same independent reference cutoff $C$ in both experiments.
The rank rule rejects exactly when the corresponding score
$S_K=U_K/s^2-CR_K$ or $S_c=U_c/s^2-CR_c$ is positive,
up to events of probability zero. By independence of $C$ from testing,
\[
 \mathbb E\{(S_K-S_c)^2\}
 \le2\{V_U(\bm v)+q_2V_R\}.
\]
To bound concentration of $S_c$ near zero, diagonalize $\bm A$
with its first coordinate along $\bm v$. The positive mean quadratic
form in $S_c$ is
$(\bm d+\sqrt c\,\bm z_0)^\top\bm A
 (\bm d+\sqrt c\,\bm z_0)$.
Its first two coordinates have Gaussian density at most
$(2\pi c)^{-1}$. The planar ellipse
$\{(b+\eta)x_1^2+bx_2^2\le u\}$ has area
$\pi u/\sqrt{b(b+\eta)}$ for $u\ge0$.
It follows that the two-coordinate quadratic form has density bounded
by $[2c\sqrt{b(b+\eta)}]^{-1}$, for every mean vector.
Adding the other independent coordinates preserves this bound.
Conditioning on all residual vectors and $C$ only translates the
positive mean quadratic form. Consequently,
\[
 \Pr(|S_c|\le t)\le\frac{t}{c\sqrt{b(b+\eta)}},\qquad t>0.
\]
If the rejection indicators differ, either $|S_c|\le t$ or
$|S_K-S_c|>t$. Thus, uniformly over unit directions,
\[
 |\pi_H(\bm v\mid\mathcal T)-\pi_c(\bm v\mid\mathcal T)|
 \le\frac{t}{c\sqrt{b(b+\eta)}}
       +\frac{2(\overline V_U+q_2V_R)}{t^2}.
\]
Minimizing over $t>0$ and truncating at one gives $\rho_H$.
Combine this bound with \eqref{eq:u-shape-power-TV} and the same
proxy optimality comparison used in Theorem~\ref{thm:u-shape-power}
to obtain the claimed regret bound.

It remains to bound $q_2$. From \eqref{eq:u-shape-envelope},
\[
 (M_{\ell,h})_+
   \le(h/\ell)Q_0\max_i V_i^{-1},\qquad
 (M_{\ell,h})_-\le p(b+\eta).
\]
The second inequality uses $A_i/V_i\le p(b+\eta)$.
Since $C^2\le\sum_{j=1}^L(M_{\ell,h}^{(j)})^2$,
independence of $Q_0$ and $V_i$ gives
\[
 q_2\le L\left[
    (h/\ell)^2\mathbb E(Q_0^2)\sum_i\mathbb E(V_i^{-2})
          +p^2(b+\eta)^2\right].
\]
Finally,
$\mathbb E(Q_0^2)=(pb+\eta)^2+2a_2$ and, because
$\nu_i V_i\sim\chi^2_{\nu_i}$ with $\nu_i>4$,
$\mathbb E(V_i^{-2})=\nu_i^2/\{(\nu_i-2)(\nu_i-4)\}$.
This proves \eqref{eq:u-shape-cutoff-moment}.
For the stated consequence as
$\epsilon=\|\bm K/c-\bm I_p\|_{\mathrm{op}}\to0$,
$e^2/c=O(\epsilon^2)$, $\|\bm d\|_2^2/c
=\|\bm\theta_c\|_2^2$, and
$\lambda_{\max}(\bm K)/c=1+O(\epsilon)$.
The assumptions bound $q_2$ and $a_k$ uniformly, so
$(\overline V_U+q_2V_R)/c^2=O(\epsilon^2)$ and
$\rho_H=O(\epsilon^{2/3})$.
\end{proof}

\section{A limitation on uniform adaptation}
\label{sec:adaptation-limit}

The oracle direction depends on the unknown alternative. Approximating its
power uniformly over all directions is a stronger requirement than retaining
power on a specified signal class. The following spherical-mixture argument
shows why a structural assumption is needed for such a uniform comparison.
It is a standard lower-bound construction, stated here for the two-sample Gaussian
experiment used in this paper. This distinction complements the different
signal regimes considered in high-dimensional mean testing
\citep{bai1996,chen2010,cai2014}.

Consider the balanced Gaussian submodel
\begin{equation}\label{eq:limit-experiment}
 \bm X_i\sim N_p(\bm\delta/2,\bm I_p),\qquad
 \bm Y_i\sim N_p(-\bm\delta/2,\bm I_p),\qquad i=1,\ldots,N,
\end{equation}
with all observations independent. The covariance matrices are known.
For a prescribed $\Lambda\ge0$, write
\begin{equation}\label{eq:limit-signal}
 \bm\delta=\sqrt{2\Lambda/N}\,\bm u,
 \qquad \|\bm u\|_2=1.
\end{equation}
The standardized contrast based on all observations is
\[
 \bm Z=\sqrt{N/2}(\overline{\bm X}-\overline{\bm Y})
 \sim N_p(\sqrt\Lambda\,\bm u,\bm I_p).
\]
Thus $\Lambda=N\|\bm\delta\|_2^2/2$ fixes the signal on the
noncentrality scale. The sample size $N$ may vary with $p$ or remain fixed.
Let $\mathbb E_{\bm u,\Lambda}$ denote expectation under
\eqref{eq:limit-experiment}--\eqref{eq:limit-signal}, and let
$\mathbb E_0$ denote expectation at $\bm\delta=\bm0$.

\begin{theorem}[A bound on worst-direction power]
\label{thm:adaptation-limit}
Let $\phi$ be any possibly randomized test based on all $2N$
observations, represented by a measurable function taking values in
$[0,1]$. Suppose $\mathbb E_0\phi\le\alpha$. Let $\bm U$ be uniform
on the unit sphere in $\mathbb R^p$. Then
\begin{equation}\label{eq:limit-average-power}
 \mathbb E_{\bm U}\mathbb E_{\bm U,\Lambda}\phi
 \le \min\left\{1,\alpha+\varepsilon_p(\Lambda)\right\},
 \qquad
 \varepsilon_p(\Lambda)
 =\frac12\left\{\exp\left(\frac{\Lambda^2}{2p}\right)-1\right\}^{1/2}.
\end{equation}
In particular,
\begin{equation}\label{eq:limit-worst-power}
 \inf_{\|\bm u\|_2=1}\mathbb E_{\bm u,\Lambda}\phi
 \le \min\left\{1,\alpha+\varepsilon_p(\Lambda)\right\}.
\end{equation}
For fixed $\Lambda$, the upper bound is $\alpha+o(1)$ as $p\to\infty$.
\end{theorem}

The theorem allows the test to use every observation for learning,
selection, and inference, provided its overall size is controlled. Hence
the limitation is not caused by a particular sample split. A test valid
under a larger composite null also satisfies the size requirement at the
null distribution used here.

To compare this bound with the unbiased direction oracle, fix integers
$N\ge n\ge2$, let $\kappa=n/N$, and give the oracle $n$ testing
observations per group. The covariance matrices are $\bm I_p$,
$s^2=2/n$, and the true direction is $\bm u$. Use $b=1$ and
$\eta=p$, and calibrate $U_{\bm u}/s^2$ at its exact known-scale
null quantile. This oracle is distinct from the feasible envelope
procedure, and its purpose is to quantify the value of knowing the axis.

\begin{corollary}[A positive gap from the unbiased direction oracle]
\label{cor:adaptation-oracle-gap}
Fix $\Lambda>0$ and $N\ge n\ge2$. Let $J_\lambda$ have law
\[
 J_\lambda=\chi_1^2(\lambda)
        -\frac{\chi_{2(n-1)}^2}{2(n-1)}
\]
with independent terms, and let $q_\alpha^J$ be the
$(1-\alpha)$ quantile of $J_0$. The exact known-scale direction-oracle
power $\pi^{\mathrm{or},U}_{p,\kappa}(\Lambda)$ satisfies
\begin{equation}\label{eq:limit-oracle-limit}
 \pi^{\mathrm{or},U}_{p,\kappa}(\Lambda)
 \longrightarrow
 \pi_\infty^U(\kappa\Lambda)
 :=\Pr\{J_{\kappa\Lambda}>q_\alpha^J\}>\alpha
 \qquad(p\to\infty).
\end{equation}
For every sequence of level-$\alpha$ tests $\phi_p$ based on all
$2N$ observations,
\begin{equation}\label{eq:limit-oracle-gap}
 \liminf_{p\to\infty}\sup_{\|\bm u\|_2=1}
 \{\pi^{\mathrm{or},U}_{p,\kappa}(\Lambda)
                  -\mathbb E_{\bm u,\Lambda}\phi_p\}
 \ge \pi_\infty^U(\kappa\Lambda)-\alpha>0.
\end{equation}
\end{corollary}

The negative chi-square term remains in the oracle limit because the
testing sample size is fixed. Replacing it by its mean would change this
finite-budget benchmark. The oracle receives the true direction, whereas
$\phi_p$ may use all the observations. Thus the limitation is not
specific to sample splitting or to a particular learning algorithm.
It concerns unrestricted directions at fixed total noncentrality;
structured or stronger alternatives remain compatible with useful
power and with the subspace improvement proved above.

\section{Proofs for the adaptation limitation}
\label{sec:proofs-limit}

\begin{proof}[Proof of Theorem~\ref{thm:adaptation-limit}]
The sample means and the within-group residuals are independent in the
Gaussian model. The residual distributions do not depend on
$\bm\delta$. Moreover,
\[
 \bm S=\sqrt{N/2}(\overline{\bm X}+\overline{\bm Y})
 \sim N_p(\bm0,\bm I_p)
\]
is independent of $\bm Z$, since their cross-covariance is zero and
they are jointly Gaussian. Consequently, the data consist of $\bm Z$
and independent parameter-free components. Integrating a test over these
components, and over any independent randomization, gives a measurable
function $\psi(\bm Z)\in[0,1]$ with the same rejection probability
under every alternative in \eqref{eq:limit-signal} and under the null.
It suffices to prove the claim for $\psi$.

Let $P_0=N_p(\bm0,\bm I_p)$ and let $P_\Lambda$ be the mixture of
$N_p(\sqrt\Lambda\,\bm u,\bm I_p)$ over a uniform unit vector
$\bm U$. Its likelihood ratio with respect to $P_0$ is
\[
 L(\bm z)=\mathbb E_{\bm U}
   \exp\{\sqrt\Lambda\,\bm U^\top\bm z-\Lambda/2\}.
\]
For independent uniform unit vectors $\bm U$ and $\bm V$, Gaussian
integration and Tonelli's theorem give
\begin{equation}\label{eq:limit-chi-square-identity}
 \chi^2(P_\Lambda,P_0)
 =\mathbb E_0L(\bm Z)^2-1
 =\mathbb E\exp(\Lambda\bm U^\top\bm V)-1.
\end{equation}
By rotational invariance, $\bm U^\top\bm V$ has the same law as
the first coordinate $U_1$ of a uniform unit vector. Its odd moments
vanish, and its even moments are
\begin{equation}\label{eq:limit-sphere-moments}
 \mathbb E(U_1^{2k})
 =\frac{1\cdot3\cdots(2k-1)}
        {p(p+2)\cdots(p+2k-2)}
 \le\frac{(2k-1)!!}{p^k},\qquad k\ge1.
\end{equation}
For $p\ge2$, the identity follows, for example, from
$U_1^2\sim\operatorname{Beta}\{1/2,(p-1)/2\}$, obtained by
normalizing a vector of independent standard normals. It also holds
directly for $p=1$. Expanding the exponential, using symmetry and
\eqref{eq:limit-sphere-moments}, yields
\begin{align*}
 \mathbb E\exp(\Lambda U_1)
 &=\sum_{k=0}^\infty
       \frac{\Lambda^{2k}}{(2k)!}\mathbb E(U_1^{2k})\\
 &\le\sum_{k=0}^\infty\frac1{k!}
                     \left(\frac{\Lambda^2}{2p}\right)^k
 =\exp\left(\frac{\Lambda^2}{2p}\right).
\end{align*}
Here the $k=0$ moment is one. Combining this inequality with
\eqref{eq:limit-chi-square-identity} and Cauchy--Schwarz gives
\[
 \|P_\Lambda-P_0\|_{\rm TV}
 =\frac12\mathbb E_0|L(\bm Z)-1|
 \le\frac12\{\chi^2(P_\Lambda,P_0)\}^{1/2}
 \le\varepsilon_p(\Lambda).
\]
Since $0\le\psi\le1$ and $\mathbb E_0\psi\le\alpha$,
\[
 \mathbb E_{P_\Lambda}\psi
 \le\mathbb E_{P_0}\psi+\|P_\Lambda-P_0\|_{\rm TV}
 \le\alpha+\varepsilon_p(\Lambda).
\]
A rejection probability is at most one, which proves
\eqref{eq:limit-average-power}. The infimum over directions is no
larger than their average, proving \eqref{eq:limit-worst-power}.
For fixed $\Lambda$, $\varepsilon_p(\Lambda)\to0$.
\end{proof}

\begin{proof}[Proof of Corollary~\ref{cor:adaptation-oracle-gap}]
Put $d=n-1$. Proposition~\ref{prop:u-reference}, applied with
$\kappa_1=\kappa_2=1/2$, gives the normalized oracle statistic as
\[
 \frac{U_{\bm u}}{p s^2}\ \overset d=
 \left(1+\frac1p\right)\chi_1^2(\kappa\Lambda)
 +\frac{\chi_{p-1}^2}{p}
 -\left(1+\frac1p\right)\frac{\chi_{2d}^2}{2d}
 -\frac{\chi_{2(p-1)d}^2}{2dp},
\]
with independent variables. Since $n$ is fixed, the second and fourth
terms both converge in probability to one and cancel in the limit.
The remaining law converges to $J_{\kappa\Lambda}$. The same argument
under the null gives convergence to $J_0$. The distribution of $J_0$
has a continuous strictly increasing distribution function on
$\mathbb R$: it is the difference of independent positive variables
whose densities are positive on $(0,\infty)$. The null quantile of $U_{\bm u}/(ps^2)$
therefore converges to $q_\alpha^J$, and the limiting alternative
has no atom at that threshold. This proves the convergence of powers.

For $\lambda>0$, the Poisson representation of
$\chi_1^2(\lambda)$ couples $J_\lambda$ with $J_0$ by adding a
nonnegative random sum that is strictly positive with positive
probability. The full support of $J_0$ makes the rejection probability
strictly larger at every finite threshold. Hence
$\Pr\{J_{\kappa\Lambda}>q_\alpha^J\}>\alpha$.

Finally the oracle power is independent of $\bm u$, and
Theorem~\ref{thm:adaptation-limit} implies
\[
 \sup_{\|\bm u\|_2=1}
 \{\pi^{\mathrm{or},U}_{p,\kappa}(\Lambda)
                  -\mathbb E_{\bm u,\Lambda}\phi_p\}
 \ge \pi^{\mathrm{or},U}_{p,\kappa}(\Lambda)
                 -\alpha-\varepsilon_p(\Lambda).
\]
Taking the lower limit proves \eqref{eq:limit-oracle-gap}.
\end{proof}

\section{Local power optimality for a general baseline}
\label{sec:u-general}

For arbitrary covariance matrices, the covariance correction in the unbiased
statistic may have a direction-dependent distribution even when the mean
quadratic has a fixed null spectrum. We therefore derive a criterion directly
for the complete unbiased statistic. This section allows
$\bm\Sigma_1$ and $\bm\Sigma_2$ to be any positive-definite matrices. Their
values are treated as known when null critical values are computed.

Let $\bm D=\overline{\bm X}_1-\overline{\bm X}_2$, and let $\bm S_i$
be the sample covariance matrix with denominator $n_i-1$. Define
\begin{equation}
 \bm M=\bm D\bm D^\top-\frac{\bm S_1}{n_1}
                         -\frac{\bm S_2}{n_2},\qquad
 U_{\eta}(\bm u)
 =\operatorname{tr}\{(\bm B+\eta\bm u\bm u^\top)\bm M\},
 \quad \|\bm u\|_2=1,
 \label{eq:u-general-statistic}
\end{equation}
where $\bm B\succ0$ and $\eta\geq0$. Thus
$U_{\eta}(\bm u)=U_{\bm B}+\eta\bm u^\top\bm M\bm u$,
including the change to both sample-covariance corrections.
Let $c_{\alpha,\eta}(\bm u)$ be the $(1-\alpha)$ null quantile and set
\[
 \pi_{\eta}(\bm u)
 =\Pr_{\bm\delta}\{U_{\eta}(\bm u)>c_{\alpha,\eta}(\bm u)\}.
\]
Write $c_0$ and $\pi_0$ for the critical value and power at $\eta=0$.
These quantities do not depend on $\bm u$.

Denote the density of $U_{\bm B}$ under mean difference $\bm\delta$ by
$f_{\bm\delta}$. The boundary conditional moment relevant to this statistic is
\begin{equation}
 \bm K_{\bm\delta}(c)
 =\mathbb E_{\bm\delta}
       [\bm M\mid U_{\bm B}=c],\qquad c\in\mathbb R.
 \label{eq:u-general-boundary}
\end{equation}
The continuous version of this conditional moment is used throughout.
The next theorem verifies its existence under explicit Gaussian sample-size
conditions. Unlike the boundary moment for a mean quadratic,
$\bm K_{\bm\delta}(c)$ includes the two residual covariance matrices
and need not be positive semidefinite.

\begin{theorem}[First variation of calibrated unbiased-statistic power]
\label{thm:u-general-variation}
Suppose the two samples are independent Gaussian samples, $n_i\geq2$,
$\bm\Sigma_i\succ0$, and
\begin{equation}
 d=p(n_1+n_2-1)>4.
 \label{eq:u-general-dimension-condition}
\end{equation}
Fix $\bm B\succ0$, $\bm\delta$, and $0<\alpha<1$.
The density $f_{\bm\delta}$ is continuous and strictly positive on
$\mathbb R$, and \eqref{eq:u-general-boundary} has a continuous version.
Define
\begin{equation}
 \bm G_{\bm\delta}
 =f_{\bm\delta}(c_0)
    \{\bm K_{\bm\delta}(c_0)-\bm K_{\bm0}(c_0)\}.
 \label{eq:u-general-gradient}
\end{equation}
There exist constants $\eta_0>0$ and $K<\infty$, depending on the fixed
model parameters, such that
\begin{equation}
 \sup_{\|\bm u\|_2=1}
 \left|\pi_{\eta}(\bm u)-\pi_0
                -\eta\bm u^\top\bm G_{\bm\delta}\bm u\right|
 \leq K\eta^2,\qquad 0\leq\eta\leq\eta_0.
 \label{eq:u-general-expansion}
\end{equation}
Moreover,
\begin{equation}
 \left.\frac{d}{d\eta}c_{\alpha,\eta}(\bm u)\right|_{\eta=0}
 =\bm u^\top\bm K_{\bm0}(c_0)\bm u.
 \label{eq:u-general-critical-derivative}
\end{equation}
\end{theorem}

Condition~\eqref{eq:u-general-dimension-condition} is automatic for
$p\geq2$ and $n_i\geq2$. It is a sufficient smoothness condition, not
a high-dimensional asymptotic assumption. The Gaussian representation
in the proof contains $p$ positive quadratic coordinates and
$p(n_1+n_2-2)$ negative quadratic coordinates. Its characteristic
function has enough decay to differentiate both the distribution and its
critical value. In particular, the result remains valid if $c_0=0$.

The subtraction of $\bm K_{\bm0}(c_0)$ accounts for null
recalibration. The residual terms also appear in the alternative boundary
moment. Omitting either contribution changes the actual-power criterion.

\begin{corollary}[Optimal local augmentation and estimation error]
\label{cor:u-general-direction}
Under the conditions of Theorem~\ref{thm:u-general-variation}, let
$\bm u_G$ be a unit leading eigenvector of $\bm G_{\bm\delta}$.
It maximizes the right derivative of calibrated power at $\eta=0$, and
\begin{equation}
 0\leq \sup_{\|\bm u\|_2=1}\pi_{\eta}(\bm u)
                       -\pi_{\eta}(\bm u_G)
 \leq2K\eta^2,\qquad 0\leq\eta\leq\eta_0.
 \label{eq:u-general-oracle-regret}
\end{equation}
If $\widehat{\bm G}$ is symmetric and $\widehat{\bm u}$ is one of its
unit leading eigenvectors, then
\begin{equation}
 0\leq\sup_{\|\bm u\|_2=1}\pi_{\eta}(\bm u)
                      -\pi_{\eta}(\widehat{\bm u})
 \leq2\eta\|\widehat{\bm G}-\bm G_{\bm\delta}\|_{\mathrm{op}}
           +2K\eta^2.
 \label{eq:u-general-estimated-regret}
\end{equation}
Furthermore,
\begin{equation}
 \operatorname{tr}(\bm B\bm G_{\bm\delta})=0.
 \label{eq:u-general-trace}
\end{equation}
Consequently, a nonzero $\bm G_{\bm\delta}$ has both positive and
negative eigenvalues.
\end{corollary}

If the matrix estimator is based on observations independent of
$(\bm D,\bm S_1,\bm S_2)$ and its operator-norm error is at most
$\epsilon$ with probability at least $1-\gamma$, conditioning gives an
unconditional power loss of at most
$2\eta\epsilon+2K\eta^2+\gamma$. The test has conditional size
$\alpha$ when its null quantile is computed using the specified
covariance matrices and the selected direction. This is an error-to-power
guarantee; it does not supply an estimator with a universal small-sample
accuracy bound.

The preceding result is local in $\eta$, with the dimension and model
parameters fixed. It makes no uniform claim over growing dimension,
nearly singular covariance matrices, or arbitrary positive augmentation
strengths. The exact orientation result for a common covariance shape
uses additional rotational structure that is absent in general.

\begin{proposition}[The covariance-matched null law need not be common]
\label{prop:u-general-noncommon-null}
Take $p=2$, $n_1=n_2=4$,
\[
 \bm\Sigma_1=\operatorname{diag}(3,1),\qquad
 \bm\Sigma_2=\operatorname{diag}(1,3).
\]
Then the mean-contrast covariance is $\bm C=\bm I_2$.
For $\bm W_{\bm v}=\bm I_2+\bm v\bm v^\top$,
the mean quadratic $\bm D^\top\bm W_{\bm v}\bm D$ has the same null
distribution for every unit $\bm v$. The corresponding unbiased
statistic does not. In particular,
\begin{equation}
 \operatorname{Var}_{\bm0}(U_{\bm e_1})=\frac{145}{12},
 \qquad
 \operatorname{Var}_{\bm0}
   \left(U_{(\bm e_1+\bm e_2)/\sqrt2}\right)=12.
 \label{eq:u-general-null-counterexample}
\end{equation}
\end{proposition}

Thus matching the baseline to $\bm C$ alone does not justify a common
critical value for the unbiased two-sample statistic under arbitrary
unequal covariance shapes. This example establishes a failure of that
reduction; it does not assert that the same direction is suboptimal for
every alternative.

\section{Proofs for a general baseline}
\label{app:u-general}

\begin{proof}[Proof of Theorem~\ref{thm:u-general-variation}]
Put $q_i=n_i-1$ and
$\bm C=\bm\Sigma_1/n_1+\bm\Sigma_2/n_2$.
Gaussian mean--residual independence gives independent standard normal
vectors $\bm Z_0,\bm Z_{ij}\in\mathbb R^p$ such that
\[
 \bm D=\bm\delta+\bm C^{1/2}\bm Z_0,\qquad
 \bm S_i\ \overset d=\
 \frac1{q_i}\sum_{j=1}^{q_i}
       \bm\Sigma_i^{1/2}\bm Z_{ij}\bm Z_{ij}^\top
                       \bm\Sigma_i^{1/2}.
\]
For a symmetric positive-definite matrix $\bm W$, define the
$d\times d$ block-diagonal matrix
\begin{equation}
 \bm A(\bm W)
 =\operatorname{diag}\left(
   \bm C^{1/2}\bm W\bm C^{1/2},
   \left\{-\frac{\bm\Sigma_1^{1/2}\bm W\bm\Sigma_1^{1/2}}
                    {n_1q_1}\right\}_{j=1}^{q_1},
   \left\{-\frac{\bm\Sigma_2^{1/2}\bm W\bm\Sigma_2^{1/2}}
                    {n_2q_2}\right\}_{j=1}^{q_2}
 \right).
 \label{eq:u-general-block-matrix}
\end{equation}
The braces indicate repeated diagonal blocks. If
$\bm Z\sim N_d(\bm m,\bm I_d)$ with
$\bm m=((\bm C^{-1/2}\bm\delta)^\top,\bm0^\top)^\top$,
then the joint representation above gives
\[
 U_{\eta}(\bm u)
 \ \overset d=\
 \bm Z^\top\bm A_{\eta,\bm u}\bm Z,\qquad
 \bm A_{\eta,\bm u}
 =\bm A(\bm B)+\eta\bm J_{\bm u},\qquad
 \bm J_{\bm u}=\bm A(\bm u\bm u^\top).
\]
The mapping $\bm A$ is linear. Its positive and negative blocks are
all nonsingular whenever $\bm B+\eta\bm u\bm u^\top\succ0$.

Allow $|\eta|\leq\lambda_{\min}(\bm B)/2$.
Uniformly over this interval and unit $\bm u$, the absolute values of
all eigenvalues of $\bm A_{\eta,\bm u}$ are bounded above and bounded
away from zero. Also $\|\bm J_{\bm u}\|_{\mathrm{op}}$ is uniformly
bounded. Write
$\bm R_t=(\bm I_d-2it\bm A_{\eta,\bm u})^{-1}$.
The characteristic function is
\begin{equation}
 \varphi_{\bm\delta}(t,\eta,\bm u)
 =\det(\bm I_d-2it\bm A_{\eta,\bm u})^{-1/2}
   \exp\{it\bm m^\top\bm A_{\eta,\bm u}\bm R_t\bm m\}.
 \label{eq:u-general-characteristic}
\end{equation}
The continuous determinant branch equals one at $t=0$.
The modulus of the exponential factor is at most one, so for a
constant $C_1$ independent of $(\eta,\bm u)$,
\begin{equation}
 |\varphi_{\bm\delta}(t,\eta,\bm u)|
 \leq C_1(1+|t|)^{-d/2}.
 \label{eq:u-general-cf-decay}
\end{equation}

Here the perturbation derivatives have the same large-$|t|$ decay.
Indeed,
\begin{align*}
 \partial_\eta\log\varphi_{\bm\delta}
 &=it\operatorname{tr}(\bm R_t\bm J_{\bm u})
    +it\bm m^\top\bm R_t\bm J_{\bm u}\bm R_t\bm m,\\
 \partial_\eta^2\log\varphi_{\bm\delta}
 &=-2t^2\operatorname{tr}
       (\bm R_t\bm J_{\bm u}\bm R_t\bm J_{\bm u})
   -4t^2\bm m^\top
       \bm R_t\bm J_{\bm u}\bm R_t
                 \bm J_{\bm u}\bm R_t\bm m .
\end{align*}
Since
$\|\bm R_t\|_{\mathrm{op}}\leq C_2\min(1,|t|^{-1})$,
both logarithmic derivatives are uniformly bounded for $|t|\geq1$.
For $|t|\leq1$, the first is $O(|t|)$ and the second is $O(t^2)$.
It follows that $\partial_\eta\varphi_{\bm\delta}$ and
$\partial_\eta^2\varphi_{\bm\delta}$ satisfy
\eqref{eq:u-general-cf-decay} for $|t|\geq1$, with possibly different
constants. Near zero they are respectively $O(|t|)$ and $O(t^2)$.
All these functions are jointly continuous in their parameters.

Let $F_{\bm\delta}(c,\eta,\bm u)$ denote the distribution function.
Fourier inversion of the integrable characteristic function gives its
continuous density. The CDF inversion formula is
\begin{equation}
 F_{\bm\delta}(c,\eta,\bm u)
 =\frac12-\frac1\pi\int_0^\infty
       \frac{\operatorname{Im}
       \{e^{-itc}\varphi_{\bm\delta}(t,\eta,\bm u)\}}{t}\,dt.
 \label{eq:u-general-cdf-inversion}
\end{equation}
The bounds just proved justify differentiation under these integrals.
In particular, $d>4$ makes
$t\varphi_{\bm\delta}$ integrable at infinity. Thus the two
$c$ derivatives, the two $\eta$ derivatives, and the mixed derivative
of $F_{\bm\delta}$ exist and are jointly continuous, uniformly on
compact $c$ intervals and the stated perturbation set.

The density is strictly positive on the real line. In the block
representation, write $U_\eta=P-N$, where $P$ is a positive-definite
noncentral Gaussian quadratic and $N$ is an independent
positive-definite central Gaussian quadratic. Both have positive
densities on $(0,\infty)$. For any $c\in\mathbb R$, integrating their
product over an interval on which $y>0$ and $c+y>0$ gives a strictly
positive contribution to the density of $P-N$. This agrees with the
continuous density obtained by inversion.

We next verify the conditional-moment formula and the derivative with
respect to the weight. For a fixed unit $\bm u$, set
$V_{\bm u}=\bm u^\top\bm M\bm u$.
The joint Gaussian representation and finite moments imply
\[
 \partial_\eta\varphi_{\bm\delta}(t,\eta,\bm u)
 =it\,\mathbb E_{\bm\delta}
        [V_{\bm u}e^{itU_\eta(\bm u)}].
\]
At $\eta=0$, the Fourier transform on the right after division by
$it$ is integrable: it is bounded near zero and is
$O(|t|^{-d/2-1})$ at infinity. Hence the finite signed measure
$A\mapsto\mathbb E_{\bm\delta}
 [V_{\bm u}\bm1\{U_{\bm B}\in A\}]$
has a continuous density $g_{\bm\delta,\bm u}(c)$.
Using the coordinate directions and their normalized pairwise sums, polarization
gives a continuous matrix-valued density $\bm g_{\bm\delta}(c)$ for
$A\mapsto\mathbb E_{\bm\delta}
 [\bm M\bm1\{U_{\bm B}\in A\}]$.
Since $f_{\bm\delta}(c)>0$, the matrix
\[
 \bm K_{\bm\delta}(c)
 =\frac{\bm g_{\bm\delta}(c)}{f_{\bm\delta}(c)}
\]
is a continuous version of
$\mathbb E_{\bm\delta}[\bm M\mid U_{\bm B}=c]$.
Differentiating \eqref{eq:u-general-cdf-inversion} yields
\begin{equation}
 \partial_\eta F_{\bm\delta}(c,0,\bm u)
 =-g_{\bm\delta,\bm u}(c)
 =-f_{\bm\delta}(c)
          \bm u^\top\bm K_{\bm\delta}(c)\bm u.
 \label{eq:u-general-cdf-derivative}
\end{equation}

Because the central density is continuous and positive at $c_0$,
choose a compact interval around $c_0$ on which it is bounded away
from zero. Joint continuity, compactness of the unit sphere, and
\eqref{eq:u-general-cdf-inversion} ensure that for all sufficiently
small $|\eta|$, uniformly in $\bm u$, the perturbed central quantile
lies in this interval and its density remains bounded away from zero.
The implicit function theorem applied to
$F_{\bm0}(c_{\alpha,\eta}(\bm u),\eta,\bm u)=1-\alpha$
therefore gives two bounded perturbation derivatives of the quantile,
uniformly over unit $\bm u$. Its first derivative at zero follows from
\eqref{eq:u-general-cdf-derivative}:
\[
 c'_{\alpha,0}(\bm u)
 =\bm u^\top\bm K_{\bm0}(c_0)\bm u.
\]
Finally,
$\pi_\eta(\bm u)=
1-F_{\bm\delta}(c_{\alpha,\eta}(\bm u),\eta,\bm u)$,
so the chain rule gives
\[
 \pi'_0(\bm u)
 =f_{\bm\delta}(c_0)
   \bm u^\top\{\bm K_{\bm\delta}(c_0)
                     -\bm K_{\bm0}(c_0)\}\bm u.
\]
The established derivative bounds also bound $\pi''_\eta(\bm u)$
uniformly over a sufficiently small perturbation interval and the
unit sphere. Taylor's theorem proves
\eqref{eq:u-general-expansion}.
\end{proof}

\begin{proof}[Proof of Corollary~\ref{cor:u-general-direction}]
The Rayleigh variational formula identifies the largest first
derivative as $\lambda_{\max}(\bm G_{\bm\delta})$.
Applying \eqref{eq:u-general-expansion} at a candidate direction and
at $\bm u_G$ proves \eqref{eq:u-general-oracle-regret}.
For the estimated direction, the leading-eigenvector property gives
\[
 \lambda_{\max}(\bm G_{\bm\delta})
 -\widehat{\bm u}^{\top}\bm G_{\bm\delta}\widehat{\bm u}
 \leq2\|\widehat{\bm G}-\bm G_{\bm\delta}\|_{\mathrm{op}}.
\]
Applying the same two power expansions proves
\eqref{eq:u-general-estimated-regret}.
If the estimator is independent of the testing observations, the
deterministic inequality holds conditionally. On an event of
operator-norm error at most $\epsilon$ it gives the asserted bound;
on the complementary event the power difference is at most one.
This proves the unconditional statement. The exact central quantile
gives conditional size $\alpha$.

Since $\operatorname{tr}(\bm B\bm M)=U_{\bm B}$ identically,
conditioning and continuity imply
$\operatorname{tr}\{\bm B\bm K_{\bm\delta}(c)\}=c$
for every real $c$. Subtracting the corresponding central identity
proves \eqref{eq:u-general-trace}. A nonzero positive-semidefinite
matrix $\bm G_{\bm\delta}$ would make
$\operatorname{tr}(\bm B\bm G_{\bm\delta})>0$, because
$\bm B\succ0$. The negative-semidefinite case is excluded by the
same argument with the opposite sign. Thus a nonzero gradient matrix
has eigenvalues of both signs.
\end{proof}

\begin{proof}[Proof of Proposition~\ref{prop:u-general-noncommon-null}]
Gaussian independence of the means and sample covariance matrices gives,
for any fixed symmetric $\bm W$ under the null,
\begin{equation}
 \operatorname{Var}_{\bm0}(U_{\bm W})
 =2\operatorname{tr}(\bm W\bm C\bm W\bm C)
  +\sum_{i=1}^2
    \frac{2\operatorname{tr}
      (\bm W\bm\Sigma_i\bm W\bm\Sigma_i)}
         {n_i^2(n_i-1)}.
 \label{eq:u-general-variance}
\end{equation}
In the specified example, $\bm C=\bm I_2$ and
$\bm W_{\bm v}$ has eigenvalues $2$ and $1$, so the first term is
$10$ for every direction. For $\bm v=\bm e_1$, the two traces in
the sum are $37$ and $13$. For
$\bm v=(\bm e_1+\bm e_2)/\sqrt2$, both traces are $24$.
Substitution into \eqref{eq:u-general-variance} gives respectively
$10+2(37+13)/(16\cdot3)=145/12$ and
$10+2(24+24)/(16\cdot3)=12$.
Their null variances differ, so their null distributions differ.
The common null law for the mean quadratic follows immediately from
its fixed eigenvalues and $\bm C=\bm I_2$.
\end{proof}

\end{vFifteenLowerHeadings}
\subsection{A general baseline for the unbiased two-sample statistic}
\label{sec:u-general-example}

We next examine the local criterion of
Theorem~\ref{thm:u-general-variation} directly for the unbiased statistic,
including its two sample-covariance corrections. Set
\[
 p=2,\qquad n_1=n_2=10,\qquad
 \bm\Sigma_1=\bm\Sigma_2=5\bm I_2,\qquad
 \bm B=\operatorname{diag}(1,6),\qquad
 \bm\delta=(0.5,1.5)^\top.
\]
Then $\bm C=\bm I_2$. For
$\bm W=\bm B+\eta\bm u(\theta)\bm u(\theta)^\top$,
where $\bm u(\theta)=(\cos\theta,\sin\theta)^\top$, let
$\lambda_j$ and $\bm e_j$ be the eigenvalues and orthonormal
eigenvectors of $\bm W$. The exact distribution used in this study is
\begin{equation}
 U_{\bm W}\ \overset d=\
 \sum_{j=1}^2\lambda_j
 \left\{\chi^2_1\bigl((\bm e_j^\top\bm\delta)^2\bigr)
              -\frac{\chi^2_{18}}{18}\right\},
 \label{eq:u-general-example-law}
\end{equation}
with independent variables throughout. Thus the numerical calculation
retains the residual contribution that distinguishes $U_{\bm W}$
from the mean quadratic.

At level $\alpha=0.05$, characteristic-function inversion gives
$c_0=17.38105919$ and baseline power $\pi_0=0.32407756$. The boundary-moment
gradient, calculated from the analytic derivative of the characteristic
function, is
\[
 \bm G_{\bm\delta}\simeq
 \begin{bmatrix}
 0.00149882&0.01528770\\
 0.01528770&-0.00024980
 \end{bmatrix}.
\]
Its leading direction has angle $43.36^\circ$, whereas the
normalized mean direction has angle $71.57^\circ$.
Table~\ref{tab:u-general-example} recomputes the null critical value
for every direction before evaluating power. At $\eta=0.20$, the
leading-eigenvector power exceeds the mean-direction power by
$0.001302$, and the trailing eigenvector reduces power below the
baseline. The numerical angle search improves on the leading direction
by approximately $7.05\times10^{-8}$ in power. These comparisons illustrate
the local criterion without asserting global finite-$\eta$ optimality.

\begin{table}[tbp]
 \centering
 \caption{Calibrated power of the unbiased two-sample statistic under a
 nonscalar baseline. The numerical comparator uses a one-degree grid
 on $[0,180^\circ)$ followed by local refinement around its best grid
 angle. The baseline power is $0.32407756$.}
 \label{tab:u-general-example}
 \small
 \begin{tabular}{rrrrrr}
 \toprule
 $\eta$ & Numerical angle & Numerical power & Leading $\bm G$
 & Mean direction & Trailing $\bm G$\\
 \midrule
0.025 & 43.33 & 0.32447363 & 0.32447363 & 0.32430364 & 0.32370954 \\
0.050 & 43.30 & 0.32486500 & 0.32486500 & 0.32452703 & 0.32333993 \\
0.100 & 43.25 & 0.32563367 & 0.32563366 & 0.32496590 & 0.32259601 \\
0.200 & 43.16 & 0.32711516 & 0.32711509 & 0.32581303 & 0.32109035 \\
 \bottomrule
 \end{tabular}
\end{table}

\begin{figure}[tbp]
 \centering
 \includegraphics[width=\textwidth]{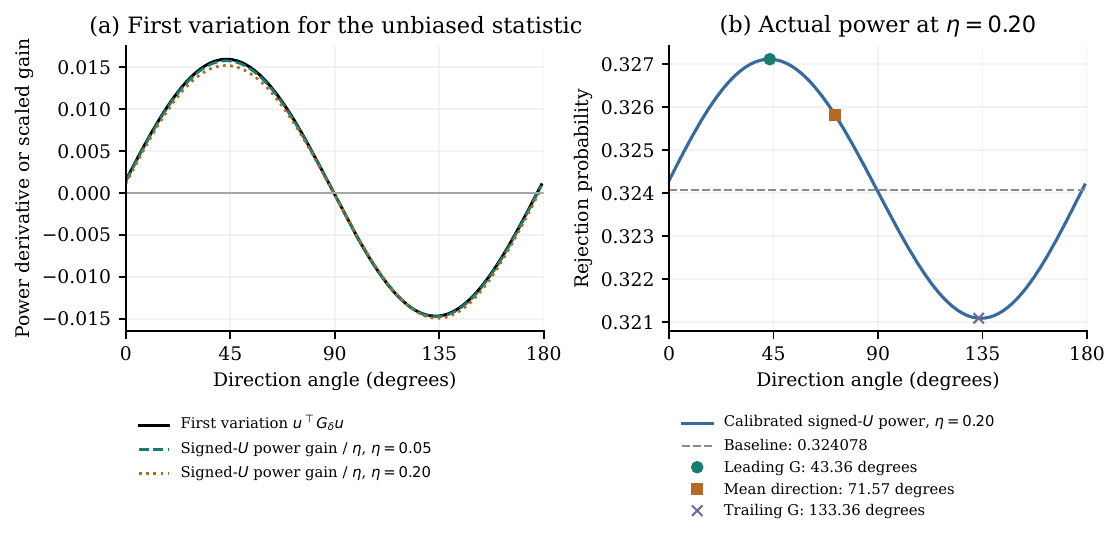}
 \caption{Local power geometry for the unbiased two-sample statistic.
 The left panel compares the analytic first variation with calibrated
 power gains divided by augmentation strength. The right panel shows
 actual rejection probabilities at $\eta=0.20$. Every power includes
 both negative Wishart terms and uses its direction-specific null cutoff.}
 \label{fig:u-general-geometry}
\end{figure}

The distribution in \eqref{eq:u-general-example-law} is evaluated by
direct characteristic-function inversion, within the Gaussian
quadratic-form framework of \citet{imhof1961}. The integral is
truncated at frequency $20$ and uses absolute and relative
finite-interval tolerances $2\times10^{-12}$. An analytic
characteristic-function bound limits the omitted CDF tail to less
than $9.87\times10^{-18}$ over the calculations. Tightening the tolerance to
$2\times10^{-13}$ and increasing the upper frequency to $24$ changes
the checked powers by less than $4.25\times10^{-13}$.
At central-difference step $10^{-4}$, the largest discrepancies
across the leading, mean, and trailing directions are
$4.68\times10^{-12}$ for the power derivative and
$3.12\times10^{-10}$ for the critical-value derivative.
The finite-interval quadrature comparisons are numerical diagnostics,
not certified integration error bounds. The accompanying
\texttt{u\_general\_example.py} reproduces the table, figure, and diagnostics.

\clearpage
\section{Leukemia application and data-based experiments}
\label{app:v20-application}

\subsection{Why the paired sign reference controls level}
\label{app:v20-sign-proof}

The role of the sign group is to preserve the entire dependence pattern
within a paired vector. Coordinatewise sign changes would require a
stronger invariance assumption. The argument below is the classical
conditional randomization argument of \citet{hemerik2018}, applied
to scores that refit an axis and diagonal variances.

\begin{proof}[Proof of Proposition~\ref{prop:v20-signflip}]
First condition on the orbit $\mathcal O(\bm D)$. Null invariance makes
the observed matrix uniform over the distinct points of this finite
orbit. Every orbit point has the same number of group elements mapping
any fixed representative to it: that number is the size of its stabilizer.

Next, conditional on the observed matrix, each independent uniform
$g_b$ maps it uniformly over the same orbit. This distribution depends
only on the orbit, not on its observed representative. Consequently
$\bm D,g_1\bm D,\ldots,g_B\bm D$ are independent and uniform
conditional on the orbit. They need not be independent unconditionally,
and exchangeability is not asserted conditional on the full observed
matrix.

Finally apply the same fitting-and-scoring map to every matrix. The
$M$ scores are conditionally exchangeable. Among any $M$ numbers, at
most $k$ have conservative upper-tail rank at most $k$; ties can only
reduce that count. Exchangeability therefore bounds the probability
that row zero has rank at most $k$ by $k/M$. Taking
$k=\lfloor M\alpha\rfloor$ gives the result. In complete enumeration,
each distinct orbit point occurs the same number of times. Counting
the group elements directly gives the corresponding $2^N$ rank bound.
\end{proof}

Independent differences centrally symmetric about zero imply the
joint invariance used in the proof. Under independent Gaussian
two-sample observations with equal means, their covariance is
$\bm\Sigma_1+\bm\Sigma_2$, with no diagonal or common-shape
restriction. The test uses only the difference matrix. This does not
assert exchangeability of the original two groups when their
covariances differ. For arbitrary non-Gaussian data, equal means alone
do not supply the required invariance; the distinction between a
parameter null and a distributional null is also central to
\citet{chung2013permutation}.

\subsection{Finite menus on the same sign orbit}
\label{app:v20-sign-menu}

Fix $K$ candidate algorithms and positive weights $w_j$. Compute all
$T_{bj}$ on the same $M$ transformed datasets, refitting each algorithm.
For each row and candidate form
\[
 R_{bj}=\sum_{a=0}^{M-1}\mathbf1\{T_{aj}\ge T_{bj}\},\qquad
 E_b=\min_{1\le j\le K}R_{bj}/w_j,\qquad
 p_{\rm menu}=\frac1M\sum_{a=0}^{M-1}\mathbf1\{E_a\le E_0\}.
\]
The map from the score matrix to $(E_b)$ is row-permutation equivariant.
Conditional exchangeability and the same rank count give level at most
$\lfloor M\alpha\rfloor/M$. Every component rank includes every row;
the observed row does not receive a separately optimized candidate
list. This is an application of the rank argument already used in
Section~\ref{sec:u-rank-selection}, not a new general combination
principle. The application below reports the prespecified components
individually, with across-pathway adjustment, and does not select the
smallest observed component $p$-value as an unadjusted test.

\subsection{Data provenance and the fixed protocol}

The public Golub expression matrices contain 38 original training
samples and 34 original test samples, with 7\,129 matching probe
identifiers. We cross-checked the downloaded raw matrices, probe
order, sample identifiers and ALL/AML labels against the official
Bioconductor \texttt{golubEsets} data objects. The archive retains
the cross-check results, source URLs, hashes and mapping table.
Only expression, sample identifier and leukemia class enter the
analysis. Auxiliary specimen and institution information is used
to describe the cohorts, not to claim covariate-adjusted effects.

Hallmark sets come from MSigDB 2024.1, with collection attribution to
\citet{liberzon2015hallmark}; the MSigDB/GSEA resource is described by
\citet{subramanian2005gsea}. The official \texttt{hu6800.db} 3.13.0
probe-to-Entrez annotation supplies the mapping. Multiple probes are
retained rather than selecting whichever shows the largest group
difference. The same annotated sets and the same fixed clipping and
logarithm are used in both cohorts. The numerical protocol was recorded
before producing rejection outcomes; this is an internal design
record, not an external preregistration.

\subsection{Score definitions and implementation conventions}

Let $d_{ij}$ be a paired difference, $\bar d_j$ its mean over $N$
pairs, and $s_j^2$ its unbiased variance. In this subsection only,
every fitted variance $s^2$ is replaced by $\max(s^2,10^{-8})$,
in squared log-expression units. With
$Y_j=\sqrt N\bar d_j/\sqrt{\max(s_j^2,10^{-8})}$, the three direct
scores on $J$ are
\[
 \frac1q\sum_{j\in J}\frac{N Y_j^2}{N-1+Y_j^2},\qquad
 \sum_{j\in J}Y_j^2,\qquad \max_{j\in J}Y_j^2.
\]
This finite numerical convention need not coincide with the
unregularized continuous-model statistic on degenerate data.
Proposition~\ref{prop:v20-signflip} only requires a consistently
defined score on every transformed matrix.

For each pair $i<k$, the LPO score estimates coordinate variances and
the standardized mean direction from the other $N-2$ paired vectors.
It uses the full set of pairs, with $b=1$, $\eta=\sqrt p$, and
the same diagonal-variance floor. A zero training contrast selects
the first supplied coordinate. The reported score is $N U$, as in
Section~\ref{sec:u-lpo}; multiplying all versions by the
same positive constant does not change their ranks. For real data
$p=q$. In the semi-synthetic experiment, the pathway-baseline and
full-baseline versions differ only in the support of their baseline;
both learn the rank-one direction within $J$.

The projection score uses the first $\lfloor N/2\rfloor$ differences
to fit the inverse-diagonal-variance mean direction on $J$. The coefficient vector is divided by its largest absolute coordinate
before applying the projected-variance floor. It then
computes the squared one-sample $t$ score on the remaining projected
differences. We recalibrate that entire split procedure by signs,
including refitting its direction on every transformation. The
application does not use a Gaussian $t$ reference for these expression
measurements. All methods share the same pairs and transformations.

Rank comparisons include scores within $10^{-10}\max(1,|T_0|)$ of
the observed score in the conservative upper tail. This can only
increase the reported $p$-value. The complete sign orbit has global
sign duplicates for these two-sided scores; these duplicates remain
in the rank count. For each method, Holm adjustment uses the three
pathway $p$-values within one cohort. It does not supply a simultaneous
claim over all methods and both cohorts.

\subsection{Noise construction and Monte Carlo uncertainty}

Let $R_1,R_2$ denote the within-class centered training residual
matrices on the fixed panel, with original class sizes $n_1,n_2$.
Set
\[
 F=\begin{pmatrix}R_1/\sqrt{n_1-1}\\R_2/\sqrt{n_2-1}\end{pmatrix},
 \quad C=F^\top F,\quad d_j=\max(C_{jj},10^{-8}),\quad
 \Sigma=0.9C+\operatorname{diag}(d_j-0.9C_{jj}).
\]
The diagonal Gaussian design uses $\operatorname{diag}(d_j)$; the
correlated Gaussian design uses $\Sigma$. Independently for each
row, multiplying a correlated Gaussian vector by
$\sqrt{3/V}$ with $V\sim\chi^2_5$ gives the third design. Its covariance
is still $\Sigma$ and its distribution is centrally symmetric.
Constant probes required the stated marginal-variance floor in the
generator. The initial protocol and a separate amendment record that
this correction was made after a deterministic zero-variance assertion
failed, before any semi-synthetic rejection outcomes were generated.
The panel, signals, sample sizes and scores were unchanged.
Only the mean is shifted between the four signal designs. The
single inside coordinate is fixed by the annotated probe order;
diffuse signals have equal marginally standardized shifts on the
specified support. No mean or covariance is fitted to favor a test.

The 500 datasets in each cell are independent. Scores share the same
data and sign matrices; the four signals within a noise regime share
the same underlying noise. For a rejection rate $\hat\pi$, the
reported Monte Carlo standard error is
$\{\hat\pi(1-\hat\pi)/500\}^{1/2}$. A paired method difference uses
the sample standard deviation of its 500 indicator differences,
divided by $\sqrt{500}$. Complete score arrays, observed rank $p$-values, indicators,
pairing identities and seeds are retained for reconstruction.

\begin{table}[tbp]
\centering\small
\caption{Unadjusted sign-reference $p$-values for the same fixed leukemia analyses. These are component results, not an adjusted choice among methods.}\label{tab:v20-real-raw}
\setlength{\tabcolsep}{4pt}
\begin{tabular}{llrrrrrr}
\toprule
Cohort & Set & $q$ & Bounded & Quadratic & Maximum & LPO & Projection\\
\midrule
Training & G2M & 156 & 0.0205 & 0.0156 & 0.0088 & 0.7227 & 0.0254\\
Training & Apoptosis & 169 & 0.0010 & 0.0010 & 0.0098 & 0.6260 & 0.0078\\
Training & IFN-$\gamma$ & 176 & 0.0010 & 0.0010 & 0.0107 & 0.3135 & 0.0020\\
\midrule
Test & G2M & 156 & 0.0110 & 0.0090 & 0.0560 & 0.3270 & 0.0385\\
Test & Apoptosis & 169 & 0.0015 & 0.0005 & 0.0010 & 0.6945 & 0.0070\\
Test & IFN-$\gamma$ & 176 & 0.0010 & 0.0015 & 0.0190 & 0.0555 & 0.0075\\
\bottomrule
\end{tabular}
\par\smallskip\begin{minipage}{.97\linewidth}\footnotesize
All methods use only the indicated pathway and the same 11 training-cohort or 14 test-cohort pairs. The references use all 2,048 signs or 1,999 random signs plus the identity, respectively. Projection uses a fixed training/testing split. Interpretation requires the symmetry assumption; smaller values do not establish higher power.
\end{minipage}
\end{table}

\begin{table}[tbp]
\centering\small
\caption{Descriptive dependence and numerical-floor use. Correlations use within-class centered residuals. These summaries do not validate central symmetry.}\label{tab:v20-floor}
\begin{tabular}{llrrrr}
\toprule
Cohort & Set & Median $|r|$ & 90th pct. $|r|$ & Full floor (\%) & LPO floor (\%)\\
\midrule
Training & G2M & 0.199 & 0.471 & 7.69 & 8.32\\
Training & Apoptosis & 0.142 & 0.359 & 10.06 & 10.96\\
Training & IFN-$\gamma$ & 0.169 & 0.427 & 7.95 & 8.51\\
\midrule
Test & G2M & 0.174 & 0.437 & 5.13 & 5.53\\
Test & Apoptosis & 0.143 & 0.382 & 6.51 & 7.81\\
Test & IFN-$\gamma$ & 0.173 & 0.418 & 6.25 & 6.56\\
\bottomrule
\end{tabular}
\par\smallskip\begin{minipage}{.95\linewidth}\footnotesize
Full floor counts all $M q$ full-sample coordinate variance calculations; LPO floor counts all $M\binom N2 q$ retained-sample coordinate variances. A constant zero difference contributes to these counts without contributing signal. Undefined correlations involving constant probes are omitted only from these descriptive summaries; every test retains all fixed probes.
\end{minipage}
\end{table}

\begin{table}[tbp]
\centering\small
\caption{Simultaneous 95\% Monte Carlo intervals for null rejection probabilities, expressed as percentages. There are 21 intervals; Bonferroni correction uses exact binomial Clopper--Pearson limits.}\label{tab:v20-null-intervals}
\begin{tabular}{lrrr}
\toprule
Method & Diagonal Gaussian & Correlated Gaussian & Correlated $t_5$\\
\midrule
Bounded & [0.95, 5.54] & [2.82, 9.18] & [4.62, 12.11]\\
Quadratic on set & [1.70, 7.15] & [3.11, 9.68] & [4.46, 11.87]\\
Maximum & [1.70, 7.15] & [3.55, 10.42] & [2.96, 9.43]\\
LPO on set & [1.84, 7.41] & [3.40, 10.17] & [4.00, 11.15]\\
LPO full baseline & [1.84, 7.41] & [3.25, 9.93] & [3.85, 10.90]\\
Quadratic full panel & [2.11, 7.92] & [3.11, 9.68] & [4.46, 11.87]\\
Split projection & [4.00, 11.15] & [1.84, 7.41] & [2.53, 8.68]\\
\bottomrule
\end{tabular}
\par\smallskip\begin{minipage}{.95\linewidth}\footnotesize
Coverage concerns the Monte Carlo probabilities under the fixed generators. The null outcomes for different methods are dependent; the Bonferroni argument does not require independence across intervals. These intervals do not establish validity for the biological data.
\end{minipage}
\end{table}

\end{document}